\documentclass[12pt]{iopart}
\usepackage{graphicx}
\usepackage{bm}
\usepackage{iopams} 

\begin{document}

\topical[Experimental Studies of the $0.7 \times 2e^{2}/h$
conductance anomaly]{What lurks below the last plateau: experimental
studies of the $0.7 \times 2e^{2}/h$ conductance anomaly in
one-dimensional systems}

\author{A P Micolich}
\address{School of Physics, University of New South Wales, Sydney NSW 2052, Australia}
\ead{adam.micolich@nanoelectronics.physics.unsw.edu.au}
\submitto{\JPCM}
\date{\today}

\begin{abstract}

The integer quantized conductance of one-dimensional electron
systems is a well understood effect of quantum confinement. A number
of fractionally quantized plateaus are also commonly observed. They
are attributed to many-body effects, but their precise origin is
still a matter of debate, having attracted considerable interest
over the past $15$ years. This review reports on experimental
studies of fractionally quantized plateaus in semiconductor quantum
point contacts and quantum wires, focussing on the $0.7 \times
2e^{2}/h$ conductance anomaly, its analogs at higher conductances,
and the zero bias peak observed in the d.c. source-drain bias for
conductances less than $2e^{2}/h$.

\end{abstract}
\maketitle

\section{Introduction}

The second half of the $20$th century saw a massive intellectual and
industrial investment in the miniaturization of semiconductor
devices to drive advances in information processing
technologies~\cite{MooreElec65, MollickIEEE06}. Despite being
dominantly an engineering endeavour, miniaturization has provided
substantial returns to fundamental semiconductor physics, the field
from which these devices originated, in the guise of methods for
realizing two, one and zero-dimensional electron systems. Research
on low-dimensional electron systems has lead to a number of
important discoveries over the past forty years including the Nobel
prize winning integer~\cite{vonKlitzingPRL80} and
fractional~\cite{TsuiPRL82} quantum Hall effects in two-dimensional
(2D) systems, conductance quantization in one-dimensional (1D)
quantum point contacts~\cite{vanWeesPRL88, WharamJPC88}, and single
electron transport in zero-dimensional (0D) quantum
dots~\cite{MeiravPRL90}.

\subsection{A brief primer on quantum point contacts and 1D
conductance}

The quantization of the 1D conductance in a quantum point contact is
an excellent demonstration of basic quantum mechanics, being a
direct result of the wave nature of the electron. Quantum point
contacts come in various designs, but their most basic/common
incarnation involves using an AlGaAs/GaAs heterostructure grown by
molecular beam epitaxy~\cite{ChoAPL71}. The heterostructure consists
of either a narrow ($\lesssim 50$~nm) AlGaAs/GaAs/AlGaAs quantum
well structure or an AlGaAs/GaAs heterojunction, buried $\sim 30 -
300$~nm beneath the heterostructure surface. At low temperature, a
thin layer of electrons accumulates in the quantum well or
immediately to the GaAs side of the heterojunction, with a density
$n_{s} \sim 10^{10} - 10^{11}$~cm$^{-2}$. These electrons are
generally supplied by a thin layer of dopants located between the
quantum well/heterointerface and the heterostructure surface. At
this density, the electron Fermi wavelength $\sim 50$~nm exceeds the
thickness of the layer the electrons are confined to, resulting in
an effectively two-dimensional electron gas (2DEG). The spatial
separation between the dopant layer and the 2DEG is significant;
combined with the high crystal quality provided by epitaxial growth,
it leads to very high electron mobilities $\sim 10^{6}$~cm$^{2}$/Vs,
and thus ballistic electron transport over length scales exceeding
several microns within the plane of the 2DEG~\cite{DingleAPL78}. To
facilitate electrical measurement, the 2DEG is usually patterned
into a Hall bar-shaped mesa structure hundreds of microns long/wide
by wet etching, with low resistance Ohmic contacts created by
deposition of small patches of NiGeAu alloy, diffused into the
heterostructure by a rapid thermal anneal step to reach the 2DEG.

Further confinement of the electrons can be achieved using
sub-micron metal gates defined by electron beam lithography and
deposited on the heterostructure surface by thermal evaporation. A
negative voltage applied to these gates electrostatically depletes
the 2DEG in the regions underneath them, directly translating the
gate pattern into the 2DEG. A short, narrow constriction
interspersed between `source' and `drain' 2DEG reservoirs, known as
a quantum point contact (QPC), can be defined using a `split-gate'
architecture consisting of a rectangular gate up to a micron long,
extending all the way across the Hall bar, and with a narrow gap
typically $200 - 1000$~nm wide in the middle~\cite{ThorntonPRL86}.

\begin{figure}
\includegraphics[width=9cm]{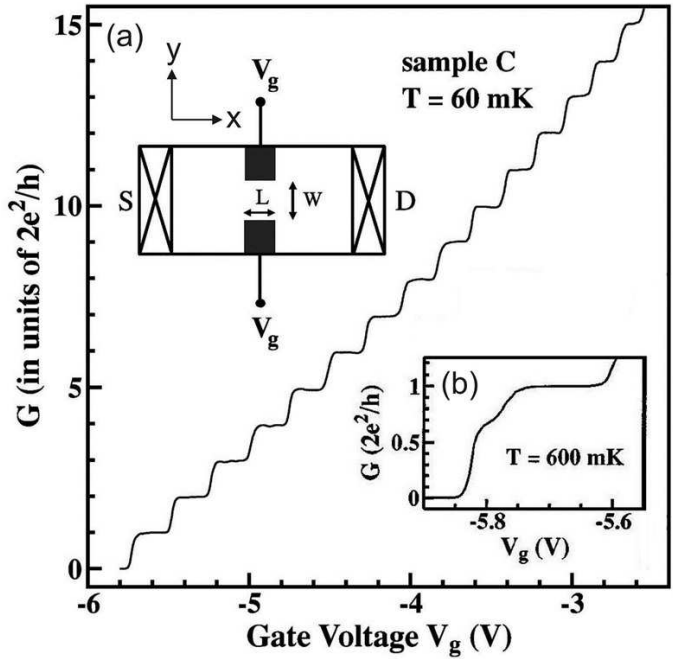}
\caption{The linear differential conductance $G = dI/dV$ vs gate
voltage $V_{g}$ from a quantum point contact with $W = 950$~nm, $L =
400$~nm at a temperature $T = 60$~mK defined in a 2DEG with a depth
of $280$~nm, an electron density $n = 1.8 \times 10^{11}$~cm$^{-2}$
and a mobility $\mu = 4.5 \times 10^{6}$~cm$^{2}$/Vs. Inset:
Schematic of a QPC of width $W$ and length $L$ defined by metal
gates biased at $V_{g}$. The two adjacent 2DEG regions connect to
source (S) and drain (D) Ohmic contacts made with annealed NiGeAu
alloy. Directions $x$ and $y$ in the plane of the 2DEG are also
indicated, with the $z$ direction pointing out of the page. (b) A
detail of the plateau at $0.7G_{0}$. Figure (a) adapted with
permission from Ref.~\cite{ThomasPRB98}. Copyright 1996 by the
American Physical Society. Figure (b) adapted with permission from
Ref.~\cite{ThomasPRL96}. Copyright 1998 by the American Physical
Society.}
\end{figure}

The width of the constriction can be tuned continuously by adjusting
the negative bias $V_{g}$ applied to the gates, leading to a
staircase of plateaus in the measured linear differential
conductance $G = dI/dV$ versus $V_{g}$, as shown in Fig.~1(a). The
staircase of plateaus occurs because the confinement in the
$y$-direction leads to quantization of the transverse wave-vector
$k_{y}$. As the constriction is narrowed by making $V_{g}$ more
negative, the allowed $k_{y}$ states, known as 1D subbands, rise up
in energy, depopulating once they exceed the Fermi energy of the
adjacent 2DEG reservoirs $E_{F} = \pi\hbar^{2}n_{s}/m^{*}$. The
plateaus occur at integer multiples of $G_{0} = 2e^{2}/h$, where $e$
is the electron charge and $h$ is Planck's constant. This
quantization of the plateau conductance can also be understood under
a simple, single-particle picture -- the conductance is dependent on
the product of the electron velocity and the 1D density of states,
each of which contain terms in $\sqrt{E}$ that fortuitously cancel
to give an equal, energy-independent conductance contribution for
each 1D subband~\cite{BeenakkerRev91}.

\subsection{Introducing the $0.7 \times 2e^{2}/h$ conductance anomaly}

With that said, it might appear that a comprehensive and complete
understanding of the 1D conductance in quantum point contacts has
been achieved. However, this is certainly not the case -- there are
several features in the conductance of QPCs that lack an accepted
explanation and are the subject of extensive
debate~\cite{FitzgeraldPT02}. Foremost is an anomalous plateau
typically observed at a conductance of $G \simeq 0.7G_{0}$, shown in
Fig.~1(b). First addressed specifically by Thomas {\it et al} in
1996~\cite{ThomasPRL96}, this feature was frequently observed in
earlier work (e.g., see Fig.~2 of Ref.~\cite{vanWeesPRL88}; Figs.~2,
6 and 7 of Ref.~\cite{vanWeesPRB91}; Fig.~3 of
Ref.~\cite{HamiltonAPL92}). Analogous non-quantized plateaus at $G >
2e^{2}/h$ for applied dc source-drain bias~\cite{KristensenPRB00}
and in-plane magnetic field~\cite{GrahamPRL03}, along with an
anomalous peak in the differential conductance versus dc
source-drain bias for $G < 2e^{2}/h$~\cite{CronenwettPRL02} known as
a `zero-bias anomaly', have also been observed in QPCs and
associated with the $0.7 \times 2e^{2}/h$ conductance anomaly.
Again, taking a quick survey pre-1996, similar features are observed
by Patel {\it et al}~\cite{PatelPRB91, PatelPRB91a}~\footnote{For
later reference, it is interesting to note the absence of a zero
bias anomaly in Fig.~4 of Ref.~\cite{PatelPRB91a}.}. A large number
of possible explanations have been offered for this effect. The two
dominant ones are a spontaneous spin-polarization~\cite{ThomasPRL96}
and Kondo-like effects~\cite{CronenwettPRL02, LindelofSPIE01,
MeirPRL02, RejecNat06}, however, others explanations include
phenomenological spin-gap models~\cite{KristensenPRB00,
BruusPhysE01, ReillyPRL02, ReillyPRB05}, subband pinning
effects~\cite{BruusPhysE01, LasslPRB07}, electron-phonon
interactions~\cite{SeeligPRL03}, singlet-triplet
effects~\cite{FlambaumPRB00, RejecPRB00}, Wigner
crystallization~\cite{SpivakPRB00, MatveevPRB04} and charge density
waves~\cite{SushkovPRB01}. Despite the diversity of explanations
offered, there is one clear point of general consensus -- the $0.7$
plateau and associated features cannot be described under a
single-particle framework, and arise from many-body effects (i.e.,
electron-electron interactions).

\subsection{Content and Structure of this Review}

This topical review focuses on experimental studies of fractionally
quantized plateaus in the 1D conductance of QPCs. My focus in
writing this review is to provide a detailed introduction for
beginners, be they new graduate students or researchers interested
in contributing to the on-going work in this area or drawing
inspiration from it. As such, I have sacrificed brevity for depth of
discussion. Experts in the field may wish to skim rather than read
or defer to the special edition of {\it Journal of Physics:
Condensed Matter} edited by Pepper and Bird~\cite{PepperJPCM08}
published in 2008. It contains a number of shorter invited reviews
of key experimental and theoretical works related to the $0.7G_{0}$
conductance anomaly and electron-electron interactions in 1D systems
and provides a more focussed coverage of specific experiments
discussed in this topical review. There is also a shorter, recent
review by Berggren and Pepper~\cite{BerggrenPTRSA10} more suited to
existing experts on 1D conductance in QPCs.

Readers seeking a general background on nanoelectronics and
low-dimensional devices can consult books by
Davies~\cite{DaviesBook} and Ferry, Goodnick and
Bird~\cite{FGBBook}. For very comprehensive reviews of earlier
studies of quantized 1D conductance, readers should consult articles
by Beenakker and van Houten~\cite{BeenakkerRev91}, van Houten,
Beenakker and van Wees~\cite{vanHoutenRev92}, and for a more recent
focus, by Clarke, Simmons and Liang~\cite{ClarkeCSST11}. Very useful
magazine-style discussions of low dimensional physics and the Kondo
effect in quantum dots can be found in articles by Berggren and
Pepper~\cite{BerggrenPW02}, and Kouwenhoven and
Glazman~\cite{KouwenhovenPW01}, respectively.

This review is divided into eleven sections and essentially two
chapters as follows: The first chapter contains five sections
focussed on establishing a foundational knowledge of the behaviour
of the $0.7$ plateau. The experiments in this part lead to the
dominant physical models considered over the lifetime of study of
the $0.7$ plateau. In Section 2 I begin with the initial
characterization of the $0.7$ plateau reported by Thomas and
coworkers in Cambridge, focussing in particular on
Refs.~\cite{ThomasPRB98, ThomasPRL96}. The aim in Section 2 is to
set a starting context for experimental investigations that
followed. Section 3 will deal with the behavior of the $0.7$ plateau
with respect to temperature, electron density and the application of
dc source-drain bias, which lead to some of the key possible
explanations for the origin of the $0.7$ plateau, namely the two
`spin-gap' models proposed by Kristensen and
Bruus~\cite{KristensenPRB00, BruusPhysE01} and Reilly {\it et
al}~\cite{ReillyPRL02, ReillyPRB05}. Section 4 discusses
experimental evidence pointing towards a Kondo-like mechanism for
the $0.7$ plateau by connecting it to the zero-bias anomaly in the
source-drain characteristics~\cite{CronenwettPRL02}. This section
includes a brief introduction to Kondo physics as it relates to
nanoscale devices, since it is an effect that many readers may be
familiar with from a different perspective, namely, metal films with
magnetic impurities. Section 5 looks at five other measurements that
provide important information regarding the $0.7$ plateau -- shot
noise, thermopower, compressibility, 1D hole systems and scanning
gate microscopy studies.

In the second chapter the focus shifts to experiments directed
towards testing the dominant explanatory models arising in the first
chapter. Section 6 begins by looking at evidence for the formation
of a quasi-bound state within the QPC as it approaches pinch-off, as
required for a Kondo-like mechanism. In Section 7 we look at
experiments testing whether or not there is a consistent and direct
connection between the $0.7$ plateau and the zero-bias anomaly. The
data here points strongly to Kondo-like phenomena in QPCs being
coincident rather than causal to the $0.7$ plateau. Section 8 shifts
focus slightly, looking at exchange mechanisms rather than Kondo,
and the link between the $0.7$ plateau and other fractional plateaus
observed in the conductance. This work is built upon in Section 9,
where we investigate the new information that can be provided via dc
conductance measurements of QPCs. These experiments point broadly
towards an exchange-driven spin-gap mechanism but with some
differences to the established models. In Section 10 we look at more
recent work focussed in a slightly different direction -- How QPCs
behave as the confinement is weakened and evidence for
exchange-driven spontaneous ordering of electrons. These experiments
provide further, albeit perhaps indirect, evidence regarding the
important role that the exchange interaction plays in QPCs. In
Section 11, I finish with some conclusions and outlook regarding
studies of fractionally quantized plateaus and spin-effects in 1D
systems.

\subsection{Conventions}

Due to different mathematical nomenclatures used by various authors,
it is necessary to redefine some variables at various points in this
Review. With few exceptions, I have deliberately chosen to use the
original nomenclature of the references where possible to avoid
confusion for readers when they consult the literature.

Unless specified otherwise, the conductance can always be assumed to
be the ac or differential conductance $G = dI/dV$. This will
sometimes be the linear conductance (i.e., with dc source-drain bias
$V_{sd} \approxeq 0$) and sometimes a non-linear conductance (i.e.,
$|V_{sd}| > 0$) depending on context. The one exception is Section
9, where the dc conductance is also discussed -- the ac and dc
conductances are referred to as $G_{ac}$ and $G_{dc}$ there. Note
that in some figures $g$ rather than $G$ is used for the
conductance; in these cases the conductance will be referred to
using $G$ in the text to avoid possible confusion with the $g$
factor $g^{*}$. All magnetic fields can be assumed to be in the
plane of the 2DEG with both $B$ and $B_{\parallel}$ often used to
indicate this. Care should be taken in Section 5.3 where
anisotropies in the in-plane $g$-factor require the in-plane field
direction relative to the QPC axis to be specified as
$B_{\parallel}$ and $B_{\perp}$. Spin-degenerate 1D subbands are
referred to by their index $n$, while the corresponding
spin-polarized 1D subbands are referred to by index and spin as
$n\uparrow$ and $n\downarrow$ for spin-up and spin-down,
respectively. Transconductance greyscales and colour-maps can be
assumed to highlight the motion of 1D subbands except in Section
5.3.

\section{Initial characterization of the $0.7 \times 2e^{2}/h$ conductance anomaly}

The $0.7G_{0}$ conductance anomaly typically presents as a clear
inflection or a weak plateau in the 1D conductance of a quantum
point contact, as shown in Figs.~1(b), 2 and 3~\cite{ThomasPRL96}.
Unlike quantum Hall plateaus, which are quantized to an accuracy of
at least parts per million~\cite{vonKlitzingPRL80}, the conductance
plateau is not fixed at precisely $0.7G_{0}$. In fact, looking
across the complete library of experimental work on this feature, it
can vary in conductance significantly, ranging in position from
$0.65$ to $0.8G_{0}$ depending on the given sample and experimental
conditions, despite being most commonly observed in close proximity
to $0.7G_{0}$.

\subsection{Ruling out impurity scattering}

Two initial studies by Thomas {\it et al}~\cite{ThomasPRB98,
ThomasPRL96} highlighted some key phenomenological properties of the
$0.7$ plateau, and eliminated some basic possible explanations for
its origin. The first and most obvious possibility is that the
plateau is a quantum interference effect caused by scattering from
an impurity local to the QPC, either directly between the two gates,
or within the electron phase coherence length of the entrance/exit
of the QPC. Impurity scattering in QPCs can generate unexpected
plateau-like and resonant features in the 1D
conductance~\cite{NixonPRB91, TekmanPRB91, McEuenSurfSci90}, and was
subsequently shown to cause suppression of plateaus below their
integer-quantized values in longer ($\sim 3-5~\mu$m) quantum
wires~\cite{LiangAPL99, LiangPRB00}. A similar suppression is
observed in $2~\mu$m long wires fabricated by cleaved-edge
overgrowth~\cite{dePicciottoPRL04, dePicciottoPRB05}, but is due
instead to coupling issues between the wire and the 2DEG
reservoirs~\cite{dePicciottoPRL00}. Returning to surface-gated
devices, the possibility that the $0.7$ plateau is an impurity
effect can be ruled out by asymmetrically biasing the gates defining
the QPC, i.e., setting the two gates to $V_{g} + \Delta V_{g}/2$ and
$V_{g} - \Delta V_{g}/2$, and monitoring $G(V_{g})$ as a function of
$\Delta V_{g}$. Asymmetric biasing shifts the QPC potential in the
2DEG along the gate-line ($y$-direction) towards the more positive
of the two gates~\cite{GlazmanSST91}. This alters the precise
location of the impurity with respect to the QPC, and hence the
amplitude of quantum interference related contributions to the
conductance. As shown in Fig.~2, the $0.7$ plateau is robust to
asymmetric biasing. Only very slight changes in its shape occur for
a $\Delta V_{g}$ range of $1.3$~V, which corresponds to an overall
lateral shift of the QPC by $80.6$~nm~\cite{ThomasPRB98}. No changes
in the conductance of the $0.7$ plateau are observed in doing this.
Similar robustness of the $0.7$ plateau to lateral shifting was
reported in Refs.~\cite{LiangPRB00, SfigakisPhysE10}.

\begin{figure}
\includegraphics[width=9cm]{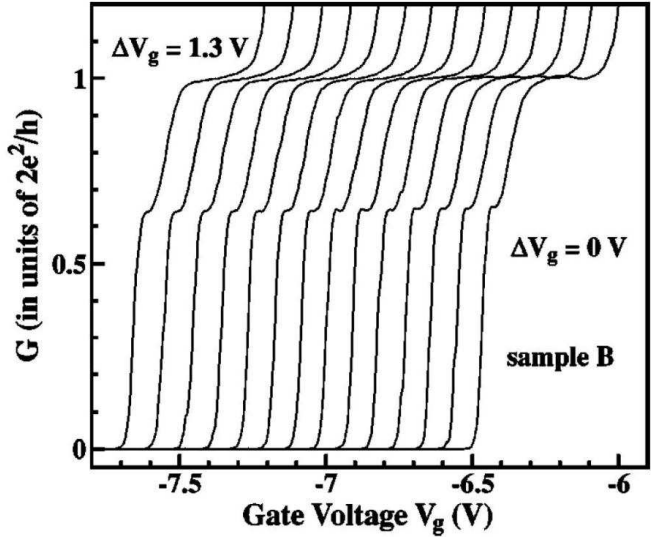}
\caption{Behaviour of the 1D conductance $G(V_{g})$ in response to
asymmetric biasing of the QPC gates to laterally shift the QPC.
Starting with symmetric biasing (right-most trace), $\Delta V_{g}$
is increased by $0.1$~V for each successive trace, and offset by
$V_{g} = -0.1$~V for clarity. Each $\Delta V_{g} = 0.1$~V step
corresponds to an additional lateral offset of the QPC by $6.2$~nm.
Data obtained from a heterojunction QPC with a 2DEG depth $z =
310$~nm, a mobility $\mu = 3.5 \times 10^{6}$~cm$^{2}$/Vs, a density
$n_{2D} = 1.4 \times 10^{11}$~cm$^{-2}$, and a split-gate separation
of $0.95~\mu$m. Figure reproduced with permission from
Ref.~\cite{ThomasPRB98}. Copyright 1998 by the American Physical
Society.}
\end{figure}

\subsection{The $0.7$ plateau is also not a transmission resonance}

Having ruled out impurity scattering, an additional possibility is
that the $0.7$ plateau is a transmission resonance due to multiple
reflection of electrons along the QPC axis (i.e., $x$
direction)~\cite{EscapaJPCM89, TekmanPRB89}. This is analogous to
Fabry-P\'{e}rot interference~\cite{SmithJPCM89a, KatinePRL97}, and
unlike scattering from an impurity inside the QPC channel, should be
relatively insensitive to lateral shifting. If the $0.7$ plateau
were a transmission effect, then it would evolve towards $0.7 \times
e^{2}/h (= 0.35G_{0})$ with applied in-plane magnetic field, rather
than the $e^{2}/h (= 0.5G_{0})$ observed in Ref.~\cite{ThomasPRB98}.
It would also result in a structure at $G = 0.49G_{0}$ for two QPCs
set to $G = 0.7G_{0}$ measured in series. Liang {\it et
al}~\cite{LiangPRB99} tested this possibility using a $0.8~\mu$m
long split-gate structure featuring three, isolated and
independently biasable $50$~nm wide `finger-gates', each located
above and running across the width of the split-gate defined QPC
channel. By setting the negative bias on the finger-gates at the two
ends of the QPC slightly higher than that in the middle, two QPCs in
series were obtained. Each of these were characterized independently
and showed clear $0.7$ plateaus. When the two QPCs were measured in
series, the integer quantized plateaus remained well quantized, as
expected for ballistic transmission through two QPCs in
series~\cite{WharamJPC88a}. There was no plateau at $0.49G_{0}$,
instead a clear $0.7$ plateau was observed~\cite{LiangPRB99}. This
not only confirms that the $0.7$ plateau is not a transmission
resonance, it shows that ballistic transport through the QPC is not
interrupted by the mechanism that causes the $0.7$ plateau.
Furthermore, this result shows that any many-body effects involved
must have a range less than $0.6~\mu$m, i.e., there is no inter-QPC
coupling between the mechanisms generating the $0.7$ plateaus in the
series measurement~\cite{LiangPRB99}.

The device configuration employed by Liang {\it et
al}~\cite{LiangPRB99} also allows the robustness of the $0.7$
plateau to changes in the QPC confining potential to be
investigated, with the $0.7$ plateau persisting despite the lateral
confinement being strengthened by a factor of two. This is
consistent with the observation of the $0.7$ plateau in a wide range
of different QPC designs and 2DEG depths. The $0.7$ plateau also
appears irrespective of the nature of the 2D confinement, being
noted for both quantum well and heterojunction-based 2DEGs by Thomas
{\it et al}~\cite{ThomasPRB98}.

\subsection{The dependence on temperature}

Further evidence mitigating explanations based on quantum
interference is provided by the dependence of the $0.7$ plateau on
temperature $T$. Figure~3(a) shows the 1D conductance measured at $T
= 70$~mK, $460$~mK, $930$~mK and $1.5$~K. Decreasing the temperature
increases the electron phase coherence length, leading to a
strengthening of all features related to quantum interference. This
is evident in the quantized plateau at $G_{0}$ in Fig.~3(a), which
becomes longer and flatter as $T$ is reduced. The $0.7$ plateau
instead strengthens with increasing $T$ in
Fig.~3(a)~\cite{ThomasPRL96}, reaching its maximum strength at $T
\sim 1.5$~K, and disappearing by $T \sim 10$~K~\cite{ThomasPRB98}.
This temperature dependence is a very important and distinctive
feature of the phenomenology of the $0.7$ plateau. We will address
the temperature dependence, as studied by other authors further in
Section 3.1, but the non-trivial temperature dependence presented by
Thomas {\it et al} demonstrates that the $0.7$ plateau does not
originate from single-particle, quantum interference effects, and
presents a significant challenge to potential explanations for the
$0.7$ plateau.

\begin{figure}
\includegraphics[width=7cm]{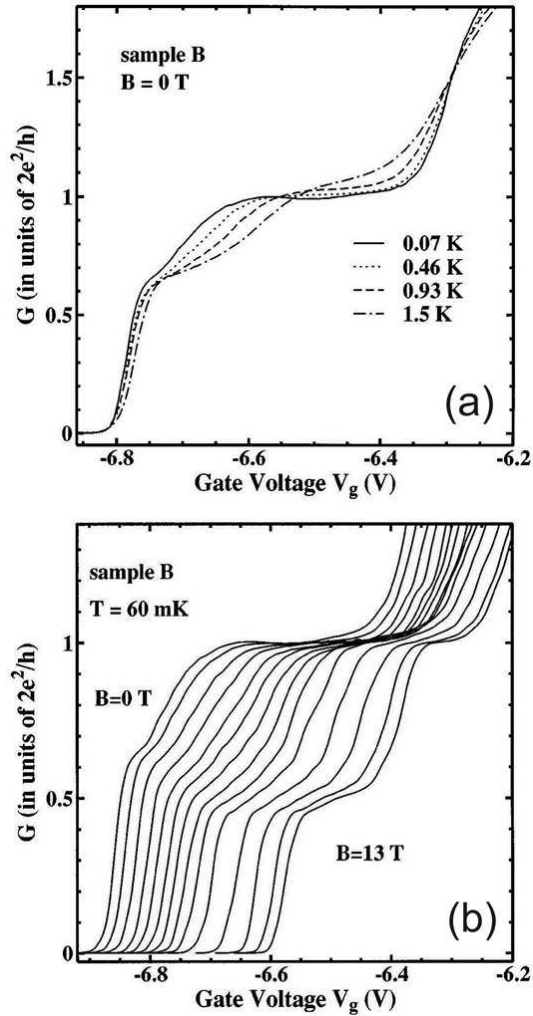}
\caption{Behaviour of the 1D conductance $G(V_{g})$ as a function of
(a) temperature $T$, and (b) an applied magnetic field $B$ in the
plane of the 2DEG, with $B$ increasing in steps of $1$~T from $B =
0$~T (left-most) to $13$~T (right-most). The temperature dependence
highlights one of the most peculiar aspects of the $0.7$ plateau,
which is that it first strengthens with increasing temperature,
becoming most apparent at $T \approx 1.5$~K, before decreasing in
strength and disappearing entirely by $T = 10$~K. With increasing
$B$, the $0.7$ plateau undergoes a smooth evolution into a plateau
at $0.5G_{0}$ corresponding to the last spin-split subband,
motivating suggestions that the $0.7$ plateau is due to a
spontaneous spin-polarization in the QPC. Figure reproduced with
permission from Ref.~\cite{ThomasPRB98}. Copyright 1998 by the
American Physical Society.}
\end{figure}

\subsection{The dependence on an in-plane magnetic field}

The dependence on a magnetic field $B$ applied in the plane of the
2DEG provides the first insight to a possible mechanism for the
$0.7$ plateau. As shown in Fig.~3(b), the $0.7$ plateau makes a
continuous downward migration to $0.5G_{0}$ with increasing $B$. The
plateau at $0.5G_{0}$ is expected, and is one of a cascade of
plateaus separated by $e^{2}/h$ rather than $2e^{2}/h$ for large $B$
due to the Zeeman splitting of the 1D subbands~\cite{WharamJPC88,
PatelPRB91}. The smooth evolution of the $0.7$ plateau into the
first spin-resolved plateau with in-plane magnetic field led to the
first hypothesis for its origin, namely the development of a
spontaneous partial spin-polarization of the electron gas within the
QPC at zero magnetic field once the last 1D subband
depopulates~\cite{ThomasPRL96}. This hypothesis initially appears
invalid due to the Lieb and Mattis theorem precluding ferromagnetic
ordering of electrons confined to one dimension~\cite{LiebPR62}.
However, this theorem holds only for an infinitely long, strictly 1D
system, and a QPC counts as neither, being only
quasi-one-dimensional, and far from infinite in length, connected to
2D reservoirs at either end. Furthermore, the second and higher 1D
subbands are not at infinite energy, as required for a pure 1D
system~\cite{GrahamPhD03}. Before discussing initial theoretical
support for spontaneous spin polarization, we address one additional
piece of evidence obtained by Thomas {\it et al}.

\subsection{Exchange-enhancement of the Land\'{e} $g$-factor}

The relationship between the Zeeman splitting $\Delta E_{z}$ and an
applied magnetic field $B$ is governed by the effective Land\'{e}
$g$-factor $g^{*}$. In the simplest linear approximation $\Delta
E_{z} = \frac{1}{2}g^{*}\mu_{B} B$, where $\mu_{B}$ is the Bohr
magneton. In a solid, the $g$-factor can differ significantly from
the free electron value $g^{*} = 2$ due to spin-orbit
effects~\cite{RothPR59}, with $g^{*} = -0.44$ in bulk
GaAs~\cite{WhiteSSC72}. The $g$-factor can also be significantly
modified by exchange interactions~\cite{JanakPR69}. Thus narrowing a
QPC should enhance the magnitude of the $g$-factor $|g^{*}|$ due to
the stronger confinement causing greater electron wavefunction
overlap. In contrast to 2D systems, $|g^{*}|$ is easily measured for
each 1D subband $n$ in QPCs using source-drain bias
spectroscopy~\cite{PatelPRB91, PatelPRB91a}. This allows the
$g$-factor to be measured as the 1D confinement is strengthened.
Thomas {\it et al} found that $|g^{*}|$ increases from the bulk
value of $0.44$ at $n = 25$ to $1.15 \pm 0.2$ at $n < 4$ with the
increase becoming more rapid as $n$ is reduced~\cite{ThomasPRB98,
ThomasPRL96}. The rapidly increasing influence of exchange
interactions and increasing $g^{*}$ provided an argument for why the
$0.7$ plateau appeared more commonly, consistently and strongly than
analogous plateaus at $G > G_{0}$ (e.g., $1.7G_{0}$, $2.7G_{0}$,
etc.) in these initial studies. Such higher non-quantized plateaus
are observed by Kristensen {\it et al}~\cite{KristensenJAP98} and
Reilly {\it et al}~\cite{ReillyPRB01} for example, and were studied
in detail by Graham {\it et al}~\cite{GrahamPRL03} as discussed in
Section 8.1. A cursory study of the dc source-drain bias dependence
of the $0.7$ plateau was also performed~\cite{ThomasPRB98}, with the
$0.7$ plateau evolving into a plateau at $0.85G_{0}$ at finite bias.
Thomas {\it et al} point out that the non-linear evolution with
source-drain bias $V_{sd}$ of the dark line separating the plateaus
at $0.85G_{0}$ and $G_{0}$ (e.g., see green dashed line in
Fig.~6(b)) indicates an occupation-dependent evolution of the lowest
1D subband. We will return to this idea in greater detail later,
with Sections 3.1, 8 and 9 presenting more in-depth studies of the
source-drain bias dependence of the $0.7$ plateau and 1D subbands.
For the purpose of closing this current Section, I will simply point
out that Thomas {\it et al}~\cite{ThomasPRB98} cite this behaviour
as further evidence for an interaction-related mechanism for the
$0.7$ plateau.

\subsection{Spin density functional calculations and spontaneous spin-polarization}

Two papers by Wang and Berggren provided initial theoretical support
for the $0.7$ plateau being related to spontaneous spin-polarization
within the QPC. Wang and Berggren performed self-consistent
calculations of the electronic structure of an infinitely long
quantum wire~\cite{WangPRB96} and a QPC~\cite{WangPRB98} using
spin-polarized density functional theory with the Kohn-Sham
approach~\cite{KohnPR65}. For the infinitely-long quantum wire, a
very strong energy splitting between spin-up ($\bf{\sigma} =
\frac{1}{2}$, $\uparrow$) and spin-down ($\bf{\sigma} =
-\frac{1}{2}$, $\downarrow$) was observed when the Fermi level
crossed through the mean (spin-unpolarized equivalent) energy for
the $n$~th 1D subband, causing it to populate/depopulate, as shown
in Fig.~4(a)~\footnote{Data calculated for an in-plane magnetic
field $B = 3$~T is shown for clarity. The same physics is observed
at $B = 0.01$~T (see Fig.~1(b) of Ref.~\cite{WangPRB96}), but the 1D
subband spacing is smaller and the spin-splitting at subband
population/depopulation is stronger, making the behaviour less
obvious.}. This spin-splitting is driven by exchange rather than the
Zeeman effect, and it leads to an intermittent spontaneous
spin-polarization within the quantum wire as the 1D subbands
depopulate (i.e., coinciding with the risers in the quantized
conductance). It also leads to the rather complex dependence of
$g^{*}$ on the 1D electron density $n_{1D}$ shown in Fig.~4(b). The
background trend of rising $g^{*}$ with decreasing $n_{1D}$ has been
observed by Thomas {\it et al}~\cite{ThomasPRL96}, as discussed in
Section 2.5. Chen {\it et al}~\cite{ChenPRB09b} claim to observe
this rise and fall in $g^{*}$ in experimental measurements of the dc
conductance of a QPC, we will return to this in Section 9.3. The
magnitude of $g^{*}$ predicted in Fig.~4(b) is over an order of
magnitude larger than reported experimentally~\cite{ThomasPRL96,
ChenPRB09b}, and Wang and Berggren point out that the finite length
of the QPC and the adjacent 2D reservoirs are likely responsible for
this~\cite{WangPRB96}.

\begin{figure}
\includegraphics[width=7cm]{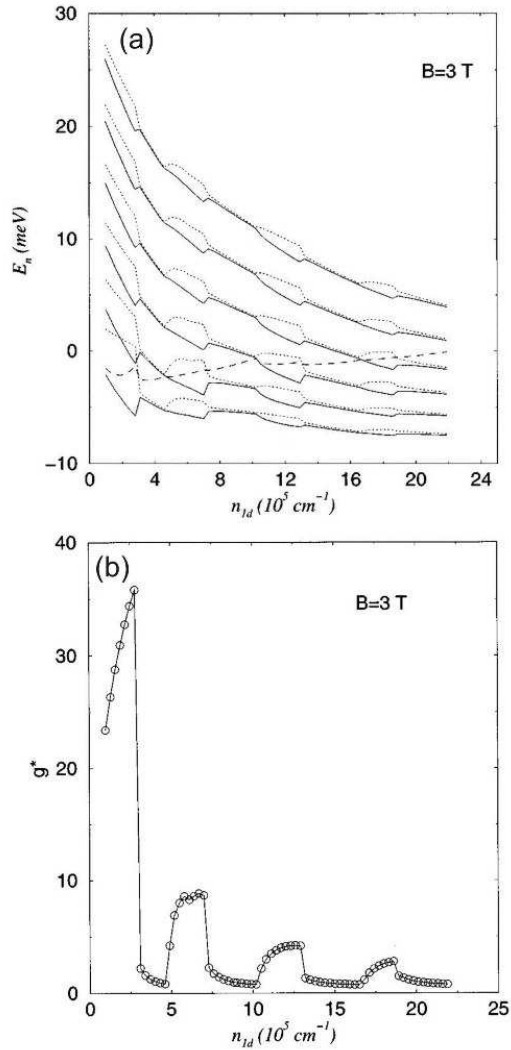}
\caption{(a) The 1D subband energies $E_{n}$ for spin-up (dotted
lines) and spin-down (solid lines) vs 1D electron density $n_{1D}$.
The dashed line shows the Fermi energy $E_{F}$. (b) Effective
Land\'{e} $g$-factor calculated vs $n_{1D}$. Data in both cases is
calculated for an in-plane magnetic field $B = 3$~T. Figure adapted
with permission from Ref.~\cite{WangPRB96}. Copyright 1996 by the
American Physical Society.}
\end{figure}

The more relevant paper for studies of QPCs is
Ref.~\cite{WangPRB98}, where the calculations are performed for a
saddle-point QPC potential~\cite{ButtikerPRB90}. Figure~5(a-d) shows
the calculated effective potential $E^{\sigma}_{1}$ for spin-up
(dotted line) and spin-down (solid line) electrons versus position
$x$ along the QPC axis with $x = 0$ corresponding to the QPC center.
The data is shown for increasing values of the potential $V_{0}$ at
the center of the QPC, which effectively translates into increasing
$V_{g}$. In each case, $E^{\sigma}_{1}$ has a maximum at $x = 0$
with $E^{\uparrow}_{1} > E^{\downarrow}_{1}$. However the height of
the spin-up barrier grows rapidly as the QPC is narrowed (i.e., more
negative $V_{g}$), eventually exceeding $E_{F}$, whereas the height
of the spin-down barrier remains approximately constant.
Qualitatively similar results were also obtained for a constriction
separating to two large quantum dot reservoirs by Bychkov {\it et
al}~\cite{BychkovNano00}. The corresponding 1D electron densities
$n^{\uparrow}_{1D}(x)$ (dotted line) and $n^{\downarrow}_{1D}(x)$
(solid line) are shown in Fig.~5(e-h). The transport of spin-up
electrons occurs by tunneling through an exchange-enhanced barrier,
while spin-down transport is largely unaffected, leading to $G <
G_{0}$. Wang and Berggren calculated the corresponding conductance
using the Landauer-B\"{u}ttiker expression (see
Ref.~\cite{BeenakkerRev91} for a discussion), and this is presented
in Fig.~5(i) versus $V_{0} \sim V_{g}$ for the non-interacting
(solid line) and interacting (dashed line) cases. A strong plateau
at $0.5G_{0}$ emerges when interactions are included, with no
feature at $0.7G_{0}$, indicative of complete spin-polarization. In
Ref.~\cite{WangPRB98}, it is argued that shortening of the QPC can
lead to onset of spin-polarization at higher $G$, leading to two
plateaus separated by $0.5G_{0}$, one located at $0 < G < 0.5G_{0}$
and the other at $0.5G_{0} < G < G_{0}$. This implies there should
be a $0.2G_{0}$ plateau attendant to the $0.7$ plateau, and although
this has been observed (e.g., by Ramvall {\it et
al}~\cite{RamvallAPL97} and de Picciotto {\it et
al}~\cite{dePicciottoPRL04}) it is certainly not a consistent
feature if one looks more broadly at available data. A subsequent
paper by Berggren and Yakimenko~\cite{BerggrenPRB02} showed that
accounting for correlation effects~\cite{TanatarPRB89} leads to
partial spin-polarization, giving a plateau at $0.7G_{0}$ rather
than at $0.5G_{0}$.

\begin{figure}
\includegraphics[width=16cm]{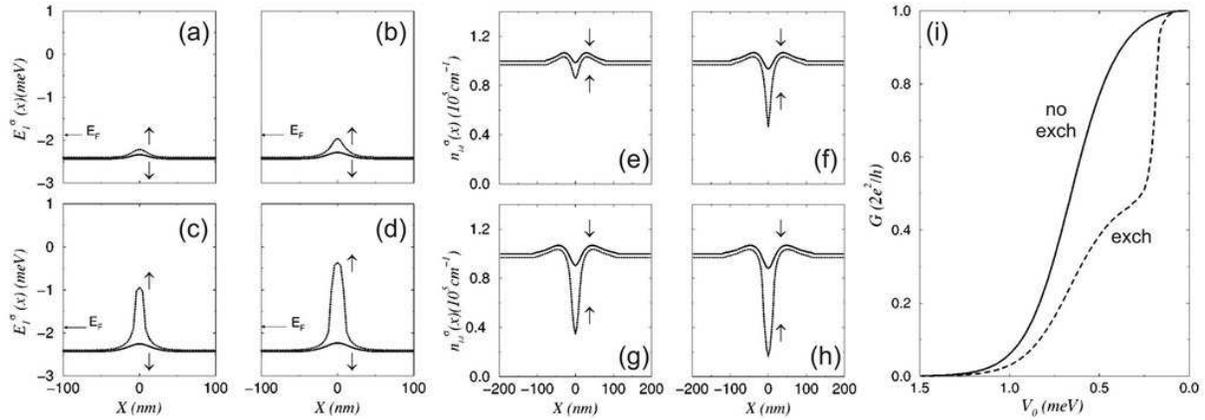}
\caption{(a-d) Effective potential barrier $E^{\sigma}_{1}(x)$ along
the QPC axis $x$ when the potential at the saddle-point $V_{0}$
takes values of (a) $0.1$~meV, (b) $0.15$~meV, (c) $0.18$~meV and
(d) $0.20$~meV. Increasing $V_{0}$ corresponds to more negative
$V_{g}$ which narrows the QPC. Calculated values for spin-up and
spin-down are represented by the dotted (upper trace) and solid
lines (lower trace), respectively, for a fixed 1D electron density
$n_{1D} = 2 \times 10^{5}$~cm$^{-1}$. The Fermi energy $E_{F}$ is
indicated in each of the four panels. (e-h) Corresponding 1D
electron density $n^{\sigma}_{1D}(x)$ vs $x$ for $V_{0}$ values of
(e) $0.1$~meV, (f) $0.15$~meV, (g) $0.18$~meV and (h) $0.20$~meV.
Calculated values for spin-up and spin-down are represented by the
dotted (lower trace) and solid lines (upper trace), respectively.
(i) Calculated conductance $G$ vs potential at the saddle-point
$V_{0}$ for cases where the exchange interaction is ignored (solid
line) and included (dashed line). Figure adapted with permission
from Ref.~\cite{WangPRB98}. Copyright 1998 by the American Physical
Society.}
\end{figure}

The $0.7$ plateau should be dependent upon electron density if it
originates from an exchange-driven spontaneous spin-polarization
within the QPC. This was examined briefly in
Ref.~\cite{ThomasPRB98}, with the $0.7$ plateau strengthening from a
weak inflection at $n_{2D} = 1.3 \times 10^{11}$~cm$^{-2}$ to a
clearly formed plateau as the density is reduced to $1.1 \times
10^{11}$~cm$^{-2}$. However, this is a small part of the real
density dependence of the $0.7$ plateau, which is discussed more
fully in Section 3.4.

\section{Studies of the temperature, electron density and dc source-drain bias dependence}

With the basic framework for the phenomenology of the $0.7$ plateau
in place, we now expand our focus to look more widely at studies of
three key experimental aspects of this problem: temperature, dc
source-drain bias and electron density.

\subsection{Temperature and dc bias}
More extended studies of the temperature and dc source-drain bias
dependence of the $0.7$ plateau were undertaken by the Copenhagen
group~\cite{KristensenPRB00, LindelofSPIE01, KristensenPhysB98,
KristensenPS02}. Kristensen {\it et al} studied six QPCs made by
shallow etching with different lengths and widths. To further vary
the 1D confinement, one of these samples had an overall metal
top-gate and two were covered with a layer of GaAlAs by MBE
regrowth. These six devices allowed the robustness to QPC geometry
and confinement potential to be investigated. Figure~6(a) shows
measurements of the differential conductance $G = dI/dV$ versus
source-drain bias $V_{sd}$ for increasing steps in gate voltage
$V_{g}$ measured for a shallow etched QPC with an overall
top-gate~\cite{KristensenPRB00}. The accumulations of traces in
Fig.~6(a) correspond to plateaus in the conductance versus gate
voltage $G(V_{g})$, such that rising along a vertical path at the
center of Fig.~6(a) corresponds to moving from left to right in
Fig.~1(a) through the population of the first four 1D subbands.
Finite source-drain bias leads to a separation between the chemical
potentials in the source $\mu_{s}$ and drain $\mu_{d}$ contacts,
$\mu_{s}-\mu_{d} = eV_{sd}$, generating plateaus at half-integer
multiples of $G_{0}$ that correspond to the chemical potential of
the source (drain) sitting above a 1D subband while the drain
(source) sits below. The $0.7$ plateau at $V_{sd} = 0$ in Fig.~6(a)
is difficult to see, as there is only a weak accumulation of traces,
but it evolves into a clear accumulation of traces (i.e.,
conductance plateau) at $G = 0.85G_{0}$. We will return to the small
peaks centered on $V_{sd} = 0$ for $G < G_{0}$ in Section 4.2. The
slight asymmetries in Fig.~6(a) with increasing $V_{sd}$ are due to
self-gating effects~\cite{KristensenPRB00}.

\begin{figure}
\includegraphics[width=7cm]{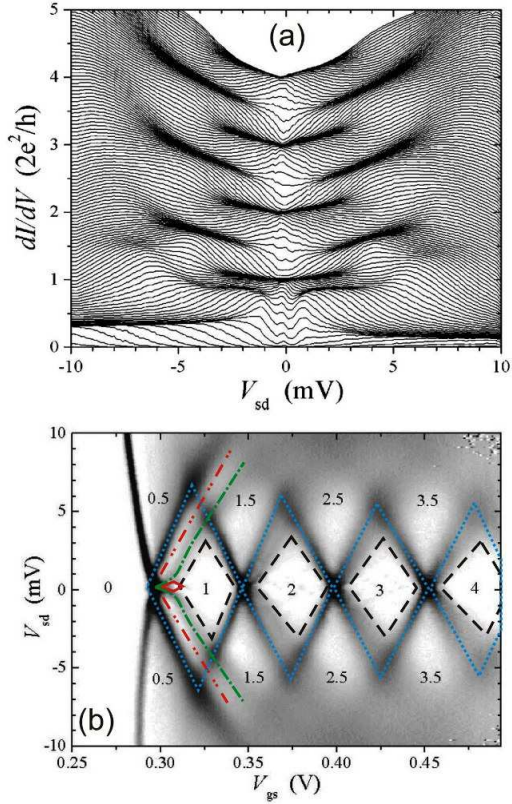}
\caption{(a) The differential conductance $G = dI/dV$ vs dc
source-drain bias $V_{sd}$. In each subsequent trace the gate
voltage $V_{g}$ is stepped by $1$~mV. (b) Transconductance greyscale
corresponding to the data in (a) with the transconductance
$dG/dV_{g}$ plotted on the greyscale axis vs $V_{sd}$ ($y$-axis) and
$V_{g}$ ($x$-axis). The light regions correspond to low
transconductance, indicating a plateau in $G(V_{g})$, while the dark
regions correspond to high $dG/dV_{g}$, indicating the risers
between plateaus. The black dashed lines are guides to the eye
highlighting the white diamonds corresponding to integer plateaus.
The blue dotted lines highlight the diamonds that correspond to
subband depopulation and thus allow the subband spacing $\Delta
E_{n,n+1}$ to be determined. The small red solid-line diamond
highlights the $0.7$ plateau region and the red dash-double-dotted
line highlights the evolution of the $0.85G_{0}$ finite bias plateau
with increasing $V_{sd}$. The green dash-dotted line highlights the
nonlinear evolution of the `anomalous' subband edge between the
$0.85G_{0}$ and $G_{0}$ plateaus. Figure adapted with permission
from Ref.~\cite{KristensenPRB00}. Copyright 2000 by the American
Physical Society.}
\end{figure}

Source-drain bias data such as that shown in Fig.~6(a) is often
plotted as a greyscale of the transconductance $dG/dV_{g}$ versus
$V_{g}$ and $V_{sd}$ on the abscissa/ordinate axes. In Fig.~6(b),
the bright regions correspond to low transconductance, which are the
plateaus in $G(V_{g})$ and the accumulations of traces in Fig.~6(a);
and the dark regions correspond to the risers between plateaus in
$G(V_{g})$ and the sparse regions in Fig.~6(a). The most informative
aspect of such a plot are the dark regions, because these provide
information about the position of the 1D subband edges relative to
the chemical potential. Starting at $V_{sd} = 0$, where $\mu_{s} =
\mu_{d} = \mu$ and moving from far left to right in Fig.~6(b), the
dark points at the apices of the light diamond-shaped structures
correspond to the edges of the first, second, third and fourth
spin-degenerate 1D subbands passing through $\mu$. Moving upwards to
finite, positive $V_{sd}$, these dark regions split into V-shaped
structures, with the left-moving (right-moving) diagonal
corresponding to a given subband edge passing through $\mu_{s}$
($\mu_{d}$). Hence moving from left to right at finite $V_{sd}$, the
picture is one of subsequent 1D subband edges dropping first through
$\mu_{s}$ and then through $\mu_{d}$, these two chemical potentials
being separated in energy by $eV_{sd}$. Note that at negative
$V_{sd}$, $\mu_{d}$ would be above $\mu_{s}$ and this attribution of
left-moving and right-moving diagonals would be reversed (i.e.,
diagonal moving down to the left corresponds to the first subband
edge crossing the drain, which is highest in energy). However,
Fig.~6(b) is symmetric to reflection about $V_{sd} = 0$, and so the
common convention is to assume that finite $V_{sd}$ raises $\mu_{s}$
above $\mu_{d}$, irrespective of the sign of $V_{sd}$ (i.e., the
attribution of source and drain is exchanged upon $V_{sd}$ sign
reversal).

At sufficient $V_{sd}$ the dark diagonals cross, giving rise to the
clear diamond-shaped structures indicated by the blue dashed lines.
These reflect one of the most useful aspects of a transconductance
greyscale -- the ability to directly read off the energy spacing
between the 1D subbands~\cite{PatelPRB91a, MartinMorenoJPCM92}. The
$V_{sd}$ at the apex of the diamond corresponding to the $n$th
plateau provides the spacing $\Delta E_{n,n+1}$, which is of order
$6.5$~meV for the first subband of the QPC studied by Kristensen
{\it et al}~\cite{KristensenPRB00, KristensenPS02}. The subband
spacing normally decreases with increasing subband index $n$, as
evident in Fig.~6(b), due to the weakening electrostatic confinement
as the gate voltage is made more positive~\cite{LauxSurfSci88}. This
allows the subband spacing to be an effective semi-quantitative
measure of confinement strength, as utilized in Section 10 (see
Fig.~41(b-d)). The transconductance greyscale in Fig.~6(b) makes the
evolution of a number of features that appear in the conductance
more clear. The flatest parts of the integer plateaus appear as
white diamonds (corresponding to $dG/dV_{g} = 0$), as highlighted by
the black dashed lines. In contrast to the higher integer plateaus,
the $G_{0}$ diamond has a small nodular extension at the low $V_{g}$
side, indicated by the small red solid-line diamond, which
corresponds to the $0.7$ plateau. The riser between the $0.7$ and
$1$ plateaus is often too short and shallow for a clear divide
between these regions to appear in a transconductance greyscale.
Another significant feature in Fig.~6(b) are the white bands
indicated by the red dash-double-dotted lines, which correspond to
the $0.85G_{0}$ and $1.85G_{0}$ plateaus at finite $V_{sd}$ (for a
clearer view of the $1.85G_{0}$ plateau, see Fig.~6(a) of
Ref.~\cite{KristensenPRB00}). The $0.85G_{0}$ plateau continues
through the left-moving dark band corresponding to the second
subband coinciding with $\mu_{d}$ to form an additional plateau at
$G \sim 1.4G_{0}$ at higher $V_{sd} \sim 5-8$~mV.

\begin{figure}
\includegraphics[width=12cm]{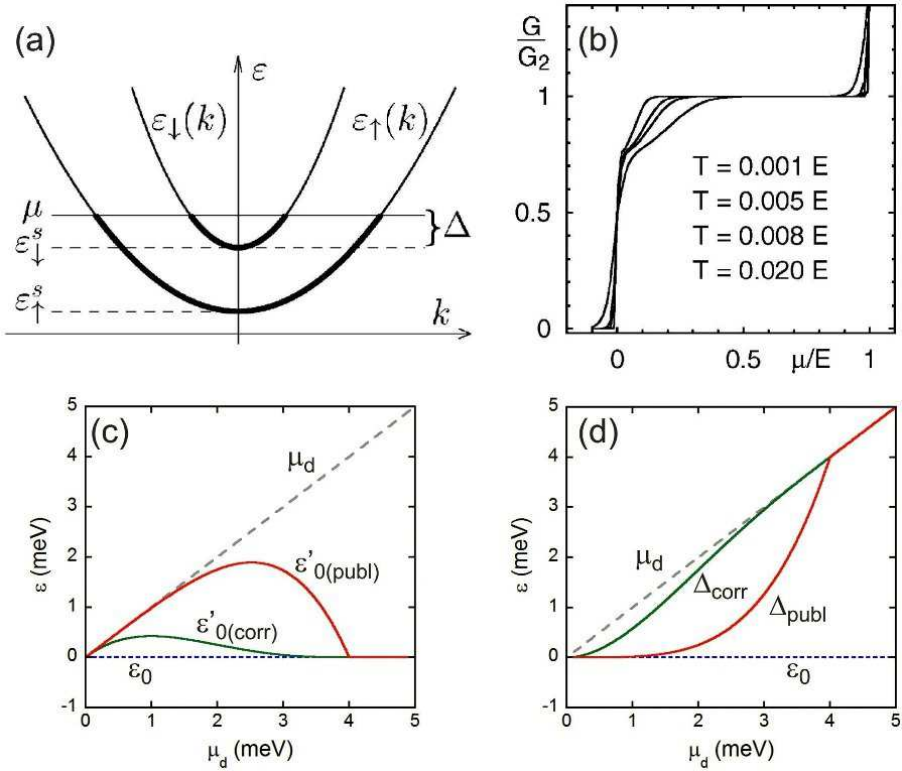}
\caption{(a) Schematic illustrating the spin-gap model proposed by
Bruus {\it et al}~\cite{BruusPhysE01}, where the lowest 1D subband
splits into spin-up $\epsilon_{\uparrow}(k)$ and spin-down
$\epsilon_{\downarrow}(k)$ components. The key energy parameter of
the model is the spin-down Fermi energy $\Delta(\mu) = \mu -
\epsilon^{s}_{\downarrow}$, where $\mu$ is the chemical potential.
(b) Simulated conductance $G$ vs $\mu$, which is proportional to
$V_{g}$ in experiments, obtained from the spin-gap model. The
normalization parameters used are $G_{2} = 2e^{2}/h$ and the
spin-degenerate 1D subband spacing $E$. The model predicts
suppression of the shoulder of the $G_{0}$ plateau to $0.75G_{0}$
for the $\mu$ range where the condition $\Delta(\mu) < k_{B}T <
\Delta_{sg} = \epsilon^{s}_{\downarrow} - \epsilon^{s}_{\uparrow}$
holds. (c) Plots of drain chemical potential $\mu_{d}$ (grey dashed
line), normal subband energy $\epsilon_{0}$ (blue dotted line), and
anomalous subband energy $\epsilon'_{0}$ using Eq.~15 in
Ref.~\cite{KristensenPS02} precisely as written (green line labelled
$\epsilon'_{0(corr)}$) and altered as per Eq.~4 to match how it is
presented in Ref.~\cite{KristensenPS02} (red line labelled
$\epsilon'_{0(publ)}$), all vs $\mu_{d}$. (d) Plots of $\Delta(\mu)
= \mu_{d} - \epsilon'_{0}$ using the corresponding $\epsilon'_{0}$
data in (c). Figures (a,b) from Ref.~\cite{BruusPhysE01} with
permission from Elsevier. Figures (c,d) prepared by the author based
on Fig.~3 and Eq.~15 of Ref.~\cite{KristensenPS02}.}
\end{figure}

The V-shaped structure that evolves from $V_{gs} = 0.3$ at $V_{sd} =
0$ in Fig.~6(b) is of particular interest. Although there is a
single left-moving branch that corresponds to a single
spin-degenerate subband passing through $\mu_{s}$, there are two
right-moving dark branches, one separating the $0.5G_{0}$ and $0.85
G_{0}$ plateaus indicated by the blue dotted line and the other
separating the $0.85G_{0}$ and $G_{0}$ plateaus indicated by the
green dash-dotted line. If one considers all transitions between
conductance plateaus as indicating the passing of a 1D subband edge
through a chemical potential, the additional dark band indicated by
the green dash-dotted line suggests that two distinct subband edges
pass through $\mu_{d}$, whereas only one passes through $\mu_{s}$
(we will return to this in Sections 8 and 9 also).

This motivated Kristensen {\it et al}~\cite{KristensenPRB00} to
suggest that the appearance of fractional plateaus, the $0.7$
plateau in particular, may be due to the presence of an anomalous
subband edge that sits above the normal subband edge during the
population of the first subband, motivating the first of two
phenomenological models that we discuss below. Note that the
attribution of `normal' comes about by association with the
right-moving diagonals at higher subbands, as is clear by looking at
the blue dotted lines surrounding each integer quantized plateau
(bright diamond) in Fig.~6(b). The fact that the two right-moving
branches for the first 1D subband evolve from the same point and are
not parallel suggests that the energy gap between them is dependent
upon source-drain bias. The fact that there is only one left-moving
diagonal suggests that the normal and anomalous subband edges are
degenerate on passing through $\mu_{s}$, i.e., the gap opens after
the first subband begins populating. How this gap opens is the key
point of difference between the two phenomenological spin-gap models
for the $0.7$ plateau. We will discuss the first now, and the second
in Section 3.5.

\subsection{The Bruus, Cheianov and Flensberg (BCF) spin-gap model}

The splitting of the first right-moving dark band in Fig.~6(b) led
Kristensen {\it et al}~\cite{KristensenPRB00} to suggest the
presence of an anomalous subband edge $\epsilon'_{0}$ that sits
slightly above a normal subband edge $\epsilon_{0}$ during
population of a subband, and that this might be responsible for the
$0.7$ plateau and other fractional plateaus at finite bias in QPCs.
Subsequent papers by Bruus, Cheianov and
Flensberg~\cite{BruusPhysE01, BruusArXiv00} and Kristensen and
Bruus~\cite{KristensenPS02} developed this suggestion further,
arguing that the normal and anomalous subbands in
Ref.~\cite{KristensenPRB00} are in fact the spin-up
$\epsilon_{\uparrow}(k)$ and spin-down $\epsilon_{\downarrow}(k)$
components of a single, spin-split 1D subband, as shown in
Fig.~7(a). Note that Bruus {\it et al}~\cite{BruusPhysE01,
BruusArXiv00} use the opposite spin-convention to most other papers
on the $0.7$ plateau -- this does not affect the physics, but needs
to be borne in mind when comparing to other works. These two
components contribute equally to a temperature-dependent
conductance:

\begin{equation}
G(T) =
\frac{1}{2}(f[\epsilon^{s}_{\uparrow}(\mu)-\mu]+f[\epsilon^{s}_{\downarrow}(\mu)-\mu])G_{0}
\end{equation}

\noindent where $\epsilon^{s}_{\uparrow}(\mu)$ and
$\epsilon^{s}_{\downarrow}(\mu)$ are the two spin subband edges, and
$f[x] = ($exp$(x/k_{B}T)+1)^{-1}$ is the Fermi-Dirac distribution.
The important energy scale in this model is the Fermi energy of the
spin-down subband $\Delta(\mu) = \mu -
\epsilon^{s}_{\downarrow}(\mu)$. If $\Delta(\mu)$ is large compared
to the spin-gap $\Delta_{sg} = \epsilon^{s}_{\downarrow} -
\epsilon^{s}_{\uparrow}$ then the system is very weakly spin
polarized, and providing $\mu$ sits below the next-highest 1D
subband then $G = G_{0}$. As $\Delta(\mu)$ becomes smaller than the
spin-gap, the spin-polarization becomes more complete, and the
conductance near the low $G$ edge of the $G_{0}$ plateau becomes
temperature dependent. At low temperatures $k_{B}T << \Delta(\mu)$
both Fermi distributions in Eq.~1 equal $1$ giving $G = G_{0}$.
However, with a slight increase in temperature to $\Delta(\mu) <
k_{B}T < \Delta_{sg}$, the first Fermi distribution falls to $0.5$,
giving $G = 0.75G_{0}$. This connects to the measurable $G(V_{g})$
via $\mu \propto V_{g}$. Hence providing that $\Delta(\mu) < k_{B}T
< \Delta_{sg}$ over some reasonable range of $\mu$ between $\mu
>> \epsilon^{s}_{\uparrow},\epsilon^{s}_{\downarrow}$ where $G =
G_{0}$ and $\mu < \epsilon^{s}_{\uparrow},
\epsilon^{s}_{\downarrow}$ where $G = 0$, then a plateau at $0.75$
should be observed. This model makes a very specific prediction
about the $0.7$ plateau, namely that will evolve as a {\it drop} in
the low $G$ shoulder of the integer plateau, as shown in
Fig.~7(b)~\cite{BruusPhysE01, BruusArXiv00}. This is indeed observed
in Fig.~8(a), as discussed in the next paragraph. Further, at finite
$V_{sd}$ the conductance should drop by only $\frac{1}{8}G_{0}$
rather than $\frac{1}{4}G_{0}$, giving rise to a finite bias plateau
at $G = 0.875G_{0}$, which is very close to the $0.85$ plateau
evident in Fig.~6(a/b). A weakness in this model is that it predicts
the evolution of a $0.5$ plateau in addition to the $0.75$ plateau
at higher temperatures $k_{B}T >> \Delta_{sg}$ (e.g., see Fig.~3 of
Ref.~\cite{BruusPhysE01}), which is inconsistent with all
experimental observations to date.

\begin{figure}
\includegraphics[width=12cm]{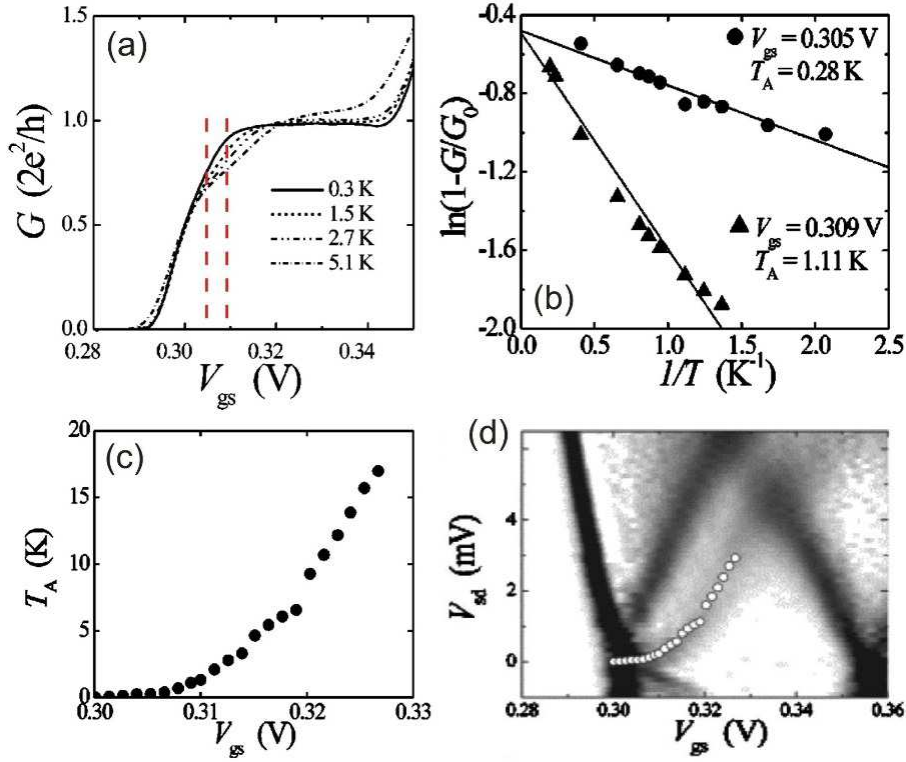}
\caption{(a) Measured conductance $G$ vs gate voltage $V_{g}$ at
four different temperatures $T$. The red dashed vertical lines
indicate the $V_{g}$ where data for (b) was obtained. (b) Arrhenius
plot of $ln(1 - G/G_{0})$ vs $1/T$, indicating thermally activated
behavior. $G$ was obtained at various $T$ between $\sim 480$~mK and
$4$~K at $V_{g} = 0.305$ and $0.309$~V. (c) Activation temperatures
$T_{A}$ vs $V_{g}$ obtained by repeating the analysis in (b) for a
number of gate voltages spanning the $0.7$ and $1$ plateaus. (d)
Data obtained in (c) converted into a corresponding source-drain
bias using $V^{*}_{sd}(V_{g}) = 2k_{B}T_{A}(V_{g})/e$ and plotted
over an expanded region of the transconductance greyscale in
Fig.~6(b). Figures adapted with permission from
Ref.~\cite{KristensenPRB00}. Copyright 2000 by the American Physical
Society.}
\end{figure}

Measurements of the evolution of the $0.7$ plateau with source-drain
bias and temperature by Kristensen {\it et
al}~\cite{KristensenPRB00} support key features of the BCF spin-gap
model~\cite{BruusPhysE01, BruusArXiv00}. Firstly, the suppression of
$G$ beneath the $G_{0}$ integer plateau was measured versus
temperature $T$ between $\sim 480$~mK and $4$~K for two gate
voltages $V_{g} = 0.305$ and $0.309$~V (vertical red dashed lines in
Fig.~8(a)) where the thermal evolution of the $0.7$ plateau is
strongest. If the $0.7$ plateau involves thermal excitation then the
data are expected to follow an Arrhenius form:

\begin{equation}
G(T)/G_{0} =  1 - C e^{(-T_{A}/T)}
\end{equation}

\noindent where $T_{A}$ is the activation temperature and $C$ is a
constant. Figure~8(b) shows the data obtained from Fig.~8(a) plotted
as ln~$(1-G/G_{0})$ versus $1/T$. The trends for both $V_{g}$ values
are clearly linear, confirming thermally activated
behaviour~\cite{KristensenPRB00}. This analysis was repeated for a
range $0.30 < V_{g} < 0.33$~V spanning the $0.7$ and $1$ plateaus,
with the corresponding activation temperature plotted against
$V_{g}$ in Fig.~8(c). This is consistent with expectations from the
BCF model~\cite{BruusPhysE01}, since $\Delta(\mu) \propto T_{A}$,
$\Delta(\mu) = \mu - \epsilon^{s}_{\downarrow}(\mu)$ and $\mu
\propto V_{g}$. A more significant analysis can be obtained by
resorting to the transconductance greyscale in Fig.~6(b), because
the path followed by the transconductance ridge between the $0.85$
and $1$ plateaus provides a direct measure of $\Delta(\mu)$. For
this purpose, $T_{A}$ can be converted into a corresponding
source-drain bias $V^{*}_{sd}(V_{g})$ using $\Delta(V_{g}) =
k_{B}T_{A}(V_{g}) = \frac{1}{2}eV^{*}_{sd}(V_{g})$. In Fig.~8(d) the
corresponding data points are plotted on an expanded section of the
transconductance greyscale from Fig.~6(b). These points sit neatly
over the transconductance ridge between the $0.85$ and $1$ plateaus.
This behavior was observed for all six QPCs studied in
Ref.~\cite{KristensenPRB00}, suggesting that a bias-dependent
spin-gap model is a good fit to the experimental phenomenology of
the $0.7$ plateau.

Before moving on, there is one last feature of the BCF model that
warrants discussion. This is how the separation between the normal
and anomalous subband edges evolves as a function of their position
relative to the chemical potential, which will be important in later
discussions, particularly relative to the second spin-gap model
introduced in Section 3.5. The fact that there is only a single
left-moving dark branch but two right-moving dark branches in
Figs.~6(b) and 8(d) suggests that the separation between the normal
and anomalous subbands (i.e., opening of the spin gap) does not
occur until after the subband edge has reached/passed the source
potential. In principle, in the BCF model the spin-gap is zero until
the subband edge reaches the drain potential, however, providing the
relationship to other energy scales is correctly accounted for, it
can potentially be finite but small after the subband edge passes
$\mu_{s}$ without adversely affecting the model. Figures 7(c) and
(d) show graphs illustrating the evolution of the normal and
anomalous subband edges, but first we need to properly define the
relevant energies involved. As mentioned earlier, $\epsilon_{0}$ and
$\epsilon'_{0}$ are the normal and anomalous subband edge energies.
It is essential to understanding Figs.~7(c/d) to know that
$\epsilon_{0} = 0$ and $\mu_{s}$ is above $\epsilon_{0}$,
$\epsilon'_{0}$ and $\mu_{d}$. In Figs.~7(c/d), $\mu_{d}$ is
increased from $0$ to $5$~meV. This corresponds to moving the drain
chemical potential from coinciding with $\epsilon_{0}$ to a position
$1.5$~meV beneath $\epsilon_{1}$ (i.e., the second subband edge).
Figure 7(c) shows four lines: the diagonal grey dashed line is
simply $\mu_{d}$ versus $\mu_{d}$; the horizontal blue dashed line
is $\epsilon_{0}$ (the normal subband edge), which is fixed at zero
as our energy reference; the red line is $\epsilon'_{0(publ)}$,
which tracks $\mu_{d}$ until $2$~meV and then drops away from the
$\mu_{d}$ diagonal, heading back to zero at $\mu_{d} = 4$~meV, where
it remains as $\mu_{d}$ is increased further (i.e., the gap closes
again); and finally the green line is $\epsilon'_{0(corr)}$, which
tracks $\mu_{d}$ briefly and then falls away back to zero.

Two lines, $\epsilon'_{0(publ)}$ and $\epsilon'_{0(corr)}$, are
presented in Fig.~7(c) due to a discrepancy in
Ref.~\cite{KristensenPS02}. Kristensen and Bruus state that the
power law dependence plotted in Fig.~3 of Ref.~\cite{KristensenPS02}
is:

\begin{equation}
\epsilon'_{0}(\mu_{d}) = \mu_{d}(1-\mu_{d}/\mu^{*})^{n}
\end{equation}

\noindent for $0 < \mu_{d} < \mu^{*}$ where $\mu^{*} = 4.0$~meV and
$n = 3$. However, if one plots this function, the green line
$\epsilon'_{0(corr)}$ in Fig.~7(c) is obtained not the red line. The
red line $\epsilon'_{0(publ)}$, which is plotted in
Ref.~\cite{KristensenPS02}, is actually the function:

\begin{equation}
\epsilon'_{0}(\mu_{d}) = \mu_{d}(1-(\mu_{d}/\mu^{*})^{n})
\end{equation}

\noindent for $0 < \mu_{d} < \mu^{*}$ where $\mu^{*} = 4.0$~meV and
$n = 3$. This discrepancy is important because the specific form of
$\epsilon'_{0}$ plotted in red in Fig.~7(c) is chosen to adapt their
model calculations to the deduction that the energy width of the
anomalous plateau is $2$~meV, based on an empirical analysis of
experimental data~\cite{KristensenPS02}. It is hard to gauge the
impact of this on the remainder of the work in this paper, however
the form presented in Eq.~3 (their Eq.~15) does feature prominently
in their subsequent calculations.

For the moment, let us ignore the problem with the exact functional
relationship between $\epsilon'_{0}$ and $\mu_{d}$, and consider the
general qualitative behaviour that is reflected in both forms,
namely that $\epsilon'_{0}$ tracks $\mu_{d}$ briefly before
returning to zero, because this is the most important aspect to
later discussion. Physically, this corresponds to the anomalous
subband edge pinning to $\mu_{d}$ and {\it not} $\mu_{s}$ (this
should be borne in mind as it is important later) over some range in
energy. The $\epsilon'_{0}$ edge eventually de-pins from $\mu_{d}$
and the degeneracy between $\epsilon_{0}$ and $\epsilon'_{0}$ is
gradually restored.  This form and choice of parameters is not built
on any particular microscopic model, it simply succeeds in
reproducing much of the observed phenomenology of the $0.7$ plateau.
Finally, for reference to earlier discussion in this section,
Fig.~7(d) plots $\Delta(\mu_{d})$, which as Fig.~7(a) shows, is the
separation between $\mu_{d}$ and the anomalous subband edge
$\epsilon'_{0}$. Again, two sets of values are plotted,
$\Delta_{publ}$ corresponding to the data shown in
Ref.~\cite{KristensenPS02}, and $\Delta_{corr}$ corresponding to
what results if their Eq.~15 is used as written. The quantity
$\Delta$ is effectively the Fermi energy of the anomalous subband. A
similar quantity can be defined for the normal subband, and it would
follow the dashed grey diagonal line marked $\mu_{d}$ in Fig.~7(d).
In other words, the normal subband edge keeps dropping below
$\mu_{d}$ unaffected as $V_{gs}$ becomes more positive, leading to a
Fermi energy difference between the normal and anomalous subbands.
Assuming these are spin-up and spin-down subbands, this results in a
partial spin-polarization and an excitation gap for flipping a spin
that, combined with thermal activation, produces the $0.7$ plateau.
It is interesting to note the discontinuity in $\Delta_{publ}$ at
$\mu_{d} = 4$~meV, as it suggests that the smooth curve of points in
Fig.~8(c) would not continue indefinitely. This discontinuity should
occur slightly beyond the points presented, and it would be
interesting to continue the Arrhenius analysis of the data in
Fig.~8(a) out to $V_{gs} > 0.328$~V to see if such a discontinuity
is observed.

\subsection{Looking for a microscopic model -- First mention of possible Kondo physics}

The BCF spin-gap model presented above is purely phenomenological,
and for any new effect in quantum devices there is a strong desire
for a microscopic model. Following on from
Ref.~\cite{KristensenPRB00}, Lindelof considered that the
self-consistent nature of the inverted parabolic potential that
occurs along the QPC axis~\cite{ButtikerPRB90} may in some cases
lead to an isolated bound-state with spin-$\frac{1}{2}$ in the
middle of the QPC~\cite{LindelofSPIE01, ReimannPRB99,
KolehmainenEPJB00}. The bound-state can have two possible
configurations with different energies that arise via a wavefunction
symmetry argument whereby the electron wavefunction has either a
maximum or minimum local density at the potential maximum at the
center of the QPC. The ground state consists of two electrons bound
around the potential maximum, and corresponds to the minimum local
density solution. The excited state consists of a single bound
electron sitting atop the potential maximum, which can act as an
isolated spin, and corresponds to the maximum local density
solution. These two solutions are expected to be separated in energy
by $\sim 0.1$~meV, with a separation that increases with electron
density~\cite{LindelofJPCM08}. Lindelof's hypothesis was that the
0.7 plateau originates from thermal fluctuations between these two
isomer states. These states are analogous to different
configurational states in a molecule, for example, the cis- and
trans-isomers of butene, in that they are not present simultaneously
but the system can alternate between the two
states~\cite{LindelofPC}. This results in the conductance switching
between values corresponding to one isomer or the
other~\cite{LindelofJPCM08}. Building a model under this basis, and
using activation energy data from Ref.~\cite{KristensenPRB00}, the
calculated conductance revealed a strengthening $0.7$ plateau with
increasing temperature, consistent with experimental
observations~\cite{ThomasPRL96, KristensenPRB00, LindelofSPIE01}.
However, a more significant insight in Lindelof's work was the
suggestion that the isolated spin-$\frac{1}{2}$ excited state might
lead to a Kondo-like resonant transmission at the expense of direct
transmission via the ground state with a transmission coefficient of
$1$, thereby producing a suppression of the conductance from $G_{0}$
to $0.75G_{0}$ to cause the $0.7$ plateau~\cite{LindelofSPIE01,
LindelofJPCM08}. Note that this is slightly different to the Kondo
effects in GaAs quantum dots~\cite{GoldhaberGordonNat98,
CronenwettSci98} and carbon nanotubes~\cite{NygardNat00}, where the
Kondo process leads to an enhancement in the conductance towards the
unitary limit $G_{0}$~\cite{vanderWielSci00}. The idea here is that
the excited state has a reduced conductance due to a much lower
density of states, which is then enhanced slightly by the Kondo
process, and that switching between the excited state and the ground
state which has conductance $G_{0}$ reduces the net conductance at
the edge of the first plateau to below $G_{0}$. The overall
conductance consists of a thermal average of the Kondo and non-Kondo
processes that results in a final value near $0.7G_{0}$ in the low
density limit (i.e., towards the low $G$ shoulder of the plateau),
thereby producing the $0.7$ plateau. A significant aspect of this
hypothesis is that it does not rely on spontaneous spin-polarization
within the QPC as proposed by Thomas {\it et al}~\cite{ThomasPRL96}
and Wang and Berggren~\cite{WangPRB98}. Lindelof suggested that the
corresponding Kondo temperature would be $T_{K} \sim 10$~K, and that
there should be some effect on the noise spectrum of the QPC for $G
< G_{0}$~\cite{LindelofSPIE01}.

Lindelof's proposal was followed up in experiments by Cronenwett and
coworkers, and we will return to this in Section 4.2. First, we will
explore one further aspect of QPCs that has significant bearing on
spin-gap models for the $0.7$ plateau -- the dependence on electron
density.

\subsection{The dependence of the $0.7$ plateau on electron density}

The strength of electron-electron interactions in nanoscale
semiconductor devices is heavily dependent upon the electron
separation and hence the electron density, making this parameter of
great interest in experimental studies of the $0.7$ plateau. Thomas
{\it et al}~\cite{ThomasPRB00} and Nuttinck {\it et
al}~\cite{NuttinckJJAP00} reported the first such studies, arriving
at the same basic result, namely that the $0.7$ plateau shifts
downwards to $0.5G_{0}$ as the electron density is lowered at $B =
0$. In the study by Thomas {\it et al}~\cite{ThomasPRB00}, this was
achieved using a QPC with an additional `mid-line' gate passing
along the channel between the two gates used to define the QPC, with
two separate devices studied. In the first, the $0.7$ feature
observed at a density $n = 1.3 \times 10^{11}$~cm$^{-2}$ evolved
downwards in a continuous manner to $0.53G_{0}$ at $n = 3 \times
10^{10}$~cm$^{-2}$. In contrast, a clear plateau emerged at $0.5
G_{0}$ once the density was reduced below $n = 4 \times
10^{10}$~cm$^{-2}$ for the second sample, without evolving
continuously from the $0.7$ plateau. The authors claim (albeit
without presenting supporting evidence) that the second sample has
more impurities, and thus the emergence of the $0.5G_{0}$ plateau is
due to a spontaneous spin polarization produced by weak
disorder~\cite{ThomasPRB00}. The study by Nuttinck {\it et
al}~\cite{NuttinckJJAP00} was performed using two devices with
n$^{+}$-GaAs back-gates. Both devices showed a clear evolution of
the $0.7$ plateau downwards to $0.5G_{0}$, but over different
density ranges: $1.1 - 2.8 \times 10^{11}$~cm$^{-2}$ in the first
and $5.0 \times 10^{10} - 1.5 \times 10^{11}$~cm$^{-2}$ in the
second. Qualitatively similar data was observed by Wirtz {\it et
al}~\cite{WirtzPRB02}, who modified their electron density using a
combination of hydrostatic pressure and illumination rather than
electrostatic gating. In contrast, both Pyshkin {\it et
al}~\cite{PyshkinPRB00} and Hashimoto {\it et
al}~\cite{HashimotoJJAP01} report evolution of the $0.7$ plateau
towards $0.5G_{0}$ in both the low {\it and} high density limits.
Hence there is clearly more to this behaviour than density alone.

\begin{figure}
\includegraphics[width=7cm]{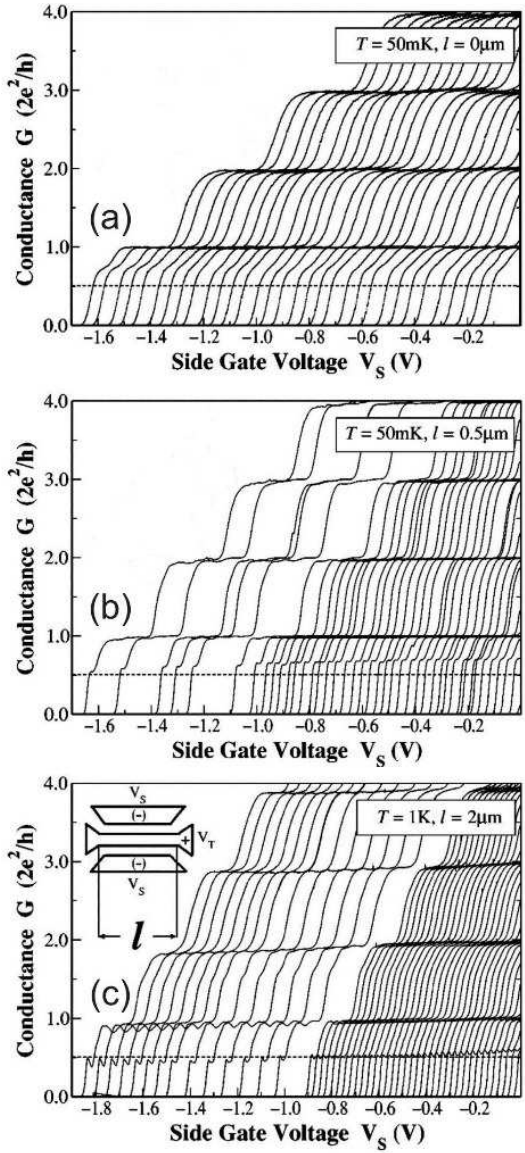}
\caption{Conductance $G$ vs side-gate voltage $V_{S}$ for doped-cap
induced 1D systems with (a) length $l = 0~\mu$m at temperature $T =
50$~mK for top-gate voltages $V_{T}$ from $172$~mV (right) to
$300$~mV (left), (b) $l = 0.5~\mu$m at $T = 50$~mK for $V_{T}$ from
$560$~mV (right) to $1500$~mV (left), and (c) $l = 2~\mu$m at $T =
1$~K for $V_{T}$ from $300$~mV (right) to $620$~mV (left). A
schematic of the device is shown inset to (c), with conduction
occurring in regions under the central gate, positively biased to
$V_{T}$ to accumulate electrons. The effective width of the central
1D region of length $l$ is tuned by the negative bias $V_{S}$
applied to the two side-gates. Figure adapted with permission from
Ref.~\cite{ReillyPRB01}. Copyright 2001 by the American Physical
Society.}
\end{figure}

A more comprehensive study by Reilly {\it et al}~\cite{ReillyPRL02,
ReillyPRB01} points to the length of the 1D system and the
difference in potential between the 1D and 2D regions playing
significant roles. This led to the development of an alternate
`spin-gap' model similar in spirit to that proposed by Bruus {\it et
al}~\cite{BruusPhysE01, BruusArXiv00} but with several important
differences. Reilly {\it et al}~\cite{ReillyPRB01} began by studying
the $0.7$ plateau in devices made on undoped AlGaAs/GaAs
heterostructures using a doped-cap architecture pioneered by Kane
{\it et al}~\cite{KaneAPL93, KaneAPL98}. This device architecture
allows the electron density to be set using a top-gate operated
independently of the two side-gates used to narrow the 1D
channel~\cite{KaneAPL98}. In practice, however, the precise density
inside the 1D region is influenced by all three gates and may differ
substantially from that generated by the top-gate alone in regions
well away from the side-gates~\cite{ReillyPRB01}. Four devices with
lengths $l = 0$, $0.5$, $2$ and $5~\mu$m were studied. In each case,
the conductance $G$ versus side-gate voltage $V_{S}$ was plotted for
a wide range of top-gate voltage $V_{T}$. This range is limited at
low $V_{T}$ by the ohmic contacts ceasing to conduct, and at high
$V_{T}$ when the top-gate begins to leak current directly to the
ohmic contacts. Figures~9(a) and (b) present $G$ versus $V_{T}$ for
the $l = 0$ and $0.5~\mu$m devices, respectively, at $T = 50$~mK.
Data for the $2~\mu$m device is shown in Fig.~9(c), and was obtained
at $T = 1$~K as this device developed significant structure
characteristic of length-resonance effects~\cite{TekmanPRB91} at $T
= 50$~mK. Migration of the $0.7$ plateau downwards in conductance
was not observed in the $l = 0~\mu$m device (Fig.~9(a)), in contrast
to Ref.~\cite{PyshkinPRB00}, however the density range explored may
have been insufficient as this device had the narrowest $V_{T}$
range of the four devices studied. The $0.5$ and $2~\mu$m long
devices show a very clear, smooth, downwards evolution of the $0.7$
feature as $V_{T}$ is reduced (i.e., moving from right to left), and
it is interesting to note that this happens much more rapidly for
the longer 1D channel (n.b., the $V_{T}$ range is $\sim 3 \times$
smaller in Fig.~9(c) compared to Fig.~9(b)). This led Reilly {\it et
al}~\cite{ReillyPRB01} to the conclusion that spin-splitting is only
fully-resolved at $B = 0$ in 1D systems above some critical length
scale and at sufficiently low density. They also note additional
structure at $G \sim 1.7G_{0}$, which was intensively investigated
by Graham and coworkers~\cite{GrahamPRL03}, and discussed in Section
8.

\subsection {Reilly's density-dependent spin-gap model}

In three subsequent papers, Reilly {\it et al} develop a
phenomenological model for the $0.7$ plateau based on a
density-dependent spin-gap in the QPC~\cite{ReillyPRL02,
ReillyPRB05, ReillyPhysE06}. Figure~10(a) shows a schematic
illustrating the proposed model. Starting at low density, the 1D
subbands $E_{j}$ are spin-degenerate and unpopulated. The density
$n$ increases as the side-gate bias $V_{s}$ is made more positive,
and a crucial aspect of this model is the non-linear rise of the 1D
Fermi energy $E_{F}^{1D}$ with $n$ due to the 1D density of states
$\rho_{1D}(E) \sim E^{-1/2}$~\cite{DaviesBook}. As soon as a 1D
subband begins to populate, $E_{F}^{1D}$ stops rising briefly due to
the very high local density of states at the subband edge, and
continues to rise approximately parabolically in $V_{s}$ as the
subband fills. The idea behind Reilly's model is that as soon as a
given 1D subband begins to populate, an energy gap $\Delta
E^{\uparrow \downarrow}$ starts to open between the spin-up
$E^{\uparrow}$ and spin-down $E^{\downarrow}$ branches of the
subband. This spin-gap is linearly dependent on density $\Delta
E_{j}^{\uparrow \downarrow} = \gamma_{j} n$, with a different
$\gamma_{j}$ for the $j$th 1D subband. These gap opening rates
$\gamma_{j}$ are the only free parameters in the model, and for
determining the behaviour below $G_{0}$, only $\gamma_{j}$ is
relevant (n.b., Ref.~\cite{ReillyPRB05} discusses $\gamma_{1}$ and
$\gamma_{2}$ however these are two different values for the opening
rate of $\Delta E_{1}^{\uparrow \downarrow}$). A helpful alternate
picture is shown in Fig.~10(b), where the schematic is altered such
that the spin branch energies $E^{\uparrow}$ and $E^{\downarrow}$
are plotted relative to $E_{F}^{1D}$ (horizontal dashed
line)~\cite{ReillyPhysE06}. The spin-up branch is only completely
empty when it is above $E_{F}^{1D}$ by much more than $k_{B}T$,
which may explain the why the $0.7$ plateau weakens with decreasing
$T$. Note also that the `floating' of the spin-up subband above the
chemical potential in Fig.~10(b) in some ways resembles a pinning of
the spin-up subband edge to $\mu_{s}$ (see Section 8.4).

This points to a key conceptual difference between the BCF and
Reilly models. In the BCF model the first subband is spin-degenerate
on passing $\mu_{s}$ and remains so until the subband reaches
$\mu_{d}$ where the anomalous subband edge (this would be spin-up in
this case) pins at $\mu_{d}$ for a brief span in $V_{g}$ (or
$V_{S}$) such that a gap opens and then closes again. In the Reilly
model the first subband is spin-degenerate on passing $\mu_{s}$
where it immediately opens and keeps opening~\cite{ReillyPhysE06},
with the relative trends of the 1D density of states and spin-up
subband giving the appearance that the spin-up subband briefly pins
above $\mu_{s}$.

\begin{figure}
\includegraphics[width=12cm]{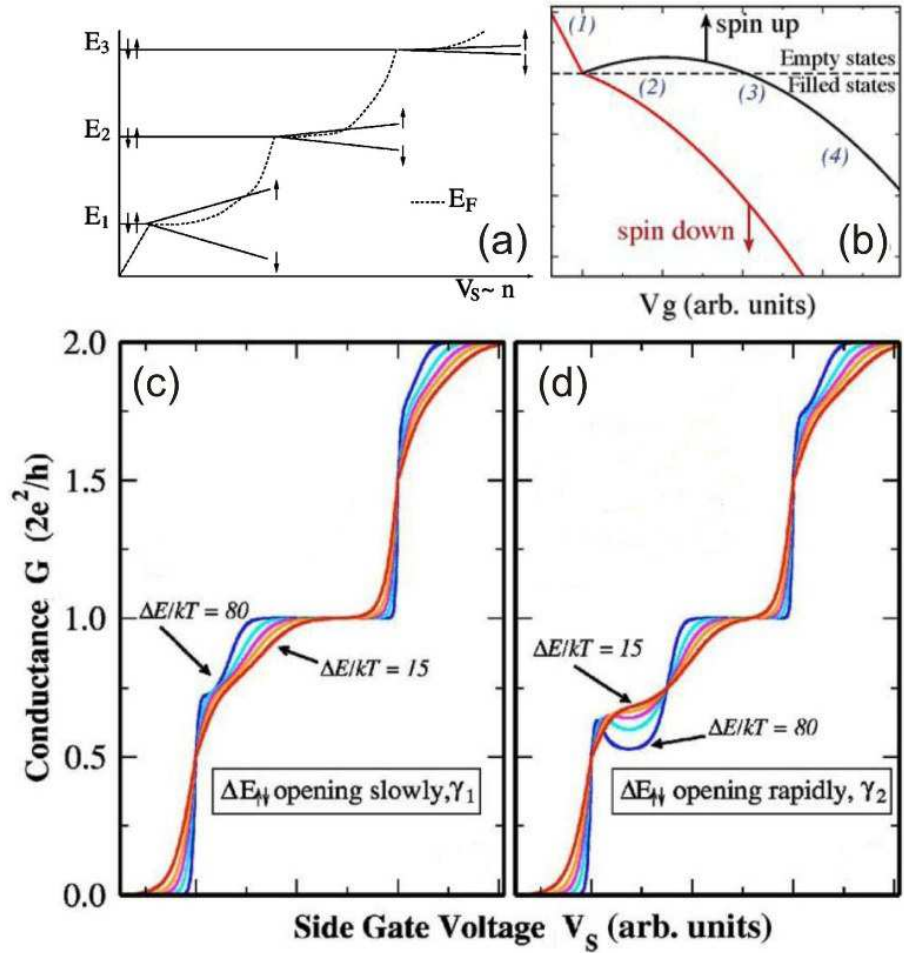}
\caption{(a) Schematic illustrating the density-dependent spin-gap
model, with the energies of the spin branches of the first three 1D
subbands (solid lines), along with the 1D Fermi energy $E_{F}$ vs
side-gate bias $V_{S}$. (b) An alternate view of the model for the
first 1D subband, with the energy of the subband edge for each
spin-branch plotted relative to the Fermi energy (dashed horizontal
line). Calculated conductance $G$ vs side-gate voltage $V_{S}$ at a
range of ratios of the 1D subband spacing $\Delta E$ to the thermal
broadening $k_{B}T$ for (c) a small and (d) a large spin-gap opening
parameter $\gamma$. Figure (a) adapted with permission from
Ref.~\cite{ReillyPRL02}. Copyright 2002 by the American Physical
Society. Figure (b) adapted from Ref.~\cite{ReillyPhysE06} with
permission from Elsevier. Figure (c/d) adapted with permission from
Ref.~\cite{ReillyPRB05}. Copyright 2005 by the American Physical
Society.}
\end{figure}

The fact that the spin-gap does not begin opening until the subband
edge reaches the chemical potential resolves two predictions of the
BCF model~\cite{BruusPhysE01} that are not observed experimentally.
These are the appearance of plateaus at {\it both} $0.7G_{0}$ and
$0.5G_{0}$, and the appearance of finite bias plateaus at $0.25
G_{0}$ and $1.25G_{0}$. The $0.25G_{0}$ and $1.25G_{0}$ plateaus are
taken care of by the spin-gap being too small at the point where
they would normally be resolved~\cite{ReillyPRL02}. Whether a
plateau is observed at $0.5G_{0}$ or $\sim 0.7G_{0}$ depends on
$\gamma$, and as we see below, both cannot occur together. If
$\Delta E_{1}^{\uparrow \downarrow}
>> k_{B}T$ then $E_{1}^{\uparrow}$ rises much faster than
$E_{F}^{1D}$ at first, staying more than $k_{B}T$ above
$E_{F}^{1D}$, until $E_{F}^{1D}$ catches up and $E_{1}^{\uparrow}$
rapidly populates. The spin-down branch on the other hand becomes
fully populated immediately, since $E_{1}^{\downarrow}$ drops below
$E_{1}$ as soon as the first subband begins to populate. The result
is plateaus at $0.5G_{0}$ and $G_{0}$ and not at $0.7G_{0}$. As
$k_{B}T$ becomes comparable to $\Delta E_{1}^{\uparrow \downarrow}$,
either because the spin-gap has decreased or the temperature has
been raised, the rising $E_{F}^{1D}$ can keep up with
$E_{1}^{\uparrow}$ such that the number of electrons thermally
excited to the spin-up branch remains roughly constant from the
moment the spin-gap opens. In this case the conductance is always
greater than $0.5G_{0}$ and a plateau is observed somewhere between
$0.5G_{0}$ and $G_{0}$, but not necessarily at $0.7G_{0}$,
consistent with experimental findings. It is also possible that no
plateau is observed below $G_{0}$, which occurs if $\Delta
E_{1}^{\uparrow \downarrow} << k_{B}T$.

\begin{figure}
\includegraphics[width=12cm]{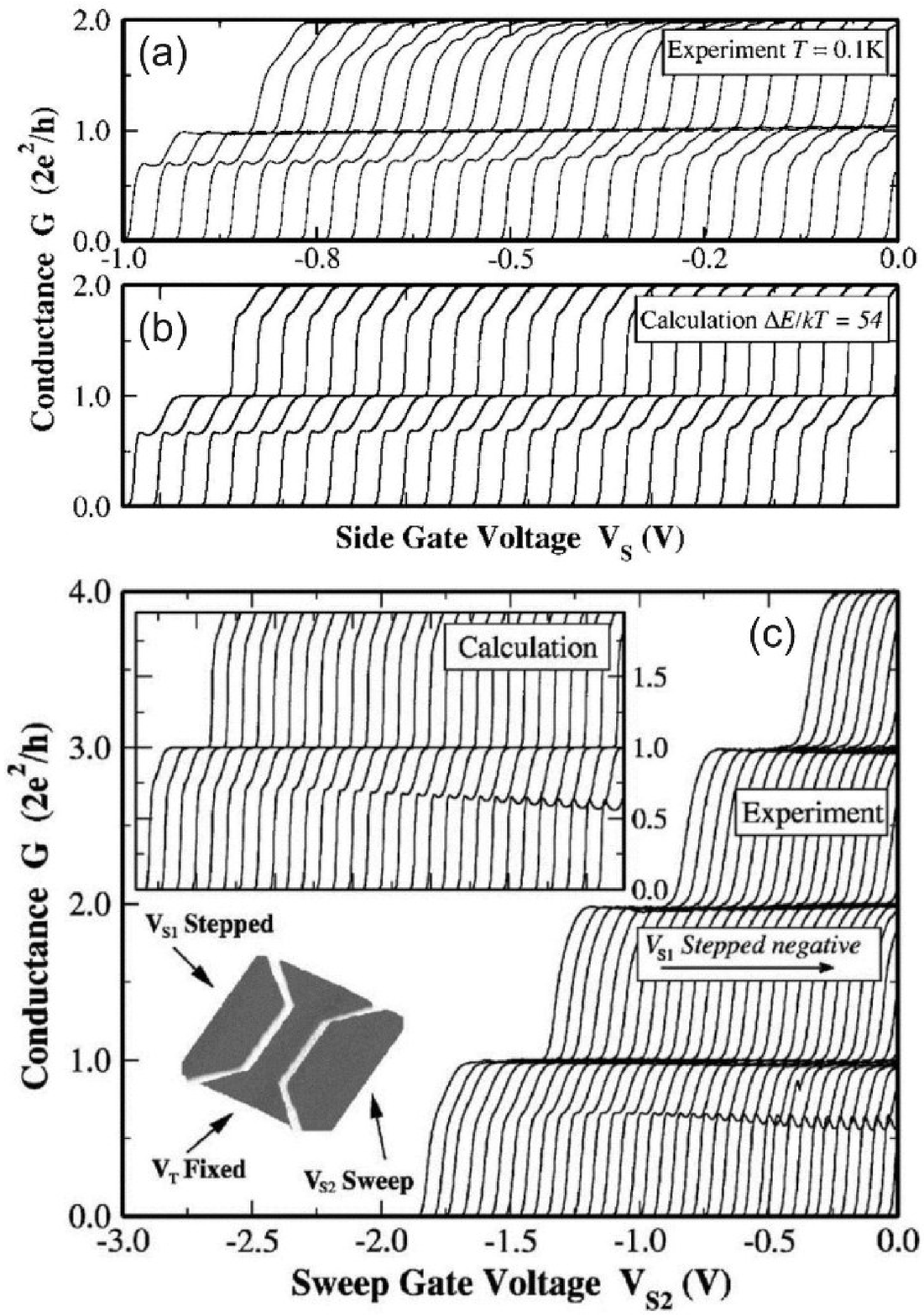}
\caption{Conductance $G$ vs side-gate voltage $V_{S}$ (a) obtained
experimentally for a doped-cap induced QPC with $l = 1~\mu$m at $T =
100$~mK for 2D densities $n$ from $2 \times 10^{11}$ (right) to $4.6
\times 10^{11}$~cm$^{-2}$ (left), and (b) calculated using the
density-dependent spin-gap model with the spin-gap opening parameter
$\gamma$ increasing gradually from right to left and $\Delta
E/k_{B}T = 54$. (c) $G$ vs bias applied to one side-gate $V_{S2}$ as
the other side-gate bias $V_{S1}$ is stepped with $V_{T}$ held
constant for a $0.5~\mu$m doped-cap induced QPC. The 2D density is
$n = 5 \times 10^{11}$~cm$^{-2}$. Corresponding spin-gap model
calculations are shown inset to (c) with $\gamma$ increasing from
right to left and $\Delta E/k_{B}T = 54$. The data in (c) suggests
that $\gamma$, the only free parameter in the spin-gap model, is
connected to the potential mismatch between the 1D region and
adjacent 2D reservoirs. Figure adapted with permission from
Ref.~\cite{ReillyPRB05}. Copyright 2005 by the American Physical
Society.}
\end{figure}

Figures 10 and 11 show some key comparisons between experimental
data and numerical predictions based on the density-dependent
spin-gap model. We start by looking at the predicted $G$ versus
$V_{S}$ obtained for a spin-gap that opens slowly (Fig.~10(c)) and
rapidly (Fig.~10(d)) for different ratios $\Delta E/k_{B}T$ of the
1D subband spacing to the thermal broadening. With the spin-gap
opening slowly, the inflection at $0.7G_{0}$ is weak at higher
temperatures (red trace) and strengthens markedly as $T$ is reduced
(blue trace), with the riser from $0.7G_{0}$ up to the $G_{0}$
plateau becoming steeper. The trends in Fig.~10(c) compare well with
what is seen experimentally, for example, in Fig.~3(a). In contrast,
when the spin-gap opens rapidly, $G$ rises up from zero to $0.7
G_{0}$, drops into a minima at $0.5G_{0}$ and then rises again to
the plateau at $G_{0}$. The reason for this is evident in
Fig.~10(b). As soon as the first subband populates, the spin-up
branch is within $k_{B}T$ of the Fermi energy. However, the spin-gap
opens quickly enough that the $1 \uparrow$ subband edge can rise
above $E_{F}~+~\sim k_{B}T$, shutting down the thermal excitation of
carriers to the spin-up branch for a short period before the Fermi
energy catches up and populates the spin-up branch, as is clear in
Fig.~10(a). Figure~11(a) shows $G$ versus $V_{S}$ obtained from a
$1~\mu$m long 1D channel at a range of $V_{T}$ corresponding to
increasing density in the 2D reservoirs adjacent to the device
moving from right to left (n.b., the density in the 1D region is not
solely a function of $V_{T}$). Corresponding calculations are shown
in Fig.~11(b), obtained by increasing $\gamma$, and clear agreement
with experimental data is evident. Reilly's model also reproduces
both the experimental evolution of the $0.7$ plateau with in-plane
magnetic field, and the 1D conductance with dc source-drain
bias~\cite{ReillyPRB05}.

However, one of the more important findings of
Ref.~\cite{ReillyPRB05} comes from comparing model calculations to
new experimental data obtained using doped-cap induced
QPCs~\cite{KaneAPL98}. The schematic in the lower left corner of
Fig.~11(c) illustrates the experiment. The density in the 2D
reservoirs either side of the 1D channel are held constant by fixing
$V_{T}$. Meanwhile $G$ is measured as a function of the bias on one
of the side-gates $V_{S2}$, while the other gate bias $V_{S1}$ is
stepped, changing the shape of the electrostatic confinement
potential in the 1D channel. The result is that the evolution of the
$0.7$ plateau is studied as a function of the potential difference
between the 1D channel and 2D reservoirs. This occurs because the 1D
channel has a saddle-point potential~\cite{ButtikerPRB90}, and
making $V_{S1}$ more negative to narrow the channel causes the
saddle-point to rise relative to the potential `floor' in the 2D
reservoirs. The experimental data is shown in the main panel of
Fig.~11(c), with the 2D-1D potential mismatch growing as $V_{S1}$
becomes more negative (i.e., moving from left to right). The
corresponding calculations are presented in the inset in the
upper-right corner of Fig.~11(c). This data is strikingly similar to
that obtained by altering the density with $V_{T}$ (c.f. Fig.
~11(a)), suggesting that $\gamma$ is linked to the 2D-1D potential
mismatch. It is interesting to note data by Lindelof and
Aagesen~\cite{LindelofJPCM08} showing the appearance of
Fabry-P\'{e}rot type diamond patterns for small $V_{sd}$ in
source-drain bias greyscales similar to that shown in Fig.~6(b).
Lindelof and Aagesen attribute these patterns to electron
localization due to non-adiabatic variation in the potential at the
QPCs causing reflections~\cite{TekmanPRB89} (i.e., 2D-1D potential
mismatch), which lends further support to the argument in
Ref.~\cite{ReillyPRB05}. The 2D-1D mismatch also offers an
explanation for why various experiments~\cite{ThomasPRB98,
ReillyPRB01, ThomasPRB00, NuttinckJJAP00, PyshkinPRB00} gave
conflicting results regarding the evolution of the $0.7$ plateau
with 2D density, since the means used to change the 2D density in
each case also influences the 1D electrostatic potential. This
differs from device to device based on gate layout and the
heterostructure used.

This is a purely phenomenological model with no connection to the
microscopic properties of the system, as with the BCF model. However
the model proposed by Reilly {\it et al} fits well with the idea of
the emergence of spontaneous spin-polarization within the QPC, as
proposed by Thomas {\it et al}~\cite{ThomasPRL96}, and
Reilly~\cite{ReillyPRB05} suggested that it may also be consistent
with a Kondo-like mechanism above the Kondo temperature $T_{K}$, in
which case the Kondo temperature would be related to $\gamma$.

\section{The possibility of Kondo physics in QPCs}

Experiments directly investigating the possibility of Kondo physics
in QPCs were performed by Cronenwett {\it et
al}~\cite{CronenwettPRL02}. Before discussing these measurements,
however, I will first divert to briefly review the basics of Kondo
Physics. An excellent basic review of Kondo physics can also be
found in a recent article by Kouwenhoven and
Glazman~\cite{KouwenhovenPW01}.

\subsection{The basics of Kondo physics}

The resistivity of a metal decreases as the temperature is lowered,
due to reduced scattering of electrons by phonons, reaching either a
saturated finite resistivity or a resistance of zero depending on
whether or not the metal enters a superconducting ground state.
However, if a small number of magnetic impurities are added, for
example $\sim 10^{-3} \%$ Fe in Cu~\cite{StarPLA69}, the resistivity
with decreasing temperature will reach a minimum at $T \sim 10$~K
and begin rising again. This rise in the low temperature resistance
is a many-body effect involving interactions between the sea of
electrons and the localized spin on the magnetic impurity. This is
known as the Kondo effect after Jun Kondo, who first explained this
phenomena in 1964~\cite{KondoPTP64}. Although the Kondo effect in
metals has been understood for many years, it has recently become
the focus of renewed attention due to the ability to produce an
analogous system using quantum dots, as investigated theoretically
by Hershfield, Davies and Wilkins~\cite{HershfieldPRL91} and Meir,
Wingreen and Lee~\cite{MeirPRL93}, and first realized experimentally
by Goldhaber-Gordon {\it et al}~\cite{GoldhaberGordonNat98}.

\begin{figure}
\includegraphics[width=12cm]{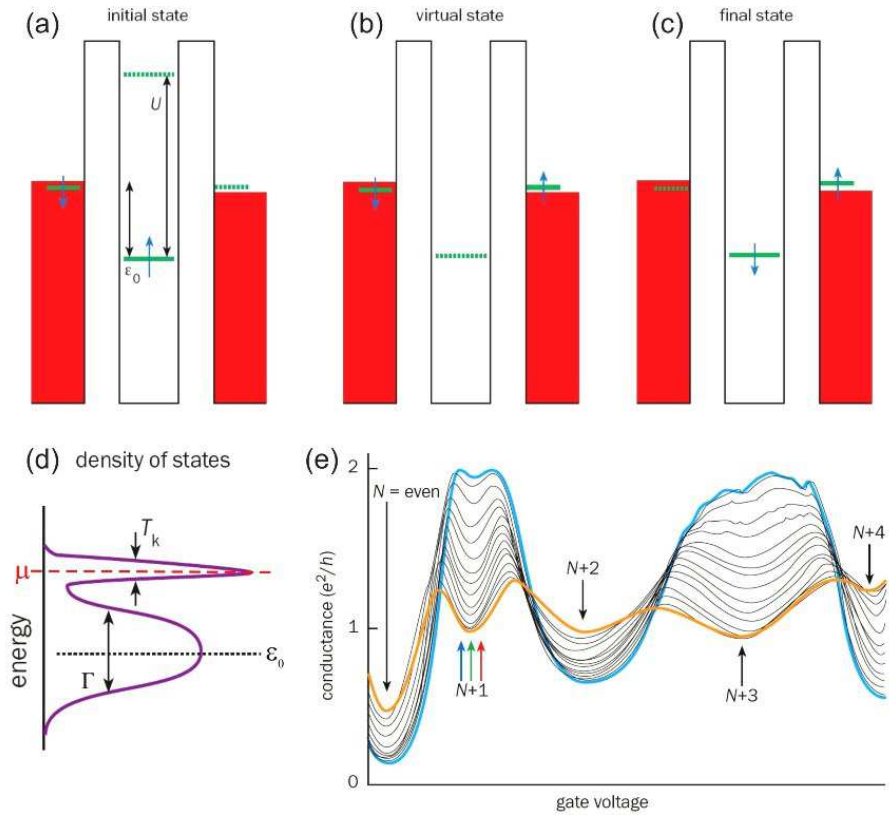}
\caption{(a)-(c) Schematics outlining the mechanism for the Kondo
effect in quantum dots. The dot has single particle energy levels
separated in energy by $\Delta E$, with the highest occupied level
$\epsilon_{0}$ below the chemical potential shown occupied by a
spin-down electron in (a). In this state, an electron in the source
(left reservoir) cannot be added to the dot, as this will raise the
level in energy by the Coulomb energy $U = e^{2}/2C$ to above the
chemical potential. Hence current cannot pass and the quantum dot is
`Coulomb blockaded'. In the Kondo mechanism, a virtual state can
arise for a short time where the spin-down electron moves to the
drain, as shown in (b). This virtual state can end in one of two
ways, the system returns to the initial state in (a), or a spin-up
electron in the source occupies the dot, as shown in (c). The latter
results in current passing through the dot, and an enhancement in
the conductance at Coulomb blockade minima where the dot contains an
odd number of electrons. (d) The quantum dot density of states vs
energy. The Kondo effect adds a second density of states peak (upper
peak) of width $k_{B}T_{K}$ centered at the chemical potential. This
is in addition to the usual density of states peaks that coincide
the dot's single-particle levels, which have a width $\Gamma$
related to the transparency of the tunnel barriers connecting the
quantum dot to the reservoirs either side. (e) Experimental data
showing the Kondo enhancement in the conductance as the temperature
is reduced when there is an odd number of electrons on the dot.
Figure adapted with permission from Ref.~\cite{KouwenhovenPW01}.
Copyright 2001, IOP Publishing.}
\end{figure}

The Kondo effect in quantum dots occurs via a higher-order,
spin-flip tunneling process mediated by a short-lived virtual state,
as outlined schematically in Fig.~12(a-c)~\cite{KouwenhovenPW01,
CronenwettSci98}. A quantum dot contains discrete energy levels
separated by $\Delta E$, with the highest occupied level sitting
$\epsilon_{0}$ beneath the chemical potential $\mu_{L} \simeq
\mu_{R}$ of the 2D reservoirs either side of the dot (note:
occupied/unoccupied levels are shown as solid/dashed green lines and
the red bars indicate the quasi-continuum of states in the source
(left) and drain (right) leads; only the highest occupied/unoccupied
level in the dot is shown in Figs.~12(a)-(c)). In this
configuration, current through the dot via first-order tunneling
processes is zero, and the dot is `Coulomb
blockaded'~\cite{MeiravPRL90}. This occurs for two reasons: Firstly
transport by the addition of a second electron to the highest
occupied level is prevented by the Coulomb repulsion energy $U =
e^{2}/C$, where $C$ is the capacitance of the dot. This is
illustrated in Fig.~12(a): the singly occupied level that sits
$\epsilon_{0}$ below the source and drain potentials would rise by
$U$ (to the green dashed line inside the dot) in the process of
becoming doubly occupied. This incurs an energy cost $U -
\epsilon_{0}$ that cannot be met unless $k_{B}T$ exceeds $U -
\epsilon_{0}$, thereby preventing the addition of an electron.
Secondly, the highest occupied level cannot be vacated if it is more
than $k_{B}T$ below the chemical potential of the leads due to the
unavailability of an empty state to tunnel into. Thus transmission
through the dot would only normally occur when the singly occupied
level aligns with the leads, which is achieved by adjusting the bias
applied to a gate that is capacitively coupled to the dot. When this
occurs the electron can leave the occupied level by resonant
tunnelling, causing the level to fall by $U$ to where it can be
reoccupied from the source or drain without an energy cost. With a
source-drain bias $V_{sd}$ applied, this results in a net current
when a dot level sits between the source and drain chemical
potentials.

However, the uncertainty principle provides a way around this second
cause for the blockade -- the electron on the dot can `borrow'
enough energy to tunnel into the right-hand reservoir, taking the
virtual state shown in Fig.~12(b), providing that this state does
not exist for a time exceeding $\hbar / |\epsilon_{0}|$. Destruction
of the virtual state can occur via either a return to the original
configuration in Fig.~12(a), or by an electron in the left reservoir
dropping down into the dot as shown in Fig.~12(c). Both processes
occur with equal probability, all that matters is that the `borrowed
energy' $\epsilon_{0}$ is returned, not where that energy comes
from. The first process produces no effective change, however, the
net result of the latter process is the transfer of an electron from
one side of the dot to the other along with a spin-flip of the
electron occupying the singly-occupied state within the dot. Note
that this process can thus only occur when the dot has a non-zero
net spin (i.e., the number of electrons on the dot $N$ is odd). This
virtual tunneling process occurs frequently with a short timescale,
and many such occurrences lead to an additional peak in the density
of states as shown in Fig.~12(d). This `Kondo peak' is centered at
the chemical potential of the leads, and sits above the peak in the
density of states corresponding to the highest occupied dot level
$\epsilon_{0}$ below. The width of the Kondo peak is determined by
the Kondo temperature $T_{K}$, which depends on $\epsilon_{0}$, $U$
and the life-time broadening of the dot state $\Gamma = \Gamma_{L} +
\Gamma_{R}$, where $\Gamma_{L}$ and $\Gamma_{R}$ are controlled by
the transparency of the tunnel barriers. The relationship between
these four parameters, the level spacing $\Delta E$ and the
temperature $T$ is vital to observing the Kondo effect in quantum
dots. The aim is to maximize $\Gamma$, since this brings $T_{K} \sim
\sqrt{U\Gamma} $exp$[-\pi(\mu - \epsilon_{0}) /
2\Gamma]$~\cite{BickersRMP87} to an experimentally accessible
temperature, but since $\Delta E > \Gamma
>> k_{B}T$ is required to observe Coulomb blockade, it means that very
small quantum dots are required~\cite{GoldhaberGordonNat98}.

The Kondo peak in the density of states leads to a number of
observable effects on the transport through the dot. The first is
clear by comparing the temperature dependence of the Coulomb
blockade minima for even and odd $N$, as shown in Fig.~12(e). As the
temperature is raised from $25$~mK (blue) to $1$~K (yellow) the even
minima rise in conductance due to thermal broadening, as expected.
In contrast, the odd minima drop in conductance as the temperature
rises towards $T_{K}$, reducing the contribution of the virtual
state process to conduction. This odd-even behaviour highlights the
need for a single localized spin for the Kondo process to occur. A
remarkable aspect of the Kondo effect in metal films is that the
ratio of the resistance $R$ to the zero temperature resistance
$R_{0}$ depends only on the ratio $T/T_{K}$~\cite{RizzutoRPP74,
BellCP75}. A similar scaling behaviour is observed in quantum dots:

\begin{equation}
G(T)/G_{0} = [1+(2^{1/s}-1)(T/T_{K})^{2}]^{-s}
\end{equation}

\noindent where $s = 0.22$ for
spin-$\frac{1}{2}$~\cite{CostiJPCM94}. This causes the $G(T)$ data
measured in the vicinity of odd $N$ to condense onto a single curve
when plotted as $G/G_{0}$ versus
$T/T_{K}$~\cite{GoldhaberGordonPRL98}. Note that Eq.~5 gives $G(T =
T_{K}) = 0.5G_{0}$. In quantum dots, another hallmark of the Kondo
effect is the zero-bias peak, a maximum in the differential
conductance versus source-drain bias centered at zero
bias~\cite{GoldhaberGordonNat98, CronenwettSci98}. This feature can
be entirely explained by reference to the density of states. Unlike
the dot energy levels, which can move up and down relative to $\mu
\simeq \mu_{L}, \mu_{R}$, the Kondo peak in the density of states is
always fixed at $\mu$, becoming two peaks centered at $\mu_{L}$ and
$\mu_{R}$ as $eV_{SD}$ exceeds $k_{B}T_{K}$~\cite{MeirPRL93} (i.e.,
half the density of states is connected to the Kondo mechanism
transferring electrons from left to right and vice versa).
Increasing $V_{SD}$ reduces the overlap between these two density of
states peaks, the conductance decreases accordingly. Finally, this
zero bias peak undergoes Zeeman splitting for an applied in-plane
magnetic field $B$, dividing into two peaks centered at finite
$V_{SD}$ separated by $2g^{*}\mu_{B}B/e$, where $\mu_{B}$ is the
Bohr magneton. It is notable that the splitting is {\it twice} that
expected for Zeeman splitting of a localized state alone, this being
a particular signature of Kondo physics in a quantum
dot~\cite{GoldhaberGordonNat98, MeirPRL93}.

The explanation for the unique $2g^{*}\mu_{B}B$ splitting is
relatively simple: At zero field the Kondo peak in the density of
states is spin degenerate, so for every spin-up/down state on the
source side there is an immediately accessible spin-down/up state on
the drain side. When a magnetic field is applied, the Kondo density
of states peak associated with source and drain sides undergoes a
Zeeman splitting of $g^{*}\mu_{B}B$ with the spin-up (spin-down)
states rising (lowering) by $\frac{1}{2}g^{*}\mu_{B}B$ relative to
$\mu_{s}$ or $\mu_{d}$. An immediate question is then why does the
differential conductance have a minimum at $V_{sd}$ if the Kondo
density of states peaks at the source and drain sides still overlap?
The reason for the suppression is that the coinciding Kondo peaks in
the source and drain now have the {\it same} spin, while the peaks
of opposite spin on either side are well separated in energy. Thus,
because the Kondo process requires a spin-flip, it is strongly
suppressed, giving a zero-bias minimum. The Kondo-enhanced
conductance can be restored by bringing the spin-up peak on one side
(i.e., source or drain) back into coincidence with the spin-down
peak on the other, and this can be achieved by applying a
source-drain bias $V_{sd} = \pm g^{*}\mu_{B}B/e$, which leads to two
peaks separated in $V_{sd}$ by the $2g^{*}\mu_{B}B$ characteristic
of the Kondo effect, one conductance peak corresponding to the
spin-up Kondo peak in the source coinciding with the spin-down Kondo
peak in the drain, and the other from the reverse situation.

\subsection{Possible Kondo physics in a QPC}

Cronenwett {\it et al}~\cite{CronenwettPRL02} performed similar
studies of a QPC for $G < 3G_{0}$, obtaining data that is strikingly
similar to that from quantum dots in the Kondo
regime~\cite{GoldhaberGordonNat98, CronenwettSci98,
GoldhaberGordonPRL98}. There are four key observations in this work
suggestive of a Kondo-like many-body state formed by an unpaired
spin localized in the vicinity of the QPC. The first appears in
Fig.~13(a), where a clear zero bias peak is observed at low
temperature for $G < G_{0}$. The zero bias peak, commonly known in
studies of the $0.7$ plateau in QPCs as the zero-bias anomaly (ZBA),
is destroyed by both an increase in temperature (see Fig.~13(b) for
data at $T \sim 600$~mK) and by large in-plane magnetic field (see
Fig.~13(c))~\cite{CronenwettPRL02}. Given the correspondence between
the evolution of the $0.7$ plateau and ZBA with temperature, the
authors argue that the $0.7$ plateau arises as the Kondo effect acts
to enhance the conductance from $0.5G_{0}$ to its unitary limit
(i.e., $G_{0}$). Note that this contrasts with the BCF model for
example, where the conductance is instead suppressed below $G_{0}$.
It is also worth noting that a weaker zero bias peak-like feature
appears for $1.5G_{0} < G < 2G_{0}$ in the same data. This second
peak is not addressed by Cronenwett {\it et al}, but shows similar
behaviour with increased $T$ and $B$.

The second key observation is related to the splitting of the ZBA
due to an in-plane magnetic field, as shown in Fig.~13(c). The rate
of this splitting is an important characteristic of the Kondo
process, and should be twice that normally
expected~\cite{MeirPRL93}, as observed in quantum
dots~\cite{GoldhaberGordonNat98, CronenwettSci98}, for reasons
explained at the end of Section 4.1. It is evident in Fig.~13(c)
that the splitting of the ZBA at different $G < G_{0}$ is not
uniform and is clearest at $G \sim 0.6 - 0.7G_{0}$. Cronenwett {\it
et al}~\cite{CronenwettPRL02} measure the $g$-factor $g^{*}_{ZBA}$
for this peak at $G \sim 0.6 - 0.7G_{0}$ and $B \sim 3$~T by
assuming that the splitting goes as $2g^{*}_{ZBA}\mu_{B}B$, and
obtain a $g$-factor of approximately $1.5$ times the magnitude of
$0.44$ found in bulk GaAs (i.e., $|g^{*}_{ZBA}| \approxeq 0.66$).
However, $g^{*} \approxeq 0.66$ is not a well established value for
the $g$-factor in a QPC, which is known to vary with subband
index~\cite{ThomasPRL96}, and may also vary with density and
confinement potential. It is difficult, with this information alone,
to argue convincingly that the prefactor for the splitting of the
ZBA is really the factor of two that is distinctive of the Kondo
effect~\cite{MeirPRL93}. However, it is possible to obtain an
independent estimate the $g$-factor $g^{*}_{1D}$ for the lowest 1D
subband using source-drain bias spectroscopy~\cite{PatelPRB91a}.
This value $g^{*}_{1D}$ can be obtained from the same device under
similar density and confinement potential conditions, and one should
obtain a ZBA energy splitting $\Delta E_{ZBA}(B) =
xg^{*}_{1D}\mu_{B}B$ where the prefactor $x = 2$ for a ZBA that is
caused by Kondo physics~\cite{KlochanPRL11}. Measurements of
$g^{*}_{1D}$ for the same device were presented by Cronenwett in
Ref.~\cite{CronenwettPhD01} using two approaches. The first was to
measure the splitting of the transconductance peaks corresponding to
the risers up to the $0.5G_{0}$ and $G_{0}$ plateaus as a function
of in-plane magnetic field, with data obtained in steps of $0.5$~T.
A 1D $g$-factor $g^{*}_{1D} = 1.12$ was obtained at $T = 80$~mK,
which fell to $0.68$ at $670$~mK, where the ZBA is suppressed, and
remained at approximately this value at higher temperatures (see
Fig.~6-9(c) of Ref.~\cite{CronenwettPhD01}). Taking the value of
$|g^{*}_{ZBA}| \approxeq 0.66$ obtained~\cite{CronenwettPRL02} for
the Zeeman split ZBA at $G \sim 0.6 - 0.7G_{0}$ and $B = 3$~T in
Fig.~13(c) gives $\Delta E_{ZBA}(B) = 0.23$~meV,~\footnote{A direct
measurement of the ZBA splitting at $G ~ 0.6 - 0.7G_{0}$ and $B =
3$~T in Fig.~13(c) suggests that $\Delta E_{ZBA}$ is closer to
$0.195$~meV, in which case $x = 0.996$, $1.468$ and $0.426$ are
obtained for $|g^{*}_{1D}| = 1.12$, $0.76$ and $2.62$,
respectively.} which corresponds to $x = 1.18$ for $g^{*}_{1D} =
1.12$. Two additional measurements of $|g^{*}_{1D}|$ were made by
reference to a source-drain bias transconductance colour map
obtained at $B = 8$~T (see Fig.~6-10 of
Ref.~\cite{CronenwettPhD01}), which gives $|g^{*}_{1D}| \approxeq
0.76$ at low $V_{sd}$ and $|g^{*}_{1D}| \approxeq 2.62$ at high
$V_{sd}$, corresponding to ZBA splitting prefactors of $x = 1.74$
and $x = 0.5$, respectively. Thus, while it is clear that the ZBA
splits with an in-plane magnetic field~\cite{CronenwettPRL02}, the
argument that this splitting has the $2g^{*}\mu_{B}B$ dependence
distinctive of Kondo physics is not so conclusive. Note also that
the splitting of the ZBA is absent for higher $G \rightarrow G_{0}$,
where the ZBA is instead suppressed with increasing $B$.  This
behaviour is unexpected given that this is where $T_{K}$ is at its
highest (see Fig.~13(e)).

\begin{figure}
\includegraphics[width=16cm]{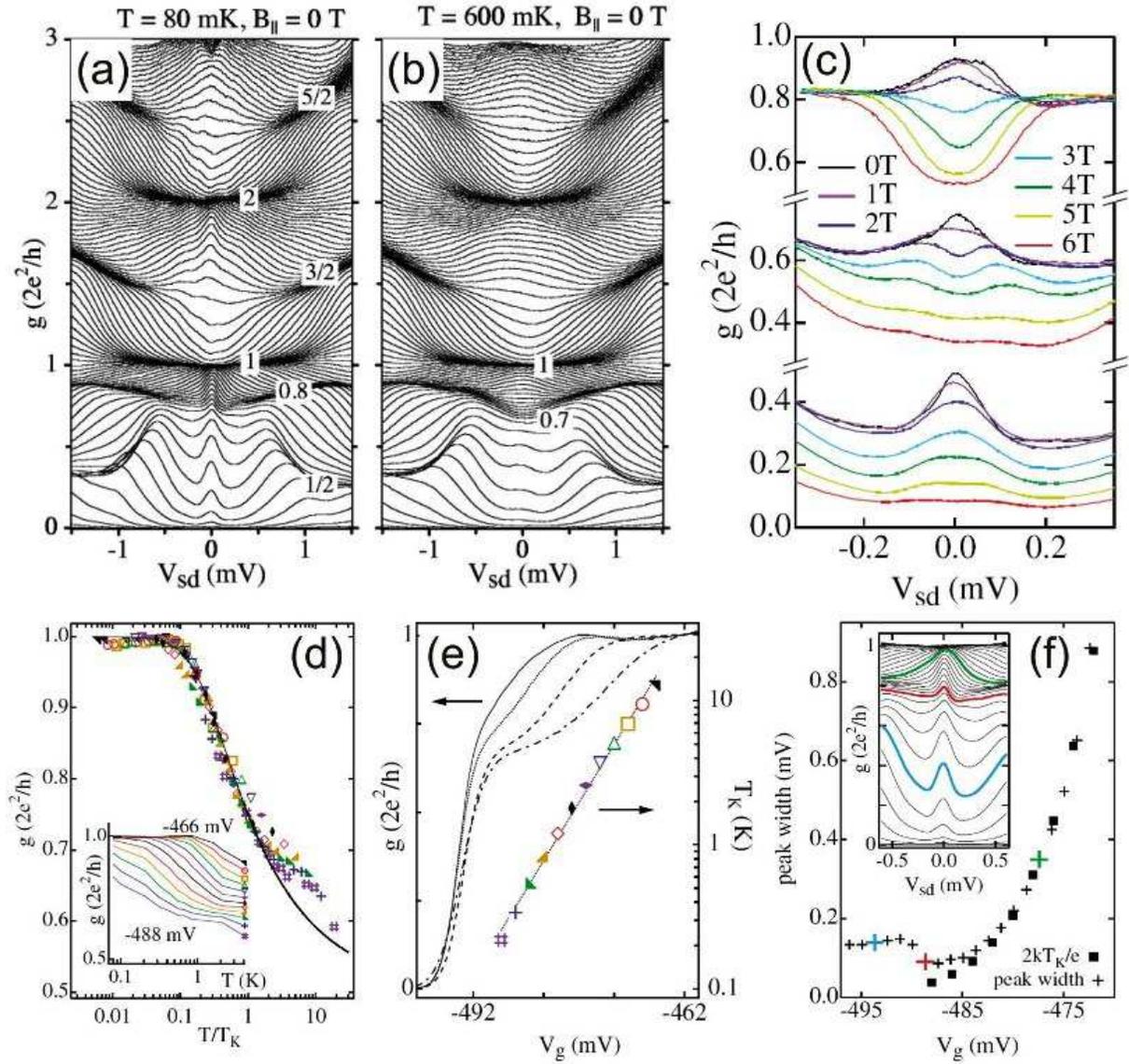}
\caption{Differential conductance $g$ vs source-drain bias $V_{sd}$
at various gate voltages $V_{g}$ at temperatures (a) $T = 80$~mK and
(b) $T = 600$~mK. (c) Selected $g$ vs $V_{sd}$ traces showing the
evolution of the zero-bias peak with in-plane magnetic field
$B_{\parallel}$. (d) A plot of $g$ vs scaled temperature $T/T_{K}$,
where the various symbols correspond to the $g$ vs $T$ traces
obtained for $-485$~mV $< V_{g} < -465$~mV shown in the inset. The
solid black line is a fit of Eq.~6 to the data. (e) Plots of $g$
(left axis) vs $V_{g}$ for $T = 80$~mK (solid line), $210$~mK
(dotted line), $560$~mK (dashed line) and $1.6$~K (dot-dashed line)
and the $T_{K}$ values (right axis) obtained from fits of Eq.~6 to
the $g(T)$ data obtained at various $V_{g}$. (f) The zero-bias peak
full-width at half-maximum (crosses) vs $V_{g}$ obtained from the
$g$ vs $V_{sd}$ data shown in the inset. The black squares show the
equivalent Kondo bias voltage $V^{K}_{sd} = k_{B}T_{K}/e$ obtained
using the data in (e). The conductance $g$ is referred to as $G$ in
the text. Figure adapted with permission from
Ref.~\cite{CronenwettPRL02}. Copyright 2002 by the American Physical
Society.}
\end{figure}

Motivated by the first observation, the third and fourth
observations relate to the universal scaling behaviour expected in
the Kondo regime~\cite{GoldhaberGordonPRL98}, namely that the
temperature dependent conductance at various locations where the
conductance is enhanced by the Kondo effect all follow a single
curve when the temperature is scaled by the Kondo temperature (i.e.,
when $G$ is plotted against $T/T_{K}$ rather than $T$). The inset to
Fig.~13(d) shows the differential conductance versus temperature for
twelve different gate voltages spanning the width of the $0.7$
plateau ($V_{g} = -488$~mV is the side closest to the riser down to
$G = 0$ and $V_{g} = -466$~mV is the side closest to the riser up to
the $G_{0}$ plateau). In each case, $G$ rises as $T$ is reduced, as
expected, saturating at the unitary limit $G_{0}$ for lower $V_{g}$.
For each gate voltage, a plot of conductance versus temperature can
be made, with an example containing data for four different gate
voltages shown in Fig.~14(a). The modified Kondo form in Eq.~6 below
can be fit to this data to extract estimates of $T_{K}$ for each
$V_{g}$~\cite{CronenwettPhD01}, which are then plotted in
Fig.~13(e). Note that a $V_{g}$-dependent $T_{K}$ is not unusual, it
is also observed in quantum dots, where ln($T_{K}$) has a quadratic
dependence on gate voltage~\cite{vanderWielSci00}. Cronenwett {\it
et al} found that $T_{K}$ depends exponentially on gate voltage (see
Fig.~13(e)), increasing from $\sim 0.2$~K at $V_{g} = -488$~mV to
$\sim 10.5$~K at $V_{g} = -466$~mV~\cite{CronenwettPRL02}. The
reduced $T_{K}$ at more negative $V_{g}$ suggests that the ZBA
should weaken at lower conductance. Although this is evident in
Fig.~2(a) of Ref.~\cite{CronenwettPRL02} and the inset to
Fig.~13(f), a wider survey of the literature suggests that this
might not be so straightforward. As we discuss further in Section
7.3, the ZBA is observed at conductances as low as $10^{-4}
G_{0}$~\cite{SarkozyPRB09, RenPRB10}, and if the trend in Fig.~13(d)
is universal, then $T_{K}$ should be inaccessibly small for $G
\lesssim 0.05G_{0}$ using existing experimental techniques.

\begin{figure}
\includegraphics[width=6cm]{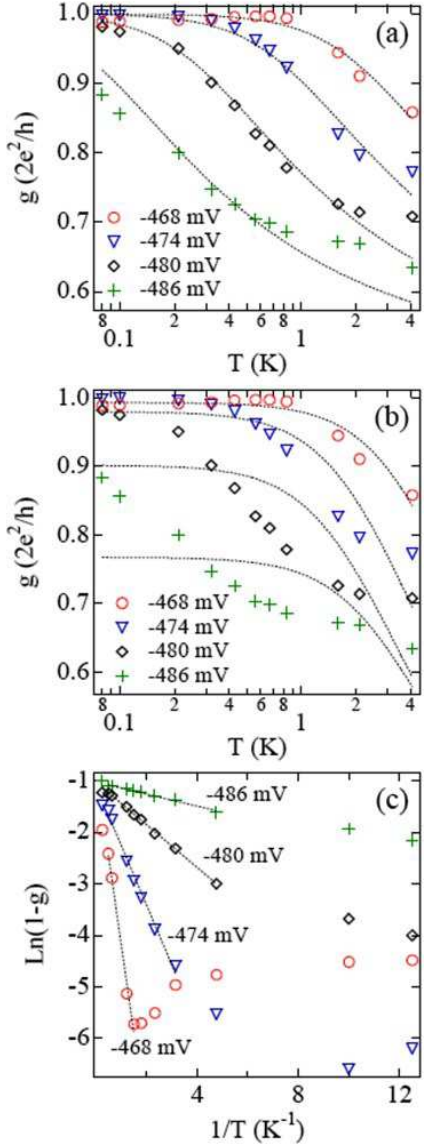}
\caption{Plots of $g$ vs $T$ with corresponding fits of (a) the
modified Kondo model (Eq.~6) and (b) the quantum dot Kondo model
(Eq.~5) for four different $V_{g}$. (c) A plot of ln $g$ vs $1/T$
with corresponding fits of the Arrhenius model (Eq.~2) for the same
four $V_{g}$ in (a) and (b). The experimental data is exactly the
same in these three plots; only the axis scaling and fits differ.
The conductance $g$ is referred to as $G$ in the text. Figure
adapted with permission from Ref.~\cite{CronenwettPhD01}.}
\end{figure}

Returning to Kondo scaling of the data, the main panel of Fig.~13(d)
shows the conductance plotted against scaled temperature $T/T_{K}$
for the data obtained at various $V_{G}$ and shown inset to
Fig.~13(d). The solid line in Fig.~13(d) is the best fit of a
modified expression for the Kondo conductance, given by:

\begin{equation}
G(T)/G_{0} = \frac{1}{2}[1+(2^{1/s}-1)(T/T_{K})^{2}]^{-s} +
\frac{1}{2}
\end{equation}

\noindent This expression differs from the expression for quantum
dots~\cite{GoldhaberGordonPRL98} -- it is half the term in Eq.~5
added to a constant of $0.5G_{0}$. In other words, whereas in
quantum dots the Kondo effect enhances the conductance from $0$ to
$G_{0}$, the fit in Fig.~13(d) implies that the Kondo effect in QPCs
instead enhances the conductance from $0.5G_{0}$ to $G_{0}$. The
choice of this modified form for the Kondo conductance is purely
empirical~\cite{CronenwettPRL02}. However, subsequent work by Meir
{\it et al}~\cite{MeirPRL02} proposed a mechanism whereby two
valence fluctuation channels contribute to the conductance via a
quasibound state formed within the QPC due to multiple reflections
from the entrance and exit of the QPC. The first channel involves
fluctuations between occupancies of 0 and 1 on the quasi-bound
state. The contribution from this process to $G$ is expected to
saturate to $0.5G_{0}$. The second channel involving fluctuations
between 1 and 2 should contribute much less at higher temperatures,
but drives the conductance towards the unitary limit of $G_{0}$ in
the limit where $T$ goes to zero. This is more in keeping with the
traditional picture of the quantum dot Kondo
effect~\cite{GoldhaberGordonNat98, CronenwettSci98, vanderWielSci00,
MeirPRL93} and contrasts with the model proposed by
Lindelof~\cite{LindelofSPIE01, LindelofJPCM08}. Returning to the
scaling behaviour in Ref.~\cite{CronenwettPRL02}, here $G(T = T_{K})
= \frac{3}{4}G_{0}$, in contrast to $G(T = T_{K}) = 0.5G_{0}$ for
quantum dots. The fourth observation is that the full width at half
maximum (FWHM) of the zero-bias peak is almost equal to
$2k_{B}T_{K}/e$ using the $V_{g}$-dependent $T_{K}$ values extracted
from fits of Eq.~6 to the $G(T)$ data in the limit where the QPC is
relatively wide open ($V_{g} > -485$~mV), as shown in Fig.~13(f).
This relationship between $T_{K}$ and the ZBA FWHM is consistent
with findings for quantum dots in carbon
nanotubes~\cite{NygardNat00}. In contrast, measurements in GaAs
quantum dots have shown that the FWHM may not provide completely
accurate estimates for $T_{K}$~\cite{vanderWielSci00}.

\begin{figure}
\includegraphics[width=12cm]{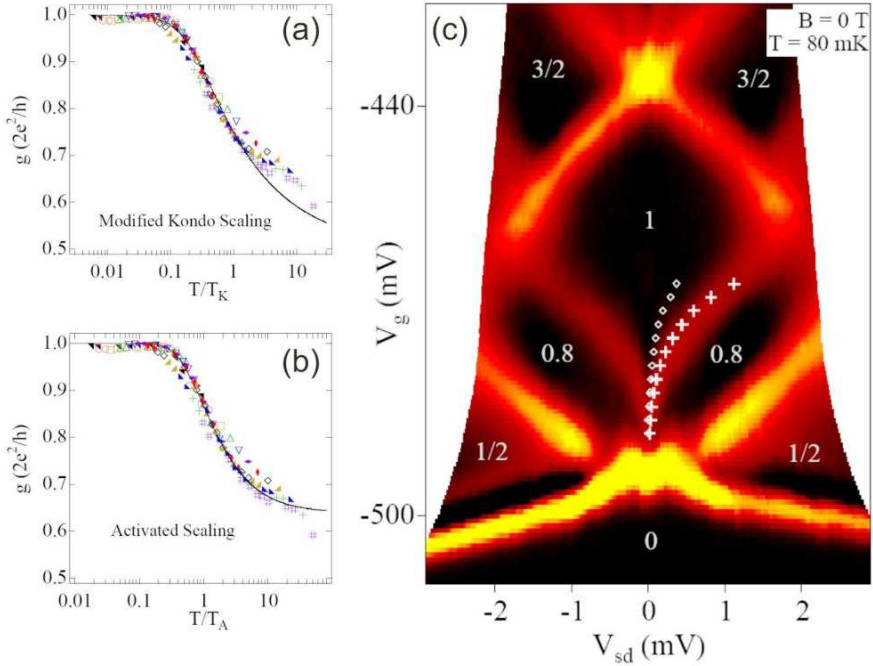}
\caption{(a) Plots of $g$ vs $T/T_{K}$ for the ten different $V_{g}$
studied, with a fit of Eq.~6 (solid line). (b) $g$ vs $T/T_{A}$ for
the ten different $V_{g}$ studied, with a fit of Eq.~2 (solid line).
The experimental data is exactly the same in these two plots; only
the axis scaling and fits differ. (c) Mapping of the equivalent
Kondo bias $V^{K}_{sd} = k_{B}T_{K}/e$ (crosses) and equivalent
Arrhenius bias $V^{A}_{sd} = k_{B}T_{A}/e$ (diamonds) superimposed
on a source-drain bias colour-map obtained at $T = 60$~mK. The
conductance $g$ is referred to as $G$ in the text. Figure adapted
with permission from Ref.~\cite{CronenwettPhD01}.}
\end{figure}

Cronenwett {\it et al}~\cite{CronenwettPRL02} convert the $T_{K}$
values they obtain into equivalent Kondo bias voltages $V^{K}_{sd} =
k_{B}T_{K}/e$, which they then plot superimposed on a source-drain
bias colourmap, as shown in Fig.~15(c), with the points closely
tracking the transconductance maximum between the $G_{0}$ plateau
and $0.8G_{0}$ finite-bias plateau. The similarity to the thermal
activation analysis by Kristensen {\it et al}~\cite{KristensenPRB00}
is remarkable (see Fig.~8(d) also), and much more insight can be
gained by undertaking a comparative analysis of the thermal
activation and Kondo models on a common data set. Although not
presented in Ref.~\cite{CronenwettPRL02}, exactly such a study was
performed by Cronenwett {\it et al} and presented in Cronenwett's
Ph.D. dissertation~\cite{CronenwettPhD01}. Figure~14(a-c) presents
fits of the modified Kondo (Eq.~6), quantum dot Kondo (Eq.~5) and
Arrhenius (Eq.~2) models to the same experimental data set. Aside
from where the QPC is widely opened (i.e., most positive $V_{g}$),
the quantum dot Kondo model is a poor fit to the data. The modified
Kondo model is a better fit, and at its best for more positive
$V_{g}$. It is also a better fit in the lower $T$ limit in each
case. In contrast, the Arrhenius model fits are at their best in the
high $T$ limit, becoming poor below $250$~mK, which interestingly,
is well below the $T \sim 1$~K where the $0.7$ feature is at its
strongest. The better quality of the Arrhenius fit in the higher
temperature limit is also apparent in comparing Figs.~15(a) and (b),
where the conductance is plotted against the scaled temperatures,
$T/T_{K}$ and $T/T_{A}$, respectively. Interestingly, for each
$V_{g}$, the best fit $T_{K} \approx 3T_{A}$. This offers the
possibility that there is more to this problem than it simply being
`either Kondo or Arrhenius' -- the data may indicate a possible
transition in the dominant mechanism at intermediate $T$, with a
transition temperature that scales with $G$, and potentially some
direct link between the two mechanisms. Indeed, the data in
Fig.~15(c) is immediately suggestive of an experiment that may
provide insight to this. In Fig.~15(c), Cronenwett plots {\it both}
$V^{K}_{sd}$ and $V^{A}_{sd} = k_{B}T_{A}/e$ onto a single
source-drain bias colour-map. As Cronenwett {\it et al} point out,
the equivalent Kondo bias data tracks the dark line in the
colour-map better than the equivalent Arrhenius bias data, but the
colour-map was obtained at $T = 60$~mK, where this would be
expected, and it would be interesting to see the two equivalent
biases plotted over a colour-map obtained at higher $T$ (perhaps
using the data from Fig.~13(b)) for comparison.

While Ref.~\cite{CronenwettPRL02} makes a strong case in favour of a
possible Kondo mechanism, it is clearly not exactly the same as that
in quantum dots (Eq.~5), and some aspects of the behaviour observed
by Cronenwett {\it et al}~\cite{CronenwettPRL02, CronenwettPhD01}
are yet to be fully understood (e.g., why the Zeeman splitting of
the ZBA does not give the prefactor of two expected for Kondo
physics). Furthermore, the behaviour reported by Cronenwett {\it et
al}~\cite{CronenwettPRL02, CronenwettPhD01} is not universally
observed in QPCs; we will return to this in Section 7. Another
important question is raised by these experiments: Does a
bound-state form in a QPC for $G < G_{0}$, and if so what is the
nature of this state? We will return to address this in detail in
Section 6. Before that, we embark on a brief discussion of five
other important experimental clues regarding the $0.7$ plateau.

\section{Five more clues: noise, thermopower, compressibility, holes and scanning gate microscopy}

\subsection{Shot-noise measurements of QPCs at $G < G_{0}$}

The granularity of electrons should lead to tiny fluctuations in the
electrical current through a device. This is known as shot noise and
was first observed in vacuum diodes~\cite{SchottkyAP18}. Shot noise
can be significant in nanoscale devices, and provide important
information beyond that obtained with conductance measurements
alone. For an in-depth discussion of shot noise in mesoscopic
conductors, see a recent review by Blanter and
B\"{u}ttiker~\cite{BlanterPhysRep00}.

Initial studies of shot noise in quantum point contacts by Reznikov
{\it et al}~\cite{ReznikovPRL95} and Kumar {\it et
al}~\cite{KumarPRL96} showed a suppression of the shot noise
coinciding with plateaus at $nG_{0}$ in the 1D conductance. The
non-equilibrium shot noise $S$ in a mesoscopic conductor is
determined by the sum of the transmission probabilities for the
conduction channels. The shot noise only takes its Poisson value
$S_{P} = 2e\langle I\rangle$, where $e$ is the electron charge and
$\langle I\rangle$ is the time-averaged current, in the limit where
the transparency is low for all channels through the
sample~\cite{BlanterPhysRep00}. At zero temperature, the Fano factor
$F$ is the ratio of the actual shot noise to the Poissonian value $F
= S/S_{P}$ and takes values between zero, where all channels have
perfect transmission, and one, where all channels have no
transmission (i.e., Poissonian noise).

Roche {\it et al}~\cite{RochePRL04} were first to undertake shot
noise measurements focussed directly on the $0.7$ plateau. Because
measurements are performed at finite temperature, there are
additional thermal contributions to the shot noise that are
difficult to fully account for. Thus the measurements presented by
Roche {\it et al}~\cite{RochePRL04} represent the upper bound
$F^{+}$ for the actual Fano factor $F$. Figure~16(a) shows
measurements of $F^{+}$ versus $G$ for in-plane magnetic fields
$B_{\parallel} = 0$, $3$ and $8$~T. Also shown is the expected
behaviour for two cases. The first behaviour is observed when there
is no spin splitting, and is that $F^{+} = 1$ at $G = 0$, with
$F^{+}$ falling linearly to zero at $G_{0}$ (upper diagonal). The
second behaviour occurs for complete spin splitting, here $F^{+}$
falls linearly from $F^{+} = 1$ at $G = 0$ to $F^{+} = 0$ at $0.5
G_{0}$ (lower diagonal), before rising to a new maximum at $F^{+}
\sim 0.2$ at $\sim 0.75G_{0}$ (`the hump') and ultimately reaching
$F^{+} = 0$ again at $G_{0}$. The measured zero-field data (solid
circles) in Fig.~16(a) drops rapidly below the upper diagonal, until
it reaches the hump at $G \sim 0.7G_{0}$, where there is a shallow
minimum before the measured $F^{+}$ follows the hump until it
reaches $F^{+} = 0$ at $G_{0}$. The fact that $F^{+}$ drops below
the upper diagonal in Fig.~16(a) shows that the two channels, one
spin-up and one spin-down, do not have the same transmission at the
$0.7$ plateau~\cite{RochePRL04}. Turning now to the data at
$B_{\parallel} = 3$ (open squares) and $8$~T (solid triangles),
these follow a similar form except that the $F^{+}$ minimum moves
closer to both $G = 0.5G_{0}$ and $F^{+} = 0$. This evolution with
increasing $B_{\parallel}$ strongly suggests that the two channels
may have different spin orientations~\cite{RochePRL04}. However, the
fact that the zero field data follows the spin-split form rather
than the spin-degenerate form implies that there is a finite
separation between the spin-up and spin-down components of the first
1D subband. Subsequent studies by Nakamura {\it et
al}~\cite{NakamuraPRB09} have shown a link between the depression in
the Fano factor at $G \sim 0.7G_{0}$ and the presence/absence of a
corresponding plateau in the $G$ versus $V_{g}$ data. Nakamura {\it
et al} also found that these features coincide with the observation
of a zero-bias anomaly in the source-drain bias characteristics at
$G < G_{0}$.

It is also interesting to note studies by Shailos {\it et
al}~\cite{ShailosJPCM06b} that aimed to mimic the situation of
transport via two spin-polarised channels indicated by the noise
experiments performed by Roche {\it et al}~\cite{RochePRL04}. This
was achieved by applying a strong magnetic field perpendicular to
the 2DEG to drive the device into the quantum Hall regime, and
tuning the gate voltage such that the QPC passes one spin-resolved
quantum Hall edge state directly and the second by tunneling across
the QPC. A clear $0.7$ plateau was observed for weaker magnetic
fields but not stronger magnetic fields, leading to the conclusion
that a small spin-gap relative to $k_{b}T$ is required for a $0.7$
plateau to develop, as predicted by Reilly {\it et
al}~\cite{ReillyPRL02}. Note that care is needed in considering this
latter result as the physics in the quantum Hall regime is different
to that at zero perpendicular magnetic field where the $0.7$ plateau
is normally observed/studied. However, it provides some
corroborating evidence for the two-channel picture proposed in
Ref.~\cite{RochePRL04}.

\begin{figure}
\includegraphics[width=16cm]{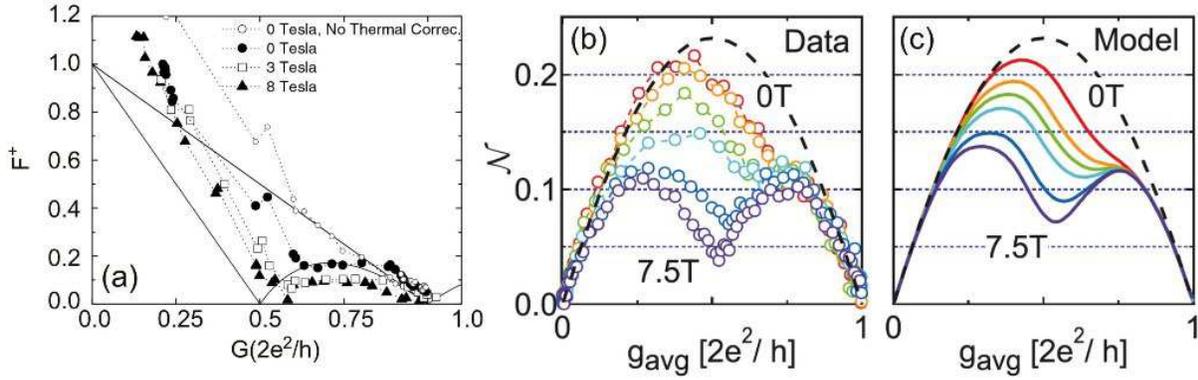}
\caption{(a) The upper bound of the Fano factor $F^{+}$ from shot
noise measurements vs conductance $G$. The two solid lines show the
expected behaviour for no spin-splitting (upper diagonal heading
directly from $F^{+} = 1$ at $G = 0$ to $F^{+} = 0$ at $G = G_{0}$)
and full spin-splitting (lower diagonal and semicircular hump to the
right of it). Data presented for three different in-plane magnetic
fields $B_{\parallel} = 0$ (open/closed symbols without/with
correction for thermal contribution to shot noise), $3$ and $8$~T.
(b) The measured noise factor $N$ and (c) calculated $N$ values
based on a density-dependent spin-gap model plotted vs the average
differential conductance $g_{avg}$ for $B_{\parallel} = 0$ (red),
$2$ (orange), $3$ (green), $4$ (cyan), $6$ (blue) and $7.5$~T
(purple). In both cases the dashed black lines indicate expected
behaviour in the absence of spin-splitting. Figure (a) adapted with
permission from Ref.~\cite{RochePRL04}. Copyright 2004 by the
American Physical Society. Figures (b,c) adapted with permission
from Ref.~\cite{DiCarloPRL06}. Copyright 2006 by the American
Physical Society.}
\end{figure}

Subsequent measurements of the shot noise as a function of
conductance, in-plane magnetic field and source-drain bias by
DiCarlo {\it et al}~\cite{DiCarloPRL06} also show interesting
behaviour in the vicinity of the $0.7$ plateau. DiCarlo {\it et al}
focus on the noise factor $N$, which is obtained from a fit of the
form: $S^{P}_{I}(V_{sd}) =
2G_{0}N[eV_{sd}$coth$(eV_{sd}/2k_{B}T_{e}) - 2k_{B}T_{e}]$ where
$T_{e}$ is the electron temperature and $S^{P}_{I}$ is the measured
partition noise, obtained from the total QPC current noise spectral
density $S_{I}$ as $S^{P}_{I}(V_{sd}) = S_{I}(V_{sd}) -
4k_{B}T_{e}G$. The noise factor $N$ relates $S^{P}_{I}$ to $V_{sd}$,
whereas the Fano factor relates $S^{P}_{I}$ to $I$ via
$S^{P}_{I}(V_{sd}) = 2eI~$coth$(eV_{sd}/2k_{B}T_{e})F(0)$ where
$F(0)$ is the Fano factor averaged over a range $k_{B}T_{e}$ around
zero energy, providing $F$ doesn't vary too rapidly with energy and
the explored energy range is less than a few $k_{B}T$. Figure~16(b)
shows the measurements of $N$ versus $G$ at various $B_{\parallel}$
between $0$ and $7.5$~T. For spin-degenerate transmission, the
expected behaviour is that $N$ becomes zero at $nG_{0}$, rising to a
maxima of $N = 1/4$ at odd multiples of $0.5G_{0}$. In contrast for
fully spin-split transmission, $N = 0$ occurs at all multiples of
$0.5G_{0}$ with maxima of $N = 1/8$ at odd multiples of $0.25G_{0}$.
The spin-degenerate case is indicated by the dashed line in
Fig.~16(b). At $B_{\parallel} = 0$, the experimental data follow the
expected behaviour until $G \sim 0.45G_{0}$, where $N$ falls below
the dashed line until $G \sim 0.8G_{0}$. For $G \sim 0.8G_{0}$ the
data intercepts the dashed line again and tracks it to $N = 0$ at
$G_{0}$. Similar behaviour is observed for $G_{0} < G < 2G_{0}$. An
inflection appears at $G \sim 0.7G_{0}$ at $B_{\parallel} = 0$,
which develops into a minimum that deepens and moves towards $G =
0.5G_{0}$ as $B_{\parallel}$ is increased. The data at
$B_{\parallel} = 7.5$~T is close to that expected for full
spin-splitting except that the noise factor at $G = 0.5G_{0}$ is not
quite zero. Figure~16(c) shows model calculations obtained using a
density-dependent spin-gap model~\cite{ReillyPRL02, ReillyPRB05},
with transmission coefficients calculated for a saddle-point QPC
potential~\cite{DiCarloPRL06}. The model curves are an excellent
match to the experimental data, both at $B_{\parallel} = 0$, and as
$B_{\parallel}$ is initially increased. However, at $B_{\parallel} =
7.5$~T, the minima is not as deep as that in the experimental data,
and the maxima to either side are not symmetric. DiCarlo {\it et
al}~\cite{DiCarloPRL06} found that this is connected to the
magnitude of the $g$-factor used. If they increase $|g^{*}|$ to
$\sim 0.6$, the model regains the minimum at $0.5G_{0}$ and becomes
symmetric. This slight increase in $|g^{*}|$ is consistent with
reports of $g$-factor enhancement in 1D systems~\cite{ThomasPRL96},
and the estimate obtained by Cronenwett {\it et al} based on Zeeman
splitting of the ZBA~\cite{CronenwettPRL02}. A suppression of the
shot noise is also predicted for the BCF model in
Ref.~\cite{BruusArXiv00}.

We conclude this section by noting theoretical calculations by Golub
{\it et al}~\cite{GolubPRL06} that reproduce the essential features
of the data in both Figs.~16(a) and (b). The agreement with the data
from DiCarlo {\it et al}~\cite{DiCarloPRL06} is striking, however,
it is worth noting that enhancement of the $g$-factor was not
required for Golub {\it et al} to obtain agreement with the data;
they used the standard bulk GaAs value of $0.44$. These calculations
were based on an extended Anderson model for a QPC, previously used
to link the $0.7$ plateau to Kondo physics~\cite{MeirPRL02}. As
Golub {\it et al} point out, the noise experiments were conducted
outside the Kondo regime, and so the data above does not imply that
the Kondo effect and the structure appearing near $G = 0.7G_{0}$ are
related~\cite{GolubPRL06}. Golub {\it et al} predict that for
temperatures and source-drain voltages (i.e., $eV_{sd}/k_{B}$) less
than the Kondo temperature, the dip in $N$ in Fig.~16(b) at $\sim
0.5 - 0.7G_{0}$ will vanish at $B_{\parallel} = 0$. This would be an
interesting aspect for future studies, with work having recently
commenced on noise studies of quantum dots in the Kondo
regime~\cite{DelattreNatPhys09}. Additionally, Lassl {\it et
al}~\cite{LasslPRB07} find the same behaviour for Hartree-Fock
calculations of the noise factor versus conductance as a function of
magnetic field. This suggests that the various spin-gap
models~\cite{BruusPhysE01,ReillyPRL02} and the Kondo
mechanism~\cite{MeirPRL02,GolubPRL06} for the $0.7$ plateau are
essentially experimentally indistinguishable as far as studies of
shot noise in QPCs are concerned, as noted in
Ref.~\cite{SfigakisArXiv09}.

\subsection{Thermopower measurements of QPCs}

A difference in temperature $\Delta T$ across a conductor leads to a
thermal current $Q$ that can flow in conjunction with an electrical
current $I$ driven by a voltage difference $\Delta V$. In the linear
response regime, these four quantities are linked by two equations:

\begin{equation}
-\Delta V = RI + S\Delta T
\end{equation}

\begin{equation}
Q = \Pi I - \kappa\Delta T
\end{equation}

\noindent where $R$ is the electrical resistance, $S$ is the
thermopower, $\Pi$ is the Peltier coefficient and $\kappa$ is the
thermal conductivity~\cite{MacDonaldBook62, DelvesRPP65}. The
Seebeck effect occurs when a current $I$ is unable to flow in Eq.~7.
Here the $\Delta V$ resulting from a finite $\Delta T$ is known as
the thermovoltage, and this along with $\kappa$ have been the focus
of studies by Appleyard {\it et al}~\cite{AppleyardPRB00} and
Chiatti {\it et al}~\cite{ChiattiPRL06}.

The Seebeck and Peltier effects are traditionally studied in
junctions between two different conductors, usually
metals~\cite{MacDonaldBook62}. However, a similar situation can be
realized in a mesoscopic system connected to reservoirs with a Fermi
sea of electrons~\cite{SivanPRB86, StredaJPCM89, ButcherJPCM90,
vanHoutenSST92}. Initial thermopower measurements of QPCs were
performed by Molenkamp {\it et al}~\cite{MolenkampPRL90}, who
observed oscillations in the thermovoltage with minima that aligned
to the 1D conductance plateaus. A very similar outcome was obtained
by Appleyard {\it et al}~\cite{AppleyardPRL98}, who used the device
structure discussed below to show that the heat loss by hot
electrons follows a $T^{5}$ dependence characteristic of acoustic
phonon emission. Subsequent work by Molenkamp {\it et
al}~\cite{MolenkampPRL92} showed similar oscillations in the Peltier
coefficient, as expected by the Kelvin-Onsager relation $\Pi =
ST$~\cite{DelvesRPP65}, as well as a direct proportionality between
the thermal and electrical conductivities consistent with the
Wiedemann-Franz relation $\kappa \approx L_{0}TG$, where $L_{0} =
(\pi k_{B})^{2}/(3e^{2})$ is the Lorenz number~\cite{FranzADP53}.

\begin{figure}
\includegraphics[width=16cm]{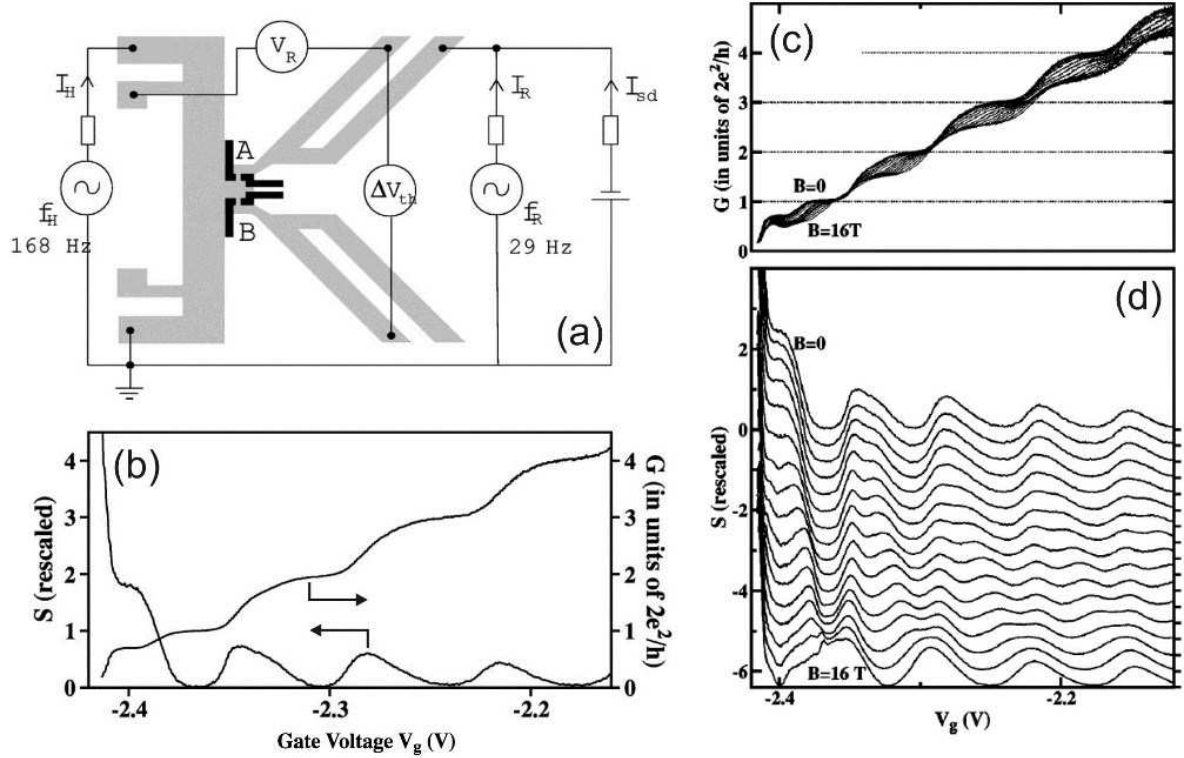}
\caption{(a) Schematic of the set-up used for thermopower
measurements by Appleyard {\it et al} and discussed in detail in the
text. (b) A plot of measured thermopower $S$ (left axis) and
conductance $G$ (right axis) vs gate-voltage $V_{g}$ showing that
minima in $S$ coincide with all plateaus except for the $0.7$
plateau, where a plateau at finite $S$ rather than a minima at $S =
0$ is observed. (c) Plot of $G$ vs $V_{g}$ for $B_{\parallel} = 0$
to $16$~T in steps of $1$~T for comparison to (d). At $B_{\parallel}
= 16$~T plateaus are observed at odd multiples of $0.5G_{0}$ and not
integer multiples of $G_{0}$ because the $n\uparrow$ and
$n+1\downarrow$ subbands cross close to this field. (d) Plot of $S$
vs $V_{g}$ for $B_{\parallel} = 0$ to $16$~T in steps of $1$~T with
traces offset for clarity, the respective zero levels for each trace
are indicated by the ticks on the right axis. The plateau in $S$
coinciding with the $0.7$ plateau at $B_{\parallel} = 0$ evolves
into a minimum coinciding with the $0.5$ plateau as $B_{\parallel}$
is increased. Figure adapted with permission from
Ref.~\cite{AppleyardPRB00}. Copyright 2000 by the American Physical
Society.}
\end{figure}

Before getting to the results of thermopower measurements near the
$0.7$ plateau, we will briefly look at the experimental method and
device structure used for these measurements. The essential aspects
are shown in Fig.~17(a), where a pair of QPCs A and B separate a
2DEG into three regions. On the left-hand side is a heating channel,
through which an ac current $I_{H}$ at frequency $f_{H}$ is passed.
This current heats the electrons to $\Delta T$ above the lattice
temperature $T_{l}$. The QPCs prevent hot electrons from reaching
the two right-hand side regions, which remain at $T_{l}$. The
thermovoltage $\Delta V_{th}$ is measured across the two QPCs in
series, and is dependent on the thermopowers $S_{A}$ and $S_{B}$ of
the two QPCs such that $\Delta V_{th} = (S_{A}-S_{B})\Delta T$. The
thermovoltage oscillates at $2f_{H}$ because the heating power goes
as $I_{H}^{2}$, and this allows thermal effects to be distinguished
from resistive effects. Simultaneous measurements of the thermopower
and conductance for QPC A can be obtained by holding QPC B fixed
while varying only QPC A, and measuring the voltage $V_{R}$ across
QPC A due to an ac current through it at a different frequency
$f_{R}$. As mentioned above, Molenkamp {\it et
al}~\cite{MolenkampPRL90} and Appleyard {\it et
al}~\cite{AppleyardPRL98} both found that the thermovoltage $-\Delta
V$ has minima that coincide with the 1D conductance plateaus. The
maxima in between become progressively higher as the 1D subband
index $n$ is reduced. The Mott relation links the thermopower to the
energy derivative of the conductance via $S =
-\frac{\pi^{2}k_{B}^{2}}{3e}\frac{\partial(\textrm{ln}
G)}{\partial\mu}\Delta T$, where $\mu$ is the chemical potential of
the source and drain relative to the 1D subbands. As such, for
constant $\Delta T$, the measured thermovoltage should go as
$-\Delta V_{th}(V_{g}) \sim d($ln$ G)/dV_{g}$. A plot of $d($ln$
G)/dV_{g}$ versus $V_{g}$ shows remarkably similar maxima and minima
to $-\Delta V$, demonstrating correspondence to the Mott
relation~\cite{AppleyardPRL98}. For more information on the Mott
relation and how it arises from the original work by Cutler and
Mott~\cite{CutlerPR69}, see an excellent discussion recently
published by Lunde and Flensberg~\cite{LundeJPCM05}.

\begin{figure}
\includegraphics[width=7cm]{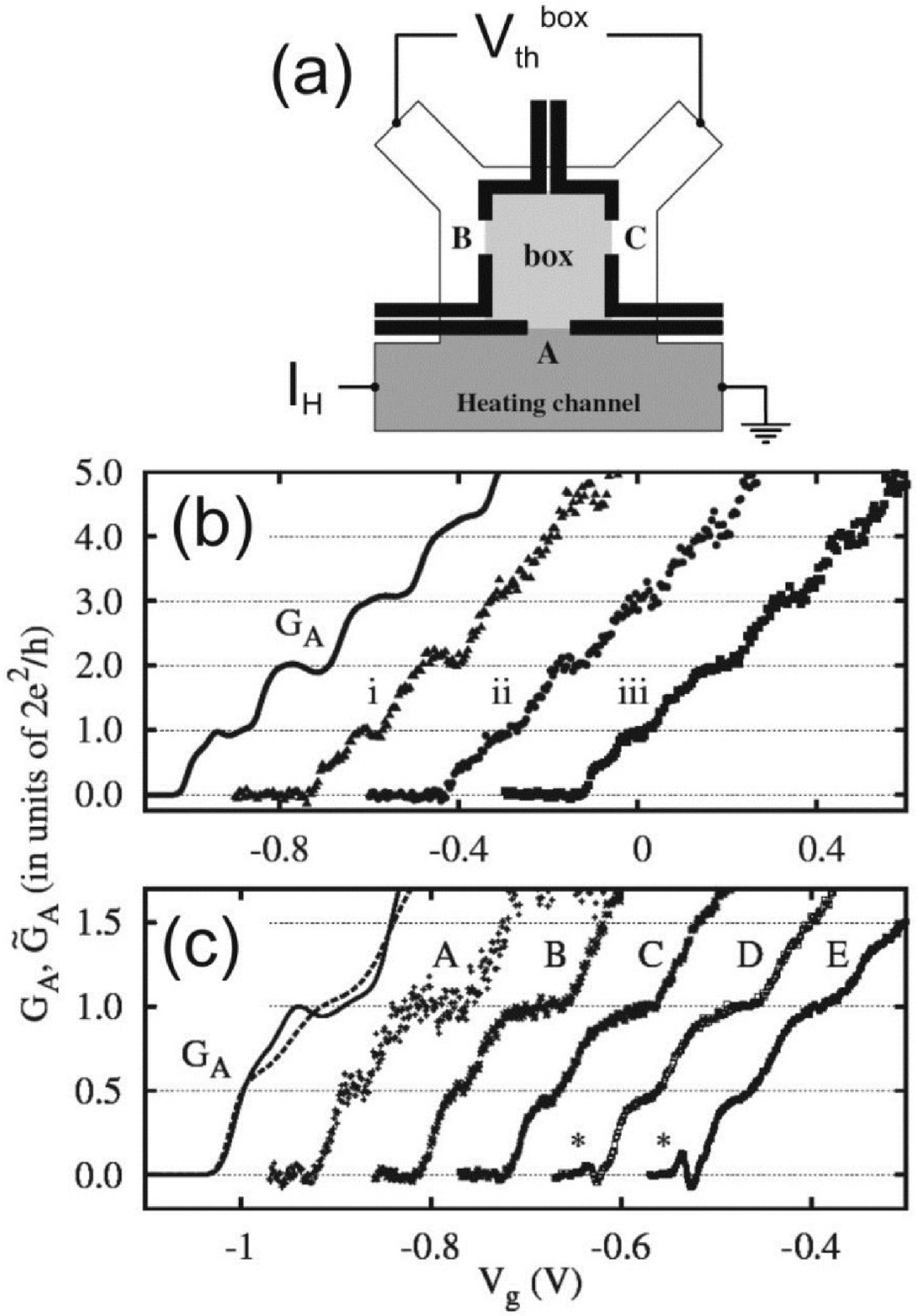}
\caption{(a) A schematic illustrating the modifications made to the
device structure in Fig.~17(a) for the thermal conductance
measurements by Chiatti {\it et al}~\cite{ChiattiPRL06} The most
notable change is the addition of QPC A to form a box between the
heating channel and the leads where the thermovoltage is measured.
(b) Plot of the electrical conductance $G_{A}$ (left trace) and
equivalent thermal conductance $\tilde{G}_{A}$ (remaining traces) vs
$V_{g}$. Traces sequentially offset horizontally by $0.3$~V for
clarity. Measurements of $\tilde{G}_{A}$ were obtained with $I_{H} =
1~\mu$A, $G_{B} = 0.5G_{0}$ and $G_{C} = 3G_{0}$, $4G_{0}$ and
$5G_{0}$ respectively for traces i-iii. (c) Close-up of $G_{A}$ and
$\tilde{G}_{A}$ vs $V_{g}$ near pinch-off. The far left traces are
$G_{A}$ obtained at lattice temperatures $T_{L}$ of $300$~mK (solid
line) and $1.2$~K (dashed line). Remaining traces, offset
sequentially by $0.1$~V for clarity, are $\tilde{G}_{A}$ at $T_{L} =
270$~mK with heating currents $I_{H} = 0.2$, $0.4$, $1$, $2$ and
$5~\mu$A for traces A-E respectively. The corresponding $\Delta T$
values are calculated to be $26$, $66$, $133$, $193$ and $234$~mK. A
and B were obtained with $G_{B} = 1.5G_{0}$ and $G_{C} = 3G_{0}$.
C-E were obtained with $G_{B} = 1.5G_{0}$ and $G_{C} = 2G_{0}$. The
small spikes in D and E indicated by * are a capacitive coupling
effect (i.e., are instrumental). Figure adapted with permission from
Ref.~\cite{ChiattiPRL06}. Copyright 2006 by the American Physical
Society.}
\end{figure}

In a subsequent paper, Appleyard {\it et al}~\cite{AppleyardPRB00}
focus more closely on the behaviour in the vicinity of the $0.7$
plateau, and in particular, whether a corresponding minimum in the
thermopower is observed. Figure~17(b) shows $S$ (left axis) and $G$
(right axis) versus $V_{g}$ for QPC A, where clear minima in $S$ are
observed coinciding with each integer plateau in $G$. The same does
not occur for the $0.7$ plateau at $V_{g} = -2.4$~V where a plateau
at finite $S$ is observed rather than a minima. Interestingly, there
is a minima in $d($ln$ G)/dV_{g}$ (see Fig.~3(c) of
Ref.~\cite{AppleyardPRB00}), which indicates a breakdown of the Mott
relation for the $0.7$ plateau. Appleyard {\it et
al}~\cite{AppleyardPRB00} attribute this breakdown to
electron-electron interactions since it is not observed for the
integer plateaus, which have single-particle origin. Delving further
into the behaviour of $S$ at the $0.7$ plateau, Figs.~17(c) and (d)
show $G$ versus $V_{g}$ and $S$ versus $V_{g}$, respectively, for
$B_{\parallel} = 0$ to $16$~T in steps of $1$~T. At $B_{\parallel} =
0$~T there are clear plateaus at $nG_{0}$ with a strong $0.7$
plateau present, whilst at $B_{\parallel} = 16$~T the plateaus
appear at odd-multiples of $0.5G_{0}$ indicating that the Zeeman
split 1D subbands cross at around this field. The $0.7$ plateau at
zero field evolves into a $0.5$ plateau at high field in agreement
with previous studies. The minima in $S$ corresponding to the
zero-field integer plateaus begin to split with increasing
$B_{\parallel}$ and evolve into minima coinciding with the plateaus
at odd integer multiples of $0.5G_{0}$ at $B_{\parallel} = 16$~T, as
expected. A corresponding minima in $S$ develops as the $0.7$
plateau moves to $0.5G_{0}$ with increasing $B_{\parallel}$.
Returning to the zero field result, Appleyard {\it et al} suggest
that one possible reason for the plateau in $S$ that coincides with
the $0.7$ plateau is that $V_{g}$ cannot change the position of the
1D subband relative to the chemical potential (i.e., the 1D subband
becomes pinned). This is in general agreement with the spin-gap
model developed by Kristensen and Bruus~\cite{KristensenPRB00,
BruusPhysE01, KristensenPS02}, and with the data obtained by Graham
{\it et al}~\cite{GrahamPRB05, GrahamPRB07} and Chen {\it et
al}~\cite{ChenPRB09b} that we will discuss in Sections 8 and 9.

Chiatti {\it et al}~\cite{ChiattiPRL06} used a similar device
structure to study the thermal conductance of a QPC near the $0.7$
plateau. This is of interest due to theoretical predictions by Kane
and Fisher~\cite{KanePRL96} and Fazio {\it et al}~\cite{FazioPRL98}
that the Wiedemann-Franz law will be violated for a one-dimensional
electron system in the presence of strong electron-electron
interactions. A schematic of the device used by Chiatti {\it et al}
is shown in Fig.~18(a), the vital change being that the two gates
running along the heating channel form a third QPC that encloses a
small quantum box containing $\sim 2 \times 10^{5}$ electrons. The
addition of the box improves on previous
studies~\cite{MolenkampPRL90, MolenkampPRL92} by providing a better
defined thermal gradient across the QPC and increased thermovoltage
for a given heating channel current $I_{H}$~\cite{ChiattiPRL06}. Hot
electrons enter the box via QPC A, the constriction for which the
thermal conductance is measured, and exit via QPCs B and C, which
allow the electron temperature $T_{box}$ inside the box to be
determined. The basic process for this is as follows. Due to the
influx of hot electrons via A, $T_{box}$ sits $\delta T$ above the
lattice temperature $T_{L}$. The increased temperature inside the
box generates a thermovoltage across both QPCs B and C. However,
following Ref.~\cite{AppleyardPRL98}, if QPC C is set so that its
conductance sits at an integer plateau, its contribution to the
measured thermovoltage $V_{th}^{box}$ (see Fig.~18(a)) becomes zero,
thus $V_{th}^{box} = S_{B}\delta T$. The thermopower $S_{B}$ can be
eliminated by knowing that $S_{B} \propto \delta T$, such that
$V_{th}^{box} = c_{B}\delta T^{2}$, where the proportionality
constant $c_{B} \approxeq 15~\mu$V/K$^{2}$ is related to the 1D
subband spacing obtained from dc source-drain bias
spectroscopy~\cite{ChiattiPRL06, AppleyardPRL98}. The final step
involves measuring the thermopower when QPC A is not defined, here
$V_{th}^{H} = c_{B}(T_{H}^{2} - T_{L}^{2}) = c_{B}\Delta T^{2}$,
allowing $\Delta T$ to be measured. At this point, the gate voltage
dependent thermal conductivity $\kappa_{A}(V_{g})$ of QPC A can be
determined by combining the measured $\delta T$ and $\Delta T$ with:
(a) the knowledge that at steady state the heat flows
$\kappa_{A}\Delta T$ into and $(\kappa_{B}+\kappa_{C})\delta T$ out
of the box are equal; (b) the assumption that for each QPC $j$ = A,
B or C, the thermal conductance $\kappa_{j} =
\alpha_{j}G_{j}\widetilde{T}$, where the three $\alpha_{j}$ are
equal to a common constant $\alpha$, $G_{j}$ is the conductance and
$\widetilde{T}$ is the average temperature across QPC $j$; and (c)
known (i.e., set) values of $G_{B}$ and $G_{C}$. Assumption (b) is
just the Wiedemann-Franz law~\cite{FranzADP53}, where $\alpha$
should equal half the Lorenz number $L_{0}$. It implies that the
quantized electrical conductance plateaus should be accompanied by
plateaus in the thermal conductance quantized in units of
$\kappa_{0} = L_{0}TG_{0} = 1.89 \times 10^{-12}T$~W/K$^{2}$, and
these are observed in both Fig.~3 of Ref.~\cite{MolenkampPRL92} and
Fig.~2 of Ref.~\cite{ChiattiPRL06}. This is also consistent with
calculations by Rego and Kirczenow~\cite{RegoPRL98} predicting the
quantization of $\kappa$.

Returning our focus to the $0.7$ plateau, Fig.~18(b) shows the
electrical conductance $G_{A}$ (left-most trace) and the
thermally-derived conductance $\tilde{G}_{A}$ (right-most three
traces) versus $V_{g}$. The thermally-derived conductance
$\tilde{G}_{A} = (G_{B} + G_{C})(V_{th}^{box}/(V_{th}^{H} -
V_{th}^{box}))$ is presented instead of $\kappa_{A}$ to facilitate
better comparison. The traces i, ii and iii in Fig.~18(a) were
obtained for $G_{B} = 1.5G_{0}$ and $G_{C} = 5G_{0}$, $4G_{0}$ and
$3G_{0}$, respectively, and demonstrate the validity of the
measurements, since $\delta T$ is different in each case, which is
naturally accounted for in obtaining $\tilde{G}_{A}$. Integer
plateaus in $\tilde{G}_{A}$ accompany those in $G_{A}$, as expected,
however they cannot be taken as definitive confirmation of the
Wiedemann-Franz law as it is unclear whether $\alpha = L_{0}/2$ in
this case~\cite{ChiattiPRL06}. The $0.7$ plateau in $G_{A}$ is not
accompanied by a $0.7$ plateau in $\tilde{G}_{A}$, a plateau at $0.5
G_{0}$ is observed instead. This is more apparent in Fig.~18(b),
which shows a close-up of $G_{A}$ and $\tilde{G}_{A}$ near
pinch-off. This suggests a breakdown of the Wiedemann-Franz relation
in the vicinity of the $0.7$ plateau, consistent with theoretical
predictions~\cite{KanePRL96, FazioPRL98}. It is important to note
that both works~\cite{KanePRL96, FazioPRL98} considered spinless
systems. In particular, the prediction by Fazio {\it et
al}~\cite{FazioPRL98} relates to charge-energy separation and not
spin-charge separation~\cite{FazioPC}, as Chiatti {\it et al} imply
in their discussion. The mechanism proposed by Fazio {\it et al} is
that energy is carried by plasmons rather than individual
electrons~\cite{FazioPRL98}. Suppose the 1D channel has
inhomogeneities on a length scale larger than the Fermi wavelength
$\lambda_{F}$; although these do not produce any backscattering for
electrons, they can backscatter plasmons with wavelength $\lambda >
\lambda_{F}$, strongly affecting the thermal conductance while the
electrical conductance remains fixed. Rejec {\it et
al}~\cite{RejecPRB02} predict structures in $\tilde{G}_{A}$ at
$0.25G_{0}$ and $0.75G_{0}$ via the Wiedemann-Franz relation, and
although these are not observed in the data in Figs.~18(b/c), their
calculations are in broad agreement with the thermopower
measurements of the $0.7$ plateau at higher temperatures. Finally,
Chiatti {\it et al} note that combining theoretical calculations of
thermal conductivity by van Houten {\it et al}~\cite{vanHoutenSST92}
with a density-dependent spin-gap model~\cite{ReillyPRL02} predicts
a plateau in $\tilde{G}_{A}$ at $0.7G_{0}$, which is inconsistent
with the experimental data they obtain. In other words, there is no
mechanism within the spin-gap models for violation of the
Wiedemann-Franz law.

\subsection{Compressibility measurements of 1D systems}

Information about the density of states is not easily accessible
through traditional transport studies, and measurements of
thermodynamic properties such as the specific
heat~\cite{GornikPRL85}, magnetization~\cite{EisensteinPRL85} and
compressibility~\cite{KaplitPRL68, SmithPRB85, EisensteinPRL92} have
been used to study the density of states in 2D electron systems.
Achieving such measurements in large-area 2D systems is a
challenging task, and extending these methods to smaller 1D electron
systems only adds to the difficulty. As a result, only the
compressibility has been successfully studied for 1D electron
systems so far~\cite{SmithPRL87, DrexlerPRB94, CastletonPhysB98,
IlaniNP06, LuscherPRL07, SmithPRL11}.

For an electron system, the compressibility $K$ is given by:

\begin{equation}
K = \frac{1}{n^{2}}\frac{\partial n}{\partial \mu}
\end{equation}

\noindent where $\frac{\partial n}{\partial \mu}$ is the
thermodynamic density of states. Thus as a physical quantity, the
compressibility reflects the change in chemical potential that
occurs in response to a given change in electron density -- for a
high (low) density of states, a given change in $n$ produces a small
(large) shift in $\mu$. The electronic compressibility is typically
measured capacitively, and in the earliest approaches, was obtained
by using a metal surface-gate and a 2DEG as the two plates in a
parallel-plate capacitor configuration~\cite{KaplitPRL68,
SmithPRB85}. The difficulty with such measurements is that the
capacitance involves two components, a geometric capacitance for the
parallel-plate configuration and the density of states contribution.
The former is substantially larger than the latter and not always
constant, making the density of states contribution difficult to
precisely measure. Eisenstein {\it et al}~\cite{EisensteinPRL92}
developed an improved method whereby the compressibility of the
upper 2DEG in a double quantum well heterostructure is measured by
using the lower 2DEG to measure the electric field that penetrates
the upper 2DEG due to an ac voltage applied to a surface gate. The
resulting signal is directly proportional to $1/K$, and thereby
mitigates the obscuring effect of the geometric capacitance
contribution.

The single 2DEG capacitance configuration was used for initial
compressibility studies of 1D systems, and given the small
capacitance changes involved, these measurements required arrays of
parallel 1D wires (i.e., grating capacitors) to obtain a viable
signal. T.P. Smith {\it et al}~\cite{SmithPRL87} studied devices
with $250 - 500$ parallel $200~\mu$m long wires formed by
electron-beam lithography, reactive-ion etching and deposition of an
overall metal top-gate. Measurements of the gate voltage derivative
of the capacitance showed oscillations as a function of gate voltage
with a period that increased with decreasing wire width. These
oscillations were attributed to the periodic 1D density of states.
While this measurement provided initial insight and proof of
concept, it was severely limited by the large geometric capacitance
contribution, which obscured the signal even in the derivative as
the wires approached the 1D limit (i.e., low density).

Drexler {\it et al}~\cite{DrexlerPRB94} subsequently studied a
device formed in a heterostructure consisting of a $20$~nm highly
doped GaAs layer used as a back-contact, a $100$~nm undoped GaAs
spacer, a $32$~nm AlAs/GaAs short period superlattice barrier layer
and a $10$~nm GaAs cap. The device consists of an array of $300$
parallel $180~\mu$m long wires defined by a pair of interdigitated
electrodes on the heterostructure surface. The wires are defined by
biasing the first electrode positively with respect to the
back-contact, and the lateral confinement of the wires can be tuned
by adjusting the bias of the second electrode relative to the first.
A key advantage of this configuration is that it reduces the
geometric capacitance contribution such that the density of states
contribution can be measured in the direct capacitance signal. The
measured capacitance showed clear oscillations corresponding to the
population of the first, second, third and fourth 1D subbands, and a
clear spin-splitting of the oscillations corresponding to the first
and second subbands in response to an applied magnetic
field~\cite{DrexlerPRB94}. A small additional maxima in the
capacitance is observed just before pinch-off, and given it is not
observed for capacitance studies of 2D systems using the same wafer,
it was argued that this feature is a characteristic of 1D systems.
Drexler {\it et al}~\cite{DrexlerPRB94} attribute this maxima to
renormalisation of the effective potential in the wires via
electron-electron interactions as the wires first begin to populate.

The first study on a single 1D wire was performed by Castleton {\it
et al}~\cite{CastletonPhysB98} using a double quantum well
heterostructure and Eisenstein's field-penetration
method~\cite{EisensteinPRL92}. This approach provided the advantage
that the conductance for the QPC could be measured along with the
compressibility, and it obviates the averaging/smearing due to
inhomogeneities in large arrays of parallel devices. Maxima in the
compressibility were observed to coincide with the risers in the
conductance, which in turn coincide with the van Hove singularities
(maxima) in the 1D density of states at each 1D subband edge. A
subsequent study of the compressibility of a single QPC was reported
by L\"{u}scher {\it et al}~\cite{LuscherPRL07} utilizing a lateral
configuration consisting of two side-by-side QPCs in a single 2DEG,
with the first QPC is used as a charge-sensitive detector for the
second `test' QPC. Each QPC is defined by one independently
controllable gate on one side, set to $V_{det}$ for the detector QPC
and $V_{QPC}$ for the test QPC, and a common $80$~nm wide gate on
the other, which is held fixed at $V_{m}$. The charge sensing
measurement involves monitoring the detector conductance $G_{det}$
as the test QPC is closed from above the third plateau to pinch-off,
and mapping $G_{det}$ to an effective bias $V_{eff}$ that would
produce the same change in $G_{det}$ if it was applied to the gate
controlling the detector QPC. During this measurement the two QPCs
share a common drain; a voltage is applied to the source of the
detector QPC with the source of the test QPC virtually grounded via
a $44$~M$\Omega$ resistor such that no intentional current flows
through the test QPC. The drain current is measured to obtain
$G_{det}$~\cite{LuscherPC}. Fourteen different, overlapping
$V_{det}$ settings were used over the full range of $V_{QPC}$ to
ensure that the detector remained at its most linear and sensitive
range $G_{det} = 0.3 - 0.5G_{0}$ throughout the measurement. The
fourteen traces were converted to a continuous effective
compressibility $D = dV_{eff}/dV_{QPC}$ that shows a linear trend in
$V_{QPC}$ for $G > 0$ in the test QPC and a series of minima
(compressibility peaks) coinciding with the risers in the test QPC
conductance, consistent with Ref.~\cite{CastletonPhysB98}.
L\"{u}scher {\it et al}~\cite{LuscherPRL07} note that the minima in
$D$ coinciding with the lowest riser (i.e., pinch-off) in the test
QPC is deeper than the others, and by comparison to density
functional theory calculations, they attribute this to
exchange-correlation effects, which should be at their strongest in
the low density environment found inside the QPC as it pinches off.
An interesting additional outcome of the calculations performed by
L\"{u}scher {\it et al} is the presence of an additional
maxima/minima in $D$ on the higher $G$ side of the minima discussed
above if the local spin density approximation (LSDA) is used.
L\"{u}scher {\it et al} argue that this additional minima would be
the signature of the formation of a magnetic moment consistent with
the Kondo scenario for the $0.7$ plateau~\cite{CronenwettPRL02,
RejecNat06}. This additional minima is not observed in the
experimental data presented by L\"{u}scher {\it et al}, however the
DFT calculations predict that this feature will be small and
possibly obscured by the spontaneously fluctuating spin polarisation
at this conductance~\cite{MeirPC}, and as a result this feature is
likely below the sensitivity limit for the measurement configuration
used in this experiment.

This possibility was very recently explored further by L.W. Smith
{\it et al}~\cite{SmithPRL11} using the configuration pioneered by
Castleton {\it et al}~\cite{CastletonPhysB98}, which provide them
with much greater sensitivity in measurements of the
compressibility. Compressibility peaks (minima in the
compressibility signal $dV_{sg}/dV_{mid}$) are observed coinciding
with the risers in the 1D conductance, consistent with earlier
work~\cite{CastletonPhysB98, LuscherPRL07}. Spin-splitting of the 1D
subbands with an applied in-plane magnetic field is also observed,
which is essentially consistent with Drexler {\it et
al}~\cite{DrexlerPRB94} who also observe spin-splitting, albeit with
the magnetic field applied perpendicular to the 2DEG, which may lead
to complications due to magnetoelectric subband
formation~\cite{BerggrenPRL86, vanWeesPRB88}. Smith {\it et al}
clearly observe the same enhanced minima (compressibility peak)
coinciding with pinch-off that is reported by L\"{u}scher {\it et
al}, but state that they do not observe the adjacent maxima/minima
predicted by the spin density functional theory calculations as a
signature of Kondo physics in Ref.~\cite{LuscherPRL07} (see
Fig.~4(b) of Ref.~\cite{SmithPRL11}). However, the data presentation
in Fig.~4(b) of Ref.~\cite{SmithPRL11} is somewhat subjective, given
the relative scaling of the left and right $y$-axes, and it could be
argued on the basis of the data scaling presented in Fig.~2(a) of
Ref.~\cite{SmithPRL11} that a Kondo feature may be present, albeit
weaker than that predicted by theoretical
calculations~\cite{LuscherPRL07}(n.b., as mentioned earlier, this
feature is expected to be very small and potentially further
diminished by other fluctuations in
spin-polarisation~\cite{MeirPC}). The predicted Kondo signature
clearly remains close to the sensitivity limit in this approach, and
so a definitive conclusion is difficult to draw.

In the analysis of their data, Smith {\it et al} focus on a
compressibility minima (peak in the compressibility signal
$dV_{sg}/dV_{mid}$) that coincides in gate voltage with the $0.7$
plateau at zero magnetic field and $0.5$ plateau at higher fields
(see arrow in Fig.~3(a) of Ref.~\cite{SmithPRL11}). This peak, which
corresponds to a relative reduction in the density of states,
increases in height and width as the magnetic field is increased and
the $0.7$ plateau migrates towards $0.5G_{0}$, as apparent in
Fig.~3(a) of Ref.~\cite{SmithPRL11} and additional
data~\cite{SmithPC}. Since the density of states for a spin-split
subband will be less than that of a spin-degenerate subband, and the
peak coincides with the $0.5$ plateau, Smith {\it et al} argue that
this peak is associated with spin-polarization of the lowest 1D
subband. Remarkably, this peak also appears to grow in height and
width with increasing temperature at $B = 0$, and Smith {\it et
al}~\cite{SmithPRL11} suggest that this might indicate a
temperature-dependent spin-gap that is driven by the exchange
interaction. While this specific hypothesis would require further
study, Smith {\it et al} argue that their data point towards
zero-field spin polarization rather than a Kondo process as the
origin of the $0.7$ plateau. It is worth noting that Drexler {\it et
al}~\cite{DrexlerPRB94} also observe a similar feature in their
data, which is the first minima after the sharp rise in capacitance
signalling the population of the first 1D subband (n.b., to compare
these two data sets directly, one needs to be mirrored about a
horizontal line before being overlaid with the other). The minima in
Ref.~\cite{DrexlerPRB94} also grows in relative height (and perhaps
width) with increasing magnetic field, albeit with the field
oriented perpendicular to the 2DEG rather than in-plane. Given also
that the device studied by Drexler {\it et al}~\cite{DrexlerPRB94}
contains 300 parallel wires, each $180~\mu$m in length, and the
device studied by Smith {\it et al}~\cite{SmithPRL11} contains a
single QPC approximately $0.5~\mu$m long, it is dangerous to assume
a causal link between these features. However, it is apparent that
further work is needed in order to draw definitive conclusions
regarding what small-scale features in compressibility measurements
imply regarding the physical mechanism responsible for the $0.7$
plateau.

\subsection{Studies of the $0.7$ plateau in 1D hole systems}

A hole is a quasiparticle corresponding to the absence of an
electron in the valence band of a semiconductor; it carries a
positive charge equal in magnitude to the electron. Considering
this, one might initially ask why the physics of the $0.7$ plateau
should be any different for holes -- The answer is two-fold:
effective mass and strong spin-orbit coupling.

The relative strength of electron-electron (or hole-hole)
interactions are commonly parameterized by the quantity $r_{s} =
\langle PE \rangle/\langle KE \rangle =
m^{*}e^{2}/4\pi^{3/2}\hbar^{2}\epsilon\sqrt{n}$, where $n$ is the 2D
electron or hole density and $\epsilon$ is the GaAs dielectric
constant. The interactions can thus be strengthened by either
lowering the density or increasing the effective mass. The hole
effective mass $m^{*}_{h} \approx 0.38 m_{e}$ is approximately five
times that for electrons $m^{*}_{e} = 0.067
m_{e}$~\cite{StormerPRL83, HirakawaPRB93,
ProskuryakovPRL02}~\footnote{The effective mass for holes is highly
anisotropic and $k$-dependent, and as a result accepted values can
vary from $0.2-0.5 m_{e}$. An in-depth discussion of the
complexities of the effective mass of holes can be found in the book
by Winkler~\cite{WinklerBook03}.}, such that at a density $n =
10^{11}$~cm$^{-2}$, $r_{s} \sim 1.8$ for electrons and $r_{s} \sim
9.9$ for holes. Since many explanations for the $0.7$ plateau are
interaction-based, being able to increase the effective mass may
provide useful insight.

The more interesting aspect, however, is afforded by the spin-orbit
interaction. The conduction band in GaAs originates from $s$-like
atomic orbitals where the orbital angular momentum $l = 0$, such
that the direct spin-orbit interaction $\sim l \cdot s = 0$.
Spin-orbit effects for electrons arise indirectly via interband
coupling~\cite{OhkawaJPSJ74}. This is enhanced in narrow band-gap
semiconductors such as InGaAs, InSb and HgCdTe, hence the interest
in these materials for spintronics
applications~\cite{AwschalomPhys09}. In contrast, the valence band
for GaAs originates from $p$-like atomic orbitals where $l = 1$,
resulting in a spin-orbit interaction so strong that $l$ and $s$ are
no longer good quantum numbers. The total orbital angular momentum
$\overrightarrow{j} = \overrightarrow{l} + \overrightarrow{s}$ must
be used for this system, with the quantum numbers $s$ and $m_{s}$
being replaced by $j$ and $m_{j}$ where $j = 1/2$ or $3/2$, and
$m_{j} = \pm 1/2$ for $j = 1/2$ and $m_{j} = \pm 3/2$ or $\pm 1/2$
for $j = 3/2$. The full ramifications of the strong spin-orbit
interaction in low-dimensional hole structures are only just
beginning to be explored, but one key aspect is that spin-orbit
interactions strongly influence the $g$-factor~\cite{RothPR59}. This
effect is complicated significantly by crystallographic
effects~\cite{WinklerBook03}, but leads to a highly anisotropic
in-plane $g$-factor in both 2D~\cite{WinklerPRL00} and
1D~\cite{KoduvayurPRL08, KlochanNJP09, ChenNJP10} hole systems; this
is not observed for electrons where the in-plane $g$-factor is
essentially isotropic~\cite{ThomasPRL96}.

Given these interesting aspects, it is remarkable that hole QPCs
were not studied earlier; however, this is not for want of trying.
Early attempts to make and measure 1D hole systems were plagued with
device instability issues~\cite{ZailerPRB94,DaneshvarPRB97}. The
problem is believed to originate from slow trapping/de-trapping
behaviour between the surface-gate/semiconductor interface and the
2D hole gas, and may or may not involve either the dopants or other
traps such as surface states~\cite{ZailerPhD94, DaneshvarPhD98,
GrbicPhD07} -- a clear picture of this problem is yet to be
established. Although other attempts using atomic force microscopy
local anodic oxidation (AFM-LAO) lithography on shallow 2DHG
heterostructures~\cite{RokhinsonSM02} and cleaved-edge
overgrowth~\cite{PfeifferAPL04} came closer to achieving clear
conductance plateaus, the stability and quality of the 1D
conductance were insufficient for detailed studies of conductance
quantization. The first report of clear, reproducible 1D conductance
plateaus with accurate quantization in a hole QPC was made by
Danneau {\it et al}~\cite{DanneauAPL06, DanneauPRL06}, who used a
modulation-doped double quantum well structure where independent
measurements of a gate-defined QPC formed in the upper and lower
quantum well could be obtained. Work in this area has more recently
shifted to undoped structures using either
semiconductor-insulator-semiconductor field-effect transistor
(SISFET) devices~\cite{ClarkeJAP06, KlochanAPL06, KlochanAPL10} or
modulation-doped devices with a HfO$_{2}$ layer deposited by atomic
layer deposition under the gates~\cite{CsontosAPL10}. For the
remainder of this section we will focus on two studies: the first is
the observation of anisotropic Zeeman splitting of the $0.7$ plateau
and associated zero bias anomaly in a hole QPC by Danneau {\it et
al}~\cite{DanneauPRL08}. The second is a study of spin polarization
in a hole QPC for $G < G_{0}$ using ballistic focussing by Rokhinson
{\it et al}~\cite{RokhinsonPRL06}. Both experiments provide
interesting conclusions regarding the link between the $0.7$
plateau/ZBA and spin.

The device studied by Danneau {\it et al}~\cite{DanneauPRL08} was
made on a (311)A-oriented double quantum well heterostructure, with
the quantum wire oriented along the higher-mobility
$[\overline{2}33]$ crystallographic direction~\cite{DaviesJCG91}.
The measured mobility along this direction was $920,000$~cm$^{2}/$Vs
at a density $p = 1.2 \times 10^{11}$~cm$^{-2}$. For these
measurements, the lower quantum well is depleted by applying a
positive bias $\sim 2.5$~V to the n$^{+}$-GaAs substrate, which acts
as a back-gate. The QPC itself is defined by three surface gates --
a pair of side-gates negatively-biased to $V_{SG}$ and a conformal
midline-gate that covers the QPC and a region extending for several
microns over the source and drain reservoirs, biased negatively to
$\sim -0.7$~V to maintain a high, homogeneous hole density within
and local to the QPC. The side-gates and midline-gate are all
surface gates, separated by a narrow ungated gap that allows them to
be biased to opposite polarities (see Ref.~\cite{DanneauAPL06} for
details). The results presented below build heavily on the key
outcome of Ref.~\cite{DanneauPRL06} that the Zeeman splitting of the
1D subbands is highly anisotropic -- for an in-plane magnetic field
$B_{\parallel}$ aligned along the QPC axis the $g$-factor is high
$g^{*}_{\parallel} \sim 0.8$, and for an in-plane magnetic field
$B_{\perp}$ aligned perpendicular to the QPC axis the $g$-factor is
low $g^{*}_{\perp} \lesssim 0.2$. This finding allowed the
connection between the $0.7$ plateau/ZBA and spin to be tested,
because if the Zeeman splitting of these features was not similarly
anisotropic, then there was clearly more to the problem than was
being considered in existing models at the time.

\begin{figure}
\includegraphics[width=16cm]{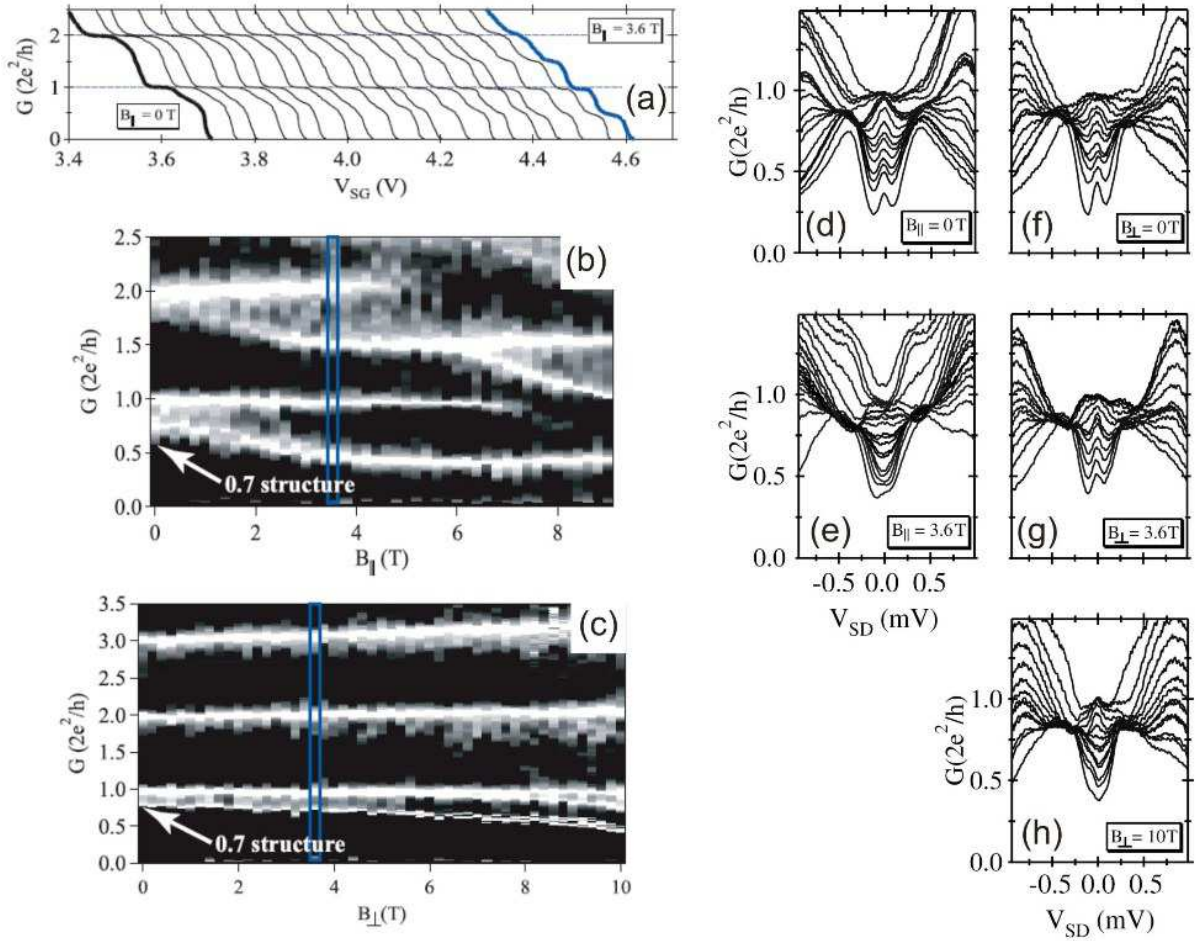}
\caption{(a) Conductance $G$ vs side-gate voltage $V_{SG}$ as a
function of in-plane magnetic field applied along the QPC axis
$B_{\parallel}$ from $0$ to $3.6$~T in steps of $0.2$~T with traces
offset horizontally for clarity. Data obtained with the back-gate at
$+2.5$~V and midline gate at $-0.225$~V. (b) Greyscale plot of the
transconductance $dG/dV_{SG}$ vs $G$ ($y$-axis) and $B_{\parallel}$
($x$-axis). (c) Similar data to that in (b) but for the in-plane
magnetic field $B_{\perp}$ aligned perpendicular to the QPC axis. In
both cases the white region corresponds to low transconductance to
emphasize the conductance plateaus in contrast to other greyscales
where emphasis is on 1D subbands. The narrow blue rectangle
coincides in field with the blue trace in (a). (d-h) Plots of
differential conductance $G$ vs source-drain bias $V_{sd}$ for a
number of gate voltage settings in the vicinity of $G(V_{sd}=0) =
G_{0}$. Data for $B_{\parallel}$ is presented in the left column
with (d) at $0$~T and (e) at $3.6$~T. Data for $B_{\perp}$ is
presented in the right column at (f) $0$~T, (g) $3.6$~T, and (h)
$10$~T. Figure adapted with permission from
Ref.~\cite{DanneauPRL06}. Copyright 2006 by the American Physical
Society.}
\end{figure}

Zeeman splitting data for the 1D conductance plateaus is presented
in Fig.~19(a-c). To best demonstrate the behaviour, in Fig.~19(b)
and (c) we show greyscale plots of transconductance $dG/dV_{SG}$
versus $G$ ($y$-axis) and $B_{\parallel}$ or $B_{\perp}$ ($x$-axis),
respectively. High transconductance appears black to emphasize the
plateaus (in contrast to many greyscales, where emphasis is on the
1D subbands). To assist with the interpretation, Fig.~19(a) shows
$G$ versus $V_{SG}$ for fields $B_{\parallel} = 0$ to $3.6$~T in
steps of $0.2$~T. The blue trace at $B_{\parallel} = 3.6$~T in
Fig.~19(a) corresponds to the narrow, blue vertical rectangle in
Figs.~19(b/c). For the data in Fig.~19(a), the $g$-factor is large,
and the 1D subbands split relatively rapidly, with plateaus at
integer multiples of $0.5G_{0}$ present by $B_{\parallel} \gtrsim
3$~T. The 1D subbands cross at $B_{\parallel} \sim 6-8$~T, indicated
by the loss of plateaus at integer multiples of $G_{0}$ around this
field range in Fig.~19(b). In both Figs.~19(b) and (c), the $0.7$
plateau is indicated and is the white band nearest to the bottom. In
the parallel case in Fig.~19(b), the $0.7$ plateau evolves to $0.5
G_{0}$ rapidly. This transition appears very similar to the
evolution of the plateau at $1.5G_{0}$ except that this higher
plateau clearly intercepts the $2G_{0}$ plateau at $B_{\parallel} =
0$~T, while the corresponding intercept for the $0.7G_{0}$ and
$G_{0}$ plateaus would occur at $B_{\parallel} < 0$~T, suggestive of
a finite energy separation between $1\uparrow$ and $1\downarrow$ at
zero field. Very different behaviour is found in the perpendicular
case in Fig.~19(c), where the integer plateaus are relatively
unaffected by the field until $B_{\perp} \sim 8$~T where the
beginnings of spin-splitting become apparent, as expected since
$g^{*}_{\perp}$ is small but finite. Accordingly, the $0.7$ plateau
just begins to move down towards $0.5G_{0}$ at the highest
$B_{\perp}$ (we return to this below). The fact that nothing unusual
happens with the $0.7$ plateau in the latter case suggests that the
$0.7$ plateau and spin are connected in a straightforward manner. In
other words there is no connection between the $0.7$ plateau and
magnetic field that cannot be explained by the Zeeman effect alone.

A similar anisotropy is apparent for the zero-bias anomaly, with the
data for $B_{\parallel}$ in the left column (Figs.~19(d/e)) and
$B_{\perp}$ in the right column (Figs.~19(f-h)). In each case, the
differential conductance $G$ versus source-drain bias $V_{SD}$ is
plotted for a range of different $V_{SG}$. The ZBA is completely
suppressed by $B_{\parallel} = 3.6$~T (Fig.~19(e)). In contrast, the
ZBA has only evolved very slightly at $B_{\perp} = 3.6$~T
(Fig.~19(g)) but is completely suppressed at $B_{\perp} = 10$~T
(Fig.~19(h)). Danneau {\it et al}~\cite{DanneauPRL08} suggest that
this strongly reinforces the hypothesis that the $0.7$ plateau and
the ZBA are linked and are spin-related phenomena. The latter is
certainly true, the fact that both show an anisotropic response to a
magnetic field that matches that of the Zeeman splitting of the 1D
subband suggests that spin lies at the heart of both phenomena. The
former is perhaps in hindsight better stated from the converse,
namely that the behaviour observed by Danneau {\it et
al}~\cite{DanneauPRL08} cannot rule out the possibility of a link
between $0.7$ plateau and the ZBA. This is because it is not
possible to distinguish between the $0.7$ plateau and ZBA being
causally linked or simply being coincident spin-related behaviors
based on the available data from experiments on 1D hole systems.
Finally, the behaviour at large $B_{\perp}$ bears some additional
comment. Danneau {\it et al}~\cite{DanneauPRL08} point out that the
drop in the $0.7$ plateau and suppression of the ZBA occur earlier
than one might expect based on the Zeeman splitting of the 1D
subbands with increasing $B_{\perp}$ -- the estimated $g^{*}$
anisotropy is greater than $4.5$ for the 1D subbands but less than
$4$ for the $0.7$ plateau and ZBA. This reduction in the anisotropy
in the 1D limit is consistent with calculations by
Z\"{u}licke~\cite{ZulickePSS06} where heavy hole - light hole mixing
with strengthening 1D confinement causes both $g^{*}_{\parallel}$
and $g^{*}_{\perp}$ to rise, but $g^{*}_{\perp}$ rises faster
causing the anisotropy to decrease. The physics of heavy hole -
light hole mixing in the 1D limit has yet to be fully explored, but
the data in Ref.~\cite{DanneauPRL08} suggests that this physics may
be accessible with some fine tuning of the 1D confinement.
Subsequent work by Csontos {\it et al}~\cite{CsontosPRB07,
CsontosPRB08} suggests that some remarkably erratic fluctuations in
$g$-factor may be observed in this regime. A study of the $0.7$
plateau in (100)-oriented hole QPCs was presented by Komijani {\it
et al}~\cite{KomijaniEPL10} and is discussed in Section 7.4.2. A
feature resembling the $0.7$ plateau has also been observed in
(100)-oriented hole QPCs by Chen {\it et al}~\cite{ChenPhysE10} and
Klochan {\it et al}~\cite{KlochanPRL11}, but requires more focussed
study. It is interesting to note that 1D exchange enhancement does
not occur for (100)-oriented hole QPCs~\cite{ChenNJP10}, making this
system of great potential interest for studying the role that
exchange effects may play in the physics of the $0.7$ plateau.

\begin{figure}
\includegraphics[width=16cm]{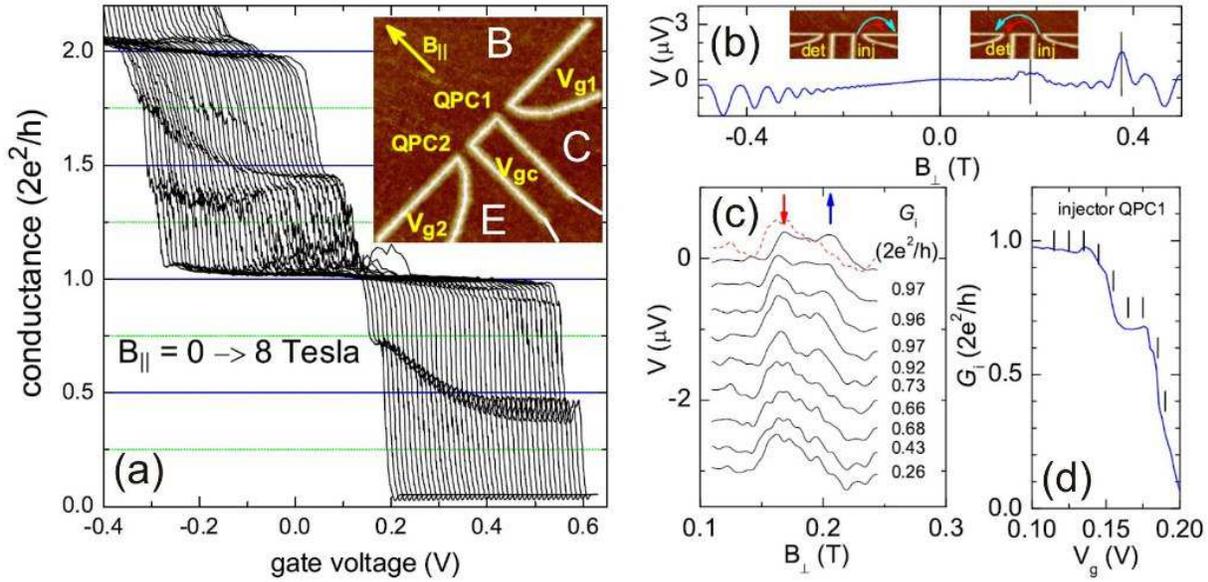}
\caption{(a) Plots of conductance vs gate voltage for QPC1 (see
inset) for different in-plane magnetic fields aligned along the QPC
axis $B_{\parallel}$ from $0$ to $8$~T in steps of $0.2$~T offset
horizontally for clarity. The inset shows an AFM image of the sample
with QPCs 1 and 2 arranged in a ballistic focussing geometry
separated by $\sim 800$~nm. The field size is $3.3 \times 3.3~\mu$m.
(b) Plot of focussing voltage $V$ vs magnetic field component
perpendicular to the 2DHG $B_{\perp}$, with the first and second
focussing peaks at positive $B_{\perp}$ indicated by the horizontal
lines. The two insets illustrate the curvature direction and why
focussing peaks only occur for positive $B_{\perp}$, while the
Shubnikov-de Haas oscillations appear on both sides. (c) Plots of
$V$ vs $B_{\perp}$ for different $G$ settings of QPC1 (as indicated
to the right of each trace), each sequentially vertically offset by
$-0.4~\mu$V for clarity. The red dashed trace is a copy of the $G =
0.66G_{0}$ trace without a vertical offset applied. QPC2 is set to
$G = G_{0}$ for all traces. The vertical arrows at the top indicate
the spin for the two focussed beams. (d) A plot of $G$ vs $V_{g}$
for QPC1. The vertical black lines correspond to the positions of
the nine traces presented in (c). Figure adapted with permission
from Ref.~\cite{RokhinsonPRL06}. Copyright 2006 by the American
Physical Society.}
\end{figure}

The device studied by Rokhinson {\it et al}~\cite{RokhinsonPRL06}
consists of two side-by-side QPCs with a center-to-center distance
of $800$~nm forming a ballistic focusing
geometry~\cite{vanHoutenPRB89}, as shown inset to Fig.~20(a). The
2DHG is located $35$~nm beneath the surface of a modulation-doped
(311)A AlGaAs-GaAs heterostructure, and the device is defined by
anodic oxidation lithography~\cite{HeldAPL97}. Despite the
shallowness of the 2DHG, the mobility is $\sim 500,000$~cm$^{2}$/Vs
corresponding to a mean free path exceeding $2.5~\mu$m. The
conductance versus gate voltage for QPC1, which is used as the
injector for the following measurements, is shown in Fig.~20(a) for
$B_{\parallel} = 0$ to $8$~T with the in-plane field aligned along
the QPC. The most notable feature of the data is the length of the
$G_{0}$ plateau compared to, for example, the $2G_{0}$ plateau. This
is not commented on in Ref.~\cite{RokhinsonPRL06}, is not observed
for other, similarly fabricated QPCs on similar heterostructures
(e.g., Ref.~\cite{KoduvayurPRL08, RokhinsonSM02}), and is a little
unusual. There are also a number of other anomalous structures in
the conductance, particularly in the range $G_{0} < G < 2G_{0}$ that
indicate that this device may be significantly affected by
disorder~\cite{CzapkiewiczEPL08}. Although the mean free path is
large enough for ballistic transport to occur within the device, the
mobility is heavily weighted against small-angle
scattering~\cite{ColeridgePRB91}, which is likely to be significant
due to the proximity between the ionized dopants and the
2DHG~\cite{MacLeodPRB09}. There are structures at $0.75$ and $1.75
G_{0}$ at $B_{\parallel} = 0$ that evolve towards $0.5$ and $1.5
G_{0}$ as the field is increased. The peak-like structure on the
high-field $0.5G_{0}$ plateau is reminiscent of that observed for
longer wires by Reilly {\it et al}~\cite{ReillyPRB01}, and may also
be indicative of disorder scattering. Rokhinson {\it et al} also
observe a ZBA in QPC1 of this device, which is suppressed without
Zeeman splitting occurring at $B_{\parallel} \sim 2$~T, or by
increasing the temperature above $\sim 350$~mK.

The focus of this experiment was not QPC1 itself but ballistic
focussing from QPC1 into QPC2, with the corresponding data presented
in Fig.~20~(b-d). These measurements are obtained by passing a
constant current of $0.5$~nA from the emitter E (see Fig.~20(a)
inset) to base B, and measuring the voltage $V$ measured at the
collector C as a function of magnetic field $B_{\perp}$ applied
perpendicular to the 2DHG to induce cyclotron motion of the holes
injected by QPC1. The collector voltage will generally be zero (or
at least small) except when a cyclotron path from QPC1 intercepts
QPC2~\cite{vanHoutenPRB89}. Figure~20(a) shows such data obtained
for $-0.5 < B_{\perp} < 0.5$~T with both QPCs set to $G = G_{0}$.
Shubnikov-de Haas oscillations develop with increasing $B_{\perp}$,
however, the symmetry afforded by the Onsager-Casimir
relations~\cite{CasimirRMP45} allows the focussing to be
distinguished despite this. Looking at the positive field side, two
focussing peaks appear in Fig.~20(b), indicated by the vertical
black lines. These two peaks correspond to zero-bounce and
single-bounce semicircular trajectories between emitter and
collector~\cite{vanHoutenPRB89}, with the peak separation $\sim
0.19$~T consistent with a cyclotron radius of $333$~nm at $p = 1.47
\times 10^{11}$~cm$^{-2}$ and $260$~nm at $p = 0.9 \times
10^{11}$~cm$^{-2}$. The first peak has a double maximum, whereas the
second has a single maximum, similar behaviour is observed in an
equivalent device in Ref.~\cite{RokhinsonPRL04} where the third and
fourth focussing peaks are also double and single maxima. The origin
of this apparent alternating single-/double-maxima behaviour in
subsequent focussing peaks is unknown, however, it is worth noting
that structure on ballistic focussing peaks is not unusual (e.g.,
see Fig.~4 of Ref.~\cite{vanHoutenPRB89}, Fig.~1(b) of
Ref.~\cite{MolenkampPRB90}, Fig.~3 of Ref.~\cite{KishenAIPC07}) and
is consistent with the effect that small-angle scattering has on
electron dynamics in focussing, as evidenced by scanning-gate
microscopy studies~\cite{AidalaNP07}. Rokhinson {\it et al} explain
the double maxima observed on the first focussing peak as a
separation of the injected beam of electrons into spin-up and
spin-down components due to the spin-orbit interaction (see
Ref.~\cite{RokhinsonPRL04} for full details). The attribution of the
maxima to particular spins was obtained by studying the evolution of
the maxima with in-plane magnetic field $B_{\parallel}$ (see
Fig.~2(a/c) of Ref.~\cite{RokhinsonPRL04}) with the QPC2 set to $G =
0.5G_{0}$ to act as a spin-selective detector~\cite{PotokPRL02}. At
zero field, the maxima are equal in height, but at $B_{\parallel}
\sim 3.0 - 4.2$~T,\footnote{n.b., Since the sample is at an angle in
this experiment, increasing $B_{\perp}$ means simultaneously
increasing $B_{\parallel}$. The field goes from $B_{\perp} = 0.16$~T
and $B_{\parallel} = 3.05$~T to $B_{\perp} = 0.22$~T and
$B_{\parallel} = 4.2$~T over the width of the focussing peak.} the
right-hand maxima is suppressed. This allows the right-hand maxima
to be attributed as the focussing peak for the spin-up component of
the injected hole beam. This is consistent with the right-most peak
(i.e., higher $B_{\perp}$ needed to bend that beam into the
detector) corresponding to the highest energy Zeeman-split spin
state, which is spin-up. The aim in Ref.~\cite{RokhinsonPRL06} is to
use this dual-peak structure to study the spin polarization of the
beam when the QPC1 (i.e., the injector) conductance $G < G_{0}$.
QPC2 is held at $G = G_{0}$ throughout this measurement so that both
spin-up and spin-down beams can be detected at the collector.

Figure~20(c) shows a series of nine ballistic focussing traces for a
range of different conductance settings for QPC1, these are
indicated by the vertical bars in the $G$ versus $V_{g}$ plot for
QPC1 in Fig.~20(d). The dashed red trace in Fig.~20(c) is a copy of
the $G = 0.66G_{0}$ trace without the offset applied. As $G$ is
reduced, the right-most peak decreases while the left-most peak
retains its amplitude. The loss of the right-most peak is most
apparent between the traces at $0.73$ and $0.66G_{0}$. Rokhinson
{\it et al}~\cite{RokhinsonPRL06} use the relative peak-heights to
estimate the spin-polarization of the injected beam, obtaining a
value of $40 \pm 15~\%$ for $G < 0.9G_{0}$. Interestingly, the
right-most peak appears to shift slightly to lower fields as this
occurs, and its disappearance coincides with a widening of the
left-most peak that may suggest a change in the spin-up subband
energy as QPC1 is pinched off. However, the emergence of periodic
structure in the focussing voltage in the vicinity of the suppressed
right-most peak for low $G$ may indicate that another effect is
responsible for this apparent behaviour. Rokhinson {\it et al}
repeat their measurements for two other samples that do not exhibit
$0.7$ plateaus. In this case similar changes in the relative
focusing peak amplitude with changes in $G$ are observed (see Fig.~4
of Ref.~\cite{RokhinsonPRL06}). The left-most peak (spin-down) drops
in one case, and the right-most peak (spin-up) drops in the other.
On the basis of this data, Rokhinson {\it et al} conclude that the
polarization of the QPC for $G < G_{0}$ is a generic property of
QPCs and that the $0.7$ plateau is an extreme indicator of the
polarization that emerges in the case where the energy gap between
spin-up and spin-down components of a 1D subband is sufficiently
large~\cite{RokhinsonPRL06}.

\subsection{Scanning gate microscopy studies of QPCs}

The development of Scanning Gate Microscopy (SGM) has provided
important new insight to the study of nanoscale electronic
devices~\cite{TopinkaPhysTod03}. By using a biased atomic force
microscope tip as a mobile, local gate situated just above the
heterostructure surface it is possible to image the flow of
electrons through both open 2DEG~\cite{TopinkaNat01, JuraNP07} and
devices structures such as QPCs~\cite{TopinkaSci00, CrookSci06},
ballistic focussing geometries~\cite{AidalaNP07} and quantum
dots~\cite{CrookPRL03, BurkePRL10}. Indeed, the technique was first
demonstrated by imaging the mode structure corresponding to the 1D
subbands in a QPC~\cite{TopinkaSci00}. A scanning probe approach
known as erasable electrostatic lithography (EEL) can also be used
to define and fine-tune the structure of nanoscale
devices~\cite{CrookNat03}. It involves depositing negative charge
into the heterostructure surface at low temperature using a scanning
probe. This charge remains trapped as long as the device remains
cold, and can deplete the 2DEG in much the same way as a
negatively-biased surface gate does, allowing device structures to
be defined.

\begin{figure}
\includegraphics[width=14cm]{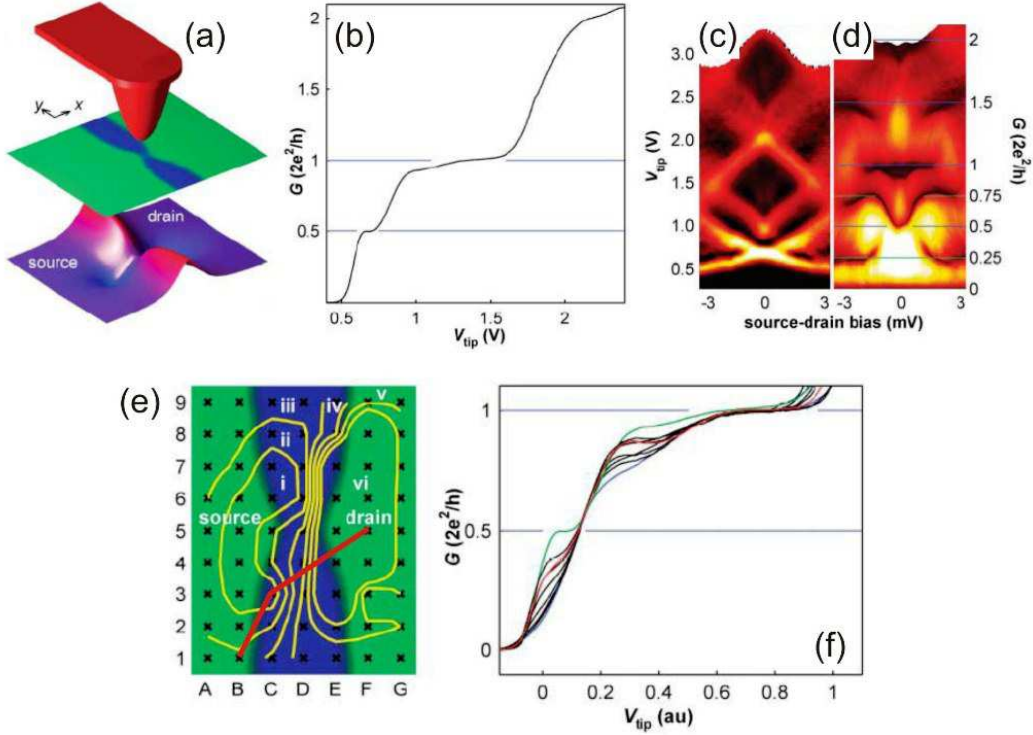}
\caption{(a) Schematic of the experimental setup showing the biased
tip (top), the heterostructure surface (middle) and the potential
profile in the 2DEG (bottom). The line of charge written by erasable
electrostatic lithography (EEL) is shown in blue and leads to a
saddle-shaped QPC potential in the 2DEG. (b) A typical plot of
conductance $G$ vs tip voltage $V_{tip}$, showing clear plateaus at
$0.5$, $1$ and $2G_{0}$ along with a weak plateau-like feature at
$0.9G_{0}$. (c) and (d) are colour maps of the transconductance
$dG/dV_{tip}$ vs source-drain bias ($x$-axis) and (c) $V_{tip}$
($y$-axis) and (d) $G$ ($y$-axis). The dark regions indicate low
transconductance (i.e., plateaus). (e) A map of tip positions
forming a $7 \times 9$ grid about the center of the QPC at which $G$
vs $V_{tip}$ was studied. The grid covers an area of $1.3 \times
1~\mu$m. The yellow contours delineate regions where the conductance
of the additional plateau at $0 < G < 0.5G_{0}$ occurs at $fG_{0}$
where i $f = 0.45$ to $0.5$, ii $f = 0.4$ to $0.45$, iii $f = 0.35$
to $0.4$, iv $f = 0.3$ to $0.35$, v $f = 0.25$ to $0.3$ and vi no
additional plateau observed. (f) Plots of $G$ vs $V_{tip}$ for a
number of positions on a path from F5 (blue trace) to B1 (red trace)
via C3 (green trace). The traces have been linearly scaled and
offset to achieve the alignment that appears in (f). Figure adapted
with permission from Ref.~\cite{CrookSci06}. Reprinted with
permission from AAAS.}
\end{figure}

Crook {\it et al}~\cite{CrookSci06} combined EEL and elements of SGM
to perform an interesting study of how conductance structures at $G
< G_{0}$ are linked to the potential landscape within a QPC. The
experimental set-up is illustrated in Fig.~21(a). The device
consists of an $8~\mu$m wide mesa structure etched into an
AlGaAs/GaAs heterostructure with a 2DEG $97$~nm beneath the surface.
A line of negative charge (blue) is drawn on the heterostructure
surface (green) by EEL to electrostatically deplete the 2DEG
underneath (lower blue/red layer), cutting it into source and drain
reservoirs, each with a set of ohmic contacts. The line of charge is
not perfectly even, and a narrow section of the line can be used as
a QPC, as shown in Fig.~21(a). The charge deposited by the EEL
technique can be imaged using Kelvin probe
microscopy~\cite{CrookPhysE06}, allowing such a section to be found.
The tip is located over this region, as shown in Fig.~21(a), and the
conductance versus tip bias $V_{tip}$ is measured, as shown in
Fig.~21(b). Quantized conductance plateaus at $G_{0}$ and $2G_{0}$
are observed, as expected, along with a strong plateau at $0.5
G_{0}$ and a weaker feature at $0.9G_{0}$ which evolves from the
$0.7$ plateau, as discussed further below.

Figure~21(c) shows a colour-map of transconductance $dG/dV_{tip}$
versus source-drain bias and $V_{tip}$. The dark regions indicate
low transconductance (i.e., plateaus), and the data has a very
similar appearance to that typically found in QPCs (c.f.,
Fig.~15(c)), demonstrating that there is nothing particularly
anomalous about using the tip bias to pinch off the QPC rather than
a set of surface split-gates. Comparing closely to Fig.~15(c)
however, there is one very notable difference in the data in
Fig.~21(c), which is that the dark bands corresponding to the $\sim
0.8G_{0}$ finite bias plateaus evolve together into a dark region at
zero source-drain bias, which is the $0.5G_{0}$ plateau in
Fig.~21(b). An interesting aspect of the finite bias plateaus is
shown in a slightly different colour-map in Fig.~21(d) where the
$y$-axis is $G$ rather than $V_{tip}$. This presentation is useful
because, unlike the plots with voltage on the $y$-axis, it
highlights the conductance of the various plateaus. Remarkably, in
this device the finite bias plateaus all appear quite accurately
quantized at integer multiples of $0.25G_{0}$ (i.e., $0.25$, $0.75$
and $1.5G_{0}$). Crook {\it et al} point out that this behaviour is
observed in quantum wires under strong in-plane magnetic
fields~\cite{CronenwettPRL02, ThomasPM98} and that combined with the
observation of a $0.5G_{0}$ plateau at zero magnetic field in
Fig.~21(b), it may indicate the presence of a relatively strong
spin-polarization close to pinch-off in this
device~\cite{CrookSci06}.

Deeper insight can be obtained by studying how $G$ versus $V_{tip}$
changes as the tip is moved to different positions near the center
of the QPC (i.e., narrow point in the EEL line). This is done for 63
different positions within a $1.3 \times 1~\mu$m window centered on
the QPC, with the sample points forming a $7 \times 9$ rectangular
grid as shown in Fig.~21(e). Figure~21(f) shows a representative set
of the $G$ versus $V_{tip}$ traces obtained for various points in
this grid and a number of features are evident. Firstly, when the
tip is located at position F5 (blue trace), the conductance takes
its normal appearance, with a clear $0.7$ plateau that is distinct
from the $G_{0}$ plateau, and no other features at $G < G_{0}$. The
tip is then moved to B1 (red trace) via C3 (green trace) and in each
case the conductance shows a plateau at $0 < G < 0.5G_{0}$, a
plateau at $0.5 < G < 1.0G_{0}$ and finally the usual integer
quantized plateau at $G_{0}$. This not only shows that the $0.9$
plateau in Fig.~21(b) evolves from the $0.7$ plateau, but that the
locations of these plateaus in conductance can be heavily dependent
upon the precise potential profile in the QPC, since this is what
the differently located tip positions will influence most. Note that
dual sub-$G_{0}$ plateaus are not commonly observed in asymmetric
biasing experiments of split-gate QPCs (see Fig.~2 for example), and
this may be an effect instead of a potential asymmetry along the QPC
axis rather than across it (i.e., along $x$ rather than $y$). This
is consistent with asymmetries in the source-drain bias
characteristics observed when the tip leads to strong asymmetry in
the QPC potential (see Fig.~2(c/d) of Ref.~\cite{CrookSci06}).

With this in mind, it is interesting to compare this with data
obtained by Shailos {\it et al}~\cite{ShailosJPCM06} who studied a
split-gate QPC with a narrow `finger' gate that extends along the
QPC axis, stopping just short of the opening of the QPC. This gate
allows an asymmetry along the QPC to be developed without disrupting
the symmetry in the QPC potential across the QPC (i.e., along the
gate line). Conductance data obtained by Shailos {\it et al} also
shows the presence of features at {\it both} $0 < G < 0.5G_{0}$ and
$0.5 < G < 1.0G_{0}$, and as in Ref.~\cite{CrookSci06}, these vary
in conductance such that when a clear plateau at $0.5G_{0}$ is
observed there is a strong feature at $\sim 0.9G_{0}$. The integer
plateaus in the data presented by Shailos {\it et al} are
comparatively weak sometimes, and although it is not clear exactly
why this is, it may be related to the action of the finger gate on
the QPC potential. A possible mechanism for this may be found in
spin-density functional theory calculations by Akis and
Ferry~\cite{AkisJPCM08}, which show that under certain circumstances
the conductance has two plateaus for $G < G_{0}$ corresponding to
the QPC passing more than one spin-down subband before the first
spin-up subband begins to conduct. This behaviour is connected to
barriers of different height for spin-up and spin-down electrons, as
found in preceding calculations~\cite{RejecNat06, BerggrenPRB02,
HirosePRL03, JakschPRB06}. The `shapes' of these barriers are
heavily dependent on device structure, electron distribution, etc.
due to their self-consistent nature, and thus are likely to be
highly sensitive to asymmetry along the QPC axis also.

We conclude this section by looking briefly at some other SGM
studies that provide evidence for the formation of a bound-state
within a QPC. Aoki {\it et al}~\cite{AokiJPCS06} conducted an SGM
study of two QPCs, the first was one of the two QPCs in an open
quantum dot, and the second was a standard QPC. Both devices were
defined by shallow wet etching of an AlGaAs/GaAs heterostructure
with a $60$~nm deep 2DEG. In both cases, ring-like structures were
observed that are consistent with localization of
electrons~\cite{SteelePRL05, SchnezPRB10}. Indeed, in
Ref.~\cite{SchnezPRB10} similar ring-like structures are observed in
the entrance and exit QPCs connecting a graphene quantum dot to its
source and drain reservoirs.

\section{Bound-state formation at $G < G_{0}$}

\subsection{Theoretical predictions of self-consistent bound-state formation in QPCs}

Early theoretical work related to the $0.7$ plateau is strongly
suggestive of the potential for the formation of a weak potential
minimum and a corresponding quasi-bound state within the QPC as it
is pinched off; for example, see Fig.~3(b) of Wang and
Berggren~\cite{WangPRB96} or Fig.~9 of work by Sushkov using
many-body Hartree-Fock calculations~\cite{SushkovPRB01}. The work by
Cronenwett {\it et al}~\cite{CronenwettPRL02} heightened interest in
this possibility since the Kondo mechanism relies entirely upon the
existence of such a state.

Wang and Berggren used a semiclassical approximation for the 1D
density $n_{1D}(x)$ along the QPC~\cite{WangPRB98} due to
limitations in computational resources. The unfortunate cost in this
approximation is a loss of the finer structure of $n_{1D}(x)$, in
particular, the Friedel oscillations (see Fig.~1 of
Ref.~\cite{WangPRB98}), which turn out to be important in terms of
the real self-consistent potential inside the QPC. Friedel
oscillations are periodic variations in electron density that emerge
from the combined effects of the electrons' wave nature and their
tendency to reorganize to screen any charge inhomogeneity that
arises within a solid (e.g., a charged impurity or a local rise in
potential, such as that occurring in a QPC)~\cite{FriedelAdvPhys54}.
The wavelength of the Friedel oscillations $\lambda_{fo} \sim
\pi/k_{F} \approx 80$~nm at $n = 10^{11}$~cm$^{-2}$ is comparable to
the length of a typical QPC ($200 - 600$~nm). Friedel oscillations
often play an important role in scattering/conduction at low
temperatures $T << E_{F}/k_{B}$, for example in 2D
systems~\cite{ProskuryakovPRL02, ZalaPRB01}. Recent work also
suggests that Friedel oscillations may lead to increases in
conductance for $G > 2e^{2}/h$ with increased temperature in
QPCs~\cite{RenardPRL08}.

\begin{figure}
\includegraphics[width=16cm]{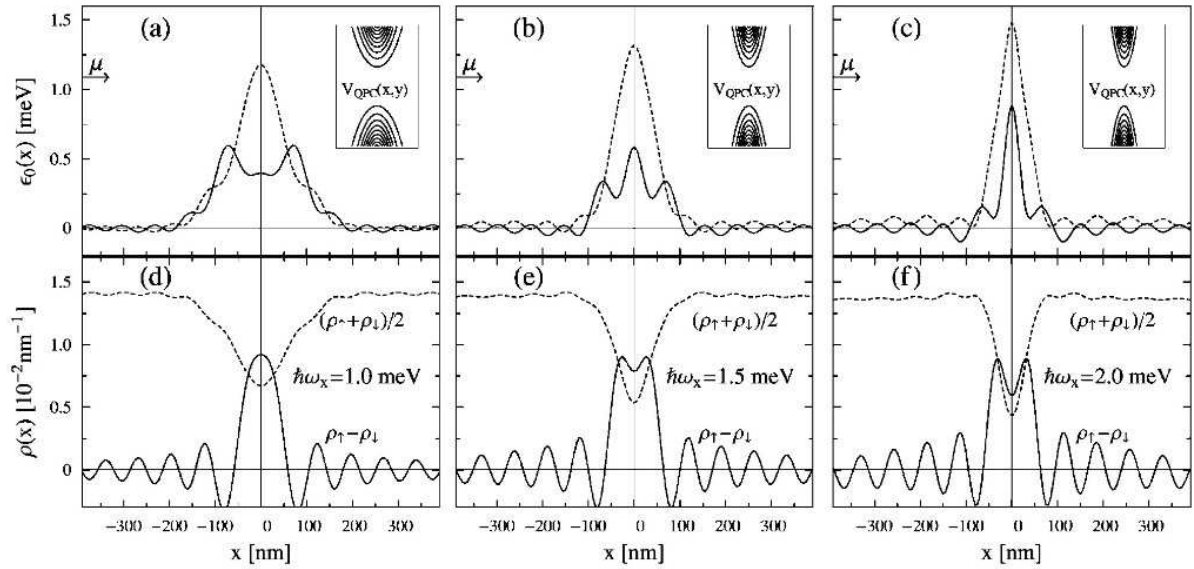}
\caption{(a-c) Self-consistent potentials $\epsilon_{0}(x)$ vs
position $x$ along the QPC axis calculated for spin-up (solid line)
and spin-down (dotted line) electrons at $T = 0.1$~K for
saddle-point potential length parameters $\hbar \omega_{x}$ of (a)
$1.0$~meV, (b) $1.5$~meV and (c) $2.0$~meV. The chemical potential
$\mu$ is indicated on the left in each panel. The insets in (a-c)
display contour plots of the QPC potential $V_{QPC}(x,y)$, which is
the standard saddle-point potential model for a
QPC~\cite{ButtikerPRB90}. (d-f) Corresponding plots of the
spin-averaged electron density
$(\rho_{\uparrow}+\rho_{\downarrow})/2$ (dotted line) and net
spin-up density $\rho_{\uparrow}-\rho_{\downarrow}$ (solid line) vs
$x$, the distance along the QPC axis from the center of the QPC.
Figure adapted with permission from Ref.~\cite{HirosePRL03}.
Copyright 2003 by the American Physical Society.}
\end{figure}

Hirose {\it et al} used a slightly different approach to solving the
Kohn-Sham Hamiltonian~\cite{HirosePRB95} that enabled them calculate
the self-consistent potential along the QPC and the resulting 1D
electron density with much greater resolution~\cite{MeirPRL02,
HirosePRL03}. Figure~22(a-c) show the calculated self-consistent
potentials for spin-up (solid lines) and spin-down electrons (dotted
lines) with three different saddle-point length parameters $\hbar
\omega_{x}$~\cite{ButtikerPRB90}. Increasing $\hbar \omega_{x}$
corresponds to shortening the QPC length. The large scale features
agree with earlier work by Wang and Berggren~\cite{WangPRB98},
however the Friedel oscillations clearly extend right into the
center of the QPC, become more pronounced than they are outside, and
cannot be safely ignored. Corresponding plots of the spin-averaged
electron density $(\rho_{\uparrow}+\rho_{\downarrow})/2$ (dotted
line) and net spin-up density $\rho_{\uparrow}-\rho_{\downarrow}$
(solid line) are shown in Fig.~22(d-f), and indicate a very large
spin-imbalance within the QPC, for example, when
$(\rho_{\uparrow}+\rho_{\downarrow})/2 =
\rho_{\uparrow}-\rho_{\downarrow}$ (i.e., the lines cross) then
$\rho_{\uparrow} = 3\rho_{\downarrow}$. The integrated net spin-up
density in all three cases corresponds to $0.85 - 0.93$ spins,
giving a local moment with spin-$\frac{1}{2}$ located within the
QPC.

\begin{figure}
\includegraphics[width=7cm]{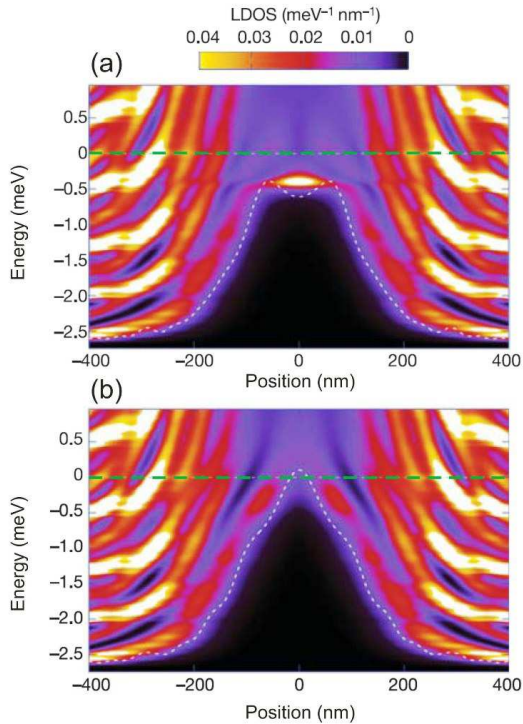}
\caption{Local density of states for (a) spin-up and (b) spin-down
obtained using spin-density functional theory with the local spin
density approximation. The spin-up potential features a quasi-bound
state located at the center of the QPCs sitting $0.5$~meV below the
Fermi energy. The spin-down potential features two weaker
quasi-bound states either side of the center of the QPC. The bright
bands to the far left and right in both cases are the 1D subbands in
the source and drain leads, these rise in energy above $E_{F}$ close
to the QPC, resulting in a QPC conductance less than $G_{0}$. In
both cases the white dashed lines are the Kohn-Sham potential and
the green horizontal dashed lines are the Fermi level. Reprinted by
permission from Macmillan Publishers Ltd, Ref.~\cite{RejecNat06},
Copyright 2006.}
\end{figure}

It is particularly interesting to note the spin-up potential in
Fig.~22(a), which has a double-barrier form that may support a
quasi-bound state. This was followed up in later calculations by
Rejec and Meir~\cite{RejecNat06}, who calculated the local density
of spin-up and spin-down states as a function of energy relative to
$E_{F}$ and position along the QPC, as shown in Fig.~23(a) and (b),
respectively. The calculations were performed using spin density
functional theory in the local spin density approximation
(LSDA)~\cite{ReimannRMP02}, with the spin densities in the 2DEG and
charge distribution in the gate electrodes obtained
self-consistently. The bright stripes to the far left and right in
Fig.~23(a/b) are the 1D subbands in the source and drain. The 1D
subbands rise in energy, exceeding $E_{F}$ close to the QPC, such
that the QPC is in the $G < G_{0}$ limit. For the spin-up case
(Fig.~23(a)), there is a clear quasi-bound state about $0.5$~meV
below $E_{F}$ as indicated by the high local density of states,
consistent with Ref.~\cite{HirosePRL03}. In the spin-down case
(Fig.~23(b)), there are two much weaker quasi-bound states either
side of the QPC center. Rejec and Meir argue that this may result in
a more complex, two-impurity Kondo effect~\cite{JonesPRL87,
JonesPRL88} influencing transport within the QPC~\cite{RejecNat06}.
The results obtained by Rejec and Meir are very length dependent --
for a very short QPC there is no bound-state formation at all,
whereas in a very long QPC, an antiferromagnetically ordered chain
of bound-states is obtained. We discuss chain formation for longer
QPCs and quantum wires further in Section 10.1.

The picture portrayed above is not a complete one, however, and
there appears to be some active debate about whether quasi-bound
states and spin polarization are robust outcomes of theoretical
investigations (i.e., emerge repeatedly despite small variations in
definition of the system, boundary conditions, treatment of exchange
and correlation terms, etc.). Calculations by Berggren {\it et
al}~\cite{BerggrenPRB02, BerggrenJPCM08} and Starikov {\it et
al}~\cite{StarikovPRB03} using the Kohn-Sham local spin-density
functional theory find that there are no quasi-bound states formed
within the QPC but spin-polarized solutions can be obtained. A
similar outcome was obtained by Havu {\it et al}~\cite{HavuPRB04}.
Jaksch {\it et al}~\cite{JakschPRB06} also perform
density-functional theory calculations; they obtain no bound states,
and instead find that there is no spin-polarization in very short
QPCs with spin-polarization developing with increasing QPC length.
Sushkov~\cite{SushkovPRB03} performed calculations using restricted
and unrestricted Hartree-Fock methods and found that the presence
and character of a bound state is dependent upon the length of the
QPC, but that the total spin of such a bound state is zero. Finally,
Ihnatsenka and Zozoulenko~\cite{IhnatsenkaPRB07} find results in
agreement with Rejec and Meir~\cite{RejecNat06}, and contradicting
Jaksch {\it et al}~\cite{JakschPRB06}, regarding the absence of spin
polarization in short QPCs and the development of spin-polarization
with an increase in length, also using spin density functional
theory. Clearly, from a theoretical perspective the generation of
quasi-bound states and spin-polarization is heavily dependent on how
the calculations are performed, and further work is needed for a
consensus to emerge. We now shift our focus back towards
experimental work, and in particular, the evidence for quasi-bound
state formation in QPCs.

\subsection{Experimental studies of bound state formation in QPCs}

Insight regarding the formation of a quasi-bound state within a QPC
as it pinches off can be obtained by studying more complex gate
geometries that define devices featuring coupled QPCs. Morimoto {\it
et al}~\cite{MorimotoAPL03} studied the geometry shown in
Fig.~24(a), featuring a pair of split-gates: the central split-gate
(gates 2 and 3) running horizontally defines a small square quantum
dot, located between a second split-gate (gates 1 and 4) running
vertically. Although all four gates can be operated independently,
gates 1,2, and 3 are held at a fixed voltage of $-1.20$~V, while
gate 4 is swept from $0$ to $-2$~V. The device contains eight ohmic
contacts, two each located in the four corners, and four-terminal
conductance measurements were obtained using the two configurations
shown in Fig.~24(b/c), meaning there are two QPCs of importance to
this study. These two QPCs are coupled via a central quantum dot
formed by gates 2 and 3. The upper QPC separates the 2D reservoirs
in the upper left and right corners, and is coupled to the quantum
dot via the opening formed by the two upper spurs on gates 2 and 3.
This QPC has a fixed potential profile and width when gates 1, 2 and
3 are held constant. On the opposite side of the quantum dot, the
lower QPC is formed by the two lower spurs on gates 2 and 3. The
width of the lower QPC is tuned by the voltage applied to gate 4,
and the conductance thus obtained for the measurement configuration
in Fig.~24(c) is presented as the solid line in Fig.~24(d). Although
there is clear step-like structure in this trace, the steps do not
sit at quantized values of $G_{0}$. This should be the case, as the
`impedance' of the quantum dot and upper QPC act as an effective
contact resistance for the measurement. If a fixed contact
resistance of $4747~\Omega$ is subtracted, then the dashed line in
Fig.~24(d) is obtained where the conductance steps do sit at
expected quantized values. This conductance quantization is due to
pinching off of the lower quantum dot entrance by the bias applied
to gate 4. The quality of the steps indicates that the upper quantum
dot entrance is relatively unaffected by gate 4.

\begin{figure}
\includegraphics[width=16cm]{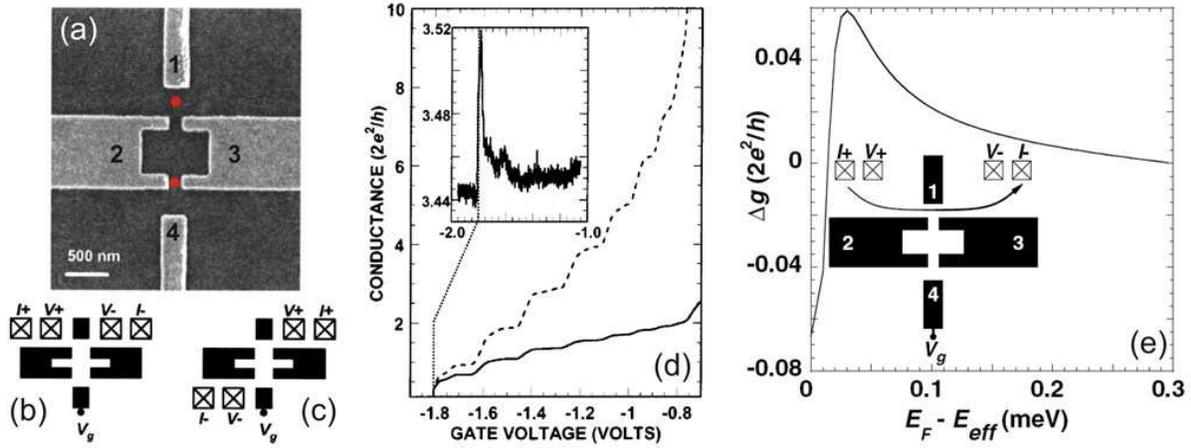}
\caption{(a) Scanning electron micrograph of the device studied by
Morimoto {\it et al} featuring a horizontal split-gate (gates 2 and
3) defining a quantum dot and a wider vertical split-gate (gates 1
and 4). The two QPCs considered in the measurement are located at
the red dots and are coupled via the quantum dot. (b) and (c) the
two measurement configurations used in the experiment. (d)
Conductance vs voltage on gate 4 with gates 1,2 and 3 held fixed at
$-1.20$~V. Data in the main panel was obtained for the configuration
in (c) and presented uncorrected (solid line) and with a fixed
contact resistance of $4747~\Omega$ removed (dashed line) to correct
for the series impedance of the quantum dot. The data in the inset
was obtained for the configuration in (b). (e) Corresponding
theoretical calculations for the data inset to (d) obtained using an
Anderson model. The plotted data is the conductance correction to
the upper QPC $\Delta g$ vs the separation between the Fermi energy
$E_{F}$ and the energy of the quasi-bound state $E_{eff}$ formed in
the lower QPC as it pinches off. Figures (a-d) adapted with
permission from Ref.~\cite{MorimotoAPL03}. Copyright 2003, American
Institute of Physics. Figure (e) adapted with permission from
Ref.~\cite{PullerPRL04}. Copyright 2004 by the American Physical
Society.}
\end{figure}

The interesting aspect of this experiment is shown in the inset to
Fig.~24(d), where the conductance through the upper QPC, obtained
using the configuration in Fig.~24(b), is plotted versus the bias on
gate 4. This configuration represents a `non-local' measurement of
the lower QPC by the upper QPC. Ignoring any interactions, the
conductance of the upper QPC should be fixed at $\sim 3.45G_{0}$,
this conductance being set by the bias on gates 1,2 and 3,
independent of the bias on gate 4 (as indicated by the subtraction
of a {\it fixed} contact resistance giving correctly quantized data
in the dashed line in Fig.~24(d)). However, a very sharp peak in the
upper QPC conductance is observed when gate 4 has a bias of
$-1.8$~V, and this coincides with the pinch off of the lower QPC.
Similar behaviour was obtained by performing the mirror image
experiment (e.g., fixing the bias on gates 2,3, and 4 and measuring
the conductance of the lower QPC as the upper QPC is pinched off
using gate 1), and for different bias settings on gates 1,2, and 3.
This includes a configuration where one or other of gates 2 and 3 is
held at ground potential, and the coupling between the two QPCs is
mediated by a large area of 2DEG~\footnote{In this configuration the
lower QPC shifts from being between the spurs on gates 2 and 3 to
between gate 2 or 3 and gate 4, with a corresponding increase in
pinch-off bias due to increased gate
separation.}~\cite{MorimotoAPL03}. The height of this peak is $0.08
G_{0}$ and subsequent work by Shailos {\it et
al}~\cite{ShailosSST04} reported that the peak height shows a small
linear increase from $0.05$ to $0.09G_{0}$ as the conductance of the
upper QPC is increased from $\sim G_{0}$ to $\sim 6G_{0}$. Morimoto
{\it et al}~\cite{MorimotoAPL03} suggest that the peak may be
connected to the presence of a localized spin in the lower QPC at
pinch-off~\cite{MeirPRL02,HirosePRL03}, and theoretical work by
Puller {\it et al}~\cite{PullerPRL04, PullerJPCM05} provides support
for this suggestion (see also Ref.~\cite{BirdSci04}).

Puller {\it et al} begin with the assumption that a net spin moment
is formed in the lower QPC at pinch-off~\cite{RejecNat06,
BerggrenPRB02, HirosePRL03}, and used an Anderson-type
model~\cite{AndersonPR61} that included a correction term to account
for the coupling between the two QPCs to calculate the conductance
of the upper QPC~\cite{PullerPRL04}. An important outcome of the
model is that the coupling between the QPCs enhances the normal 1D
density of states in the upper QPC at the energy $E_{eff}$ of the
quasi-bound state formed in the lower QPC. Since the upper QPC
conductance depends on the derivative of the upper QPC density of
states, this generates a correction to the upper QPC conductance
$\Delta g$, which is plotted versus the separation $E_{F} - E_{eff}$
between the Fermi energy and the energy of the bound state in the
lower QPC in Fig.~24(e). This correction bears a clear resemblance
to the experimentally observed peak inset to Fig.~24(d), suggesting
that the upper QPC can be used as a detector for local
magnetic-moment formation in a nearby QPC. It is worth noting that
under this model the conductance correction is proportional to the
tunneling probability between the QPCs. This is dominated by the
height of the tunnel barrier in the lower QPC for $G < G_{0}$ and is
relatively independent of the conductance of the upper QPC,
explaining the result obtained by Shailos {\it et
al}~\cite{ShailosSST04}. Furthermore, the model makes no reliance on
the nature of the energy level structure in the region mediating the
coupling, explaining why the peak remains if either gate 2 or 3 is
left unbiased~\cite{MorimotoAPL03, ShailosSST04}.

Subsequent work by Mourokh {\it et al}~\cite{MourokhAPL05} suggested
that the signal inset to Fig.~24(d) is actually a Fano resonance,
and further evidence of bound-state formation in QPCs was obtained
in a series of experiments by Yoon {\it et al}~\cite{YoonPRL07,
YoonPRB09, YoonAPL09}, showing the presence of a Fano resonance in
the conductance of a `detector' QPC coupled to a QPC that is swept
towards pinch-off. Prior to discussing these experiments, I will
make a brief interlude to introduce some essentials regarding Fano
physics.

\subsection{A short primer on Fano physics}

The Fano effect has its roots in atomic physics, beginning with a
seminal paper by Ugo Fano in 1935~\cite{FanoNC35} explaining the
strange lineshapes observed in the absorption spectra of noble
gases~\cite{BeutlerZP35}. The lineshapes occur when a discrete state
in the spectrum interferes with a continuum of states amongst which
it resides. A subsequent paper by Fano in 1961~\cite{FanoPR61},
covering the same concept but to greater theoretical depth, has been
cited over 5800 times and has become one of the most influential
papers in the history of {\it The Physical Review}. The `classic'
Fano lineshape obtained experimentally from absorption studies of He
is shown in Fig.~25(a). Moving from left to right, it consists of a
sharp, almost asymptotic rise from an initial background, an even
sharper drop to a minima, and a second asymptotic rise back towards
the background level~\cite{FanoPR65}. However, this lineshape is but
one of a family of such lineshapes predicted by the Fano lineshape
formula:

\begin{equation}
\sigma(\epsilon) =
\sigma_{a}[(q+\epsilon)^{2}/(1+\epsilon^{2})]+\sigma_{b}
\end{equation}

\noindent where $\sigma(\epsilon)$ is the absorption cross-section
for incident photons of energy $E$ and $\epsilon = (E -
E_{r})/\frac{1}{2}\Gamma$ is the separation between $E$ and a
resonance energy $E_{r}$ due to a discrete auto-ionizing level in
the atom, with $\Gamma$ being the lifetime broadening of the
discrete level. The prefactors $\sigma_{a}$ and $\sigma_{b}$
represent the two components of the spectrum corresponding to states
in the continuum that do and do not interact with the discrete
level. Finally, the Fano parameter $q$ characterizes the lineshape,
which starts out as a symmetric minima or antiresonance for $q = 0$,
becoming highly asymmetric at $q = 1$, as shown in Fig.~25(b), and
ultimately takes the Lorentzian Breit-Wigner
lineshape~\cite{BreitPR36} (not shown) as $q \rightarrow \infty$.

\begin{figure}
\includegraphics[width=7cm]{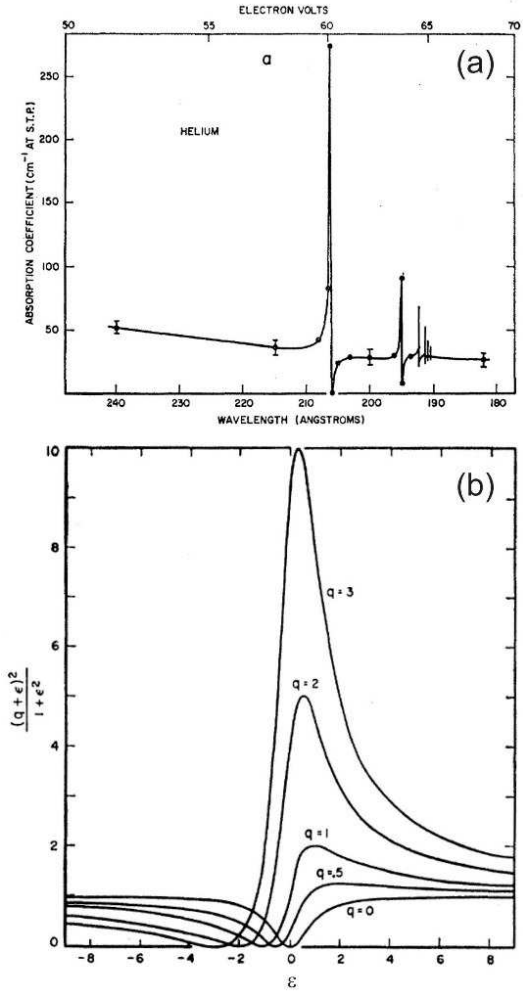}
\caption{(a) Absorption spectrum for He near the $n = 2$ level of
He$^{+}$ obtained by Madden and Codling. (b) Calculated Fano
lineshapes for different values of $q$. Figure (a) adapted with
publisher's permission from Ref.~\cite{FanoPR65}. Copyright 1965 by
the American Physical Society. Figure (b) adapted with publisher's
permission from Ref.~\cite{FanoPR61}. Copyright 1961 by the American
Physical Society.}
\end{figure}

The great significance in Fano's work lies in its ubiquity;
interaction between discrete and continuous states extends well
beyond atomic absorption spectra, having been observed in electron
and neutron scattering~\cite{AdairPR49, SimpsonPRL63}, organic
impurities in rare gas solids~\cite{PyshPRL65}, Raman
scattering~\cite{CerdeiraPRB73}, quantum well optical
absorption~\cite{FaistNat97}, STM studies of single magnetic atoms
on surfaces~\cite{MadhavanSci98}, and most recently in semiconductor
single-electron transistor devices~\cite{GoresPRB00}. The latter was
the first in a number of notable observations of Fano resonances in
the conductance of mesoscopic semiconductor devices that have seen
the phenomenon gradually evolve into a method for the study of the
interaction between a continuum of states in a 1D system and the
discrete energy level states in a quantum dot. The advantage of
performing such studies in these devices is the unprecedented
opportunity to tune the properties of the device to study a range of
Fano physics behaviours~\cite{NockelPRB94}.

\begin{figure}
\includegraphics[width=16cm]{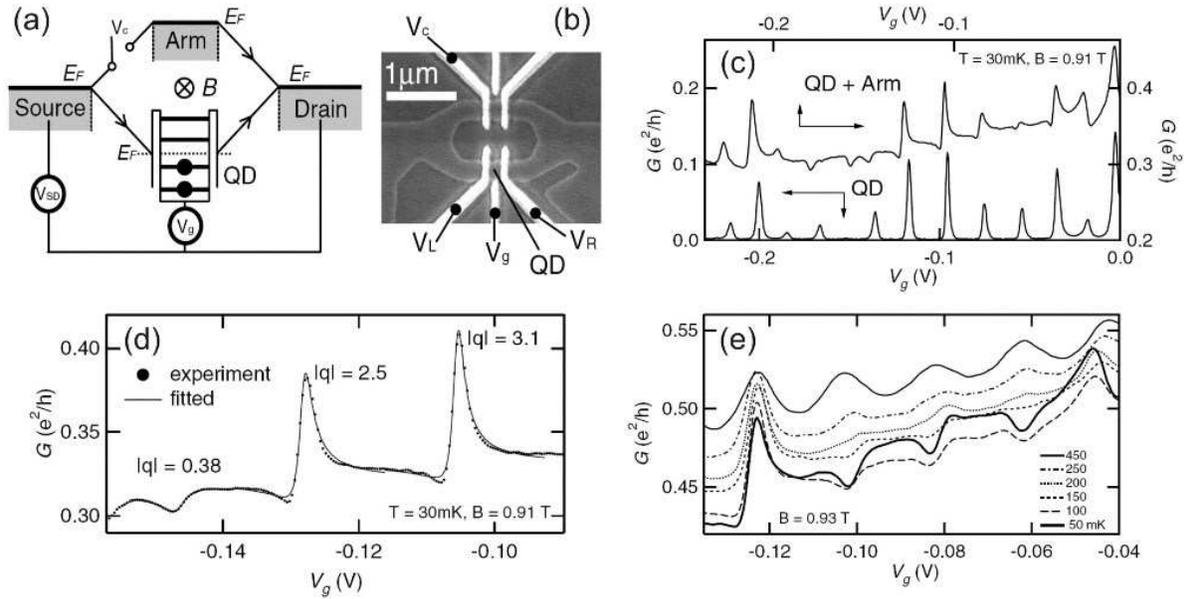}
\caption{(a) Schematic and (b) scanning electron micrograph of an
Aharonov-Bohm ring interferometer with a quantum dot embedded in one
arm, as studied by Kobayashi {\it et al}~\cite{KobayashiPRL02}. (c)
Conductance $G$ vs gate voltage $V_{g}$ for the quantum dot with the
other arm of the interferometer severed (left axis - lower trace)
and with both arms conducting (right axis - upper trace), showing
the emergence of Fano resonances due to interference between the
continuum of states in the arm and discrete states in the quantum
dot. (d) Fits of the Fano lineshape to three selected peaks from
(c). (e) A section of the data in (c) plotted at six different
temperatures demonstrating that the Fano effect gradually disappears
as the temperature is raised. Figure adapted with permission from
Ref.~\cite{KobayashiPRL02}. Copyright 2002 by the American Physical
Society.}
\end{figure}

A key experiment in the study of the Fano effect in mesoscopic
systems was performed by Kobayashi {\it et
al}~\cite{KobayashiPRL02}, who studied an Aharonov-Bohm
interferometer~\cite{AharonovPR59, TimpPRL87} with a quantum dot
embedded in one of the arms. The device used is shown schematically
in Fig.~26(a), with a scanning electron micrograph of the device
shown in Fig.~26(b). The continuum of states is supplied by the
upper arm of the interferometer. This arm can be cut by applying a
negative bias $V_{c}$ to a gate running over it, allowing the
quantum dot to be measured independently. The quantum dot is
operated in the Coulomb blockade regime and provides the discrete
state required to observe the Fano effect. The conductance of the
quantum dot versus gate voltage $V_{g}$ is shown in the lower trace
in Fig.~26(c), and exhibits clear Coulomb blockade oscillations
indicative of strongly resolved discrete energy level structure
within the dot~\cite{MeiravPRL90}. Reconnecting the arm to the
circuit leads to the conductance shown in the upper trace in
Fig.~26(c). Here, interference occurs between the continuum states
in the arm and each discrete level of the dot when it passes through
a window of width $\sim k_{B}T$ centered at the Fermi energy,
causing the characteristic asymmetric Fano lineshape to appear where
the Coulomb blockade peaks were previously located. Figure~26(c)
shows the quality of fit that the Fano form (Eq.~10 with
cross-sections $\sigma$ replaced by conductances $G$ or Eq.~11)
makes to the conductance, and highlights that the Fano factor $q$
can evolve considerably with the specifics of the coupling. Finally,
Fig.~26(e) demonstrates the importance of coherence in the Fano
effect, with the asymmetric Fano lineshapes observed at low
temperature evolving into the Breit-Wigner (Lorentzian) lineshape
for $|q| \rightarrow \infty$ as the temperature is increased.

\begin{figure}
\includegraphics[width=12cm]{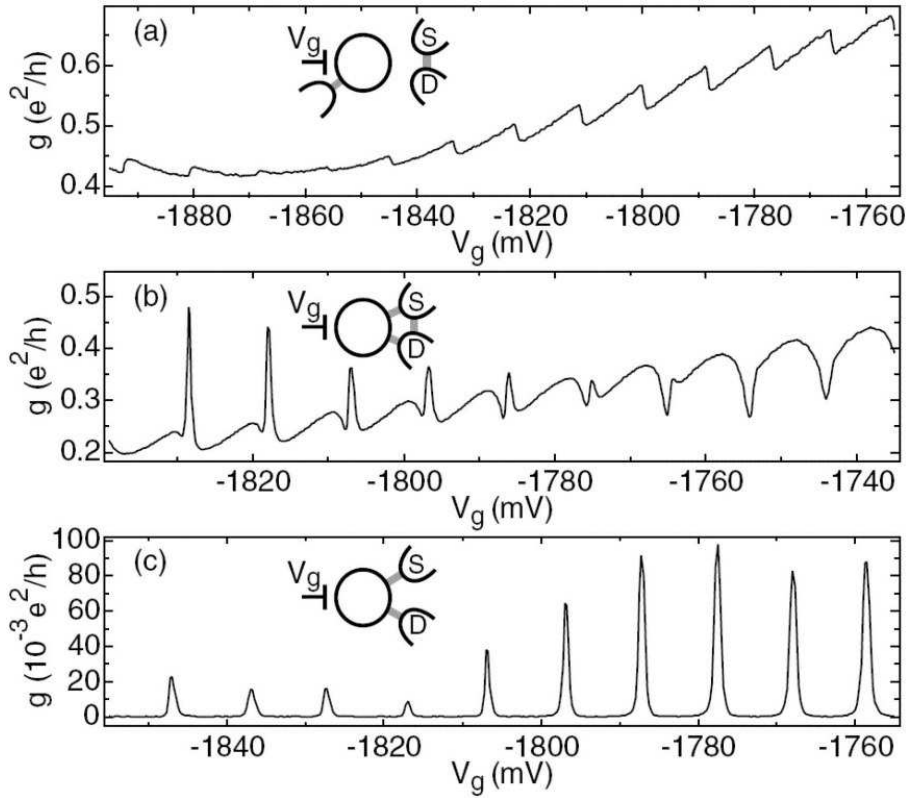}
\caption{Conductance $g$ vs gate voltage $V_{g}$ for three different
configurations of a device consisting of a 1D channel with a quantum
dot coupled into one of the side-walls: (a) current passes from
source to drain via the 1D channel only, which is capacitively
coupled to the quantum dot; (b) current passes via both the 1D
channel and quantum dot, the resulting interference producing Fano
resonances in the conductance; and (c) current passes only via the
quantum dot, resulting in Coulomb blockade oscillations. Figure
adapted with permission from Ref.~\cite{JohnsonPRL04}. Copyright
2004 by the American Physical Society.}
\end{figure}

The experiment by Kobayashi {\it et al}~\cite{KobayashiPRL02}
demonstrates the interesting possibility that the conductance though
a 1D system can be sensitive to coupling to a nearby discrete state
participating in the transport. However, the configuration need not
be as contrived as an Aharonov-Bohm interferometer with an embedded
quantum dot for this to occur. For example, Johnson {\it et
al}~\cite{JohnsonPRL04} investigated a 1D channel with a quantum dot
coupled to the side in such a way that with appropriate
configuration of the gates, current can flow from source to drain
via the 1D channel alone with the dot isolated, via the dot alone,
and via both. When current passes via the 1D channel alone, a
sawtooth signal is observed as shown in Fig.~27(a). This signal
arises due to capacitive coupling, with the 1D system acting as a
charge sensor for the dot~\cite{FieldPRL93}. When current passes
only via the dot, standard Coulomb blockade oscillations are
observed, as shown in Fig.~27(b). However, when both paths are
active, strong Fano resonance structure is observed as shown in
Fig.~27(c).

\subsection{Experimental studies of bound-state formation in QPCs - Continued}

Yoon {\it et al}~\cite{YoonPRL07} studied the device shown in
Fig.~28(a), which consists of eight surface gates G$_{1}$ to G$_{8}$
forming a cross-like pattern with a pair of ohmic contacts located
in each corner. The device is operated in various configurations
with one common theme -- one pair of gates is used to define a
`detector' QPC in close proximity to a second `swept' QPC, and their
conductances $G_{d}$ and $G_{s}$, respectively, are monitored as the
swept QPC is pushed towards pinch-off. Comparing back to Fig.~24,
the detector and swept QPCs in this experiment correspond to the
upper and lower QPCs, respectively. Figures~28(b-e) show $G_{d}$
(red trace) and $G_{s}$ (black trace) obtained versus the gate
voltage $V_{g}$ applied to the pair of gates indicated in red that
define the swept QPC, for four different device
configurations~\cite{YoonPRL07}. In each case $G_{s}$ falls toward
zero as $V_{g}$ is made more negative, showing a clear $0.7$
plateau, as highlighted by the horizontal black arrows. The data
shown in Fig.~28 was obtained at $T = 4.2$~K, which is why integer
plateaus are not observed. In contrast, $G_{d}$ decreases
approximately linearly with $V_{g}$ due to capacitive coupling, but
in each case $G_{d}$ shows a sharp peak that occurs just after
pinch-off in the swept QPC. The observation of this behaviour in
four separate pairs of QPCs confirms that it is not due to random
impurities. The sharp peak observed in $G_{d}$ is consistent with
that obtained by Morimoto {\it et al}~\cite{MorimotoAPL03}. As found
in Ref.~\cite{MorimotoAPL03}, the general appearance of the peak is
relatively independent of the conductance through the detector QPC.

\begin{figure}
\includegraphics[width=14cm]{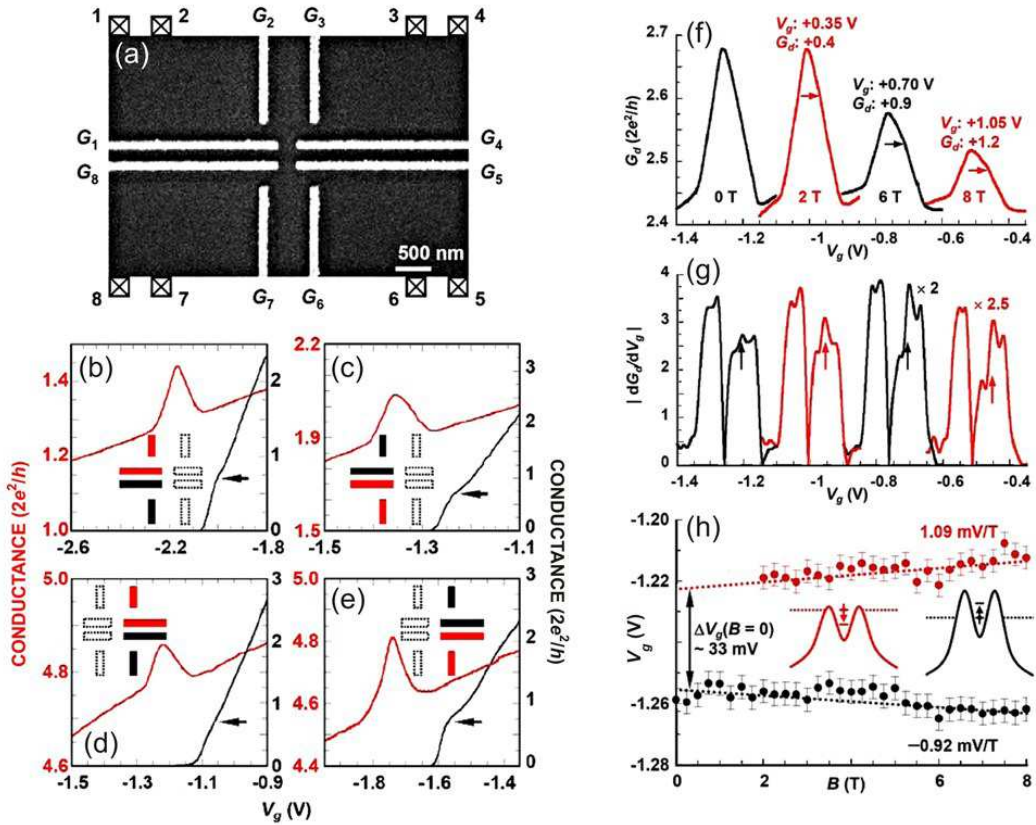}
\caption{(a) Scanning electron micrograph of the device studied by
Yoon {\it et al} (b)-(e) Swept QPC conductance $G_{s}$ (black trace
- right axis) and detector QPC conductance $G_{d}$ (red trace - left
axis) vs gate voltage $V_{g}$ applied to the swept QPC with the
detector bias held constant for four different configurations of
gates defining nominally identical coupled QPC pairs (detector QPC
in red, swept QPC in black, remaining four gates grounded). In each
case a clear $0.7$ plateau is observed in $G_{s}$, and a peak is
observed in $G_{d}$ just after the swept QPC pinches off,
demonstrating that the effect is not connected to random impurity
scattering or disorder. (f/g) Plots of (f) conductance $G_{d}$ and
(g) its derivative $|dG_{d}/dV_{g}|$ vs swept QPC gate voltage
$V_{g}$ for in-plane magnetic fields $B = 0, 2, 6$ and $8$~T,
respectively. The traces for $B > 0$~T have been offset in $V_{g}$
by increments of $+0.35$~V, and offset in $G_{d}$ or multiplied in
$|dG_{d}/dV_{g}|$ as annotated in the figure for clarity. The arrows
in both case highlight the weak, right-most shoulder due to
population of the upper Zeeman branch. (h) The $V_{g}$ position of
the left-most (black) and right-most (red) shoulders of the $G_{d}$
peak vs $B$. The peaks have a finite splitting at $B = 0$~T. The two
insets are schematics illustrating population of the upper (left)
and lower (right) branches of the quasi-bound state formed as the
QPC pinches off. Figure adapted with permission from
Ref.~\cite{YoonPRL07}. Copyright 2007 by the American Physical
Society. Society.}
\end{figure}

Figures 28(f-h) present the results of a study of how the $G_{d}$
peak responds to an in-plane magnetic field, which would induce a
Zeeman splitting $g^{*}\mu_{B}B$ in the discrete quasi-bound state
formed in the swept QPC. The evolution of the peak itself is shown
in Fig.~28(f), with each peak at successively higher $B$ offset by
$+0.35$~V in $V_{g}$ for clarity. Although at first sight the peak
appears to simply broaden, the peak actually divides, with a strong
shoulder moving to more negative $V_{g}$ and a much weaker shoulder,
highlighted by the horizontal arrows, moving to more positive
$V_{g}$. The presence of twin peaks is evident in Fig.~28(g), where
the derivative $|dG_{d}/dV_{g}|$ is plotted versus $V_{g}$ for each
of the four peaks in Fig.~28(f), and can be explained by the two
schematics shown inset to Fig.~28(h). The right inset (black) in
Fig.~28(h) shows the lower Zeeman branch of the one-electron
ground-state, which moves down in energy with increasing $B$,
shifting the corresponding left-most shoulder in the $G_{d}$ peak to
more negative $V_{g}$. Correspondingly, the right-most shoulder
moves to more positive $V_{g}$ as the upper Zeeman branch moves up
in energy with increasing $B$. The lower occupation probability for
the upper Zeeman branch explains why the right-most shoulder is much
weaker than its counterpart. Figure~28(h) shows the peak location
$V_{g}$ for the left-most (black) and right-most (red) peaks versus
$B$, and interestingly these peaks do not converge when extrapolated
to $B = 0$, suggestive of a spontaneous spin-polarization of the
quasi-bound state at zero field, consistent with suggestions by
Thomas {\it et al}~\cite{ThomasPRL96} and Wang and
Berggren~\cite{WangPRB98}. Based on the residual splitting, and
assuming the bulk $g^{*}$ for GaAs, this corresponds to a zero-field
spin-splitting of $\sim 0.8$~meV. However, this is likely an
under-estimate due to the 1D enhancement of
$g^{*}$~\cite{ThomasPRL96, WangPRB96}. A more precise measurement
can be obtained using non-local source-drain bias
spectroscopy~\cite{YoonAPL09}. Using these measurements, Yoon {\it
et al} found an effective $g$-factor of $|g^{*}| = 2.0 - 2.7$
indicating a zero field spin-splitting of as much as $4$~meV.
Although this energy is relatively large, it is consistent with
temperature dependence studies of the $G_{d}$ peak showing that it
survives to temperatures above $35$~K~\cite{YoonPRL07, YoonPRB09}.
This zero-field spin-splitting is inconsistent with a Kondo
model~\cite{MeirPRL02, HirosePRL03}, but in accordance with models
predicting a static spin-polarization within the QPC at
pinch-off~\cite{BerggrenPRB02, StarikovPRB03, HavuPRB04}.

In subsequent work, Yoon {\it et al}~\cite{YoonPRB09} investigated
the influence of the coupling between the two QPCs in more detail.
The separation and relative geometries of the QPC pair were varied
by activating different combinations of the eight available gates in
the device, as shown in the insets to Fig.~29(a-e), with the
detector QPC gates in red and the swept QPC in blue. At maximal
separation, the peak in $G_{d}$ that occurs as the swept QPC pinches
off, as shown in Fig.~29(a), has a similar appearance to that shown
in Figs.~28(b-e)~\cite{YoonPRL07}. However, as the QPCs become more
proximal, the $G_{d}$ peak becomes considerably more asymmetric,
taking a form that is very reminiscent of the Fano lineshape,
consistent with predictions by Mourokh {\it et
al}~\cite{MourokhAPL05}. The red lines in Figs.~29(a-c) are fits of
the Fano form for the conductance:

\begin{equation}
G_{d}(\epsilon) \propto (q + \epsilon)^{2}/(1+\epsilon^{2})
\end{equation}

\noindent where $\epsilon = 2(V_{S} - V_{0})/\Gamma$, with $V_{0}$
and $\Gamma$ the bound-state potential and width, and $V_{S}$ the
swept QPC gate voltage. The Fano factor $q$ is a free parameter for
the fit, and takes best fit values of $-20$, $-9$ and $-1$
respectively for the data in Fig.~29(a), (b) and (c). This trend in
$q$ is directly related to the separation between the two QPCs, as
shown in Fig.~29(d), and it is interesting to compare this data with
the work by Johnson {\it et al}~\cite{JohnsonPRL04} in Fig.~27.
There a strongly asymmetric lineshape in the conductance is observed
when transport occurs via {\it both} the 1D channel and quantum dot,
and is lost as the connection of the transport path to the 1D
channel or quantum dot is broken. Working from right to left in
Fig.~29(a-c) the same occurs, and the question becomes one of how
the interference between the discrete state and continuum is broken
in the process. The sawtooth-like appearance of the $G_{d}$ peak in
Fig.~29(a) at first suggests evolution to the charge sensing
arrangement in Fig.~27(a). The presence of capacitive coupling can
be tested by biasing a fifth gate, as shown inset to Fig.~29(f), to
sever the direct connection between the two QPCs via the electron
gas. A comparison of $G_{d}$ under identical conditions with the
fifth gate grounded and defined is shown as the red traces in
Figs.~29(e) and (f), respectively. The disappearance of the $G_{d}$
peak indicates that direct interference between the two QPCs is
vital to this feature.

\begin{figure}
\includegraphics[width=14cm]{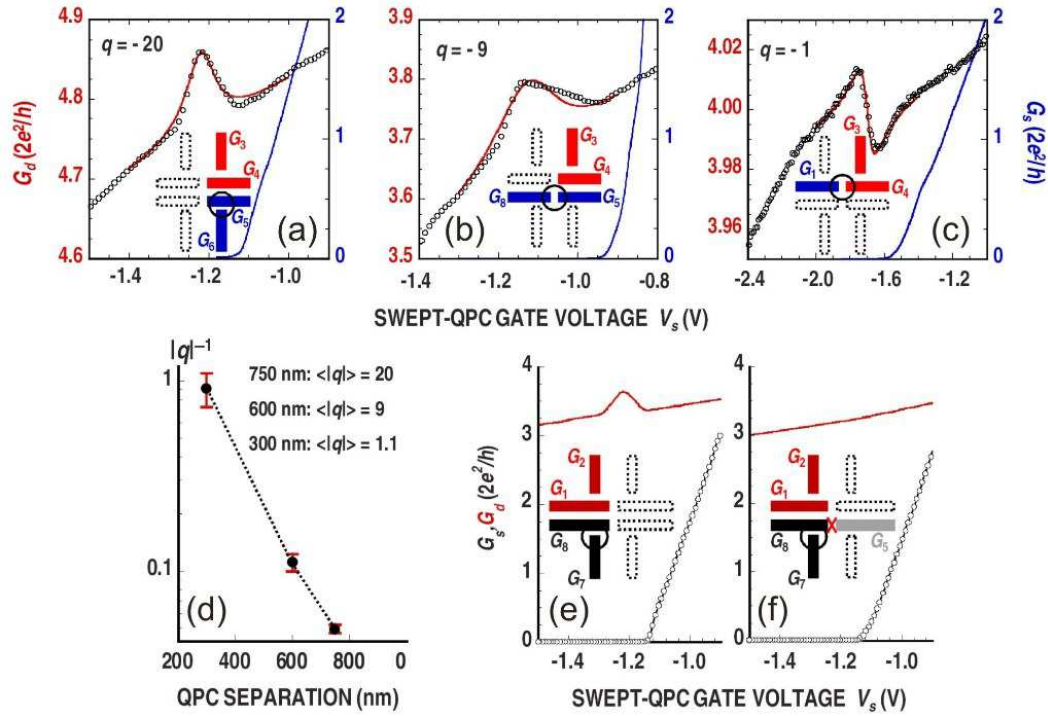}
\caption{(a) - (c) Plots of $G_{s}$ (blue trace - right axis) and
$G_{d}$ (red trace and data points - left axis) vs swept QPC gate
voltage $V_{S}$ for three different configurations giving different
separations between the swept and detector QPCs. The red lines are
fits of the Fano lineshape to the data points with the best fit Fano
factor $q$ listed in each panel. (d) Plot of inverse Fano factor
$|q|^{-1}$ vs QPC separation for the data presented in (a) - (c).
(e) and (f) $G_{s}$ (black trace/lower) and $G_{d}$ (red
trace/upper) vs $V_{S}$ with gate 5 grounded (d) and defined (e).
The disappearance of the peak in $G_{d}$ demonstrates that the
coupling between the QPCs is mediated via the electron gas rather
than simply being due to capacitive coupling, as in
Ref.~\cite{FieldPRL93}. Figure adapted with permission from
Ref.~\cite{YoonPRB09}. Copyright 2009 by the American Physical
Society.}
\end{figure}

To obtain further insight into the separation dependence, it is
worth considering the physical origin of the Fano factor $q$. In
Ref.~\cite{FanoPR61}:

\begin{equation}
q = \frac{(\Phi|T|i)}{\pi V_{E}^{*}(\psi_{E}|T|i)}
\end{equation}

\noindent where $(\Phi|T|i)$ and $(\psi_{E}|T|i)$ are the
transmission coefficients to go from the initial state $i$ to the
discrete state $\Phi$ and continuum $\psi_{E}$, respectively. These
interfere with opposite phase on opposite sides of the resonance.
The term $V_{E}^{*} = (\Phi|H|\psi_{E})$, where $H$ is the
Hamiltonian for the system, represents the strength of the
interaction between the discrete state and continuum, and is related
to the discrete state width $\Gamma =
2\pi|V_{E}^{2}|$~\cite{FanoPR61}. The result in Fig.~29(e/f)
suggests that the denominator terms in Eq.~12 are responsible for
the increasing $q$ as the QPCs are separated, and this may be a
signal of either weaker coupling between the two channels (i.e.,
reduced ($\Phi|H|\psi_{E}$)) or a suppressed continuum contribution
for the Fano process (i.e., reduced ($\psi_{E}|T|i$)) or perhaps
some combination of the two~\cite{BirdPC}. By comparison to Johnson
{\it et al}~\cite{JohnsonPRL04}, this suggests that the $G_{d}$ peak
evolves into a Coulomb blockade peak with increased separation
between the two QPCs. This corresponds to the increasing $q$, which
drives the asymmetric Fano resonance towards being a symmetric
Lorentzian peak. Further studies into this process would be
interesting, however, this work~\cite{MorimotoAPL03, YoonPRL07,
YoonPRB09, YoonAPL09} provides clear experimental evidence for the
formation of a self-consistent bound-state within a QPC as it is
pinched off, as predicted by many theoretical
studies~\cite{MeirPRL02, RejecNat06, HirosePRL03}. As a final note
on this topic, very recent work by Wu {\it et al}~\cite{WuArXiv10}
arrives at a similar conclusion, albeit with a different
experimental signature and a slightly different explanation for the
origin of the bound-state. Wu {\it et al} study a QPC with a strong
lateral asymmetry (i.e., along the $y$-direction) and observe
suppression of the $G_{0}$ plateau along with strong
resonant-structures on the first riser from zero conductance. Both
signatures are predicted to arise from formation of a quasi-bound
state within the QPC in theory calculations by Bardarson {\it et
al}~\cite{BardarsonPRB04} using a $T$-matrix Lippmann-Schwinger
approach. Wu {\it et al}~\cite{WuArXiv10} suggest that their
bound-state originates from a momentum-mismatch between the 1D and
2D regions, consistent with Lindelof and
Aagesen~\cite{LindelofJPCM08}, which is accentuated by the
asymmetry.

This brings us to an interesting point in this review, where there
is a growing conflict regarding the origin of the $0.7$ plateau. On
the one hand, there is much work heavily favoring a {\it static}
spin polarization forming within the QPC~\cite{ThomasPRB98,
ThomasPRL96, KristensenPRB00, ThomasPRB00, RochePRL04,
RokhinsonPRL06, CrookSci06}, and numerous theoretical calculations
using spin-density functional theory argue {\it against} the
formation of a quasi-bound state within the QPC~\cite{WangPRB98,
BerggrenPRB02, BerggrenJPCM08, StarikovPRB03, HavuPRB04}. On the
other hand, other works using variants of the {\it same} theory
argue strongly {\it for} the formation of a bound
state~\cite{MeirPRL02, RejecNat06, HirosePRL03, IhnatsenkaPRB07,
MeirJPCM08}, favouring a Kondo-like mechanism for the $0.7$ plateau
that involves a {\it dynamic} spin-polarization process driven by
the Coulomb charging energy of the quasi-bound state with no static
spin-polarization ~\cite{CronenwettPRL02, MeirPRL02,
CronenwettPhD01}. The work by Morimoto {\it et al} and Yoon {\it et
al} fuels this conflict further, as it provides evidence for the
presence of a quasi-bound state, but with a {\it static}
spin-polarization present at zero magnetic
field~\cite{MorimotoAPL03, YoonPRL07, YoonPRB09, YoonAPL09}. While
this scenario is inconsistent with the traditional picture for a
Kondo-like origin for the $0.7$ plateau, it finds support in very
recent numerical calculations using an exact diagonalization
technique~\cite{SongPRL11}. Song and Ahn~\cite{SongPRL11} suggest
that localized states produced by non-adiabaticity in the confining
potential at the QPC openings~\cite{TekmanPRB89} lead to the
existence of ferromagnetically coupled magnetic impurities within a
QPC. An outcome of this model is the coexistence of Kondo
correlations and static spin-polarization. Although there is
evidence to support the formation of such a localized
state~\cite{LindelofJPCM08}, this picture is very new and further
calculations are clearly required. That said, the possibility raised
by Song and Ahn~\cite{SongPRL11} is an interesting avenue to explore
given that the wider experimental literature on the $0.7$ plateau in
QPCs repeatedly shows signatures of both Kondo-related and
spin-polarization-related processes.

Additional insight into quasi-bound state formation in a QPC is
difficult to obtain and not yet available, but further information
regarding the role that a Kondo-like mechanism might play at $G <
G_{0}$ can be obtained via more in-depth transport studies,
focussing in particular on a key experimental signature of the Kondo
effect when it is observed in quantum dots -- the zero bias anomaly.

\section{The zero-bias anomaly and its relationship to the $0.7$ plateau}

As discussed in Section 4, a key feature of the Kondo effect in
quantum dots is a peak in the differential conductance versus dc
source-drain bias~\cite{GoldhaberGordonNat98, CronenwettSci98}. The
observation and properties of a similar peak for $G < G_{0}$ in the
source-drain bias characteristics of QPCs was a key motivator for
the suggestion by Cronenwett {\it et al}~\cite{CronenwettPRL02,
CronenwettPhD01} that the $0.7$ plateau might be connected to Kondo
physics. The properties of this zero-bias anomaly have received
significant experimental interest recently~\cite{SfigakisPhysE10,
KlochanPRL11, SarkozyPRB09, RenPRB10, SfigakisPRL08, ChenPRB09,
LiuPRB10}, focussed on establishing whether it is fully consistent
with a Kondo model, and whether it is causally linked to the $0.7$
plateau or simply a separate but coincident effect in a QPC near
pinch off.

\begin{figure}
\includegraphics[width=16cm]{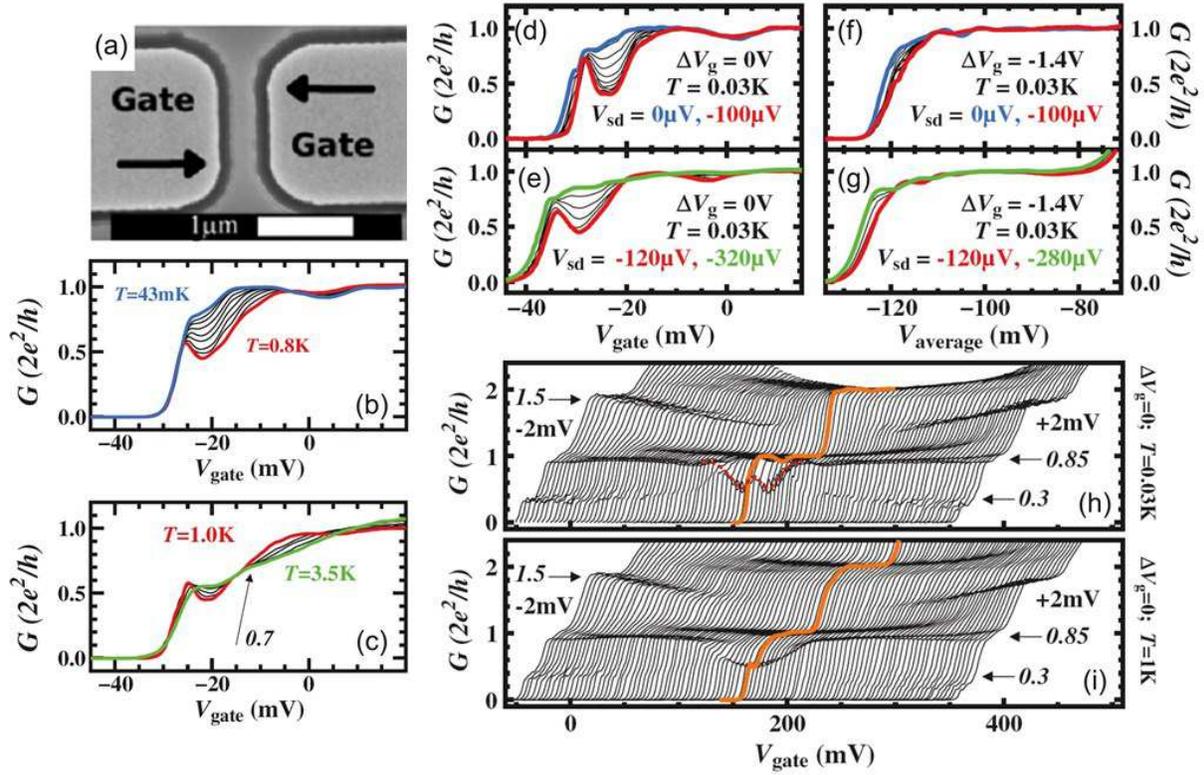}
\caption{(a) Scanning electron micrograph of a QPC featuring two
microconstrictions, indicated by the arrows, which generate a
shallow bound-state within the QPC. (a) and (b) show the evolution
of the conductance $G$ vs gate voltage $V_{gate}$ for temperatures
$T$ in the range (b) $47-800$~mK and (c) $1.0-3.5$~K. (d) and (e)
show the evolution of $G$ vs $V_{gate}$ with changing source-drain
bias $V_{sd}$ for (d) $0-100~\mu$V and (e) $120-320~\mu$V for a
symmetrically biased QPC $\Delta V_{g} = 0$~V. (f) and (g) show the
evolution of $G$ vs the average bias $V_{average}$ of the
asymmetrically biased QPC gates with changing source-drain bias
$V_{sd}$ for (f) $0-100~\mu$V and (g) $120-320~\mu$V for an
asymmetrically biased QPC $\Delta V_{g} = -1.4$~V. (h) and (i) show
source-drain bias `waterfall' plots of $G$ vs $V_{average}$ for
$40~\mu$V steps of $V_{sd}$ between $-2$ and $+2$~mV for $V_{g} =
0$~V and (h)$T = 30$~mK and (i) $T = 1$~K. The red dashed line in
(h) is a guide to the eye. In the text both $V_{gate}$ and
$V_{average}$ are referred to as $V_{g}$ when discussing this
figure. Figure adapted with permission from
Ref.~\cite{SfigakisPRL08}. Copyright 2008 by the American Physical
Society.}
\end{figure}

\subsection{The ZBA in a QPC supporting a clear shallow bound-state}

Sfigakis {\it et al}~\cite{SfigakisPRL08} studied a QPC featuring a
pair of etch defects, one on each gate, as indicated by the arrows
in Fig.~30(a). These etch defects act as microconstrictions,
generating a double barrier potential along the length of the QPC
that supports a shallow bound-state. The bound-state produces a
Coulomb blockade (CB) peak in the conductance versus gate voltage
trace at $G < G_{0}$ (see red trace in Fig.~30(b)), similar to that
previously observed in a QPC containing an impurity by McEuen {\it
et al}~\cite{McEuenSurfSci90} and a QPC with three very narrow
transverse top-gates crossing it by Liang {\it et
al}~\cite{LiangPRL98}. The shallow bound state is unstable to
asymmetric biasing of the QPCs, with the CB peak disappearing for
$|\Delta V_{g}| > 1.0$~V (see Fig.~1(b) of Ref.~\cite{SfigakisPRL08}
or Fig.~6.14 of Ref.~\cite{SfigakisPhD05}). The temperature
dependence of the CB peak is shown over the ranges $47 - 800$~mK and
$1.0 - 3.5$~K in Fig.~30(b) and (c), respectively. As Fig.~30(b)
shows, the conductance minima to the right of the CB peak is
strongly temperature dependent, with the conductance enhanced as the
temperature is reduced. This behaviour is consistent with the Kondo
effect in a quantum dot containing an odd number of electrons (c.f.
Fig.~12(e))~\cite{KouwenhovenPW01}. For $T > 1$~K in Fig.~30(c), the
conductance of the CB minima begins to rise again, indicating that
the Kondo effect has become suppressed, which is suggestive of a
Kondo temperature $T_{K} \sim 1$~K. Additionally, the CB peak itself
starts to diminish due to thermal smearing, whilst the $0.7$ plateau
strengthens substantially, highlighting that the $0.7$ plateau is
strongest at a very different temperature range to that where the
Kondo effect is dominant. The Arrhenius (Eq.~2), quantum dot Kondo
(Eq.~5) and modified Kondo (Eq.~6) models were fitted to the $G$
versus $T$ from Fig.~30(b) for a number of fixed gate voltages $-8
\leq V_{g} \leq -22$~mV. The quantum dot Kondo model was an
excellent fit to the data over the entire $V_{g}$ and $T$ range of
this data, with a $V_{g}$-dependent Kondo temperature that varied
linearly from $\sim 0.5$~K at $-22$~mV to almost $20$~K at $-8$~mV.
In contrast, the Arrhenius model was only a good fit in the high $T$
limit, and the modified Kondo model did not fit the data at
all~\cite{SfigakisPRL08}.

Figures~30(d-i) present a study of the behaviour of the CB peak and
associated minima with dc source-drain bias. Figures~30(d) and (e)
show the evolution of the conductance with bias at $G < G_{0}$ for a
symmetrically biased QPC at $30$~mK for the low and high $V_{sd}$
regimes, respectively. This evolution appears very similar to that
observed with changing temperature in Figs.~30(b/c), and leads to
the w-shaped trend highlighted by the red dashed line in Fig.~30(h),
the central peak of which is the ZBA. This data clearly indicates
the quenching of the Kondo effect by $V_{sd} \sim 100~\mu$V,
consistent with observations in quantum
dots~\cite{GoldhaberGordonNat98, CronenwettSci98}, and with a $T_{K}
\sim 1$~K. Figure~30(i) shows the same measurement as Fig.~30(h)
obtained at higher $T = 1$~K, the most notable change being the loss
of the ZBA consistent with the quenching of the Kondo effect above
$T = 800$~mK in Figs.~30(b/c). Note that the finite bias plateaus at
$0.3G_{0}$ and $0.85G_{0}$ appear in both Figs.~30(h/i) indicating
that they do not arise from the Kondo effect, which is quenched by
both temperature and source-drain bias in Fig.~30(i).
Figures~30(f/g) show the same data as in Figs.~30(d/e) obtained with
an asymmetric bias $\Delta V_{g} = -1.4$~V applied to the gates to
disrupt the confinement imposed by the microconstrictions. The CB
peak that was observed under symmetric gate bias (c.f. Fig.~30(d/e))
is accordingly destroyed by the asymmetric bias. The shoulder of the
$G_{0}$ plateau (i.e., the vicinity of the lost CB peak structure)
shows considerably less evolution with $V_{sd}$ in Figs.~30(f/g),
confirming that Kondo-like behaviour is lost with the CB peak when
the shallow bound-state is destroyed by asymmetric biasing. Further,
a comparison of Figs.~30(e) and (g) confirms that the finite bias
plateau at $0.85G_{0}$ still arises well after any Kondo behaviour
is suppressed, corroborating the finding from Fig.~30(i) above. The
fact that these devices show a strong Kondo effect consistent with
that in quantum dots~\cite{KouwenhovenPW01, GoldhaberGordonNat98,
CronenwettSci98} is perhaps not surprising given the shallow bound
state established by the microconstrictions. However the
co-existence of this behaviour with all the usual hallmarks of the
$0.7$ plateau suggests that the Kondo-like effect in 1D channels
reported by Cronenwett {\it et al}~\cite{CronenwettPRL02} and the
$0.7$ plateau are separate and distinct effects.

\subsection{The ZBA in QPCs on modulation doped heterostructures}

Moving back to more conventional QPCs (i.e., without
microconstrictions), further work by Chen {\it et
al}~\cite{ChenPRB09} focussed on the response of the ZBA to in-plane
magnetic fields up to $10$~T. Chen {\it et al} studied more than
fifteen devices~\cite{ChenPC} and in all but two the ZBA was
observed to remain as a single, unsplit peak centered at $V_{sd} =
0$~V at $B \gtrsim 6$~T. The 1D electron system was clearly
spin-polarized at this field, as indicated by a strong plateau at
$0.5G_{0}$ and a strong corresponding accumulation of $G$ versus
$V_{sd}$ traces at $0.5G_{0}$ in the source-drain bias data (e.g.,
see Fig.~1 of Ref.~\cite{ChenPRB09}). The spin-flip processes
required for the Kondo effect are blocked in this regime, making it
unlikely that the Kondo effect is responsible for the ZBA they
observe. Instead, Chen {\it et al}~\cite{ChenPRB09} present an
alternate phenomenological model for the ZBA that relies on the
energy of the 1D subbands rising with dc source-drain bias. A small
linear rise of only $0.33$~meV/mV is required to cause a sharp ZBA
over the entire range $0 < G < G_{0}$. This phenomenological model
has no reliance on spin, and would thus show no field-induced
splitting of the ZBA. This would allow the ZBA to remain visible
even in the fully spin-polarized transport regime, as Chen {\it et
al}~\cite{ChenPRB09} observed experimentally. It is interesting to
note that both the Arrhenius (Eq.~2) and modified Kondo (Eq.~5)
models are good fits to the data obtained by Chen {\it et al}, while
the quantum dot Kondo model (Eq.~6) is not. This is consistent with
the findings by Cronenwett {\it et al}~\cite{CronenwettPRL02}, and
suggests there is nothing particularly different about the devices
Chen {\it et al} study that would invalidate a direct comparison
with earlier data. In particular, it suggests that the ZBA in these
devices does not arise from the formation of a quantum dot like
bound-state, as in the device studied by Sfigakis {\it et
al}~\cite{SfigakisPRL08}, pointing to the possibility that there is
more than one mechanism that can result in a ZBA in QPCs. The
$V_{g}$-dependent Kondo temperatures obtained by Chen {\it et al}
are in general agreement with earlier
results~\cite{CronenwettPRL02}, ruling out insufficient magnetic
field (i.e., $g^{*}\mu_{B}B < k_{B}T_{K}$) as a reason for the
splitting of the ZBA not being observed in this experiment.

Chen {\it et al}~\cite{ChenPRB09} also highlight a significant
problem in studies of the ZBA in QPCs, namely the effect of
disorder. In two of their devices, Chen {\it et al} found that the
ZBA does split as the magnetic field is increased, as expected under
a Kondo model~\cite{MeirPRL02}. However, this splitting was strongly
influenced by laterally shifting the QPC by asymmetric biasing of
the gates -- the splitting was enhanced for $\Delta V_{g} = +0.3$~V,
while the peaks merge into a single asymmetric peak at $\Delta V_{g}
= -0.3$~V, which becomes sharper and more symmetric at $\Delta V_{g}
= -0.6$~V. This behaviour conflicts with that described above, and
such inconsistencies in ZBA behaviour are frequent in QPCs formed in
modulation-doped heterostructures. For example, Liu {\it et
al}~\cite{LiuPRB10} find that the ZBA is suppressed by increasing
the QPC length, while Koop {\it et al}~\cite{KoopJSNM07} find the
opposite, with no length dependence of the ZBA at all. Both Liu {\it
et al}~\cite{LiuPRB10} and Ren {\it et al}~\cite{RenPRB10} report
that the shape, height and width of the ZBA vary from device to
device, and even between subsequent cool-downs on the same device.
This naturally leads to the question of whether the ZBA is simply
due to scattering from the background potential produced by the
ionised dopants, since the spatial distribution of the ionised
fraction of the dopants in the AlGaAs modulation-doping layer is
known to change randomly between devices and subsequent cool-downs
of the same device~\cite{BervenPRB94, ScannellArXiv11}.

\subsection{The ZBA in QPCs on undoped heterostructures}

The impact of disorder can be tested by studying devices made using
heterostructures where the modulation doping has been removed. This
was done by Sarkozy {\it et al}~\cite{SarkozyPRB09}, who studied ten
separate electron QPCs made using undoped AlGaAs/GaAs
heterostructures where the 2DEG is populated electrostatically using
a metal top-gate biased positively to $V_{top}$, and isolated from
the semiconductor and QPC surface gates, which are biased negatively
to $V_{gate}$ (or $V_{g}$), by a thin layer of
polyimide~\cite{SarkozyECS07}. The lack of any intentional doping in
the heterostructures results in an average impurity spacing $D =
0.6~\mu$m, based on a measurement of the mobility and analysis of
the dominant sources of scattering in this
system~\cite{SarkozyPRB09}. For a QPC of length $L = 0.4~\mu$m, this
gives a probability $P = 1 - e^{(-L/D)} \approx 48 \%$~\footnote{The
minus sign is missing in the exponent in Ref.~\cite{SarkozyPRB09},
this is a typographical error.} of finding an impurity within the
QPC. A symmetric, unsplit ZBA was observed for $G < 0.8G_{0}$ in all
ten devices studied, and using the probability argument above it is
extremely unlikely (i.e., $\sim P^{10} \sim 7 \times 10^{-4}$) that
all of these ZBAs are due to an impurity within the QPC.

\begin{figure}
\includegraphics[width=10cm]{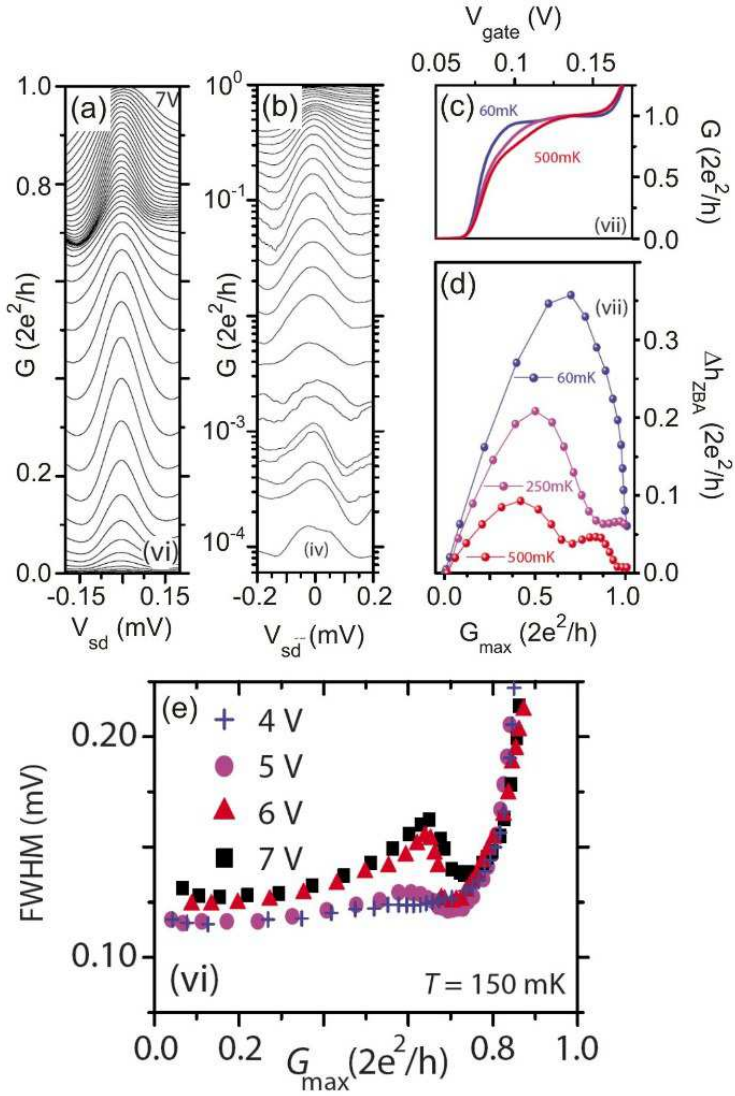}
\caption{Conductance $G$ vs source-drain bias $V_{sd}$ at increments
of $V_{g}$ for the range $0 < G < G_{0}$  plotted on (a) linear and
(b) logarithmic scales of $G$ for two of the ten studied, designated
(vi) and (iv), in Ref.~\cite{SarkozyPRB09}, respectively. (c) $G$ vs
$V_{g}$ and (d) the corresponding ZBA amplitude $\Delta h_{ZBA}$ vs
$G$ from device (vii) at temperatures $T = 60$, $250$ and $500$~mK.
(e) The full-width at half maximum (FWHM) of the ZBA vs $G$ obtained
from device (vi) at four different top-gate voltages $V_{top}$
spanning the density range from $n = 8 \times 10^{10}$~cm$^{-2}$ at
$V_{top} = +4$~V to $n = 1.6 \times 10^{11}$~cm$^{-2}$ at $V_{top} =
+7$~V. In the text $V_{gate}$ will be referred to as $V_{g}$ when
discussing this figure. Figure adapted with permission from
Ref.~\cite{SarkozyPRB09}. Copyright 2009 by the American Physical
Society.}
\end{figure}

We now explore the data obtained by Sarkozy {\it et al} in some
detail to provide an interesting counterpoint for the paper
discussed in Section 7.4.1. Figures~31(a/b) show $G$ versus $V_{sd}$
for incremented $V_{g}$ over the range $0 < G < G_{0}$. The
clustering of traces near $0.7G_{0}$, particularly evident in
Fig.~31(a), indicates the presence of the $0.7$ plateau. Yet,
despite this, the ZBA is observed over the {\it entire} range of $G$
from $G_{0}$ down to zero, remaining as a clear peak at $G \sim
10^{-4}G_{0}$. Similar behaviour was also reported by Ren {\it et
al}~\cite{RenPRB10}. Figure~31(d) shows the amplitude of the ZBA
$\Delta h_{ZBA}$ versus $G$ at temperatures $T = 60$, $250$ and
$500$~mK, with the corresponding $G$ versus $V_{g}$ traces shown in
Fig.~31(c). The amplitude $\Delta h_{ZBA}$ is measured as the
conductance at the peak minus the average of the conductance at the
minima on either side of the ZBA. At $T = 60$~mK, where the $0.7$
feature is weakest, $\Delta h_{max}$ versus $G$ shows a clear single
peak at $\sim 0.8G_{0}$. However, as the temperature is increased
and the $0.7$ plateau strengthens, the peak $\Delta h_{max}$ drops
substantially and moves to lower $G$ highlighting the very different
temperature dependencies of the $0.7$ plateau and ZBA noted by
Cronenwett {\it et al}~\cite{CronenwettPRL02}. Note also that a
clear minimum in $\Delta h_{ZBA}$ versus $G$ develops with
increasing $T$. This minima sits at $\sim 0.75G_{0}$, shifts to
lower $G$ as the temperature is raised, and can be very strong,
corresponding to a $50 \%$ reduction in ZBA amplitude compared to
its maximum at $\sim 0.5G_{0}$ (see Fig.~3(c) of
Ref.~\cite{SarkozyPRB09}). This trend is largely independent of the
2D density. The full-width at half maximum (FWHM) for the ZBA versus
$G$ is shown in Fig.~31(e) at four different 2D densities between $n
= 1.6 \times 10^{11}$~cm$^{-2}$ at $V_{top} = +7$~V and $n = 8
\times 10^{10}$~cm$^{-2}$ at $V_{top} = +4$~V. The background trend
in Fig.~31(e) is in general agreement with the FWHM versus $V_{g}$
data presented by Cronenwett {\it et al}~\cite{CronenwettPRL02} (see
Fig.~13(f)), in particular the very sharp rise in FWHM as $G$
approaches $G_{0}$. However, Sarkozy {\it et al} also observe a very
pronounced dip in FWHM at $0.7 - 0.8G_{0}$, which is particularly
strong at higher density. Note that the FWHM of the ZBA is directly
linked to the Kondo temperature, and hence this directly contradicts
theoretical predictions of a clean exponential $T_{K}$ versus
$V_{g}$ dependence made by the 1D Kondo model~\cite{MeirPRL02}. With
the benefit of hindsight, the dip in FWHM is clearly evident in the
data obtained by Cronenwett {\it et al}~\cite{CronenwettPRL02} (see
vicinity of $V_{g} = -485$~mV in Fig.~13(f)), and its strength is
not inconsistent in terms of the density dependence in Fig.~31(e).
The data in Fig.~13(f) was obtained at a density of $1.1 \times
10^{11}$~cm$^{-2}$~\cite{CronenwettPRL02}, which would correspond to
a top-gate bias of $+5.125$~V in the experiment performed by Sarkozy
{\it et al}~\cite{SarkozyPRB09}. Similar data to that in
Figs.~31(d/e) was also observed by Ren {\it et al}~\cite{RenPRB10}
at a density of $1.11 \times 10^{11}$~cm$^{-2}$.

\begin{figure}
\includegraphics[width=16cm]{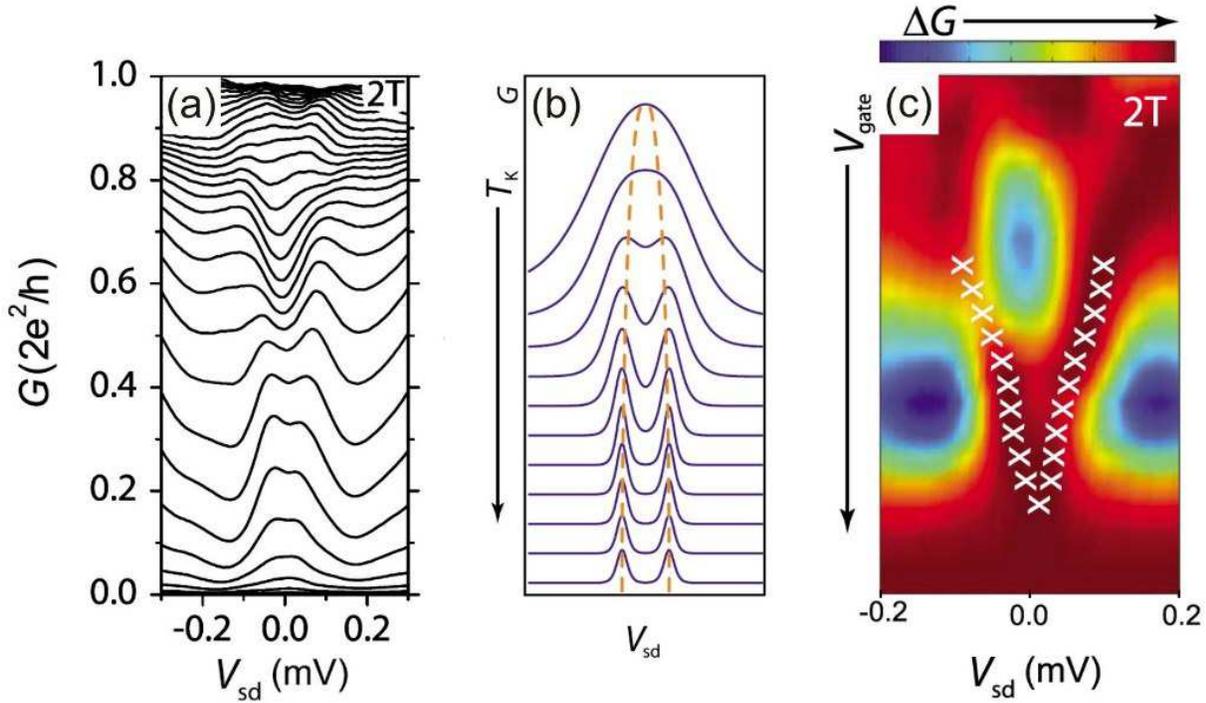}
\caption{(a) $G$ vs $V_{sd}$ for incremented $V_{g}$ with an
in-plane magnetic field $B = 2$~T applied. (b) Schematic of the
expected Zeeman splitting of the ZBA at constant $B$ and $T$ for the
singlet Kondo effect as the Kondo temperature $T_{K}$ is decreased
moving from top to bottom (traces offset vertically for clarity).
Note that both Cronenwett {\it et al}~\cite{CronenwettPRL02} and
Chen {\it et al}~\cite{ChenPRB09} find that $T_{K}$ decreases as
$V_{G}$ is made more negative and $G$ heads towards zero. (c)
Colour-map plotting the fluctuation from mean conductance $\Delta G$
for a given $G(V_{sd},V_{g})$ trace (colour axis) vs $V_{sd}$
($x$-axis) and $V_{g}$ ($y$-axis) at $B = 2$~T. The superimposed
white crosses highlight the increasing Zeeman splitting of the ZBA
as $V_{G}$ is made more positive. Figure adapted with permission
from Ref.~\cite{SarkozyPRB09}. Copyright 2009 by the American
Physical Society.}
\end{figure}

Figure~32(a) shows $G$ versus $V_{sd}$ with incremented $V_{g}$ for
an in-plane magnetic field $B = 2$~T, and although Zeeman splitting
of the ZBA is clearly evident in the vicinity of $0.5G_{0}$, this
behaviour is certainly not constant or consistent over the entire $0
< G < G_{0}$ range. As discussed in Section 4.1, one of the defining
features of the Kondo effect, as it is observed in quantum dots, is
the Zeeman splitting of the ZBA by {\it twice} the amount normally
expected, i.e., the ZBA splitting $e\Delta V_{ZBA} = 2g^{*}\mu_{B}B$
rather than $g^{*}\mu_{B}B$~\cite{CronenwettSci98, MeirPRL93}. To
investigate the Zeeman splitting behaviour in Fig.~32(a) further, we
will now focus our attention on Figs.~32(b) and (c) together.
Figure~32(b) is a schematic illustrating how the splitting of the
ZBA should evolve with $T_{K}$ at fixed $B$ and $T$. Figure~32(c)
shows a colour map of the deviation $\Delta G$ from average
conductance across the $V_{sd}$ range measured at a given $V_{g}$
versus $V_{sd}$ on the $x$-axis and $V_{g}$ on the $y$-axis, it is
the same data as that in Fig.~32(a) but with a different
presentation. The interpretation of Figs.~32(b/c) is complex, so we
will do it in two steps. First, it is important to note that the
traces in Fig.~32(b) are offset vertically such that moving
downwards corresponds to the evolution of the ZBA with decreasing
$T_{K}$ -- at large $T_{K}$ (top) the Zeeman splitting cannot be
resolved, for intermediate $T_{K}$ such that $g^{*}\mu_{B}B <
k_{B}T_{K} < k_{B}T$ (middle) the ZBA is strong and the splitting
resolved, and finally at small $T_{K}$ (bottom) the ZBA begins to
diminish. The second step involves noting that the $T_{K}$ values
obtained from fits of the modified Kondo model (Eq.~6) are
$V_{g}$-dependent and decrease as $V_{g}$ becomes more negative (see
Fig.~13(e))~\cite{CronenwettPRL02, ChenPRB09}, which in turn
corresponds to reduced $G$. Thus, moving from the top downwards in
Fig.~32(b) corresponds to a set of traces ordered in decreasing
conductance, allowing direct comparison with the data in
Figs.~32(a/c). This comparison reveals a significant discrepancy
because the splitting of the ZBA should remain constant but become
better resolved as we move to lower $G$, as shown in Fig.~32(b).
However, the splitting actually evolves with $G$ and is most clearly
resolved at intermediate $G$, as shown in Figs.~32(a/c). Note that
there is an `apparent' change in ZBA splitting in the high $T_{K}$
limit in Fig.~32(b), as indicated by the orange dashed lines. This
is a resolution effect rather than a real change in the Zeeman
splitting, and would produce an apparent reduction in the splitting
at high $G$. The data in Fig.~32(c) clearly shows that this effect
is negligible for $G < 0.7G_{0}$. The splitting behaviour in Fig.~32
is inconsistent with theoretical frameworks for both
Kondo~\cite{MeirPRL02} and spin-polarization models~\cite{WangPRB96,
WangPRB98, BerggrenJPCM08}. A possibility considered by Sarkozy {\it
et al}~\cite{SarkozyPRB09} (and discussed by Sfigakis {\it et
al}~\cite{SfigakisPhysE10}) is that this behaviour might be
connected to higher-order Kondo phenomena, such as the two-impurity
Kondo effect~\cite{JonesPRL87} or competition between Kondo
processes and the Ruderman-Kittel-Kasuya-Yosida (RKKY)
interaction~\cite{RudermanPR54, KasuyaPTP56, YosidaPR57}. This is
motivated by spin DFT calculations suggesting that two bound spins
can be confined within the QPC at pinch-off~\cite{RejecNat06,
BerggrenJPCM08}. However, this should produce a ZBA that has a
finite splitting at zero magnetic field~\cite{AguadoPRL00,
DiasDaSilvaPRL06}, as observed in experiments by Jeong {\it et
al}~\cite{JeongSci01} with two series-coupled quantum dots. In the
data obtained by Jeong {\it et al} the initially spin-split ZBA
peaks converge with increasing $B$, merge into a single maxima at
zero bias at some finite field, and and then diverge as $B$ is
increased further. The data presented by Sarkozy {\it et
al}~\cite{SarkozyPRB09} shows no evidence of zero field ZBA
splitting, suggesting that a two-impurity Kondo
model~\cite{JonesPRL87, JonesPRL88} cannot explain the observed
splitting in Figs.~32(a/c). As a final note, follow-up work on
similar devices by Sfigakis {\it et al}~\cite{SfigakisPhysE10}
demonstrates that the magnetic field dependence of the ZBA is
different from that of the $0.7$ plateau, drawing additionally on
findings by Koop {\it et al}~\cite{KoopJSNM07}. The evidence from
undoped QPCs discussed above adds further weight to the claim that
the $0.7$ plateau and the ZBA are separate and distinct phenomena
that coexist in the QPC at $G < G_{0}$~\cite{SfigakisPRL08}, and
ultimately Sarkozy {\it et al}~\cite{SfigakisPhysE10, SarkozyPRB09}
argue that the ZBA is a fundamental property of quantum wires that
cannot be explained by either 1D Kondo physics or spin-polarization
models alone.

\begin{figure}
\includegraphics[width=16cm]{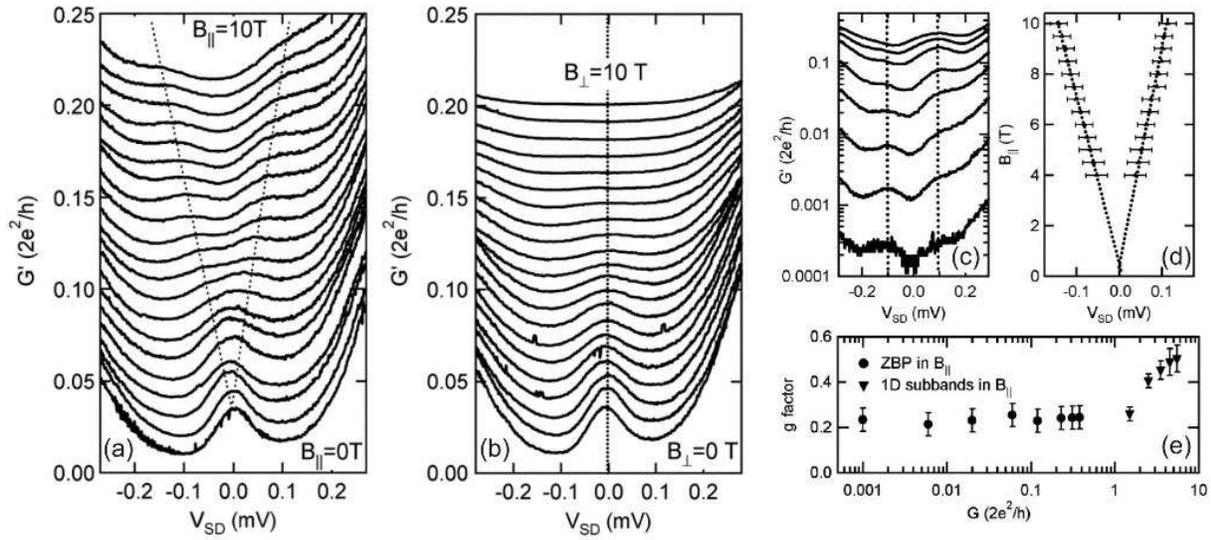}
\caption{Plots of differential conductance $G^{\prime}$ vs
source-drain bias $V_{SD}$ as a function of in-plane magnetic field
applied (a) along the QPC axis $B_{\parallel}$, and (b)
perpendicular to the QPC axis $B_{\perp}$ obtained at a linear
conductance $G \sim 0.03G_{0}$ with a side-gate voltage of $V_{SG} =
1.105$~V. Each successive trace is incremented in field by $0.5$~T
and offset upwards by $0.01G_{0}$ for clarity. (c) Plot of
$G^{\prime}$ vs $V_{SD}$ at a range of $V_{SG}$ at fixed
$B_{\parallel} = 8$~T demonstrating that the peak splitting is
independent of gate voltage. (d) Plot of the center of the Zeeman
split zero-bias peaks in $V_{SD}$ ($x$-axis) vs $B_{\parallel}$
($y$-axis) for the data from (a). (e) Effective Land\'{e}
$g$-factors $g^{*}$ obtained for the ZBPs using the Kondo expression
$2g^{*}\mu_{B}B$ (circles) and the 1D subbands using the usual
Zeeman expression $g^{*}\mu_{B}B$ for $B_{\parallel}$ (triangles)
plotted against $G$. Figure adapted with permission from
Ref.~\cite{KlochanPRL11}. Copyright 2011 by the American Physical
Society.}
\end{figure}

\subsection{The ZBA in hole QPCs}

\subsubsection{The ZBA in a hole QPC with a clear shallow bound-state}

We now consider some very recent data obtained by Klochan {\it et
al}~\cite{KlochanPRL11} in an undoped hole quantum dot. The dot
studied in this experiment is formed spontaneously in a quantum
wire, presumably due to some combination of roughness in the
confining potential, as in the device studied by Sfigakis {\it et
al}~\cite{SfigakisPRL08}, and self-consistent electrostatic
effects~\cite{YoonPRL07}. Evidence for dot formation is provided by
resonant structures that emerge in the linear conductance versus
gate voltage characteristics~\cite{McEuenSurfSci90, KomijaniEPL10,
LiangPRL98}. The zero-bias peak (ZBP) observed in the differential
conductance $G^{\prime}$ versus source-drain bias $V_{SD}$ is robust
to lateral shifting~\cite{GlazmanSST91} of the 1D wire by asymmetric
biasing (see Fig.~2 of Ref.~\cite{KlochanPRL11}), indicating that
the features Klochan {\it et al} observe are not a disorder effect.
In contrast to Sarkozy {\it et al}~\cite{SarkozyPRB09}, the undoped
devices studied by Klochan {\it et al} were produced following the
degenerately-doped-cap approach pioneered by Kane {\it et
al}~\cite{KaneAPL98}. However, the structure was adapted to use a
p$^{+}$-doped cap to reduce the biases required for inducing a 2D
hole gas~\cite{ClarkeJAP06}. This approach results in hole QPCs that
show very stable 1D quantized conductance
plateaus~\cite{KlochanAPL06}, and which have been used to
demonstrate the strongly anisotropic Zeeman spin-splitting in 1D
hole systems on both (311)-oriented~\cite{KlochanNJP09} and
(100)-oriented~\cite{ChenNJP10} heterostructures. The device studied
by Klochan {\it et al} is a $400$~nm long quantum wire oriented
along the $[01\overline{1}]$ direction of a (100)-oriented
heterostructure. It was measured in a dilution refrigerator
featuring an {\it in-situ} piezo-rotator mechanism~\cite{YeohRSI10}
to enable the in-plane magnetic field to be applied both along and
perpendicular to the QPC axis, denoted as $B_{\parallel}$ and
$B_{\perp}$ respectively for this discussion, with the sample
remaining at a temperature below $200$~mK during the entire rotation
process. Previous studies have shown that $B_{\parallel}$ and
$B_{\perp}$ correspond to the high and low $g^{*}$ directions for a
QPC on a (100)-oriented heterostructure~\cite{ChenNJP10}.
Figures~33(a) and (b) show plots of $G^{\prime}$ versus $V_{SD}$
illustrating the evolution of the zero-bias peak obtained at a
linear conductance $G = G^{\prime}(V_{SD} = 0) \sim 0.03G_{0}$ as a
function of $B_{\parallel}$ and $B_{\perp}$, respectively. In the
parallel orientation, the ZBP clearly splits into two peaks for
$B_{\parallel} > 4$~T. In contrast, for the perpendicular
orientation, the ZBP is suppressed for $B_{\perp} > 7$~T with no
sign of any Zeeman splitting. This anisotropy is consistent with
expectations from the known anisotropy of the 1D hole subbands in a
(100) heterostructure~\cite{ChenNJP10}.

Two hallmarks of the Kondo effect in quantum dots are that the
Zeeman splitting of the ZBP is independent of gate-voltage, and that
the splitting goes as $2g^{*}\mu_{B}B$ rather than the usual
$g^{*}\mu_{B}B$~\cite{GoldhaberGordonNat98, CronenwettSci98,
MeirPRL93}. Figure~33(c) shows $G^{\prime}$ versus $V_{SD}$ obtained
at $B_{\parallel} = 8$~T over a range of gate voltages such that the
ZBP spans the range $\sim 10^{-4}G_{0} < G < 0.5G_{0}$. As the
dashed lines show, the ZBP splitting is constant over this range, as
expected for the quantum dot Kondo effect~\cite{CronenwettSci98}.
Figure~33(d) shows the zero bias peak positions in $V_{SD}$
($x$-axis) plotted against $B_{\parallel}$ ($y$-axis) for all the
traces in Fig.~33(a) where the Zeeman splitting of the ZBP can be
resolved. The splitting of the ZBP is linear in $B_{\parallel}$ as
the dotted lines forming a V-shape in Fig.~33(d) highlight, allowing
an estimate of the $g$-factor to be obtained for that particular
conductance. Klochan {\it et al} repeated this analysis for ZBPs
obtained at a range of $G$, with the corresponding $g$-factors
plotted as the circles in Fig.~33(e).

At this point, there is one important detail that remains to be
addressed -- which form of the splitting should be used, the
standard Zeeman form $g^{*}\mu_{B}B$ or the Kondo form
$2g^{*}\mu_{B}B$? Since there was strong evidence for the presence
of a bound-state (i.e., quantum dot) in this device, Klochan {\it et
al} assumed the Kondo form in obtaining the $g^{*}$ from the ZBP. We
will refer to this $g$-factor as $g^{*}_{ZBP}$ to distinguish it
from the $g$-factor $g^{*}_{1D}$ that Klochan {\it et al} obtained
independently from the 1D subbands using source-drain bias
spectroscopy~\cite{PatelPRB91a, DanneauPRL06}. We will return to
$g^{*}_{1D}$ in a moment. As shown in Fig.~33(a), $g^{*}_{ZBP}$ is
relatively constant in $G$, as expected based on the data in
Fig.~33(c), and takes an average value $g^{*}_{ZBP} = 0.236 \pm
0.012$. Klochan {\it et al} obtain estimates of $g^{*}_{1D}$ from
the 1D subbands using the procedure outlined by Danneau {\it et
al}~\cite{DanneauPRL06} assuming the usual Zeeman form
$g^{*}\mu_{B}B$. These are plotted as triangles in Fig.~33(e). The
decrease in $g^{*}_{1D}$ with reduced $G$ is consistent with
previous results obtained by Chen {\it et al}~\cite{ChenNJP10}; the
absence of exchange enhancement~\cite{ThomasPRL96} is a particular
feature of (100)-oriented hole systems~\cite{WinklerPRB05}. Klochan
{\it et al} obtained $g^{*}_{1D} = 0.25 \pm 0.03$ for the lowest 1D
subband, in excellent agreement with the constant $g^{*}_{ZBP}$
value obtained assuming the Kondo form of the Zeeman splitting for
the ZBP. This observation provides conclusive evidence for the
presence of quantum dot Kondo physics in this device.

There are a couple of points of significance to this result. First,
as Klochan {\it et al}~\cite{KlochanPRL11} point out, it represents
the first observation of Kondo physics for a hole quantum dot, which
is interesting due to the strong spin-orbit interaction and
spin-$\frac{3}{2}$ nature of holes in GaAs. But considering it in
terms of the other studies of zero-bias maxima in the differential
conductance in QPCs above, the system studied by Klochan {\it et al}
presents a clear case of what the Kondo physics should look like for
an unintentional quantum dot embedded within a 1D system.
Remarkably, there appears to be no significant difference to the
behaviour observed for a quantum dot embedded in a 2D
system~\cite{GoldhaberGordonNat98, CronenwettSci98}. In particular,
the behaviour of the $g$-factor obtained from the zero-bias peak
matches that of the underlying system in which it resides. For the
device studied by Klochan {\it et al}~\cite{KlochanPRL11} the
directional anisotropy of $g^{*}_{ZBP}$ matches that of
$g^{*}_{1D}$. And for the quantum dots studied by Cronenwett {\it et
al}~\cite{CronenwettSci98} and Nyg{\aa}rd {\it et
al}~\cite{NygardNat00}, $g^{*}_{ZBP}$ matches the bulk $g$-factor
for GaAs and the known values for carbon nanotubes~\cite{TansNat97,
CobdenPRL98}, respectively. The findings by Klochan {\it et
al}~\cite{KlochanPRL11} strongly suggest that many of the phenomena
normally associated with Kondo-like physics in QPCs may instead have
a different origin.

\subsubsection{The ZBA in a modulation-doped hole QPC under perpendicular magnetic fields}

We finish Section 7 by noting that similar resonant-like structures
were observed in the conductance versus gate voltage of a hole QPC
by Komijani {\it et al}~\cite{KomijaniEPL10}. In the absence of a
magnetic field, they observe a strong $0.7$ plateau at temperatures
between $300$~mK and $1.84$~K. At lower temperatures, the $0.7$
plateau rises up to become part of the $G_{0}$ plateau.
Additionally, a clear ZBA is observed over the entire range $0 < G <
G_{0}$ at $T = 100$~mK, this does not occur at $800$~mK. This
behaviour is essentially the same as that reported by Cronenwett
{\it et al}~\cite{CronenwettPRL02} for electrons except perhaps that
the low temperature disappearance of the $0.7$ plateau is more
evident here; a weak inflection or rounding of the plateau shoulder
remains in the data obtained at $80$~mK by Cronenwett {\it et
al}~\cite{CronenwettPRL02}. Komijani {\it et al} performed their
study using a magnetic field $B_{\perp}$ oriented {\it
perpendicular} to the 2DHG, which is unusual. Studies are normally
conducted using in-plane magnetic fields to avoid Landau
quantization and edge-state transport from affecting the data. On
increasing the perpendicular magnetic field, the $0.7G_{0}$ and
$G_{0}$ plateaus both develop into a plateau at $0.5G_{0}$ (see
Figs.~2(a/b) of Ref.~\cite{KomijaniEPL10}, the latter is peculiar as
the $G_{0}$ plateau would not fall to $0.5G_{0}$ for a similar
magnitude in-plane magnetic field (c.f., Fig.~3(c), for example).
Additionally, a very strong resonant-like structure develops on the
riser from $G = 0$ to the $0.5G_{0}$ plateau. The resonant structure
is first observed at $B_{\perp} \sim 2$~T and becomes very clear for
$B_{\perp} > 5$~T. Komijani {\it et al}~\cite{KomijaniEPL10} argue
that this structure indicates a quasi-localized state within the
QPC, consistent with a Kondo-like model for the $0.7$ plateau.
However, the peak cannot be conclusively ascribed to a
quasi-localized state formed self-consistently by Friedel
oscillations as predicted theoretically~\cite{MeirPRL02, RejecNat06,
HirosePRL03}. Although an impurity-based origin for the resonant
structure~\cite{McEuenSurfSci90} can be ruled out because the peak
is robust to asymmetric biasing of the QPC, there is another
possible cause for such a structure. Tunneling between edge-states
has been predicted to produce a resonant structure in
QPCs~\cite{JainPRL88}, and although this has been
observed~\cite{vanWeesPRB91}, these structures in earlier studies
appear much weaker than the structure observed by Komijani {\it et
al}~\cite{KomijaniEPL10}. Similar structures have also been observed
for tunneling between edges-states across a fully depleted
region~\cite{HaugSST93}, which may also be relevant given the
proximity to pinch-off in the data presented by Komijani {\it et
al}~\cite{KomijaniEPL10}. Looking towards a more quantitative
assessment, let us assume that a QPC in the single-mode limit has a
width comparable to the Fermi wavelength $\lambda_{F} \sim 40$~nm at
a 2D hole density $p = 4 \times 10^{11}$~cm$^{-2}$ as in
Ref.~\cite{KomijaniEPL10}. The transition to edge-state transport
through a QPC commences once $2r_{cyc} < W$, where $r_{cyc} =
\hbar/eB_{\perp} \times \sqrt{2\pi p}$ is the cyclotron radius and
$W$ is the width of the QPC~\cite{DaviesBook}. This transition
occurs at $B_{\perp} \sim 2.6$~T, a field not much different to the
onset of the resonant feature that Komijani {\it et al} observe.
Such a structure would, in principle, also be robust to asymmetric
gating of the QPC because unlike an impurity, which is fixed, the
edge-states will simply shift laterally along with the confinement
potential. Although the structure reported in
Ref.~\cite{KomijaniEPL10} cannot be given a definitive origin, its
occurrence is interesting and warrants further study.

Putting all of the results in this section together, it is evident
that there may be more than one effect that can give rise to a ZBA
in a QPC. The first is formation of a shallow bound-state, which
produces a ZBA that behaves very similarly to the ZBA caused by the
Kondo effect in quantum dots, with the data fitting a quantum dot
Kondo model (Eq.~5), and showing a gate voltage independent Zeeman
splitting with the prefactor of two characteristic of the Kondo
effect~\cite{KlochanPRL11, SfigakisPRL08}. The second is still of
undetermined origin but behaves differently to the ZBA in quantum
dots, with the data instead fitting a modified Kondo (Eq.~6) or
Arrhenius (Eq.~2) model, and showing a ZBA that survives at large
in-plane fields with a Zeeman splitting that is approximately linear
in gate voltage~\cite{CronenwettPRL02, SarkozyPRB09, ChenPRB09}. The
origin of this latter ZBA is unlikely to be the Kondo effect, at
least in its most basic manifestation, and may be related to more
complicated many-body effects as the QPC is pinched off. An
interesting possibility is electron-phonon coupling, which can also
produce a ZBA in the differential conductance, along with other
characteristics of the $0.7$ plateau, as discussed by Seelig and
Matveev~\cite{SeeligPRL03}. Ultimately, a full understanding of the
ZBA in QPCs will entail more work, however a strong case can be made
for the absence of a direct causal link between the ZBA and the
$0.7$ plateau, and much of the data points towards them being
competing effects. Hence, we will now leave our discussion of the
ZBA and move on to another set of features often observed with the
$0.7$ plateau, the `$0.7$ analogs'.

\section{Revisiting the exchange interaction: $0.7$ plateau analogs above $G_{0}$}

The $0.7$ plateau is not the only non-quantized feature that appears
in the 1D conductance, similarly behaved plateaus are commonly
observed well above $G_{0}$ (e.g., Fig.~1 of
Ref.~\cite{KristensenJAP98} and Fig.~2 of Ref.~\cite{ReillyPRB01}),
and have been extensively studied by Graham {\it et
al}~\cite{GrahamPRL03, GrahamPRB05, GrahamPRB07}. These plateaus,
known as `$0.7$ analogs', are associated with the crossing of
Zeeman-split 1D subband edges with differing subband index $n$ and
opposite spin. Not only do they highlight the important role that
the exchange interaction plays in the physics of 1D electron systems
at low electron density, but they provide significant insight into
the physics of the $0.7$ plateau itself.

\subsection{The `0.7 analog' and `0.7 complement' structures}

Figure 34(a) shows a greyscale plot of the transconductance
$dG/dV_{g}$ versus gate voltage $V_{g}$ ($x$-axis) and in-plane
magnetic field $B_{\parallel}$ ($y$-axis)~\cite{GrahamPRL03}. Low
transconductance regions are white, with the first two integer
conductance plateaus indicated by the numbers 1 and 2 and the $0.7$
plateau indicated by $\alpha_{0}$. The interspersed dark regions
indicate the risers between plateaus, and correspond to the 1D
subband edges crossing the Fermi energy. Accordingly, as
$B_{\parallel}$ is increased, the 1D subbands, which are
spin-degenerate at zero field, divide forming V-shaped structures.
The spin-down/up components form the left/right branches of each V,
and these begin to cross at $B_{\parallel} \gtrsim 10$~T. In the
vicinity of these crossings, additional non-quantized plateaus
appear in the conductance, two of which are highlighted in
Fig.~34(b). These coincide with the crossing of the $1\uparrow$ and
$2\downarrow$ subband edges at $\alpha_{1}$ in Fig.~34(a). The first
is a plateau that falls smoothly from $1.5G_{0}$ to $G_{0}$ with
increasing $B_{\parallel}$ as indicated by the red dashed line. The
evolution of this plateau closely resembles that of the $0.7$
plateau with both $B_{\parallel}$ and temperature (c.f. Fig. 4 of
Ref.~\cite{GrahamPRL03} with Fig.~3(a) of this review and Fig.~1 of
Ref.~\cite{ThomasPhysE02}), and hence this feature is known as a
`$0.7$ analog' (marked A in Fig.~34(a/b)).

\begin{figure}
\includegraphics[width=7cm]{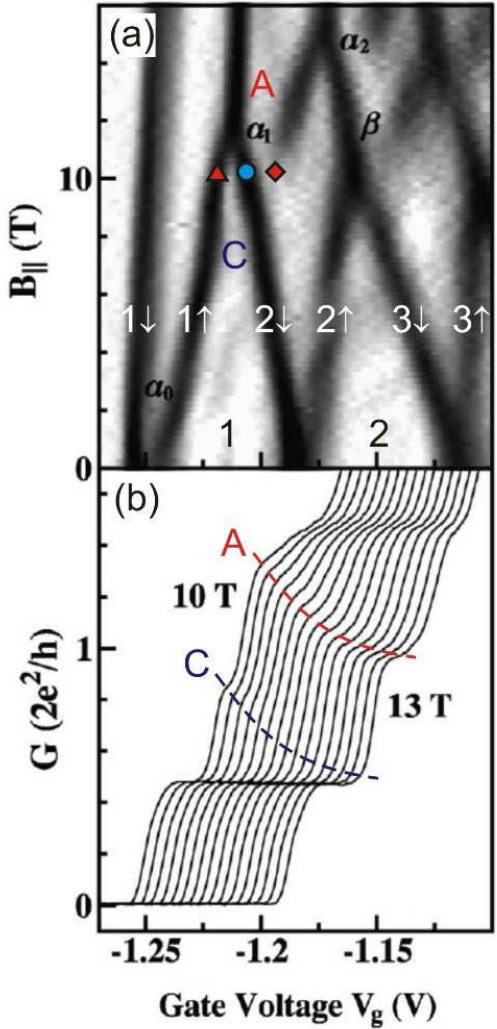}
\caption{(a) Greyscale plot of the transconductance $dG/dV_{g}$ vs
gate voltage $V_{g}$ ($x$-axis) and in-plane magnetic field
$B_{\parallel}$ ($y$-axis) showing the Zeeman splitting of the 1D
subbands, indicated by the white numbers. The black 1 and 2 indicate
the quantized conductance plateaus at $G_{0}$ and $2G_{0}$. (b)
Conductance $G$ vs $V_{g}$ for $B_{\parallel} = 10$ to $13$~T in
steps of $0.2$~T. The red dashed line marked A, and the blue dashed
line marked C indicate the $0.7$ analog and complement structures,
respectively, in (a) and (b). The triangle, circle and square in (a)
facilitate connection to Fig.~36(d), all other annotations are
referred to in the text. Figure adapted with permission from
Ref.~\cite{GrahamPRL03}. Copyright 2003 by the American Physical
Society.}
\end{figure}

The analog is accompanied by a second plateau that falls from
$G_{0}$ to $0.5G_{0}$ with increasing $B_{\parallel}$, indicated by
the dashed blue line, known as a `0.7 complement' (marked C in
Fig.~34(a/b))~\cite{GrahamPRB07}. Since the analog and its
corresponding complement are separated by a single spin-polarized
subband (see Fig.~34(a)), they have a constant separation in
conductance of $0.5G_{0}$. Similar structures are also observed at
higher conductances and for higher order crossings (e.g., the
crossing between $1\uparrow$ and $3\downarrow$ at $\alpha_{2}$ in
Fig.~34(a)), however these are typically weaker conductance
structures due to the higher electron densities at which they occur.
The most significant aspect of the $1\uparrow$/$2\downarrow$
crossing at $\alpha_{1}$ in Fig.~34(a) is the discontinuity in the
evolution of the subband edges in $V_{g}$. This is particularly
evident for the $1\uparrow$ subband edge, which approaches the
intersection from below at $V_{g} = -1.225$~V, vanishes, and then
reemerges above the intersection at $V_{g} = -1.2$~V. A similar
discontinuity is observed at the $2\uparrow$/$3\downarrow$ crossing
at $\beta$ in Fig.~34(a). Theoretical calculations by Berggren {\it
et al}~\cite{BerggrenPRB05} using the Kohn-Sham
spin-density-functional method reproduce these discontinuities
observed experimentally by Graham {\it et al}~\cite{GrahamPRL03}
remarkably well. We now turn to dc source-drain bias spectroscopy to
understand the physics of these crossings in more detail.

\begin{figure}
\includegraphics[width=10cm]{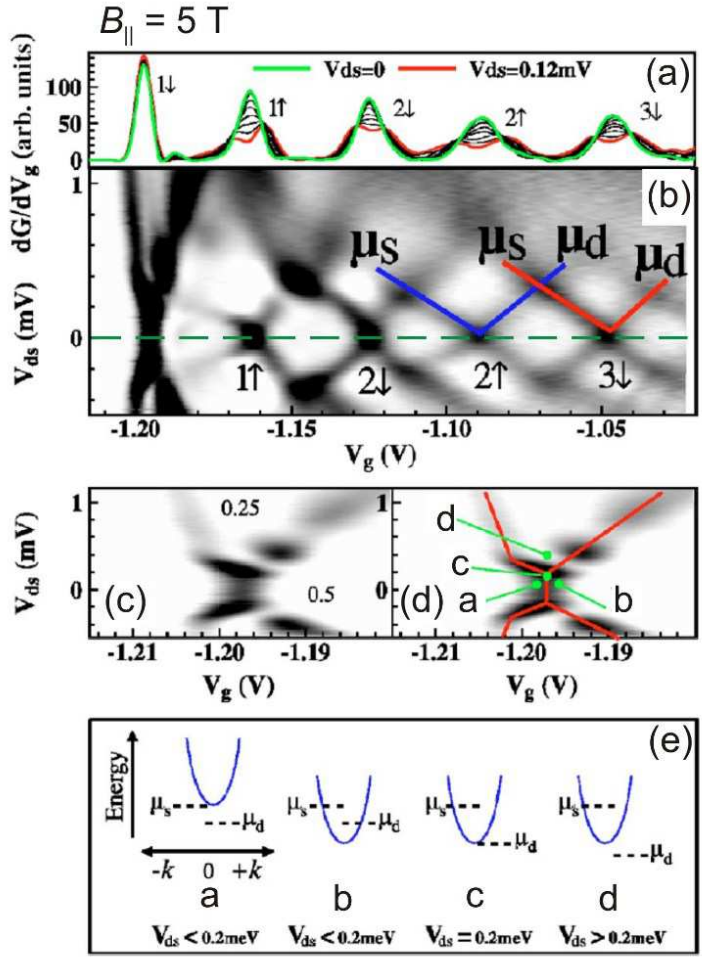}
\caption{(a) $dG/dV_{g}$ vs $V_{g}$ for dc source-drain bias
$V_{ds}$ from $0$~mV to $0.12$~mV at $B_{\parallel} = 5$~T,
corresponding to horizontal sections through (b). (b) Greyscale plot
of $dG/dV_{g}$ vs $V_{g}$ ($x$-axis) and $V_{ds}$ ($y$-axis) at
$B_{\parallel} = 5$~T. The horizontal green dashed line indicates
$V_{ds} = 0$~mV and corresponds to the green trace in (a). The red
and blue V-shaped structures indicate the separation of the source
$\mu_{s}$ and drain $\mu_{d}$ chemical potentials with $V_{ds}$ such
that the left (right) branches correspond to the $2\uparrow$ and
$3\downarrow$ subband edges coinciding with $\mu_{s}$ ($\mu_{d}$),
respectively. (c) Close-up of (b) in the vicinity of the $1\uparrow$
subband edge. (d) Duplicate of (c) with annotations that refer to
the schematic in (e). In each case a-d, the location of the
$1\uparrow$ subband (blue parabola) is shown relative to $\mu_{s}$
and $\mu_{d}$. Figure adapted with permission from
Ref.~\cite{GrahamPRB05}. Copyright 2005 by the American Physical
Society.}
\end{figure}

\subsection{Population-induced energy lowering of spin-down 1D subbands}

Figure~35(b) shows a greyscale plot of transconductance versus
$V_{g}$ ($x$-axis) and dc source-drain bias $V_{ds}$ ($y$-axis),
with $dG/dV_{g}$ versus $V_{g}$ for several specific $V_{ds}$ (i.e.,
horizontal slices through the greyscale) shown in Fig.~35(a),
obtained at $B_{\parallel} = 5$~T~\cite{GrahamPRB05}. Starting from
the left at $V_{ds} = 0$ (horizontal dashed green line), the dark
regions correspond to the first five spin-resolved 1D subband edges
$1\downarrow$, $1\uparrow$, $2\downarrow$, $2\uparrow$ and
$3\downarrow$. The intervening white regions correspond to
half-integer quantized conductance plateaus, with the $0.7$ plateau
having evolved to $0.5G_{0}$ by $B_{\parallel} = 5$~T. With
increasing $|V_{ds}|$, each of these dark regions splits into
V-shaped structures, as highlighted by the blue and red lines for
the $2\uparrow$ and $3\downarrow$ subbands at positive $V_{ds}$. In
each case, the left/right-moving branch corresponds to the subband
edge crossing $\mu_{s}$/$\mu_{d}$, which are separated in energy by
$\mu_{s} - \mu_{d} = eV_{ds}$. For ease of discussion, we will refer
to this as `bias-splitting' of a given subband edge, but it is
important to note that the subband edge itself does not actually
split in energy, it remains as a single entity. Instead the
splitting reflects the change in $V_{g}$ required to take the
subband edge from coinciding with $\mu_{s}$ to coinciding with
$\mu_{d}$. Physically, this gives some indication of the rate at
which a subband edge moves in energy between the two chemical
potentials; this is an important theme in Section 9.

For the $1\uparrow$ and higher subband edges, the evolution with
increasing $V_{ds}$ is straightforward -- it is clearly linear in
Fig.~35(b), and as shown in Fig.~35(a), the corresponding
transconductance peaks are immediately broadened by $V_{ds}$, with
the bias-split peaks becoming resolved at about $V_{ds} = 0.12$~mV.
However, the $1\downarrow$ subband edge behaves differently. In
Fig.~35(b), the splitting for $1 \downarrow$ does not commence until
$V_{ds} \sim 0.2$~mV, and then the rate of splitting is lower than
for the other subbands. This is also evident in Fig.~35(a), where
the transconductance peak corresponding to $1\downarrow$ shows no
visible broadening up to $V_{ds} = 0.12$~mV. The lack of splitting
for $0< |V_{ds}| < 0.2$~mV suggests that the $1\downarrow$ subband
edge passes through both $\mu_{s}$ and $\mu_{d}$ within an
irresolvably small change in $V_{g}$, despite there being a
separation of up to $0.2$~meV between $\mu_{s}$ and $\mu_{d}$.
Graham {\it et al}~\cite{GrahamPRB05} interpreted this as the
$1\downarrow$ subband edge falling rapidly in energy, downwards
through both $\mu_{s}$ and $\mu_{d}$, as soon as the $1\downarrow$
subband begins to populate.

To explore the anomalous behaviour for the $1 \downarrow$ subband
further, Fig.~35(c) shows a close-up of the relevant region of
Fig.~35(b); this data is repeated in Fig.~35(d) where the points a-d
in the overlay correspond to the four schematics in Fig.~35(e). At
point a in Fig.~35(d), the $1\downarrow$ subband edge is just above
$\mu_{s}$, as shown in schematic a in Fig.~35(e). Reducing $V_{g}$
weakens the 1D confinement, pushing the subband lower in energy.
This corresponds to a horizontal motion towards point b in
Fig.~35(d). In moving from a to b, the conductance rises from $0$ to
$0.5G_{0}$, and this can only occur if the $1\downarrow$ edge falls
rapidly downwards through $\mu_{s}$ and $\mu_{d}$ as illustrated in
schematic b. The extent of this population-induced drop in
$1\downarrow$ can be determined by locating the vertex of the
bias-induced splitting, i.e., point c in Fig.~35(d). Here, when
$1\downarrow$ populates the $1\downarrow$ edge immediately drops
down to coincide with $\mu_{d}$ as shown in Fig.~35(e). Finally,
point d represents a return to behaviour characteristic of the other
subbands, with a plateau at $0.25G_{0}$ appearing in addition to
plateaus at $0$ and $0.5G_{0}$ in the conductance. Here, the
population-induced drop in $1\downarrow$ still occurs, it is just
insufficient for the $1\downarrow$ edge to pass through $\mu_{d}$
also. This behaviour is not restricted to the $1\downarrow$ subband
alone, it can occur for higher spin-down subbands, despite becoming
weaker with increasing subband index due to increased electron
density (see Fig.~2(c-f,h) of Ref.~\cite{GrahamPRB05} for this
data).

\subsection{Pinning of the spin-up 1D subbands to the chemical potential}

The behaviour described above for the $1\downarrow$ subband at
$B_{\parallel} = 5$~T also occurs at $B_{\parallel} = 0$, with
Fig.~36(a/b) showing data obtained at $B_{\parallel} = 0$
corresponding to that in Figs.~34(d,e) obtained at $B_{\parallel} =
5$~T. Here, moving from left to right along $V_{ds} = 0$ (green
dashed horizontal line) in Fig.~36(a), the white regions correspond
to plateaus at $0$, $0.7G_{0}$, $G_{0}$ and $2G_{0}$. Figure~36(b)
shows a close-up in the vicinity of the $0.7$ plateau, with the
evolution of the $1\downarrow$ and $1\uparrow$ subband edges
highlighted by the red solid and blue dotted lines, respectively. A
corresponding schematic of the subband edge evolution is presented
in Fig.~36(c). Consistent with the data at finite magnetic field,
the bias-splitting of the $1\downarrow$ subband edge begins at a
substantial finite bias $|V_{ds}| \sim 0.5$~mV (Fig.~36(b)),
indicating that the $1\downarrow$ subband drops sharply in energy
once it starts populating. Interestingly, the $1\uparrow$ subband
edge also shows peculiar behaviour in the vicinity of the $0.7$
plateau. As shown in Fig.~36(c), the $1\uparrow$ subband edge should
bias-split from point c into two branches corresponding to crossings
with $\mu_{s}$ (thin dashed left-moving diagonals) and $\mu_{d}$
(thick solid right-moving curves). However, the left-moving branch
of $1\uparrow$ is clearly missing in the data in Fig.~36(b). This
indicates that the $1\uparrow$ subband edge becomes pinned at $\mu$
when it populates, as predicted by Kristensen {\it et
al}~\cite{KristensenPRB00}. To better understand the pinning of the
$1\uparrow$ edge, we briefly return to data obtained at
$B_{\parallel} = 9$~T.

\begin{figure}
\includegraphics[width=16cm]{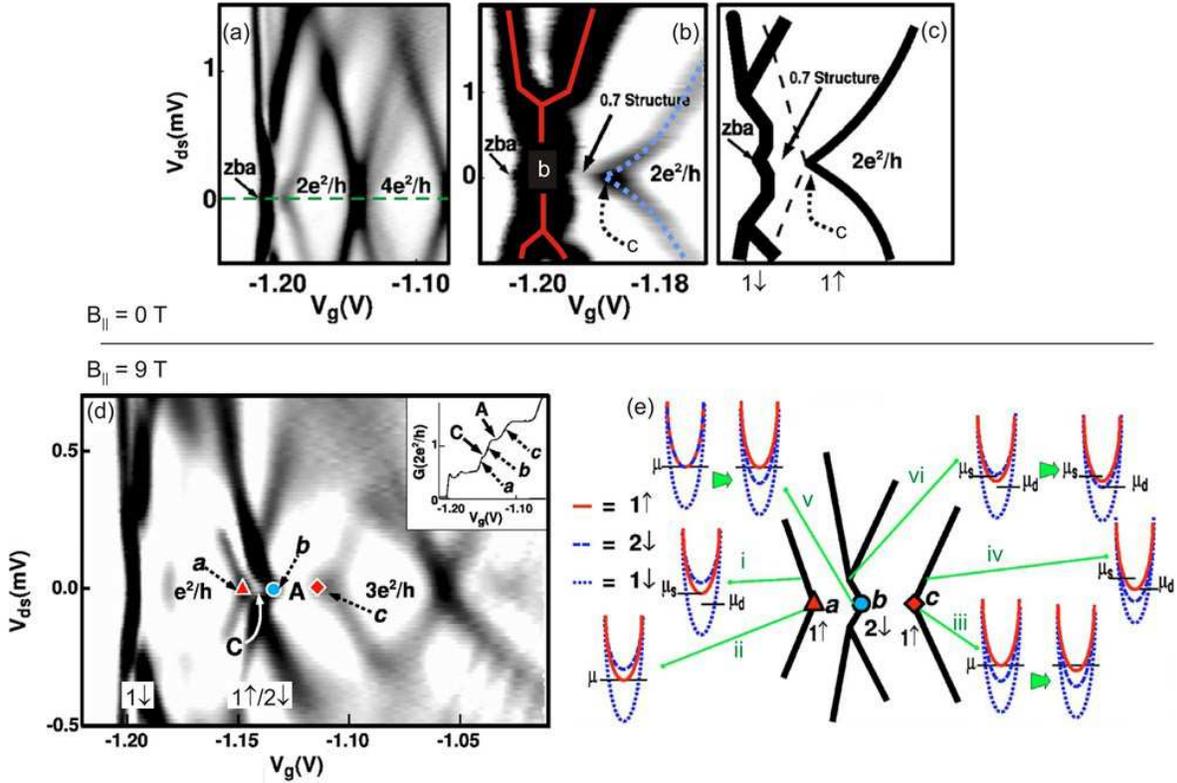}
\caption{(a) Greyscale plot of transconductance $dG/dV_{g}$ vs gate
voltage $V_{g}$ ($x$-axis) and dc source-drain bias $V_{ds}$
($y$-axis) at $B_{\parallel} = 0$~T. The green dashed horizontal
line indicates $V_{ds} = 0$. (b) A close up of the data in (a)
focussing on the bias-splitting of the $1\uparrow$ (red solid lines)
and $1\downarrow$ (blue dashed lines) subband edges. (c) A schematic
of the subband edge evolution in (b). The missing left branch of the
bias-splitting for the $1\downarrow$ subband edge is indicated by
the thin dashed line. Panels (a)-(c) were obtained at $B_{\parallel}
= 0$~T, panels (d)-(e) were obtained at $B_{\parallel} = 9$~T. (d)
Greyscale plot of $dG/dV_{g}$ vs $V_{g}$ ($x$-axis) and $V_{ds}$
($y$-axis) at $B_{\parallel} = 9$~T focussing on the crossing of the
$1\uparrow$ and $2\downarrow$ subband edges. The corresponding $G$
vs $V_{g}$ trace at $V_{ds} = 0$~mV is shown inset upper right. The
A and C indicate the analog and complement structures, respectively.
The red triangle, blue circle and red diamond correspond to those in
(e) and Fig.~34(a). Since this data is obtained at the magnetic
field where the edges of adjacent subbands of opposite spin cross,
only plateaus at odd multiples of $G_{0}/2$ are observed in the
conductance, the white regions in the main panel corresponding to
the plateaus at $G_{0}/2$ and $3G_{0}/2$ are indicated. (e)
Schematics i-vi indicate the positions of the $1\downarrow$ (blue
dotted parabola), $1\uparrow$ (red solid parabola) and $2\downarrow$
(blue dashed parabola) subbands relative to the source and drain
chemical potentials $\mu_{s}$ and $\mu_{d}$, or common chemical
potential $\mu$ if $V_{ds} = 0$. Figure adapted with permission from
Ref.~\cite{GrahamPRB07}. Copyright 2007 by the American Physical
Society.}
\end{figure}

Figure~36(d) shows a source-drain bias greyscale obtained in the
vicinity of the $1\uparrow/2\downarrow$ subband edge crossing in
Fig.~34(a). The triangle, circle and diamond in Fig.~36(d)
correspond to those in Fig.~34(a), with the analog (A) and
complement (C) plateaus in the conductance, shown inset to
Fig.~36(d), appearing immediately to the right and left of the blue
circle in both figures. Figure~36(e) illustrates the physics behind
the high transconductance (dark) regions in Fig.~36(d), where the
evolution of both the $1\uparrow$ and $2\downarrow$ subband edges
differ markedly from what would be expected in the absence of
electron-electron interactions. The evolution of the $2\downarrow$
edge occurs as explained in Section 8.2. However, the evolution of
the $1\uparrow$ edge is very different to what we have previously
seen for spin-up subbands (c.f. Fig.~35(b)). Although it initially
appears that the $1\uparrow$ subband populates twice as $V_{g}$ is
reduced, the situation is more complex, as evident from schematics i
- vi in Fig.~36(e). The behaviour is best understood by approaching
the figure from the left- and right-hand sides. First, moving
horizontally from the left edge of Fig.~36(e) to location i
corresponds to reducing $V_{g}$ at constant $V_{ds}$. This causes
the $1\uparrow$ subband edge to drop in energy, reaching $\mu_{s}$
at i, whereupon the $1\uparrow$ subband begins to populate. Moving
diagonally from i to ii (point a), $\mu_{s}$ and $\mu_{d}$ come back
together and the $1\uparrow$ edge tracks $\mu_{s}$ to sit at
$\mu_{s} = \mu_{d} = \mu$. Moving over to the right edge of
Fig~36(e) and heading horizontally left towards iv, the $1\uparrow$
edge rises up to meet $\mu_{d}$ and the $1\uparrow$ subband begins
to depopulate. Moving diagonally from iv to iii (point c), we also
arrive at the $1\uparrow$ edge coinciding with $\mu$. Thus
$1\uparrow$ edge starts at point a, coinciding with $\mu$, and ends
up at point c, still coinciding with $\mu$. The reason for this
pinning of $1\uparrow$ edge at $\mu$ is that this subband populates
very slowly, in contrast to the $2\downarrow$ subband, which
populates rapidly as $V_{g}$ is changed. With this in mind, the
structure of the two subband edges in the dc bias greyscale makes
sense. The rapid population of $2\downarrow$ pushes the left-moving
and right-moving branches of the V-shaped structure that would be
expected in the absence of interactions closer together in $V_{g}$.
This makes the two branches coincide at point b and not diverge
until finite $V_{ds}$ (i.e., vi in Fig.~36(e)). The gradual
population of $1\uparrow$ instead stretches the V-shaped
bias-splitting along $V_{g}$ (i.e., horizontally), giving the
appearance in Fig.~36(e) that $1\uparrow$ has bias-split such that
the subband edge coincides with a chemical potential both before and
after the $2\downarrow$ edge passes through the same two chemical
potentials (i.e., source and drain, which are separated in energy if
$V_{ds} \neq 0$). Most importantly, the different population rates
for $1\uparrow$ and $2\downarrow$ mean that these subbands rearrange
in energy in moving from point a to point c. This is evident at ii
in Fig.~36(e), where $2\downarrow$ is above $1\uparrow$, and iv,
where $2\downarrow$ is below $1\uparrow$, with the $1\uparrow$ edge
remaining pinned at $\mu$ throughout. This behaviour is reminiscent
of exchange-induced phase transitions at Landau-level
crossings~\cite{GiulianiPRB85, PiazzaNat99}, and is addressed in
more detail in Ref.~\cite{GrahamPRL08}. The vast difference in
population rates between spin-up and spin-down subbands also
explains why the analog and complement structures are non-quantized
but separated by a fixed value of $0.5G_{0}$. At $T > 0$, the
pinning of the $1\uparrow$ edge at $\mu$ places it where the Fermi
function $0 < f(\mu,T) < 1$, giving a conductance $G_{complement} =
G_{1\downarrow} + G_{1\uparrow} = 0.5(1 + f)G_{0} < G_{0}$. In
contrast, the $2\downarrow$ edge is rapidly driven well below $\mu$
to where $f \sim 1$ when it populates, giving $G_{analog} =
G_{1\downarrow} + G_{1\uparrow} + G_{2\downarrow} = 0.5(1 + f +
1)G_{0} = G_{complement} + 0.5G_{0} < 1.5G_{0}$.

\subsection{Connection to the $0.7$ plateau and the spin-gap phenomenological models}

Graham {\it et al}~\cite{GrahamPRB07} argue that by direct
comparison of Figs.~36(c) and (e), it is evident that the $0.7$
plateau is simply a special case of the more general physics related
to the analog and complement structures. Here the relevant subbands
are $1\downarrow$ and $1\uparrow$, with the $0.7$ plateau being the
corresponding analog structure. The corresponding complement would
be expected at $0.2G_{0}$, and although such structures have been
reported (e.g., Ref.~\cite{RamvallAPL97, dePicciottoPRL04}), they
are not consistently observed. Within the analog framework presented
by Graham {\it et al}, the $0.7$ plateau is caused by pinning of the
$1\uparrow$ subband edge at the chemical potential $\mu$, coincident
with a sudden drop in $1\downarrow$, upon population of the first
subband. The sudden drop in $1\downarrow$ while $1\uparrow$ remains
pinned represents the almost immediate opening of a spin-gap $\Delta
E^{\uparrow \downarrow} = E_{1\uparrow} - E_{1\downarrow}$ with a
magnitude of order $0.1 - 0.5$~meV. Hence the experimental data is
in good agreement with the various spin-gap phenomenological models
for the $0.7$ plateau~\cite{KristensenPRB00, BruusPhysE01,
ReillyPRL02, ReillyPRB05} as well as numerous calculations
predicting the development of a spin-polarization within the
QPC~\cite{LasslPRB07, WangPRB96, JakschPRB06, LindPRB11}, however
two aspects warrant further discussion.

The first is the nature of the opening of the spin-gap. In
Ref.~\cite{GrahamPRB05}, Graham {\it et al} state that the abrupt
drop in $1\downarrow$ indicated by their data is not in agreement
with the gradual and linear opening in spin-gap with gate voltage in
the Reilly model. Considering the typical schematic for this model
in Fig.~10(a)~\cite{ReillyPRL02, ReillyPRB05}, it might initially
seem that Reilly's spin-gap model would not explain the data in
Ref.~\cite{GrahamPRB05}. However it is important to bear in mind
that a linear opening of the spin-gap with more positive $V_{g}$ is
only half of the picture. Ignoring spin for a moment, the other half
of the picture is that a more positive $V_{g}$ leads to filling of
the 1D subbands, which results in a non-linear rise in the chemical
potential due to the form of the 1D density of
states~\cite{ReillyPRL02}. Putting both of these together one gets a
picture more like that in Fig.~10(b) where, viewed from the
perspective of how the subband edges move relative to the chemical
potential, the $1 \downarrow$ subband edge does drop
abruptly~\cite{ReillyPhysE06}. It is not clear whether this drop is
abrupt enough to reproduce the experimental results in
Ref.~\cite{GrahamPRB05}, however, it is worth noting that the Reilly
model produces $0.7$ analogs at the $1\uparrow/2\downarrow$ and
$2\uparrow/3\downarrow$ subband edge crossings in addition to the
$0.7$ plateau, as is clear in Fig.~4 of Ref.~\cite{ReillyPhysE06}.
Note also that even in the BCF model~\cite{KristensenPRB00,
BruusPhysE01} there is no abrupt or discontinuous drop in the
majority spin subband. Additionally, density functional theory
calculations by Jaksch {\it et al}~\cite{JakschPRB06} also support a
linear approximation for the the opening of the spin-gap with
$V_{g}$ upon population of a subband, and do not show any remarkably
abrupt opening of this gap. Similar behaviour has also been observed
in calculations by Lind {\it et al}~\cite{LindPRB11}. While
Hartree-Fock calculations by Lassl {\it et al}~\cite{LasslPRB07} do
predict an abrupt drop in $1\downarrow$, this occurs only very
briefly before the drop slows and ultimately, the spin-gap closes
(see Fig.~4 of Ref.~\cite{LasslPRB07}).

The second aspect is the pinning of the $1\uparrow$ edge to the
chemical potential. The BCF model explicitly includes pinning of the
minority spin subband edge to the chemical potential, in agreement
with Graham {\it et al}~\cite{GrahamPRB07}; the Reilly model makes
no such stipulation. While at first this might appear as an
inconsistency, Fig.~10(a/b) offer a possible resolution -- although
the $1\uparrow$ subband edge rises in energy as $V_{g}$ is made more
positive, so does the Fermi energy. However $E_{F}$ does so
nonlinearly such that for some interval of $V_{g}$, the $1 \uparrow$
subband edge remains very close to $\mu$, resulting in a
coincidental `quasi-pinning' of the $1\uparrow$ edge to $\mu$.
Ultimately, in view of Fig.~10(b), the correspondence between the
experimental data obtained by Graham {\it et al}~\cite{GrahamPRB05,
GrahamPRB07} and the Reilly model~\cite{ReillyPRL02, ReillyPRB05,
ReillyPhysE06} is stronger. Finally, it is worth noting that
theoretical support for pinning of the spin-up subband extends
beyond the phenomenological models -- Hartree-Fock calculations by
Lassl {\it et al}~\cite{LasslPRB07}, where the screened Coulomb
interaction between electrons is approximated by a repulsive contact
potential, lead to the pinning of the spin-up subband edge at $\mu$
and the opening of a spin-gap as the spin-down subband edge drops
below the chemical potential.

\section{dc Conductivity measurements of 1D subband population behaviour}

Additional information about the motion of the 1D subbands in energy
as they populate is provided by the dc conductance measurements
obtained by Chen {\it et al} ~\cite{ChenPRB09b, ChenAPL08,
ChenPhysE08, ChenNL10}. Before discussing these measurements, we
will briefly look at how the dc conductance compares to the more
commonly measured ac conductance, since the difference is subtle but
important. A more comprehensive discussion of dc conductance
measurements can be found in a recent article by Micolich and
Z\"{u}licke~\cite{MicolichJPCM11}.

\subsection{The practicalities of dc conductance measurements of QPCs}

Up to this point, all the measurements of the 1D conductance
discussed in this review have involved the ac or differential
conductance. These measurements are obtained by applying a small ac
bias $V^{ac}_{sd}$, typically $10-100~\mu$V in rms amplitude at
$5-200$~Hz, to the source. This bias is held constant and is
separate to the dc source-drain bias $V_{sd}$ referred to
previously. The resulting ac drain current $I^{ac}_{d}$, typically
$1-100$~nA, is measured using a lock-in amplifier, giving the
conductance $G$ used so far. Although this conductance is obtained
practically as $G_{ac} = I^{ac}_{d}/V^{ac}_{sd}$ it is really a
differential conductance $G_{ac} = dI/dV_{sd}$, the reason becomes
apparent on considering how the dc bias measurements are made.

At each gate voltage $V_{g}$, the device has an particular $I$-$V$
characteristic, and $G_{ac}$ measures only a small component of
this. The ac conductance is obtained by applying a finite
$V^{ac}_{sd}$ and measuring the resulting current $I^{ac}_{d}$ at
the same frequency. Physically, this corresponds to taking source
and drain potentials, which are on average equal, and periodically
slightly raising and lowering one relative to the other, causing an
oscillatory current. The resulting conductance is thus the average
slope of the $I$-$V$ curve within the small bias window,
$V^{ac}_{sd}$ wide, centered at $V_{sd} = 0$. In practice, a dc
source-drain bias is applied using an `ac/dc adder' circuit,
resulting in an excitation to the source consisting of the small ac
bias riding on top of the dc bias (i.e., with the drain at ground,
the source potential is $V_{sd} + V^{ac}_{sd}$~\footnote{Note that
an implicit sign reversal is assumed here, because a positive bias
applied to one reservoir lowers it in energy with respect to the
other. The convention above is used in keeping with the literature
where it is generally thought of as a positive $V_{sd}$ raising the
source above the drain.}). In this case, $G_{ac}$ gives the slope of
the $I$-$V$ curve within a $V^{ac}_{sd}$ wide window centered on
$V_{sd}$, and in this light the differential nature of this
conductance becomes clear. In contrast, the dc conductance $G_{dc}$
is obtained from the dc drain current $I^{dc}_{d}$ obtained at a
given $V_{sd}$, with any superimposed ac bias or noise removed
through low-pass filtering. The dc conductance is the integral of
the ac conductance over the range from $V = 0$ to $V = V_{sd}$, and
is sensitive to motion of the subband edge through the entire range
between $\mu_{s}$ and $\mu_{d}$. The ac and dc conductances can be
measured simultaneously by using a current preamplifier, which
converts the broadband drain current (i.e., dc + all ac components)
into a broadband voltage with a known $I:V$ conversion/gain, and
provides a virtual ground at the drain. This voltage can be
duplicated, with one copy sent to a lock-in amplifier to extract
$I^{ac}_{d}$, and the other via a low-pass filter to a multimeter to
extract $I^{dc}_{d}$.

An understanding of the difference between these two conductance
measurements is important to interpreting what $G_{ac}$ and $G_{dc}$
say about the motion of the 1D subbands relative to the chemical
potential. Starting with $G_{ac}$; imagine a 1D system with finite
$V_{sd}$ where the $n$~th subband edge is below $\mu_{d}$ and the
$n+1$~th subband edge is above $\mu_{s}$. Making $V_{g}$ more
positive lowers these subbands in energy, but there will be no
change in the conductance until the $n+1$~th subband edge comes
within the $V^{ac}_{sd}$ bias window centered at $\mu_{s}$. The
conductance will then rise by $0.25G_{0}$ or $0.5G_{0}$ depending on
whether the subband is spin-polarized or spin degenerate. Once the
subband edge drops below the bottom of this window, $G_{ac}$ remains
constant until it reaches the bias window centered at $\mu_{d}$,
when $G_{ac}$ rises again. The outcome of the same process is
different for the dc conductance. Again, nothing changes with
$G_{dc}$ until the $n+1$~th subband edge reaches $\mu_{s}$, but now
it enters a bias window that extends all the way down to $\mu_{d}$.
As the subband moves downwards, it brings states that can contribute
to the conduction into the window such that $G_{dc}$ rises
continuously and gradually until the subband edge reaches $\mu_{d}$.
At this point, $G_{dc}$ takes a quantized value, stepping up by
$0.25G_{0}$ or $0.5G_{0}$ from the value $G_{dc}$ had before the
subband edge reached $\mu_{s}$, depending on whether the subband is
spin-polarized or spin degenerate. One might ask why the ac
conductance does not change gradually between plateaus as it does
for the dc conductance. It does, but only within the narrow ac bias
windows local to $\mu_{d}$ and $\mu_{s}$, where it simply broadens
the risers, lowering the corresponding transconductance. Thus
$G_{dc}$ is useful as it enables the motion of the subband edges to
be tracked over the entire range between $\mu_{s}$ and $\mu_{d}$ as
$V_{g}$ is changed.

\subsection{Additional evidence for rapid drop of spin-down and pinning of spin-up subbands}

Figures~37(a) and (b) show transconductance colour-maps obtained
using the ac and dc conductances, respectively, at $B_{\parallel} =
16$~T. In both cases, the subbands are spin-resolved due to the high
magnetic field and the plateaus (indicated by dark regions) at
$V_{sd} = 0$ occur at integer multiples of $0.5G_{0}$. The ac
transconductance takes its characteristic form (c.f., Fig.~35(b)),
and shows a strong odd-even behaviour -- the bright regions at
$V_{sd} = 0$ are smaller for the $\uparrow$ subbands than the
$\downarrow$ subbands. This corresponds to a faster drop in
transconductance with $V_{sd}$ indicating that the $\uparrow$
subband edges move more slowly through $\mu_{s}$ and $\mu_{d}$ than
the $\downarrow$ subband edges. This behaviour is more strikingly
evident in Fig.~2 of Ref.~\cite{ChenPRB09b} where plots of
$dG_{ac}/dV_{g}$ versus $V_{g}$ as a function of $T$ and $V_{sd}$
are shown. The onset of bias-splitting for the $\downarrow$ subband
edges also occurs at a slightly higher $V_{sd}$ than for $\uparrow$
subband edges, as found by Graham {\it et al}~\cite{GrahamPRB05}
(n.b., this bias-splitting is not an actually energy splitting of
the 1D subbands, it is an effect of the separation in $\mu_{s}$ and
$\mu_{d}$ with $V_{sd}$ -- see Section 8.2).

\begin{figure}
\includegraphics[width=12cm]{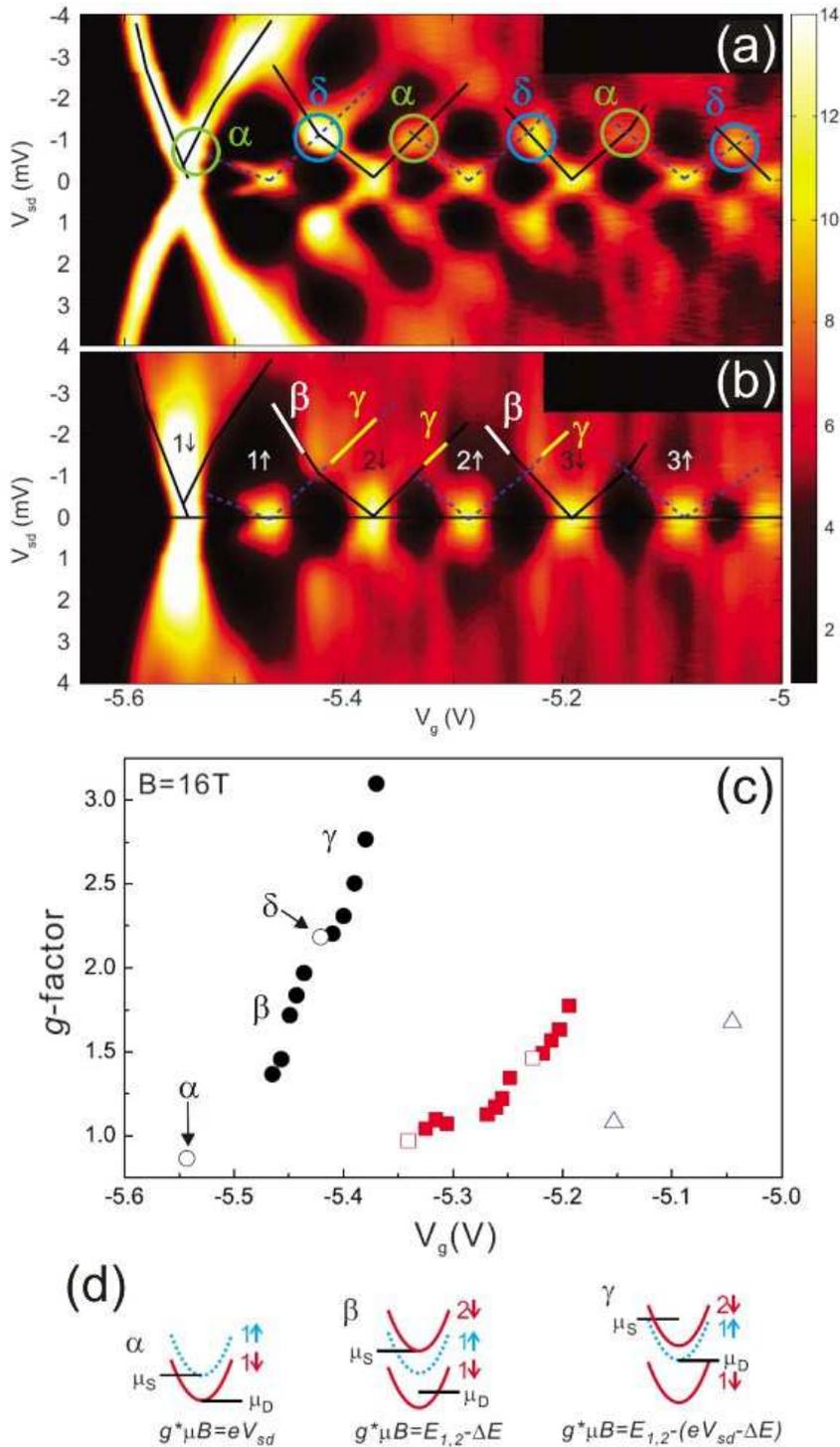}
\caption{Colour maps of (a) the ac transconductance $dG_{ac}/dV_{g}$
and (b) the dc transconductance $dG_{dc}/dV_{g}$ vs gate voltage
$V_{g}$ ($x$-axis) and dc source-drain bias $V_{ds}$ ($y$-axis)
obtained simultaneously at $B_{\parallel} = 16$~T and $T = 130$~mK.
The green and blue circles marked $\alpha$ and $\delta$ in (a)
indicate the locations where the open symbols in (c) are obtained.
The white and yellow lines marked $\beta$ and $\gamma$ indicate
locations where the solid symbols in (c) are obtained. (c)
Measurements of effective $g$-factor $g^{*}$ vs $V_{g}$, the open
(closed) symbols indicate data obtained using ac (dc) conductance.
(d) Schematics illustrating the locations of subbands and chemical
potentials for the $\alpha$ point and the $\beta$ and $\gamma$
branches, $\delta$ is not shown, but is similar to $\alpha$ except
that the $2\downarrow$ edge coincides with $\mu_{s}$ and the
$1\uparrow$ edge coincides with $\mu_{d}$. Figures (a-d) adapted
with permission from Ref.~\cite{ChenPRB09b}. Copyright 2009 by the
American Physical Society.}
\end{figure}

The corresponding dc transconductance in Fig.~37(b) has a slightly
different appearance and interpretation. We start with the
horizontal line at $V_{sd} = 0$, along which appear alternating
bright and dark regions, exactly as in the ac transconductance in
Fig.~37(a). This occurs because for $V_{sd} \lesssim V^{ac}_{sd}$
the two methods are approximately equivalent -- any subband edge
passing through the $eV_{sd}$ gap between $\mu_{s}$ and $\mu_{d}$
will necessarily be within either of the $V^{ac}_{sd}$ wide,
overlapping bias windows centered on $\mu_{s}$ and $\mu_{d}$. In
considering what happens as $V_{sd}$ is increased, one needs to
remember that the ac transconductance is only non-zero when the
subband edge crosses $\mu_{s}$ and $\mu_{d}$, while the dc
transconductance is nonzero for the entire transition of the subband
edge between $\mu_{s}$ and $\mu_{d}$. Hence the V-shaped
bias-splitting features in the ac transconductance should appear as
bright triangular regions, as for the $2\downarrow$ subband in
Figs.~37(a/b). Here the brightness corresponds to the rate in
$V_{g}$ at which the subband edge moves between $\mu_{s}$ and
$\mu_{d}$. In Fig.~37(b), the brightness of the bias-splitting
triangles alternates, appearing brighter for $\downarrow$ subbands
and much darker for $\uparrow$ subbands. For the $1\uparrow$ subband
in particular, the bias-splitting triangle is entirely invisible.
The much lower transconductance for the $\uparrow$ subbands
indicates that they fill more slowly than the $\downarrow$ subbands.
The $1\downarrow$ subband has a particularly high dc
transconductance, consistent with the $\downarrow$ subbands dropping
rapidly in energy upon populating~\cite{GrahamPRB05}.

\subsection{Evolution of the Land\'{e} $g$-factor with subband filling}

Chen {\it et al}~\cite{ChenPRB09} used the dc conductance to track
the motion of the subband edges over a wide energy range and thereby
investigate how the effective $g$-factor evolves as the 1D subbands
populate. Figure~37(c) shows measurements of $g^{*}$ versus $V_{g}$,
with the regions $\alpha$, $\beta$ and $\gamma$ corresponding to the
adjacent energy level schematics. The open symbols represent $g^{*}$
values obtained from the ac transconductance by measuring the
source-drain bias where adjacent subband edges of opposite spin
coincide in energy~\cite{PatelPRB91a}. The green circles marked
$\alpha$ in Fig.~37(a) correspond to the $n\downarrow$ ($n\uparrow$)
edge coinciding with $\mu_{d}$ ($\mu_{s}$) allowing $g^{*}$ to be
obtained directly from the $V_{sd}$ at the crossing. The blue
circles marked $\delta$ correspond instead to the $n\uparrow$
($n+1\downarrow$) edge coinciding with $\mu_{d}$ ($\mu_{s}$). This
situation is similar to $\alpha$ except that energy separations of
the form $E_{1,2} = E_{2\downarrow} - E_{1\downarrow}$ are needed to
extract $g^{*}$, and these can be obtained at higher $V_{sd}$
subband edge intersections, as indicated by the grey/white circles
in Fig.~37(a). In this sense, the method used for obtaining the
$\alpha$ and $\delta$ points is no different to that used in
previous studies of the spin-splitting of 1D
subbands~\cite{ThomasPRL96, PatelPRB91, DaneshvarPRB97,
DanneauPRL06, MartinAPL08, MartinPRB10}.

The dc conductance can be used to obtain additional $g^{*}$ values
at intermediate $V_{g}$, which can be divided into two `strands'
corresponding to schematics $\beta$ and $\gamma$. Note however, that
these are just two in a set of four more general strands that can be
defined and measured~\cite{MicolichJPCM11}. The dc conductance data
needed to measure $g^{*}$ is obtained along a zig-zag path that runs
between the lowest and second-lowest subband crossings in $V_{sd}$
-- the exact regions used to obtain the data in Fig.~37(c) are
indicated by the white and yellow lines in Fig.~37(b). As the two
corresponding schematics show, the key to the measurement is $\Delta
V$, which is obtained from $G_{dc}$. As the $n\downarrow$ edge moves
from $\mu_{s}$ to $\mu_{d}$, $G_{dc}$ will rise by $2e^{2}/h$, and
so the fraction of that $2e^{2}/h$ that has been added to the
conductance measured for $\beta$ in Fig.~37(d) gives $\Delta
V/V_{sd}$ (i.e. $G_{dc} = (m + \Delta V/V_{sd}) e^{2}/h$, where $m$
is the number of spin-polarised 1D subbands beneath $\mu_{d}$).
Similar arguments hold for $\gamma$ in Fig.~37(d). In both cases,
the energy separation $E_{1,2}$ is needed to extract $g^{*}$ for the
first subband (and $E_{2,3} = E_{3\downarrow} - E_{2\downarrow}$ for
the second subband). This points to a significant source of error in
the $g^{*}$ values, because $E_{1,2}$ and $E_{2,3}$ necessarily
depend on $g^{*}$ themselves (see Fig.~2(d) of
Ref.~\cite{MicolichJPCM11}). This error is small for $V_{g}$ close
to where $E_{1,2}$ and $E_{2,3}$ are measured in Fig.~37(a), but
become significant further away, for example, at the top of the
$\gamma$ branch.

Indeed, the measured $\gamma$ branch data has an additional
implication. Chen {\it et al} state that $|g^{*}|$ oscillates as
each subsequent 1D subband begins to populate, pointing out that
this is similar to the oscillatory $|g^{*}|$ in quantum Hall systems
as the Landau levels are filled~\cite{AndoJPSJ74} and to the 1D
theoretical prediction by Wang and Berggren~\cite{WangPRB96}. The
theoretical prediction that Chen {\it et al} refer to is shown in
Fig.~4(b), and care needs to be taken in comparing experimental data
to this. What Wang and Berggren plotted is $g^{*}_{1}$ monitored
{\it continuously} as the higher subbands are gradually filled. The
second subband begins to occupy at $n_{1D} = 5 \times 10^{5}$
cm$^{-1}$ (see Fig.~4(a), which coincides with the rise out of the
first minima in Fig.~4(b)). However, the measurement made by Chen
{\it et al} is insensitive to $g^{*}_{1}$ for $V_{g} \gtrsim
-5.38$~V (i.e., to the right of the end of the first $\gamma$ branch
in Fig.~37(c))~\cite{MicolichJPCM11}. More crucially however, the
$1\uparrow$ edge reaches $\mu_{d}$ at the first $\delta$ point, and
it is here that Fig.~4(a) predicts a precipitous drop in
$g^{*}$~\cite{WangPRB96}, in stark contrast to the continued rise
measured by Chen {\it et al}~\cite{ChenPRB09b} in Fig.~37(c).
Instead, the observed drop in $g^{*}$ at the end of the $\gamma$
branch in Fig.~37(c) occurs because the gap being measured changes
from that between the $1\downarrow$ and $1\uparrow$ edges to that
between the $2\downarrow$ and $2\uparrow$ edges (see Fig.~1 of
Ref.~\cite{MicolichJPCM11}, note that the $\gamma$ branch in
Ref.~\cite{ChenPRB09b} corresponds to the $\delta\gamma_{1,2}$
branch in Ref.~\cite{MicolichJPCM11}). In this light, the data in
Fig.~37(c) appear more supportive of the Reilly
model~\cite{ReillyPRL02, ReillyPRB05}, where the spin-gap for each
subband opens linearly once that subband begins to populate, but
remains small until population commences.

In a follow-up to the work by Chen {\it et al}, Micolich and
Z\"{u}licke~\cite{MicolichJPCM11} have reanalysed the data in
Ref.~\cite{ChenPRB09b} to focus instead on the motion of the subband
edges with $V_{g}$. The reanalysis shows that the $2\downarrow$
subband edge drops rapidly in energy upon populating, consistent
with Ref.~\cite{GrahamPRB05}, however this subband edge appears to
momentarily track $\mu_{s}$, which rises relative to $\mu_{d}$ due
to increasing $V_{sd}$, as the increasing $V_{g}$ attempts to drive
the $2\downarrow$ edge below $\mu_{s}$~\cite{MicolichJPCM11}. This
occurrence appears in the $g^{*}$ data as a suppression of the
left-most data point in the $\gamma$ branch below the otherwise more
linear rise that this branch takes further to the right in
Fig.~37(c). A similar albeit weaker trend is observed for the
$3\downarrow$ subband edge. This behaviour may indicate that initial
population of the $\downarrow$ subbands involves an interplay
between competing energy contributions. It is interesting to note
also that the left-most parts of the $\beta$ branch sit below the
line joining $\alpha$ and $\delta$. Further data for how this branch
evolves closer to $\alpha$ would be a useful contribution. The
population of the $2\uparrow$ subband occurs at a slower rate,
consistent with Ref.~\cite{GrahamPRB07}. However, as the $2\uparrow$
subband edge approaches $\mu_{d}$, the population rate becomes
comparable to that of the $2\downarrow$ subband when the
$2\downarrow$ edge approaches $\mu_{d}$. This suggests that the
spin-gap between $2\downarrow$ and $2\uparrow$ may close again, or
at least stop opening, as the $\uparrow$ subband approaches and
passes through $\mu_{d}$, roughly consistent with recent
calculations~\cite{LasslPRB07, JakschPRB06, LindPRB11}.

Additionally, Chen {\it et al}~\cite{ChenNL10} report the
observation of a non-quantized plateau in $G_{dc}$ at $\sim 0.7
G_{0}$ that occurs at finite source-drain bias $V_{sd} \sim 0.5$~mV,
and drops smoothly to $0.5G_{0}$ as the in-plane magnetic field is
raised to $14$~T (see Fig.~1(a) of Ref.~\cite{ChenNL10}). As
discussed above, in a single-particle picture, plateaus at $G_{dc} =
nG_{0}$ only occur when the bias window contains no subband edges. A
non-quantized plateau indicates that a subband edge within the bias
window has stopped moving in energy as $V_{g}$ is changed, and the
$G_{dc}$ value can be used to identify where this occurs. An
analysis of this plateau suggests that the $1\uparrow$ subband stops
populating for a short range of $V_{g}$ around when the
$1\downarrow$ subband edge reaches $\mu_{d}$ and begins passing
current in both directions. It is interesting to note that the
reanalysis in Ref.~\cite{MicolichJPCM11} indicates that similar
behaviour occurs for the $2\uparrow$ subband, which populates
relatively slowly immediately after the $2\downarrow$ edge reaches
$\mu_{d}$, compared to the more rapid population of $2\uparrow$ as
it reaches $\mu_{d}$, as discussed in the preceding paragraph.

Combined, this data suggests that the population rate of a 1D
subband can vary significantly with gate voltage, and it is clear
that further studies using the dc conductance techniques pioneered
by Chen {\it et al}~\cite{ChenPRB09b, ChenAPL08, ChenPhysE08,
ChenNL10} may offer new insight regarding the physics of 1D subband
population.

\subsection{Finite bias plateaus at $0.25G_{0}$ and $0.85G_{0}$}

In two separate papers, Chen {\it et al} also discuss dc conductance
studies of plateaus observed at $0.25G_{0}$~\cite{ChenAPL08} and
$0.85G_{0}$~\cite{ChenPhysE08} in $G_{ac}$ versus $V_{g}$ as a
function of finite $V_{sd}$ (see Fig.~38(a)). The interpretation of
these plateaus and associated features in the dc bias greyscale
plots is intricate, and we will start by looking at the $0.25G_{0}$
plateau observed at large $V_{sd} \gtrsim 2$~mV.

\begin{figure}
\includegraphics[width=10cm]{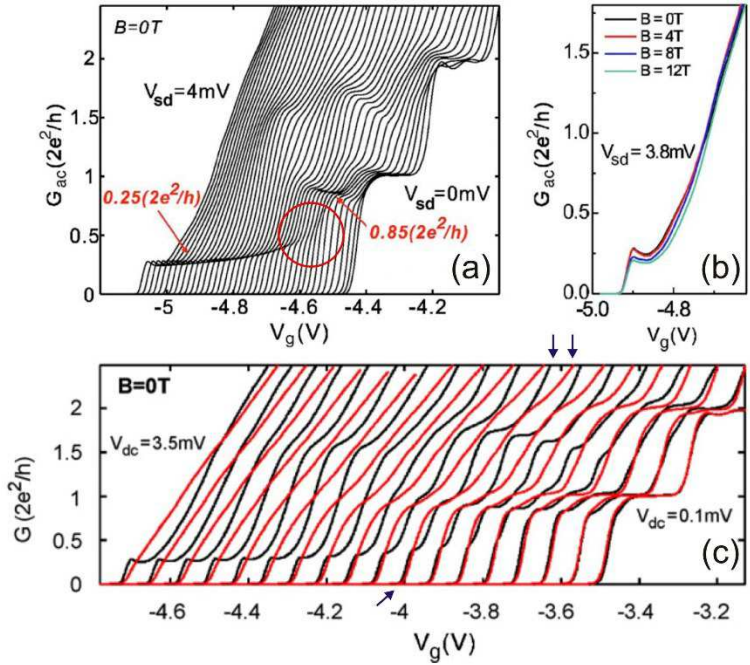}
\caption{(a) Plots of the ac conductance $G_{ac}$ vs $V_{g}$ for
different $V_{sd}$ increasing from $0$ (right) to $4$~mV (left) in
steps of $0.1$~mV. Traces are sequentially offset horizontally for
clarity by an unspecified amount. The $0.25$ and $0.85G_{0}$
plateaus are indicated. (b) $G_{ac}$ vs $V_{g}$ at $V_{sd} = 3.8$~mV
for in-plane magnetic fields $B_{\parallel} = 0$, $4$, $8$ and
$12$~T, showing that the $0.25$ plateau is relatively unaffected by
$B_{\parallel}$. (c) Plots of the ac (black) and dc (red)
conductance vs $V_{g}$ for different $V_{dc} (= V_{sd})$ increasing
from $0.1$ (right) to $3.5$~mV in steps of $0.2$~mV. Traces are
offset horizontally for clarity by an unspecified amount. The blue
arrows at top/bottom indicate the $V_{sd} = 1.5$~mV trace, which
corresponds to the horizontal line in Fig.~39(a) along which
schematics a-e from Fig.~39(g) occur. Figures (a,b) adapted with
permission from Ref.~\cite{ChenAPL08}. Copyright 2008, American
Institute of Physics. Figure (c) adapted from
Ref.~\cite{ChenPhysE08} with permission from Elsevier.}
\end{figure}

The fact that a plateau is observed at $0.25G_{0}$ immediately
suggests that the first subband is already spin-polarized, because a
spin-degenerate first subband crossing $\mu_{s}$ would instead give
a plateau at $0.5G_{0}$. One way to confirm this is to see if there
is a Zeeman splitting associated with the $0.25$ plateau at large
in-plane magnetic field $B_{\parallel}$. Figure~38(b) shows that the
$0.25$ plateau is relatively unaffected for $B_{\parallel}$ as large
as $12$~T, and the data in Fig.~2 of Ref.~\cite{ChenAPL08} provides
further evidence for the absence of Zeeman splitting for this
plateau. If we now look above the $0.25$ plateau a very interesting
conundrum eventually emerges.

\begin{figure}
\includegraphics[width=10cm]{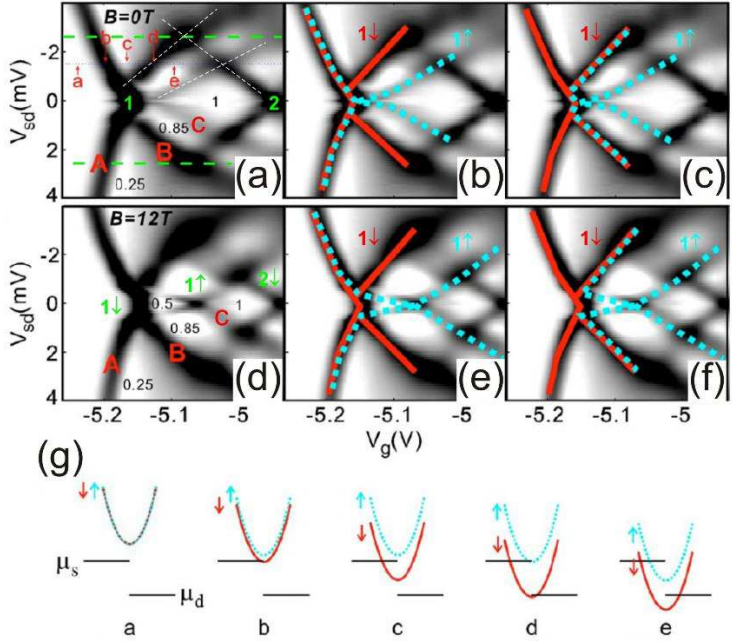}
\caption{(a/d) Greyscale plot of ac transconductance
$dG_{ac}/dV_{g}$ vs $V_{g}$ ($x$-axis) and $V_{sd}$ ($y$-axis) at
(a) $B_{\parallel} = 0$ and (d) $B_{\parallel} = 12$~T. The
annotations a-e correspond to the five schematics in (g) and A, B
and C are discussed in the text, the black numbers indicate the
conductance of given plateaus (white regions) and the green numbers
indicate the relevant 1D subbands. (b) and (e) contain the same data
as (a) and (d) respectively but show overlays indicating the
$1\downarrow$ (red solid line) and $1\uparrow$ (blue dashed line)
subband edges in the traditional picture of subband edge evolution
with dc bias. (c) and (f) also contain the same data as (a) and (d)
respectively but show overlays indicating the evolution of the
$1\downarrow$ (red solid line) and $1\uparrow$ (blue dashed line)
subband edges proposed by Chen {\it et al}~\cite{ChenPRB09b}. (g)
Schematics illustrating the locations of the $1\downarrow$ (red
solid line) and $1\uparrow$ (blue dashed line) subbands relative to
$\mu_{s}$ and $\mu_{d}$ for the five positions marked a-e in (a).
Figure adapted with permission from Ref.~\cite{ChenAPL08}. Copyright
2008, American Institute of Physics.}
\end{figure}

Figure~38(c) shows the ac (black) and dc (red) conductances measured
simultaneously for $0.1 < V_{sd} < 3.5$~mV at $B_{\parallel} =
0$~\cite{ChenPhysE08} (n.b., Fig.~2(c) in Ref.~\cite{ChenAPL08} is
from the same data set covering the range $1.6 < V_{sd} < 3.8$~mV
instead). For the moment we focus our attention on the high $V_{sd}$
limit towards the left-hand side of Fig.~38(c). The first item of
note is that the beginning of the riser above the $0.25$ plateau in
$G_{ac}$ always coincides closely with the point where $G_{dc}$
passes through $0.5G_{0}$ This is also shown and emphasized in
Fig.~2(c) of Ref.~\cite{ChenAPL08}. The value of $0.5G_{0}$ is
exactly half the change in $G_{dc}$ expected for a spin-degenerate
subband edge having passed through the dc bias window, and can only
occur in one of two ways: either a spin-degenerate subband edge has
moved to a point halfway between $\mu_{s}$ and $\mu_{d}$ or a
spin-resolved subband edge has reached $\mu_{d}$ while its opposite
spin counterpart remains above $\mu_{s}$. Only the latter would lead
to a $G_{ac}$ plateau at $0.25G_{0}$. The second item of note is the
absence of any plateaus at $0.5G_{0}$ in either $G_{ac}$ or
$G_{dc}$. Indeed, at the highest $V_{sd}$ there are no clear
plateaus observed at higher $G$, we will explain why below, but if
we reduce $V_{sd}$ (i.e., move to the right in Fig.~38(c)) the next
plateau that emerges above the $\sim 0.25G_{0}$ plateau occurs at
$\sim 0.85G_{0}$ and develops at $V_{sd} \sim 0.5 - 1.0$~mV. Chen
{\it et al} argue that the absence of a $0.5(2e^{2}/h)$ feature in
both $G_{dc}$ and $G_{ac}$ implies that to get from $0.25(2e^{2}/h)$
to $0.85(2e^{2}/h)$, the $1\uparrow$ edge passes through $\mu_{s}$
simultaneously with the $1\downarrow$ edge passing through
$\mu_{d}$. This also implies that the $1\uparrow$ band is unable to
populate until transport can occur in both directions via
$1\downarrow$, resulting in a transition from complete to partial
spin polarization. Under this argument, plateaus in $G_{ac}$ should
occur at $0.25G_{0}$ and $0.75G_{0}$. Both plateaus in the measured
$G_{ac}$ data are elevated above these expected values and Chen {\it
et al} attribute this to $V_{sd}$-dependence of the subband energy,
which leads to an added current contribution that breaks the
expected quantization of the plateaus in Ref.~\cite{ChenPhysE08},
but later comment that the nonlinear population of the spin-up
subband suggested by the observation of a non-quantized plateau in
$G_{dc}$ might be responsible for the enhanced conductance of the
$0.85$ plateau~\cite{ChenNL10}. Note, however, that finite bias
plateaus accurately quantized at $0.25$ and $0.75G_{0}$ have been
reported for an InGaAs QPC by Simmonds {\it et
al}~\cite{SimmondsAPL08}. This data was obtained at a temperature of
$1.5$~K in a weak out-of-plane magnetic field of $\sim 0.6$~T, and
in a device that showed the evolution of a resonant peak in response
to lateral shifting of the 1D channel, which makes it difficult to
put in proper context with the work by Chen {\it et al} above. It
would be interesting to pursue this material system further,
particularly given the much higher $g$-factor due to spin-orbit
interactions and ease in achieving large 1D subband spacings due to
the reduced effective mass~\cite{RamvallAPL97, MartinAPL08,
MartinPRB10, SimmondsAPL08, SchapersAPL07}.

The conundrum mentioned in paragraph before last emerges when one
puts this result in the context of the dc bias greyscale plots, such
as those shown in Figs.~39(a-f). Looking at Fig.~39(a) there is only
one left-moving diagonal for the first subband, and this would
normally mean that the $1\uparrow$ and $1\downarrow$ edges {\it
both} cross through $\mu_{s}$ simultaneously. This conventional
scenario is shown at $B_{\parallel} = 0$ and $12$~T in Figs.~39(b)
and (e), respectively, with the red solid (blue dashed) lines
indicating the $1\downarrow$ ($1\uparrow$) subband edge. The only
way the $1\uparrow$ edge can pass $\mu_{s}$ {\it after} the
$1\downarrow$ edge passes $\mu_{d}$ is if the $1\uparrow$ and
$1\downarrow$ subbands are separated in energy, as occurs due to
Zeeman splitting at high $B_{\parallel}$ in Fig.~39(d). This is
certainly not the case in Fig.~39(a) at $B_{\parallel} = 0$, where
the left-moving diagonal at $V_{g} \sim -5$~V corresponds to the
spin-degenerate second subband edge, rather than the $1\uparrow$
edge, coinciding with $\mu_{s}$. Chen {\it et al} suggest instead
that the $1\uparrow$ subband edge coincides with $\mu_{s}$ along the
higher-slope right-moving diagonal and $\mu_{d}$ along the
lower-slope right-moving diagonal as shown in Figs.~39(c/f), rather
than the left-moving and lower-slope right-moving diagonal,
respectively, as in Figs.~39(b/e). This is unconventional, but as
Chen {\it et al}~\cite{ChenAPL08} point out, the scenario in
Figs.~39(c/f) explains the data in Figs.~38(a/c), whereas that in
Figs.~39(b/e) cannot.

The implication of the scenario in Figs.~39(c/f) is that a spin-gap
between $1\downarrow$ and $1\uparrow$ begins to open before the
first subband reaches $\mu_{s}$, to look more closely at how this
would work, Fig.~39(g) shows five schematics a-e indicating the
positions of the $1\downarrow$ (red) and $1\uparrow$ (blue) subbands
relative to $\mu_{s}$ and $\mu_{d}$. The corresponding locations in
the dc bias greyscale are indicated in Fig.~39(a). At position b
only the $1\downarrow$ edge passes through $\mu_{d}$. The
$1\downarrow$ subband then drops rapidly in energy while the
$1\uparrow$ subband edge remains above $\mu_{s}$. This would lead to
a plateau at $0.25G_{0}$ in $G_{ac}$ as found in Fig.~38(a). At
position d on the higher-slope right-moving diagonal, $1\downarrow$
crosses $\mu_{d}$ simultaneously with $1\uparrow$ crossing
$\mu_{s}$, which would nominally give a plateau at $0.75G_{0}$ in
$G_{ac}$. The lower-slope right-moving diagonal then corresponds to
$1\uparrow$ crossing $\mu_{d}$ leading to a plateau at $G_{ac} =
G_{0}$. This would mean that at $V_{sd} \sim 1.5$~mV there should be
three plateaus in $G_{ac}$ at $0.25G_{0}$, $0.75G_{0}$ and $G_{0}$.
The corresponding trace is 11th from left and 8th from right in
Fig.~38(c) and it is clear there are only two $G_{ac}$ plateaus
below $G_{0}$, one at $\sim 0.35G_{0}$ and one at $\sim 0.9G_{0}$.
Hence it is not clear that the data conclusively demonstrates the
scenario proposed by Chen {\it et al}~\cite{ChenAPL08}, and it would
be an interesting avenue for further experiments.

We now return to some further discussion about the data in the high
$V_{sd}$ limit in Fig.~38(c). Considering the upper sections of
Fig.~39(a), it is clear that the argument about $1\uparrow$ passing
through $\mu_{s}$ simultaneously with $1\downarrow$ passing through
$\mu_{d}$ causing the riser above the $0.25$ plateau only holds for
$|V_{sd}| \lesssim 2.65$~mV (below the dashed horizontal green line
in Fig.~38(a)). The left-most trace in Fig.~38(c) is obtained at
$V_{sd} = 3.5$~mV (the left-most trace in Fig.~2(c) of
Ref.~\cite{ChenAPL08} is at $3.8$~mV), and if you follow a
horizontal path at this $V_{sd}$ in Fig.~39(a) (essentially the very
top of this greyscale), it is clear that as we make $V_{g}$ more
positive, we would intercept the left-moving diagonal corresponding
to the second subband coinciding with $\mu_{s}$ {\it before}
intercepting either of the right-moving diagonals corresponding to
the $1\downarrow$ and $1\uparrow$ subbands coinciding with
$\mu_{d}$. This explains why there are no plateaus in {\it any} of
the corresponding $G_{dc}$ vs $V_{g}$ traces obtained at $|V_{sd}|
\gtrsim 3$~mV. The bias window is so large that there is never any
point where there are no subband edges within the bias window.
According to Fig.~39(a), the $1\downarrow$ subband will reach
$\mu_{d}$ before the second subband reaches $\mu_{s}$ providing
$|V_{sd}| < 2.65$~mV (i.e., between the horizontal green dashed
lines). This corresponds to the fifth and sixth traces from the left
in Fig.~38(c), where a plateau begins to form just below $G_{0}$ in
$G_{dc}$. The $0.85$ plateau does not appear in Figs.~38(a/c) until
$V_{sd} < 1.7$~mV, and this is close to where the right-moving
$1\downarrow$ diagonal intercepts the left-moving $2$ diagonal,
which occurs at $V_{sd} \sim 1.44$~mV. This is the first point where
the both edges of the first subband (i.e., $1\uparrow$ and
$1\downarrow$) have moved out of the dc bias window before either of
the second subband edges enter. It is also the first place where a
single subband edge is within the bias-window at all times and
$G_{dc}$ is a straightforward measurement. This does not invalidate
the argument that the $0.25$ plateau in $G_{dc}$ indicates that
current only flows through the $1\downarrow$ subband in
Ref.~\cite{ChenAPL08}, but it does highlight that care is needed in
interpreting the data at large $V_{sd}$.

As a final point before moving on, it is also worth noting that
Kothari {\it et al}~\cite{KothariJAP08} report bunching of
conductance traces in the vicinity of $0.25G_{0}$, and attribute the
origin of this effect to the highly asymmetric drop of a relatively
large source-drain bias across a QPC when it is near pinch-off.
Additionally, theoretical calculations by Ihnatsenka and
Zozoulenko~\cite{IhnatsenkaPRB09} suggest that nonlinear screening
and the self-consistent nature of the QPC potential can produce a
finite bias plateau at $0.25G_{0}$, and argue that the feature
observed by Chen {\it et al}~\cite{ChenAPL08} is not spin related.

Reference~\cite{ChenPhysE08} explores the $0.85G_{0}$ plateau
further. Figures~39(a) and (d) present ac transconductance
greyscales obtained at $B_{\parallel} = 0$ and $12$~T, respectively,
with the numbers over the white regions indicating the conductance
in units of $G_{0}$ for the corresponding plateau (note that these
are not dissimilar to the data in Figs.~6(b) and 15(c)). As in
earlier greyscales in this review, the diagonal dark regions
correspond to a subband edge coinciding with a chemical potential.
Note that the first subband edge in Fig.~39(a) has two such
diagonals, labelled B and C, extending outwards to the right, but
only one heading towards the left labelled A. Ignoring any many-body
behaviour, there should only be one diagonal heading in each
direction at $B_{\parallel} = 0$ corresponding to the
spin-degenerate subband crossing $\mu_{s}$ (left-moving) and
$\mu_{d}$ (right-moving). The $0.85$ plateau sits between these two
right-moving diagonals, implying that a subband edge already sits
below $\mu_{d}$, while another passes through $\mu_{d}$ to take
$G_{ac}$ up onto the $G_{0}$ plateau. Increasing $B_{\parallel}$ to
$12$~T breaks the spin-degeneracy, and as Fig.~39(d) shows, the
$0.85$ plateau remains and the origin of the diagonals B and C
becomes clear -- they correspond to $1\downarrow$ and $1\uparrow$
crossing $\mu_{d}$. Hence at the $0.85$ plateau $1\downarrow$ is
below $\mu_{d}$ while $1\uparrow$ is between $\mu_{d}$ and
$\mu_{s}$, and thus must be contributing to $G_{dc}$. This still
holds if $1\uparrow$ coinciding with $\mu_{s}$ corresponds to the
right-moving diagonal B rather than the left-moving diagonal as
usual (i.e., the scenario in Fig.~39(e) rather than Fig.~39(f)).

We conclude by returning to Fig.~38(c) to connect back to the
discussion at the end of Section 9.3. As $V_{sd}$ is increased from
zero, a weak plateau/shoulder at $\sim 0.8G_{0}$ in $G_{dc}$ begins
to emerge, disappearing again along with the $0.85$ plateau in
$G_{ac}$ once $V_{sd}$ exceeds $1.1$~mV. The plateau in $G_{dc}$
indicates that the motion of the $1\uparrow$ subband in energy with
$V_{g}$ is not constant~\cite{ChenPhysE08}, a point that is
elaborated on in Ref.~\cite{ChenNL10}. The fact that this $G_{dc}$
plateau is non-quantized suggests that the $1\uparrow$ edge pins at
neither $\mu_{s}$ as proposed by Graham {\it et
al}~\cite{GrahamPRB07} nor $\mu_{d}$ as proposed by Kristensen {\it
et al}~\cite{KristensenPRB00}, but instead populates at a slower
rate somewhere in between. It is interesting to compare this to
Fig.~3 of Ref.~\cite{MicolichJPCM11}, where the same behaviour is
shown to occur for the $2\uparrow$ subband edge in the data obtained
by Chen {\it et al} in Ref.~\cite{ChenPRB09}. Note that in
Fig.~38(c) the $0.7$ plateau at $V_{sd} = 0$ evolves smoothly into
the $0.85$ plateau at finite bias, and under this basis, Chen {\it
et al}~\cite{ChenNL10} argue that the $0.7$ plateau is also caused
by the $1\uparrow$ subband populating very slowly, consistent with
earlier experiments suggesting a finite zero-field spin-gap as the
origin of this effect.

\subsection{Some final comments on the dc conductance measurements}

Figure~1(d) of Ref.~\cite{ChenPhysE08} shows the data in Fig.~39(a)
over an extended $V_{g}$ range that encompasses the $2G_{0}$
plateau. From this additional data it is clear that the one
left-moving/two right-moving diagonal structure also occurs for the
second subband. Under the normal picture for bias-splitting of the
1D subbands, the single left-moving branch suggests that for each 1D
subband the zero-field spin-gap is zero or irresolvably small when
the subband first begins to populate. The spin-gap must open on
population, and the shape of the $1\uparrow$ right-moving diagonal
(i.e., C in Fig.~39(a)) gives some clue to this. Imagine two
V-shaped bias-splitting structures in a greyscale, one that opens as
a wide-V (i.e, the diagonal moves rapidly along $V_{g}$ for a small
increase in $V_{sd}$) and one as a narrow-V. The wide-V case says
that the $V_{g}$ shift needed to move the subband edge from
$\mu_{d}$ to $\mu_{s}$ is large, and hence the subband moves (i.e.,
populates) very slowly in between. A narrow-V says the opposite,
that the subband populates rapidly, as we saw in Section 8.2. The
right-moving branch C begins very shallow and then turns upwards
sharply to rise at almost the same rate as branch B (see also
Figs.~6(b) and 15(c)). Note that there is white space between the C
branches at positive and negative $V_{sd}$ that extends right back
to the first subband edge at $V_{sd} = 0$ (i.e., the two C branches
at positive and negative $V_{sd}$ bound a white diamond that has its
left-most $V_{sd} = 0$ vertex at the point where the spin-degenerate
first subband crosses $\mu$). This suggests that the $1\uparrow$
edge does not pin precisely to $\mu_{s}$ but rather drops beneath it
and then populates very slowly, before populating more quickly. Note
that this also happens for the $2\uparrow$ subband in Fig.~3 of
Ref.~\cite{MicolichJPCM11}.

How this picture alters under the framework where the higher-slope
right-moving diagonal corresponds to the $1\uparrow$ edge coinciding
with $\mu_{s}$ is interesting. Under the framework given in
Fig.~39(c) the spin-gap at population of the first subband is zero
if $V_{sd} = 0$. However, the spin-gap at population of the first
subband becomes finite for $|V_{sd}| > 0.7$~mV (i.e., where the
left-moving branch A splits from the higher-slope right-moving
branch B) and increases roughly linearly with increasing $V_{sd}$.
This implies a bias-dependent spin-gap, and this is in many ways
similar to the BCF model~\cite{BruusPhysE01} except that the subband
edge pinning and gap opening is tied to $\mu_{s}$ rather than
$\mu_{d}$. Another aspect to note for the framework provided by Chen
{\it et al} in Fig.~39(c) is that the spin-gap at $|V_{sd}|
> 0.7$~mV decreases as the first 1D subband populates. This is
evident in Fig.~39(c) by considering the horizontal separation
between (a) the left-moving red diagonal and the higher-slope
right-moving blue diagonal, and (b) the right-moving red diagonal
and the lower-slope right-moving blue diagonal, at some fixed
$V_{sd}$ (n.b. the 1D subband spacing also changes with $V_{g}$ so
this observation should be considered to be qualitative). The full
implications of such a spin-gap scenario as a phenomenological model
are yet to be considered and would be an interesting contribution.

As a final comment, there is much that remains to be understood in
the dc conductance data that could benefit from further measurement
and study. To give one example, consider the evolution of the $0.25$
plateau in Fig.~38(a) with decreasing $V_{sd}$. This strong feature
rises smoothly at first, and eventually becomes a much weaker
plateau at $0.5G_{0}$ (see red circle in Fig.~38(a)), whereafter it
rises much more rapidly with $V_{sd}$ to intercept the $0.85G_{0}$
plateau. In parallel, the $0.85$ plateau decreases slightly with
decreasing $V_{sd}$ towards the intersection point with the rising
$0.25$ plateau, but if this is traced back towards higher $V_{sd}$
it would reach $G_{0}$ (if it did not diminish in strength and
disappear) at the same $V_{sd}$ where a plateau structure at $1.5
G_{0}$ appears. Note also that at the point where the $0.25$ plateau
has reached $0.5G_{0}$ there is also a plateau at $1.5G_{0}$, which
does not appear at lower $V_{sd}$ and which rises towards $1.75
G_{0}$ as $V_{sd}$ is increased, falling back down to become the
high $V_{sd}$ plateau at $1.5G_{0}$ mentioned in the preceding
sentence. In particular, at $V_{sd} = 0.6$~mV, the ac conductance
has five plateau-like structures at $0.5$, $~0.7$, $1$, $1.5$ and
$2G_{0}$. I point this out not because I think there is an
explanation, but because I think the dc conductance measurements so
far have only just scratched the surface regarding the connection
between conductance plateaus and the relative motion of 1D subbands
in energy. Further study is clearly needed, and use of the dc
conductance to more closely track the 1D
subbands~\cite{MicolichJPCM11} might provide new insight into how
many-body effects influence the energetics of 1D subbands as they
populate.

\section{Spontaneous electron ordering in 1D systems}

Sections $4$ and $6$ presented compelling evidence for bound-state
formation in a QPC near pinch-off, and a natural extension is to ask
whether more complex organized electron states may also occur in
such cases. This possibility is evident in the density functional
theory calculations presented in Section $6.1$, and in Fig.~23 in
particular, where the spin-up bound-state at the center of the QPC
has adjoining spin-down quasi-bound states along the QPC
axis~\cite{RejecNat06}. In this section we explore this idea in more
detail, beginning with brief coverage of key theoretical papers in
this direction, and following with some very recent experimental
studies presenting initial evidence for the formation of ordered
electron transport in QPCs in the weakly confined limit.

\subsection{Theoretical studies on electron ordering in 1D systems}

In the low density limit, the potential energy dominates over the
kinetic energy for a gas of electrons, and it was predicted by
Wigner that the electrons will form an ordered crystal lattice to
minimise their energy~\cite{WignerPR34}. The long-range order in
this lattice is destroyed by quantum
fluctuations~\cite{MatveevPRB04}, however it is expected that the
remaining short-range order can have significant effects on
transport through a QPC or quantum wire. In such a case, the system
can be treated as an antiferromagnetic Heisenberg spin chain with an
exponentially small exchange coupling parameter $J$. This causes 1D
conductance suppression for $J >> k_{B}T$, and in the limit $J <<
k_{B}T$, leads to an additional quantized plateau at $0.5
G_{0}$~\cite{MatveevPRB04, MatveevPRL04}. This spin-chain model has
been extended by Klironomos {\it et al}~\cite{KlironomosEPL06,
KlironomosPRB07} to consider the case where the system is no longer
purely one-dimensional and is instead confined to a quantum wire
with a parabolic confining potential. At low density the electrons
form an ordered 1D lattice, but as the density is increased, a
transition to a zig-zag chain occurs once the electron separation
becomes less than a characteristic length-scale for the 1D
confinement $r_{0} = (2e^{2}/\epsilon m^{*}\Omega^{2})^{1/3}$, where
$\Omega$ is the harmonic frequency for the confining
potential~\cite{KlironomosEPL06}. To parameterize this, Klironomos
{\it et al} invoke a dimensionless density $\nu = nr_{0}$, where $n$
is the 1D density in the wire. They find that the linear 1D crystal
is stable for $\nu < 0.78$, while the zig-zag chain occurs for $0.78
< \nu < 1.75$. The zig-zag chain takes an equilateral configuration
at $\nu = 1.46$, where the nearest neighbour separation equals the
next nearest neighbour separation. While only nearest-neighbour
exchange is relevant for the purely 1D chain, in the zig-zag case,
both the nearest neighbour exchange $J_{1}$ and next-nearest
neighbour exchange $J_{2}$ become important. The zig-zag geometry
also allows more complicated multi-electron exchange interactions
that lead to the possibility of ferromagnetic states being
formed~\cite{KlironomosEPL06, KlironomosPRB07}. In this case the
deviation from one-dimensionality afforded by adopting the zig-zag
structure avoids any problems with the Lieb-Mattis
theorem~\cite{LiebPR62}. Structures with three or more rows are
obtained at higher $\nu$, as discussed by Piacente {\it et
al}~\cite{PiacentePRB04}, and it is interesting to note that in
between the regions where two and three row states are stable, there
is a small range of $\nu$ where a four row state is stable (e.g.,
see Fig.~2 of Ref.~\cite{PiacentePRB04}). It would be interesting to
explore if this `two row to four row to three row' structural
pattern could produce a distinct experimental signature in the
conductance. Furthermore, Piacente {\it et al}~\cite{PiacentePRB04}
predict that the higher transitions are first order phase
transitions while the one row to two row transition is second order,
which may also provide a route to more definitive evidence for
electron ordering in QPCs.

\begin{figure}
\includegraphics[width=7cm]{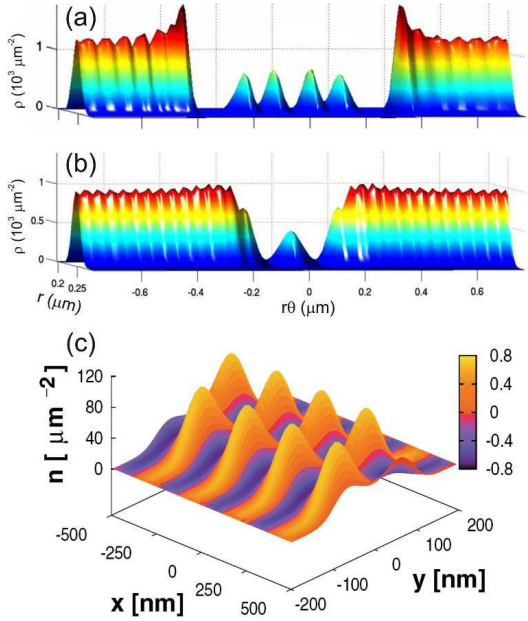}
\caption{(a,b) The two-dimensional ground state density $\rho$ for
(a) a long and (b) a short constriction in a narrow 2D system
obtained using quantum Monte Carlo calculations. (c) Electron
density $n(x,y)$ for a quantum wire in the local spin density
approximation, showing a zig-zag formation of localized electrons.
Figures (a,b) adapted with permission from Ref.~\cite{GucluPRB09}.
Copyright 2009 by the American Physical Society. Figure (c) adapted
with permission from Ref.~\cite{WelanderPRB10}. Copyright 2010 by
the American Physical Society.}
\end{figure}

Electron ordering is also born out in numerical calculations by
G\"{u}\c{c}l\"{u} {\it et al}~\cite{GucluPRB09} and Welander {\it et
al}~\cite{WelanderPRB10}. G\"{u}\c{c}l\"{u} {\it et al} performed
quantum Monte Carlo calculations for a model consisting of a narrow
two-dimensional quantum ring with a constriction, essentially
resembling a QPC where the source and drain leads are connected
together. The constriction is formed by a potential barrier with a
hyperbolic tangent form of specified width and sharpness. As shown
in Fig.~40(a), numerous electrons can be localized in a row for a
longer constriction, with a clear gap present between the electron
lattice in the constriction and the electron liquid in the leads.
The width of this gap is dependent on the sharpness of the barrier.
The lattice is isolated and in the Coulomb blockade regime for a
sharp barrier, becoming isolated as the barrier is softened, and
ultimately developing into a continuation of the Friedel
oscillations in the leads that extends through the constriction with
an attendant small drop in electron density. This smooth evolution
from liquid to localized chain and back fits well with the
theoretical model studied by Matveev~\cite{MatveevPRB04,
MatveevPRL04}. As the constriction is made very short, a single
electron can be localized, as shown in Fig.~40(b), and
G\"{u}\c{c}l\"{u} {\it et al} note that this may be consistent with
the Kondo mechanism for the $0.7$ plateau~\cite{CronenwettPRL02,
MeirPRL02, RejecNat06}, a point disputed by Welander {\it et
al}~\cite{WelanderPRB10} as discussed below. G\"{u}\c{c}l\"{u} {\it
et al} also obtained good agreement between the quantum Monte Carlo
calculations and those obtained using spin density functional theory
under the local spin density approximation
(LSDA)~\cite{ReimannRMP02}(see Fig.~6 of Ref.~\cite{GucluPRB09}),
indicating that the localization of electrons within a QPC close to
pinch-off is a robust behaviour of the theoretical models. A
ferromagnetic ground state is not found in any of the calculations
performed by G\"{u}\c{c}l\"{u} {\it et al}~\cite{GucluPRB09}.

In more recent work, Welander {\it et al}~\cite{WelanderPRB10} have
used the different theoretical framework of spin density functional
calculations to approach the same problem with an almost identical
geometric configuration/potential. As in Ref.~\cite{GucluPRB09},
Welander {\it et al} find a single electron bound state, however
this state is very fragile, with an increase in electron density at
fixed barrier sharpness or a slight softening of the barrier at
fixed density leading to a delocalization of the electron. Due to
the instability of this localized state, Welander {\it et al}
suggest that it is more likely related to the Fano resonance
structures reported by Yoon {\it et al}~\cite{YoonAPL09}, as
discussed in Section 6.4, as opposed to producing the $0.7$ plateau
via a Kondo-like mechanism. The numerical calculations discussed
above have all been obtained with the narrow quantum ring being held
in the single-mode limit by a strong transverse harmonic potential.
Welander {\it et al} investigate what happens as this transverse
confinement is relaxed. As shown in Fig.~40(c), the system first
evolves to a zig-zag chain of electrons, similar to the model
studied by Klironomos {\it et al}~\cite{KlironomosEPL06,
KlironomosPRB07}, and even into three-row lattice structures as the
number of electrons in the ring is increased~\cite{WelanderPRB10}.
These multichain structures form under weak transverse confinement
because the electrons have more space available than they would
under tight confinement to localize and reduce their interaction
energy without a severe penalty in terms of kinetic and transverse
confinement energies. Welander {\it et al} predict that these
multichain structures should lead to unusual crossovers of the
conductance plateaus in 1D systems.

\subsection{Experimental studies of weakly confined QPCs}

In a series of four papers, Hew {\it et al}~\cite{HewPRL08,
HewPRL09, HewPhysE10} and Smith {\it et al}~\cite{SmithPRB09} study
transport in a weakly confined QPC. The device studied by Hew {\it
et al} consists of a pair of split-gates forming a channel
$0.4~\mu$m long deposited directly onto a modulation-doped
AlGaAs/GaAs heterostructure, with a top-gate located directly above
the split-gates. The top gate is $1~\mu$m wide (i.e., it extends to
$300$~nm beyond the entrance and exit of the QPC in the
$x$-direction) and extends all the way across the Hall bar in the
$y$-direction. The top gate is separated from the split-gates by a
$200$~nm thick layer of cross-linked polymethylmethacrylate
electron-beam lithography resist. The top-gate can be swept
independently over a range $-1.5 < V_{TG} < 0.6$~V while the
side-gates are held at a fixed bias in the range $-2.0 < V_{SG} <
-0.52$~V, allowing the first four to five quantized conductance
plateaus to be measured as a function of confinement strength.

\subsubsection{Formation of a spin-incoherent state in a weakly confined QPC}

Figure~41(a) shows a series of top-gate sweeps obtained at different
side-gate biases. Starting at the right hand side, the device is
much like a typical QPC -- the electron density is relatively high
and the side-gates are strongly negative resulting in a strong,
sharp 1D confinement potential. Moving towards the left, the
side-gate voltage becomes less negative, as does the top-gate
voltage. This reduces the density and weakens the confinement,
transferring the QPC from the strong confinement (sc) regime,
through an intermediate confinement (ic) regime, and ultimately, to
a weak confinement (wc) regime, which is the region of particular
interest in this section. It can be established that the combination
of $V_{TG}$ and $V_{SG}$ have this effect by measuring the 1D
subband spacing by dc bias spectroscopy~\cite{HewPhD09}. The ac
transconductance $dG/dV_{tg}$ greyscales corresponding to the wc, ic
and sc regimes are shown in Figs.~41(b), (c) and (d). As the
confinement becomes weaker, the width of the diamonds decreases
indicating that the 1D subband spacing is reduced, consistent with
the weaker confinement.

In the strongly confined regime, the 1D conductance takes its usual
appearance, with clear plateaus accurately quantized in integer
multiples of $G_{0}$. As the confinement is weakened, the odd
integer plateaus in particular begin to drop below their quantized
values. Near the transition between the ic and wc regimes, indicated
by the upward-pointing arrow in Fig.~41(a), a plateau at $0.5G_{0}$
develops and continues to grow stronger as the confinement is
weakened further. This coincides with suppression of the $G_{0}$
plateau, but not the higher plateaus. Note that to the far left in
Fig.~41(a), the $0.5G_{0}$ and suppressed $2G_{0}$ plateaus cease
dropping as the confinement is weakened, unlike the suppressed $3
G_{0}$ plateau, which continues to drop at an ever growing rate.
This behaviour is inconsistent with the conductance suppression
being due to the well-known trend for the electron mean free path to
drop rapidly as the density is reduced in modulation-doped
heterostructures~\cite{StormerAPL81}. If this were the case, all
plateaus should suppress evenly, and no new quantized plateaus are
expected to emerge/vanish when this occurs~\cite{LiangPRB00,
CzapkiewiczEPL08}.

\begin{figure}
\includegraphics[width=16cm]{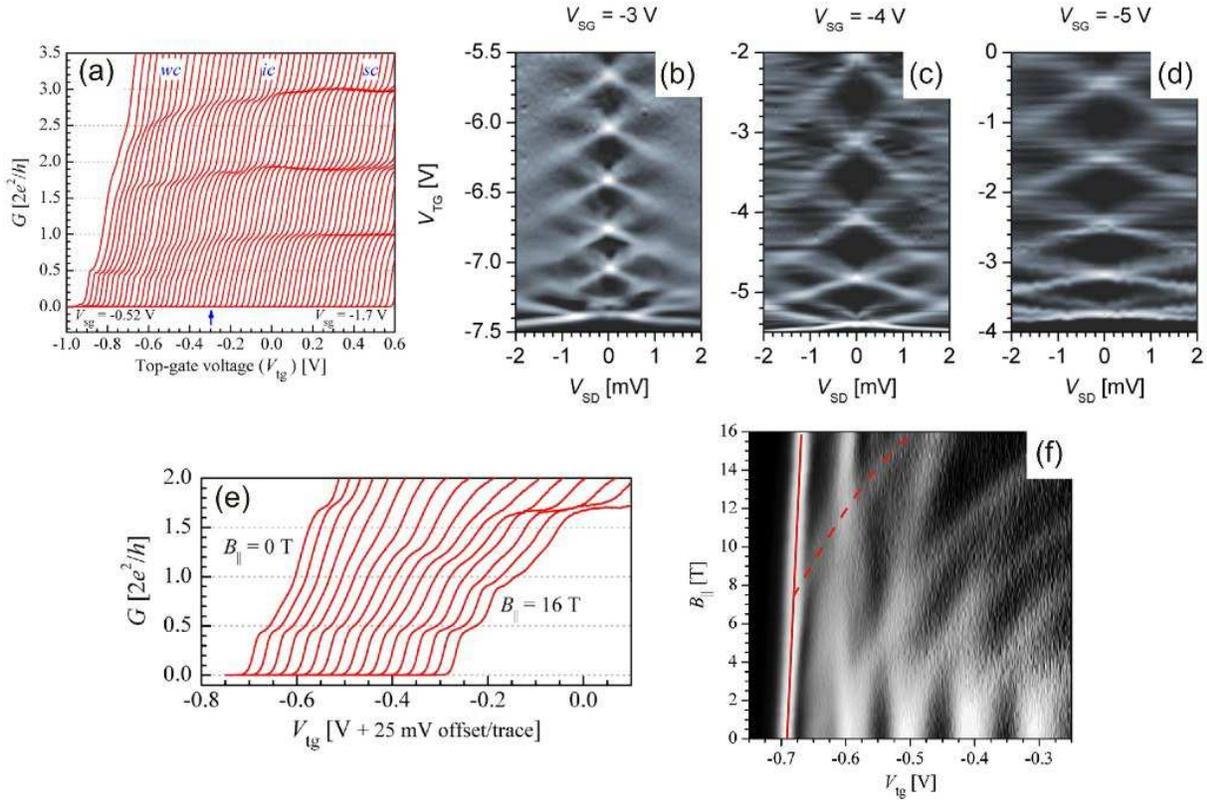}
\caption{(a) Conductance $G$ vs top-gate voltage $V_{tg}$ for a
range of side-gate voltages $-0.52 > V_{sg} > -1.7$~V. The QPC is in
the strong confinement (sc) regime on the right, and as $V_{sg}$
becomes more positive and $V_{tg}$ more negative, the confinement is
weakened and the QPC passes through intermediate confinement (ic) to
the weak confinement (wc) regime on the left. The vertical arrow at
the bottom marks the point where the $0.5G_{0}$ plateau first
appears. (b-d) Greyscale plots of ac transconductance $dG/dV_{tg}$
vs source-drain bias $V_{sd}$ ($x$-axis) and top-gate voltage
$V_{tg}$ ($y$-axis) for three different side-gate voltages $V_{sg}$,
which represent the (b) weak, (c) intermediate and (d) strong
confinement regimes. The width of the diamonds is proportional to
the subband spacing, with increased subband spacing indicative of
stronger confinement. This data was obtained from a separate device,
hence the mismatch in voltages, however the concept is the same. (e)
$G$ vs $V_{tg}$ for in-plane magnetic fields $B_{\parallel}$ from
$0$ to $16$~T in steps of $1$~T for a trace in the wc regime. Traces
are offset horizontally for clarity. The $0.5G_{0}$ plateau weakens
at around $9$~T before strengthening again. (f) Greyscale plot of
$dG/dV_{tg}$ vs $V_{tg}$ ($x$-axis) and $B_{\parallel}$ ($y$-axis).
On the right hand side, the four white V-shaped structures indicate
the Zeeman splitting of the second to fifth 1D subbands. At the left
hand side, the sharp line (with superimposed solid red line) and the
foggy line to its right correspond to the riser up to the $0.5
G_{0}$ plateau and the weak gradient change between the $0.5G_{0}$
and $G_{0}$ plateaus. The sharp line splits at $B_{\parallel} \sim
8$~T, as highlighted by the superimposed red dashed line and
discussed in the text. Figures (a,e,f) adapted with permission from
Ref.~\cite{HewPRL08}. Copyright 2008 by the American Physical
Society. Figures (b-d) adapted with author's permission from
Ref.~\cite{HewPhD09}.}
\end{figure}

The behaviour in both the ic and wc regimes is interesting. As
discussed in the previous section, calculations by
Matveev~\cite{MatveevPRB04, MatveevPRL04} predict conductance
suppression when the exchange coupling $J >> k_{B}T$ and the
appearance of an extra plateau at $0.5G_{0}$ for $J << k_{B}T$.
Based on magnetic depopulation measurements~\cite{BerggrenPRL86},
Hew {\it et al} estimate a 1D density $n_{1D} \sim 1 \times
10^{5}$~cm$^{-1}$ in the wc regime, which gives $J/k_{B} \sim
2.2$~mK and $E_{F}/k_{B} \sim 1.6$~K consistent with $J << k_{B}T <<
E_{F}$. The exchange coupling rises exponentially as the separation
between electrons decreases, explaining why in the intermediate
confinement regime the $0.5G_{0}$ plateau is lost and only
conductance suppression is observed. This was tested by illumination
of the device, which increases the electron density through the
persistent photoconductivity effect~\cite{NelsonAPL77}. After an
initial illumination, the $0.5G_{0}$ plateau is weaker and its onset
shifts from $V_{tg} = -0.3$~V down to $V_{tg} = -0.7$~V, with
further illumination eliminating this plateau entirely and restoring
integer quantized plateaus over the entire available $V_{tg}$
range~\cite{HewPhD09}.

Figure~41(e) shows the evolution of the 1D conductance in the wc
regime with increasing $B_{\parallel}$ from $0$ to $16$~T. The $0.5
G_{0}$ plateau is present throughout, but undergoes a distinct
weakening at approximately $9$~T. Figure~41(f) shows a corresponding
greyscale of transconductance $dG/dV_{tg}$ versus $B_{\parallel}$
($y$-axis) and $V_{tg}$ ($x$-axis) that reveals what is occurring at
$B_{\parallel} = 9$~T in Fig.~41(e). Looking to the right of
Fig.~41(f), first there are four white V-shaped structures that
curve slightly towards the right with increasing field. These are
the risers above the $G_{0}$ plateau (white indicates high
transconductance) and correspond to the Zeeman splitting of the
second and higher 1D subbands. This splitting is first resolved at
$B_{\parallel} \sim 3$~T. Additionally, there are a sharp line at
$V_{tg} \sim -0.69$~V and a `foggy' line at $V_{tg} \sim -0.64$~V in
Fig.~41(f). These correspond to the riser up to the $0.5G_{0}$
plateau, and the weak slope change between the $0.5G_{0}$ and
$G_{0}$ plateaus in Figs.~41(a/e). The sharp line shows no visible
spin-splitting until $B_{\parallel} = 9$~T, where a branch emerges
moving rapidly to the right. This behaviour is not consistent with
the $0.5G_{0}$ plateau arising from a ferromagnetic state, which
should show no Zeeman splitting with $B_{\parallel}$. Instead, Hew
{\it et al} attribute the $0.5G_{0}$ plateau in the wc regime to the
formation of a spin-incoherent Luttinger liquid~\cite{CheianovPRL04,
FietePRL04, FieteRMP07}. The weakening of the plateau at
$B_{\parallel} = 9$~T is attributed to destruction of the
spin-incoherent Luttinger liquid~\cite{HewPRL08}. The mechanism for
this involves competition between spin-ordering and thermal
randomization of the spins. At low fields, there is only the
exchange interaction to impose order, and since $J << k_{B}T$ due to
the low density, the thermal energy destroys any spin-ordering. The
in-plane magnetic field can also impose an ordering effect, and as
$B_{\parallel}$ increases, the Zeeman energy $g^{*}\mu_{B}B$
eventually exceeds both $J$ and $k_{B}T$, imposing spin-order and
destroying the spin-incoherent Luttinger liquid state. At this
point, the system reverts to having a spin-polarized first subband,
which is consistent with the appearance of plateaus at both $0.5
G_{0}$ and $G_{0}$ for $B_{\parallel} \gtrsim 13$~T. Finally, Hew
{\it et al} note the presence of a zero-bias anomaly in their data,
which shows no Zeeman splitting and is fully suppressed for
$B_{\parallel} > 7$~T. This is inconsistent with the ZBA behaviour
expected if a Kondo-like mechanism was
involved~\cite{CronenwettPRL02}.

\subsubsection{Experimental evidence of electron ordering in weakly confined QPCs}

In a subsequent paper, Hew {\it et al}~\cite{HewPRL09} used the same
device structure to explore the possibility of the electrons forming
a zig-zag chain in a weakly confined quantum wire, as proposed by
Klironomos {\it et al}~\cite{KlironomosEPL06, KlironomosPRB07} and
Welander {\it et al}~\cite{WelanderPRB10}. Figure~42(a) shows data
similar to that in Fig.~41(a) obtained for this device, where the 1D
density is estimated to be approximately three times higher ($\sim 3
\times 10^{5}$~cm$^{-1}$). As a result, the $0.5G_{0}$ plateau is no
longer observed in the weakly confined limit, however, as in
Fig.~41(a), suppression of the plateaus still occurs as the
confinement is weakened. This suppression happens first for the
$G_{0}$ plateau, which vanishes as it approaches $0.5G_{0}$. Once
this plateau vanishes, the conductance jumps directly to a plateau
at $2G_{0} = 4e^{2}/h$ from $G = 0$ when the QPC populates; this has
been also observed in a second, similar device~\cite{HewPRL09}. Note
that the $G_{0}$ plateau reappears in the left-most traces in
Fig.~41(a), demonstrating that the disappearance of the $G_{0}$
plateau is not due to thermal or disorder effects. In the presence
of normal 1D subbands, population of each spin-degenerate subband
can contribute at most $2e^{2}/h$ to the conductance, and this
direct jump to $4e^{2}/h$ indicates a breakdown of normal 1D subband
behaviour. Hew {\it et al}~\cite{HewPRL09} argue that this doubled
conductance contribution is due to Coulomb interactions driving this
system into a double-row configuration where there are effectively
two channels contributing to the conduction. This is consistent with
experiments by C.G. Smith {\it et al}~\cite{SmithJPCM89} where two
closely-spaced QPCs were measured in a parallel configuration and
gave a 1D conductance quantized in units of $4e^{2}/h$ rather than
the usual $2e^{2}/h$ for a single QPC.

\begin{figure}
\includegraphics[width=14cm]{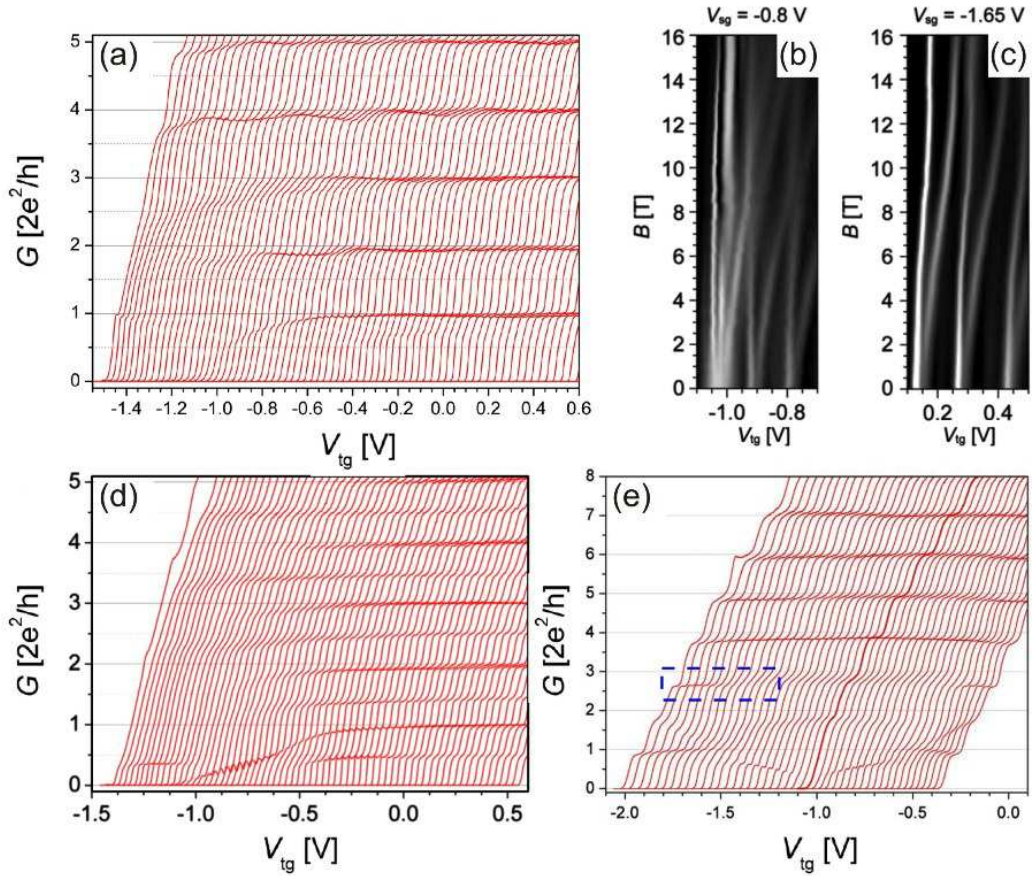}
\caption{(a) Conductance $G$ vs top-gate voltage $V_{tg}$ at various
fixed side-gate voltages $V_{sg}$ incremented from $-0.52$~V (left)
to $-2$~V (right) in steps of $20$~mV. Moving to the left
corresponds to a weakening of the 1D confinement. Note in particular
the drop and disappearance of the $G_{0}$ plateau such that for
slightly weaker confinement the first plateau occurs at $2G_{0}$.
(b) and (c) Greyscale plots of transconductance $dG/dV_{tg}$ vs
$V_{tg}$ ($x$-axis) and in-plane magnetic field $B_{\parallel}$
($y$-axis) for weak and strong confinement, respectively. Normal
Zeeman splitting is observed for strong confinement, but for weak
confinement, the first subband splits into more than two branches
with increasing field. (d) Similar data to (a) obtained at
$B_{\parallel} = 7$~T. Note that the $G_{0}$ plateau drops downwards
dividing the $0.5G_{0}$ plateau into two segments, one in the strong
confinement limit and one in the weak confinement limit, as the
confinement is weakened. (e) $G$ vs $V_{tg}$ at $V_{sg} \approxeq
-1.64$~V as a function of asymmetric bias $\Delta V_{sg}$ applied to
the side-gates. The bold trace corresponds to $\Delta V_{sg} = 0$,
with the maximum asymmetric bias $\Delta V_{sg} = \pm 0.5$~V
corresponding to a lateral displacement of approximately $350$~nm.
The blue dashed box highlights the evolution of the $3G_{0}$ plateau
with $\Delta V_{sg}$. The behaviour of the lowest three plateaus is
complex and discussed in the text. Figures (a-d) adapted with
permission from Ref.~\cite{HewPRL09}. Copyright 2009 by the American
Physical Society. Figure (e) adapted from Ref.~\cite{HewPhysE10}
with permission from Elsevier.}
\end{figure}

Further studies in this region with an in-plane magnetic field
applied reveal some interesting results. Firstly, the normal twofold
Zeeman splitting of the 1D subbands that occurs at stronger
confinement (see Fig.~42(b)), evolves into a band with more than two
(Hew {\it et al} suggest four) components that are degenerate at
zero field at $V_{sg} = -0.8$~V, where the direct jump to a $2G_{0}$
plateau in Fig.~42(a) occurs. Figure~42(d) shows similar data to
that in Fig.~42(a) obtained at $B_{\parallel} = 7$~T. In the strong
confinement regime, conductance plateaus accurately quantized in
units of $0.5G_{0}$ are observed as expected for Zeeman-split 1D
subbands. Interestingly, as the confinement is weakened (moving left
in Fig.~42(d)), the plateaus at $1.5G_{0}$, $2G_{0}$, $2.5G_{0}$,
etc. vanish right at around the point where the plateau that started
out at $G_{0}$ in the strong confinement regime reaches $G = 0$. The
$0.5G_{0}$ peak reappears at this point, only to vanish again at the
very weakest confinements measured. This behaviour is strange --
normally the $0.5G_{0}$ plateau corresponds to the $1\downarrow$
subband. Since $1\downarrow$ is the lowest energy 1D subband and it
decreases in energy with magnetic field, it should remain the lowest
subband (i.e., in a conventional 1D subband energy diagram
$1\downarrow$ never crosses another subband edge within a measurable
range of $V_{sd}$)~\cite{HewPhD09}. Hence the `interruption' of the
$0.5G_{0}$ plateau by the plateau dropping down from $G_{0}$
suggests that physics beyond that normally observed for 1D subbands
takes place. During the continuous suppression of the $G_{0}$
plateau towards zero, a resonant structure appears on the plateau,
reaching its maximum strength as the plateau passes through $0.5
G_{0}$ before diminishing again. This resonant structure becomes
considerably stronger at $B_{\parallel} = 16$~T, attaining an
amplitude $\Delta G$ of almost $0.25G_{0}$. Hew {\it et
al}~\cite{HewPhD09} suggest that this feature is due to the approach
of the $1\downarrow$ and $2\downarrow$ subband edges as the
weakening confinement reduces the 1D subband spacing, and propose
that this effect signals the threshold of a structural bifurcation
between single- and double-row electron formations, which would be
nearly degenerate at this point.

Hew {\it et al} probe this structure further by asymmetric biasing
of the side-gates. The maximum asymmetry studied was $\Delta V_{sg}
= \pm 0.9$~V for an initial symmetric bias $V_{sg} = -1$~V at
$B_{\parallel} = 16$~T (see Fig.~4(b) of Ref.~\cite{HewPRL09}). The
sharp, deep resonant peak observed at $\Delta V_{sg} = 0$ decays
rapidly, returning to a relatively sharp plateau. Figure~42(e) shows
similar data obtained at $B_{\parallel} = 0$~T. The central bold
trace is obtained with the side-gates symmetrically biased at
$V_{sg} = -0.8$~V. The maximum asymmetry applied $\Delta V_{sg} =
\pm 0.5$~V is estimated to give a lateral displacement of
approximately $350$~nm~\cite{HewPhD09}. In the central bold trace,
the direct rise towards the first plateau at $\sim 4e^{2}/h$ is
clear, with a normal sequence of integer quantized plateaus
appearing above it. The plateaus at $4G_{0}$ and above are
remarkably unaffected by the asymmetric biasing compared to the ones
below. The $4e^{2}/h$ plateau makes a gradual descent with
increasing $\Delta V_{sg}$ and this coincides with a rising plateau
that appears at $e^{2}/h$. These two plateaus appear to be heading
towards a crossing at large $\Delta V_{sg}$. However, both weaken
and vanish, and a plateau like structure appears at $2e^{2}/h$,
which appears to split into a rapidly rising and a gradually
lowering plateau as $\Delta V_{sg}$ is increased further. This data
is hard to interpret without additional studies, and Hew {\it et al}
only comment that it suggests that the double-row state is fragile
and easily destroyed by changes in the transverse confining
potential~\cite{HewPRL09, HewPhysE10}. An interesting feature of the
data in Fig~42(e), however, is the behaviour of the $6e^{2}/h$
plateau. This plateau does not move upwards or downwards with
asymmetric biasing as the plateaus beneath it do. Instead it is
weakened compared to the plateaus above. Remarkably, this weakening
of the $6e^{2}/h$ plateau begins to diminish when the descending
$4e^{2}/h$ and ascending $e^{2}/h$ plateaus vanish (highlighted by
the blue dashed box in Fig.~42(e)), and recovers the same strength
as the plateaus above it at the point where the splitting of the
$2e^{2}/h$ plateau becomes resolved. This suggests that the state
that forms at low density in weakly confined QPCs has effects up
into at least the third subband, and may be more complex than merely
the appearance of double-row structure reported in
Ref.~\cite{HewPRL09}. One possibility that might be considered is
that it represents one of the higher-row states predicted by
Piacente {\it et al}~\cite{PiacentePRB04}. This scenario would be
consistent with the recovery in strength of the $6e^{2}/h$ plateau
around where the $4e^{2}/h$ and $2e^{2}/h$ plateaus are restored in
Fig.~42(e), since all of these row-states should be suppressed by
breaking the transverse symmetry of the 1D channel. Any further
suggestions on the mechanism at this point would be speculation, but
this is certainly an interesting place for experimenters to start
towards a deeper understanding, perhaps by coupling dc source-drain
bias spectroscopy~\cite{ChenPRB09} with these asymmetric gate-bias
measurements, or using thermodynamic measurements such as the
compressibility~\cite{LuscherPRL07, SmithPRL11}.

We conclude this discussion with data obtained by Smith {\it et
al}~\cite{SmithPRB09} looking more closely at the possibility of
double-row formation in weakly confined QPCs. The device studied is
nominally identical, albeit with slightly wider split-gates to
improve control over the confinement potential, and a constant
asymmetric bias of $\Delta V_{sg} = 2$~V applied throughout the
measurements. The asymmetric bias was applied to laterally shift the
channel to a clean region due to disorder effects observed in the
conductance at zero offset~\cite{SmithPC2}. To ensure that the
offset was not responsible for the observed behaviour, the
experiment was repeated in a second device, with the data published
in Ref.~\cite{SmithPhysE10}. Figure~43(a) shows similar data to that
presented in Fig.~41(a) for the device studied by Smith {\it et al}.
Once again, the $G_{0}$ plateau is lost as the confinement weakens,
before recovering at the weakest confinement at the left of
Fig.~43(a). Very similar behaviour is observed in the $0.5G_{0}$
plateau in measurements obtained at $B_{\parallel} = 16$~T (see
Fig.~1(a) of Ref.~\cite{SmithPRB09}). In both cases, the weakening
of the plateau coincides with what appears to be an anticrossing in
a plot of transconductance $dG/dV_{tg}$ versus $V_{sg}$ and
$V_{tg}$, with the $B_{\parallel} = 0$~T case shown in Fig.~43(b).
The dark diagonal bands correspond to regions of high
transconductance (risers) which typically indicate 1D subband edges
moving through the chemical potential, as in the strongly confined
regime at the far right in Fig.~43(b). In the intermediate
confinement regime at the left, there is a clear modulation of the
dark lines, and an apparent anticrossing of the two upper-most
lines, as highlighted by the red arrow in Fig.~43(b). Smith {\it et
al} argue that this anticrossing occurs because there is no
mechanism in the standard subband model for the energy of the first
excited state (i.e., the second subband) to fall below that of the
ground state (i.e., the first subband). As the two levels approach
degeneracy, they hybridise into bonding and antibonding states,
represented as $1$ (solid line) and $1^{*}$ (dashed line) in the
schematic shown upper left inset to Fig.~43(b), which anticross.

\begin{figure}
\includegraphics[width=16cm]{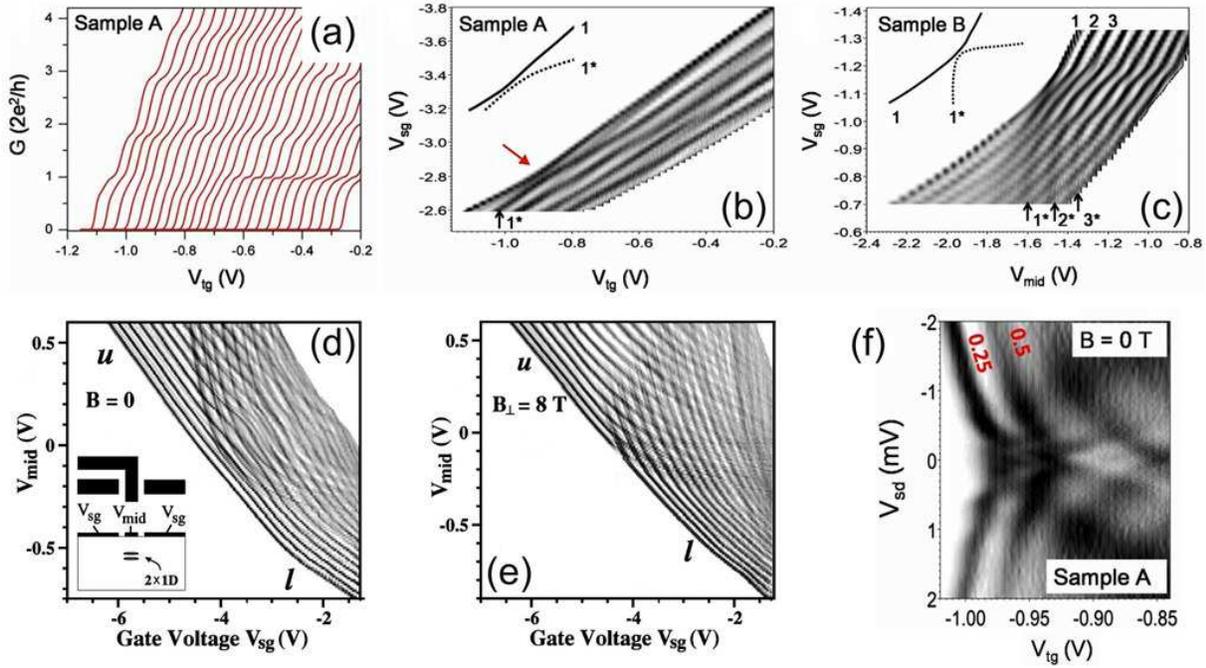}
\caption{(a) Conductance $G$ vs top-gate voltage $V_{tg}$ at fixed
side-gate voltage $V_{sg}$. The data is obtained with a constant
asymmetric bias $2$~V to laterally shift the channel away from an
impurity~\cite{SmithPC}, and $V_{sg}$ is incremented from
$-2.6,-0.6$~V (left) to $-3.7,-1.7$~V (right). (b) Greyscale plot of
transconductance $dG/dV_{tg}$ vs $V_{tg}$ ($x$-axis) and $V_{sg}$
($y$-axis) with dark regions indicating high transconductance (i.e.,
risers). The first bonding $1$ and antibonding $1^{*}$ states are
illustrated in the schematic upper left. The red arrow points to the
anticrossing between $1$ and $1^{*}$. (c) Similar data to (b) with
mid-line gate voltage $V_{mid}$ ($x$-axis) for Sample B. The first,
second and third bonding and antibonding states are indicated. Both
(b) and (c) are data obtained at $B_{\parallel} = 0$~T. (d) and (e)
are transconductance $dG/dV_{sg}$ vs $V_{sg}$ ($x$-axis) and
$V_{mid}$ ($y$-axis) obtained from strongly-coupled 1D systems
formed in a double quantum well heterostructure at $B = 0$~T and
$B_{\perp} = 8$~T, respectively. The inset in (d) contains a
schematic of the gate pattern from above (top) and a schematic of
the device from the side (bottom) used to obtain data in (d) and
(e). (f) Greyscale plot of $dG/dV_{tg}$ vs $V_{tg}$ ($x$-axis) and
source-drain bias $V_{sd}$ ($y$-axis) for sample A at $B = 0$~T in
the weak confinement regime. There appears to be a `shifted
duplicate' of the structure normally associated with the first
subband coinciding with the source chemical potential $\mu_{s}$.
This is not yet understood, highlighting the need for additional
measurements in the weak confinement regime. Figures (a-c,f) adapted
with permission from Ref.~\cite{SmithPRB09}. Copyright 2009 by the
American Physical Society. Figures (d,e) adapted with permission
from Ref.~\cite{ThomasPRB99}. Copyright 1999 by the American
Physical Society.}
\end{figure}

The behaviour in Fig.~43(b) bears strong resemblance to that
observed in strongly-coupled QPCs formed using a double-quantum well
heterostructure~\cite{CastletonPhysB98, ThomasPRB99}. In these
studies, the upper and lower QPCs are controlled by a pair of
side-gates and a mid-line gate, which allow the system to be driven
all the way from where only the upper QPC is populated, through a
regime where both conduct, and to where only the lower QPC is
populated. Providing the two quantum wells are sufficiently close
(separation $\lesssim 10$~nm) the two QPCs interact strongly. This
coupling can also be quenched by a strong magnetic field
$B_{\parallel},B_{\perp} \gtrsim 8$~T~\cite{ThomasPRB99}.
Figure~43(d) shows a plot of transconductance $dG/dV_{sg}$ versus
mid-line $V_{mid}$ and side-gate biases $V_{sg}$ obtained for these
vertically coupled QPCs. The parallel sets of dark bands at the
upper left and lower right corners correspond to where the upper and
lower QPCs are the only device populated. Moving to the right from
the upper left corner, or upwards from the lower right corner,
crossings of the higher subband edges of one QPC with the lower
subband edges of the other are apparent. Anticrossings do not form
here due to the vastly different subband index, however, the
quantization of the corresponding plateaus is significantly affected
(see Fig.~1 of Ref.~\cite{ThomasPRB99}). Following a diagonal
heading towards the upper right near the middle of the band, we see
a change towards two families of states -- flatter ones running from
upper left to lower right and more parabolic ones running from the
top downwards before turning to head to the right side. These are
bonding and anti-bonding states corresponding to symmetric and
antisymmetric linear combinations (i.e., hybridisation) of
wavefunctions with similar index $n$ due to coupling between the two
QPCs. A careful inspection of Fig.~43(d) reveals a number of
anticrossings and significant bending of the dark lines where this
occurs. As a counterpoint, Fig.~43(e) shows an example of how
Fig.~43(d) looks when the coupling is suppressed with a strong
magnetic field, which confirms that the anticrossings are due to
coupling between the two 1D systems.

Figure~43(c) shows an intermediate position between the devices
studied in Figs.~43(b) and (d) -- this device has a mid-line gate
like that studied by Thomas {\it et al}~\cite{ThomasPRB99} with no
top-gate and only a single 2DEG. The midline gate is $1.1~\mu$m
wide, with the split-gate separation increased to $1.9~\mu$m to
accommodate it. Here the data has a very similar character to that
in Fig.~43(d) albeit rotated clockwise by $90^{\circ}$ due to choice
of axes. However, care is needed in interpreting it, because this
gate configuration can produce a transverse double-well potential
that may support two coupled 1D systems, oriented horizontally
rather than vertically as in Ref.~\cite{ThomasPRB99}. As such, this
data is less suggestive itself of the formation of a double-row
state, but it adds confidence to the interpretation of the feature
in Fig.~43(b) as signalling the formation of such a double-row
state. Note that there only appears to be one parabolic anti-bonding
state running through the data in Fig.~43(b), which suggests that
unlike Fig.~43(d) where there are always two 1D systems present,
`coupled' behaviour in sample A only occurs in the limited
confinement range where the double-row state is stable. The
symmetric-antisymmetric gap was estimated by Smith {\it et
al}~\cite{SmithPRB09} to be of order $0.2$~meV for sample A,
approximately five times smaller than the double-2DEG system studied
by Thomas {\it et al}~\cite{ThomasPRB99}.

Finally, Fig.~43(f) shows a source-drain bias greyscale obtained in
the weak confinement regime, with a structure that resembles a
`shifted duplication' of the structure normally associated with the
first subband edge coinciding with $\mu_{s}$, resulting in finite
bias plateaus at $0.25$ and $0.5G_{0}$. This is not yet well
understood, and I present this final piece of data because it is
another example highlighting that there is still plenty to study and
understand about the nature of conduction in 1D systems at $G
\lesssim 2e^{2}/h$.

\section{Conclusions and Outlook}

The $0.7$ plateau and associated structures in QPCs have been the
subject of extensive experimental and theoretical studies since the
first paper on the topic by Thomas {\it et al}~\cite{ThomasPRL96} in
1996. The two questions in closing this review are what have we
learned and where do we go from here?

{\it What we have learned:} The quantization of conductance in a QPC
is a striking demonstration of the wave nature of electrons. This is
a single particle effect, however, and the observation of the $0.7$
plateau and associated features are a reminder of the important
contribution that interactions between electrons make to conduction
in nanoscale quantum devices. The rich behaviour of these features
compared to the integer quantized plateaus highlight the complexity
of these many-body interactions.

The key defining characteristics of the $0.7$ plateau are its
unusual temperature dependence, with the plateau becoming most
prominent at intermediate temperatures $T \sim 1$~K, and a clear and
consistent drop to $0.5G_{0}$ with an applied in-plane magnetic
field~\cite{ThomasPRB98, ThomasPRL96}. The latter indicates that
spin plays a vital role, as in more commonly known interaction
phenomena ranging from magnetism through to the Kondo effect. A
comparative study of the $0.7$ plateau and the Land\'{e} effective
$g$-factor provides some important clues. Enhancement of $g^{*}$ is
commonly observed as the 1D confinement is strengthened in electron
QPCs~\cite{ThomasPRL96, PatelPRB91, DaneshvarPRB97, DanneauPRL06,
MartinAPL08, MartinPRB10}, pointing strongly to the growing
importance of the exchange interaction in the 1D
limit~\cite{WangPRB96, WangPRB98}. In hole QPCs on (311)A-oriented
heterostructures, the directional dependence of the rate at which
the $0.7$ plateau drops to $0.5G_{0}$ with increasing in-plane
magnetic field matches that of the anisotropic
$g^{*}$~\cite{DanneauPRL08}, pointing clearly to spin playing a role
in the appearance of the $0.7$ plateau. Interestingly, the $0.7$
plateau also appears to be present in hole QPCs on (100)-oriented
heterostructures~\cite{KlochanPRL11, KomijaniEPL10, ChenPhysE10},
where 1D exchange enhancement is suppressed~\cite{ChenNJP10,
WinklerPRB05}. This does not mean that the $0.7$ plateau is not an
exchange effect, but it suggests that a comparative study of
electron and hole QPCs on (100)-oriented heterostructures may
provide further insight into the role that exchange effects play in
the appearance of the $0.7$ plateau.

A broad spectrum of explanations have been offered for the $0.7$
plateau, but the two dominant ones in terms of focussed experimental
studies have been the development of an exchange-driven static
spontaneous spin polarization within the QPC, as originally proposed
by Thomas {\it et al}~\cite{ThomasPRL96} and first indicated in
theoretical calculations by Wang and Berggren~\cite{WangPRB96,
WangPRB98}; and a Kondo-like effect as proposed by
Lindelof~\cite{LindelofSPIE01} and Cronenwett {\it et
al}~\cite{CronenwettPRL02}, and formalized by Meir {\it et
al}~\cite{MeirPRL02}. I will address the latter first.

The Kondo scenario draws inspiration from the discovery of a Kondo
effect in quantum dots~\cite{GoldhaberGordonNat98, CronenwettSci98},
where for odd occupancy of the dot, interactions between a single
localized spin and the sea of electrons in the source and drain
reservoirs can lead to enhancement of the conductance through the
dot. Initial evidence in favour of a Kondo-like origin for the $0.7$
plateau was primarily based on the presence of an anomalous peak in
the source-drain bias characteristics of a QPC at $G < G_{0} =
2e^{2}/h$ known as the `zero-bias anomaly'. A similar feature is
observed with the Kondo effect in quantum
dots~\cite{GoldhaberGordonNat98, CronenwettSci98}. Additionally, a
temperature-dependent scaling behaviour of the conductance is
observed~\cite{CronenwettPRL02}, however this scaling follows a
different dependence to that observed in quantum
dots~\cite{GoldhaberGordonPRL98}, and the form (see Eq.~6) is
somewhat empirical involving the addition of a constant $0.5G_{0}$
to the conductance and fixing of the prefactor to the temperature
dependence to $0.5G_{0}$ also. Although a theoretical justification
for this modified form of Kondo scaling has been
provided~\cite{MeirPRL02}, the validity of this modified form is yet
to be firmly established. For example, the modified Kondo form in
Eq.~6 is incompatible with the appearance of the ZBA at $G < 0.5
G_{0}$, which is concerning, as the ZBA is often very strong at $G <
0.5G_{0}$ in QPCs, and has been reported at conductances of $10^{-3}
G_{0}$ and below~\cite{KlochanPRL11, SarkozyPRB09, RenPRB10,
SfigakisPRL08}. Additionally, it is not always possible to fit this
modified form to data obtained from QPCs~\cite{KlochanPRL11,
SfigakisPRL08}, although it is a reasonable fit in some
instances~\cite{CronenwettPRL02, ChenPRB09}.

Subsequent experiments related to the Kondo mechanism focussed on
two important directions. The first is whether and how there is a
bound-state within a QPC that can support a single-spin and thereby
mediate the Kondo process. The second is whether the $0.7$ plateau
and the zero-bias anomaly are directly related or merely coincident
but separate phenomena in QPCs.

Regarding the possibility of bound-state formation in QPCs,
theoretical arguments both for~\cite{RejecNat06, HirosePRL03,
MeirJPCM08, WelanderPRB10} and against~\cite{JakschPRB06,
BerggrenJPCM08, StarikovPRB03} have been made on the basis of
spin-density functional theory. Recent quantum Monte Carlo
simulations also point to bound-state formation~\cite{GucluPRB09}.
While the theoretical picture is still a matter of some
disagreement, the experimental evidence is more compelling.
Experiments on coupling effects between nearby QPCs by Morimoto {\it
et al}~\cite{MorimotoAPL03} and Yoon {\it et al}~\cite{YoonPRL07,
YoonPRB09, YoonAPL09} indicate the presence of a quasi-bound state
within a QPC for $G < G_{0}$. In particular, the detection of Fano
resonances by Yoon {\it et al}~\cite{YoonPRB09} is compelling
evidence for discrete state formation within a QPC near pinch-off. A
remaining question, however, is whether such a bound-state forms
purely from self-consistent distribution of the electrons, or due to
other causes for localization within the QPC such as
impurities/disorder~\cite{McEuenSurfSci90}, roughness in the
definition of the QPC~\cite{SfigakisPRL08} or 2D-1D density mismatch
between the QPC and reservoirs~\cite{LindelofJPCM08}.

Regarding whether the $0.7$ plateau and the zero-bias anomaly (ZBA)
are coincident or causally linked phenomena, the main studies have
focussed in two directions. The first is on QPCs where a clear
bound-state is formed, as evidenced by a Coulomb-blockade peak at $G
< G_{0}$. The device studied by Sfigakis {\it et
al}~\cite{SfigakisPRL08} features two micro-constrictions that lead
to both a very strong Kondo effect and a clear $0.7$ plateau. The
ZBA and $0.7$ plateau have very different temperature dependencies
in this device -- the $0.7$ plateau is strongest and the ZBA is
completely suppressed for $T > 1$~K whereas the ZBA is at its
strongest in the low temperature limit, where the $0.7$ plateau is
obscured by the Coulomb blockade peak caused by the bound-state
within the QPC. The modified Kondo model (Eq.~6) could not be fit to
the data obtained by Sfigakis {\it et al}~\cite{SfigakisPRL08},
however, at low temperatures where the $0.7$ plateau is at its
weakest, the quantum dot Kondo model was a good fit to the data
instead. Additionally, the $\sim 0.25$ and $\sim 0.85G_{0}$ plateaus
at finite source-drain bias, features normally linked with the $0.7$
plateau in QPCs, appear well after the Kondo effect has been
suppressed and are oblivious to whether or not the temperature
exceeds the Kondo temperature. A similar situation to that reported
by Sfigakis {\it et al} was obtained in a hole QPC by Klochan {\it
et al}~\cite{KlochanPRL11}, with a bound-state supporting the
quantum dot Kondo effect observed. The ZBA observed in this device
bears all of the definitive hallmarks of the quantum dot Kondo
effect including a Zeeman splitting that goes as $2g^{*}\mu_{B}B$
rather than $g^{*}\mu_{B}B$~\cite{MeirPRL93}. Although a feature
resembling the $0.7$ plateau is apparent in
Ref.~\cite{KlochanPRL11}, it is not actively discussed due to
concerns regarding the obscuring effect of Coulomb blockade features
present due to the quantum dot formed within the QPC. In addition to
representing the first observation of the quantum dot Kondo effect
for spin-$\frac{3}{2}$ holes, the data presented by Klochan {\it et
al} demonstrate that true quantum dot behaviour can be obtained from
a device that is nominally an open QPC (i.e., with no features
intended to deliberately produce a bound-state as in
Refs.~\cite{SfigakisPRL08, LiangPRB99, LiangPRL98}). In order to
avoid disorder issues and establish a clearer picture regarding the
link between the $0.7$ plateau and the ZBA, Sarkozy {\it et
al}~\cite{SarkozyPRB09} studied ten different devices made on
undoped heterostructures. Firstly, the ZBA is observed at all
conductances $G < G_{0}$ to as low as $10^{-5}G_{0}$ (ZBAs at very
low conductance are also seen in Refs.~\cite{KlochanPRL11, RenPRB10,
SfigakisPRL08}). The fact that the ZBA is still evident well away
from $0.7G_{0}$ strongly suggests that the two phenomena are
distinct. Further, the ZBA in these devices show numerous behaviours
that are {\it not} consistent with expectations from the quantum dot
Kondo effect. This includes a Kondo temperature $T_{K}$ that
increases non-monotonically with gate voltage, and a finite
source-drain bias splitting of the ZBA that opens linearly with gate
voltage at fixed in-plane magnetic field, rather than remaining
constant. This latter data by Sarkozy {\it et al} suggests that
there may be more than one manifestation of the zero-bias anomaly in
QPCs: one that obeys the usual behaviour of the quantum dot Kondo
effect when a clear bound-state is formed in the
device~\cite{KlochanPRL11, SfigakisPRL08}, and one that is a more
generic feature of QPCs that is not yet understood and cannot be
explained by existing spin-polarization or Kondo effect
models~\cite{SarkozyPRB09}. At the very least, a broad survey of the
data suggests that the $0.7$ plateau is not simply a new
manifestation of the quantum dot Kondo effect, however, it is
difficult (and dangerous) at this stage to rule out the possibility
that Kondo-like correlations contribute in some way to the various
phenomena observed at $G < G_{0}$ in QPCs.

A separate line of study by Graham {\it et al}~\cite{GrahamPRL03}
has focussed on other non-quantized plateaus in the 1D conductance,
denoted as `0.7 analog' and `0.7 complement' structures. These
structures arise due to a spontaneous splitting that occurs at
crossings between spin-split 1D subband edges of opposite spin and
adjacent subband index (e.g., $1\downarrow$ and $2\uparrow$) under
strong in-plane magnetic field. These structures show many
characteristics of the $0.7$ plateau. The spontaneous splitting
observed at the subband edge crossings is reproduced remarkably well
by spin-density functional theory calculations~\cite{BerggrenPRB05}
and similar features can be obtained from a density-dependent
spin-gap model~\cite{ReillyPhysE06}. Subsequent work focussing on
source-drain bias spectroscopy studies of 1D subband population
under strong in-plane fields revealed that the spin-down subbands
tend to drop rapidly in energy upon populating~\cite{GrahamPRB05}
whereas the spin-up subband edges pin to the chemical potential upon
populating~\cite{GrahamPRB07}. Studies of the dc conductance of QPCs
provide additional support for these behaviours~\cite{ChenPRB09b,
ChenAPL08, ChenNL10, MicolichJPCM11}. These behaviours lead to the
opening of an energy-gap between spin-up and spin-down subbands, in
broad agreement with phenomenological spin-gap models proposed by
Kristensen {\it et al}~\cite{KristensenPRB00, KristensenPS02},
Bruus, Cheianov and Flensberg~\cite{BruusPhysE01, BruusArXiv00} and
Reilly {\it et al}~\cite{ReillyPRL02, ReillyPRB05, ReillyPhysE06}.
Although the precise details differ between these models, they all
rely on thermal excitation across an energy gap that exists between
spin-up and spin-down components of a 1D subband whilst it is near
the chemical potential, and reproduce the observed behaviour of the
$0.7$ plateau with remarkable accuracy. While these phenomenological
models make no discussion of the possible microscopic mechanisms
involved, recent theoretical calculations based on spin-density
functional theory~\cite{JakschPRB06} and Hartree-Fock
calculations~\cite{LasslPRB07, LindPRB11} arrive at results that
show qualitative agreement with the phenomenological models. Studies
of shot-noise in QPCs~\cite{RochePRL04,DiCarloPRL06} point to the
$0.7$ plateau involving conduction via two separate channels, and
yield good agreement with predictions based on a spin-gap
model~\cite{DiCarloPRL06}. A similar finding is obtained from
studies performed under strong perpendicular fields where
spin-polarized edge-states were used to mimic two-channel conduction
through a QPC~\cite{ShailosJPCM06b}. Finally, studies using scanning
gate microscopy also point towards spin-polarization causing plateau
structures at $G < G_{0}$~\cite{CrookSci06}. Combined together, this
work strongly suggests that exchange-driven spin-splitting of the 1D
subbands plays a significant role in both the $0.7$ plateau and
other non-quantised plateaus observed in the conductance of QPCs.

The most recent studies have focussed on the potential for
exchange-driven spontaneous ordering of electrons in QPCs and
quantum wires~\cite{HewPRL08, HewPRL09, HewPhysE10, SmithPRB09}.
Analytical calculations have shown that under appropriate conditions
of electron density and interaction strength, the electrons in a
quantum wire can enter a ferromagnetic ground-state with a zig-zag
chain structure~\cite{KlironomosEPL06, KlironomosPRB07}. Similar
possibilities have also been predicted by quantum Monte
Carlo~\cite{GucluPRB09} and spin-density functional
theory~\cite{WelanderPRB10} calculations. Experiments by Hew {\it et
al}~\cite{HewPRL08, HewPRL09} and Smith {\it et
al}~\cite{SmithPRB09, SmithPhysE10} have recently explored this
possibility in QPCs with an additional top-gate such that 1D
conductance can be studied for $G \lesssim 5G_{0}$ as the strength
of the 1D confinement is varied. While the behaviour for strong
confinement is similar to that commonly observed in QPCs, some
remarkable behaviour is observed for weak confinement including
disappearance of the $G_{0}$ plateau~\cite{HewPRL09} and
anticrossings of 1D subband edges~\cite{SmithPRB09} that signal a
possible transition to a double-row or zig-zag structure
ground-state.

{\it Where we go from here:} Ultimately, looking across the full
spectrum of data, it is clear that a conclusive origin for the $0.7$
plateau still remains to be determined and there are a number of
open questions regarding both the physics of fractionally quantized
conductance plateaus and the effects of many-body interactions in 1D
systems more broadly. There are several aspects where contradictory
results are obtained, the clearest example being the precise
behaviour of the the zero-bias anomaly in the differential
conductance of QPCs. At present there appear to be two separate sets
of behaviour obeyed by this feature, one following that normally
observed in the quantum dot Kondo effect~\cite{KlochanPRL11,
SfigakisPRL08} and another with different
properties~\cite{CronenwettPRL02, SarkozyPRB09, ChenPRB09}, and it
would be useful to investigate the latter further to understand the
underlying mechanism. The use of dc conductance measurements to
track the 1D subband edges also has considerable potential to shed
further light regarding the physics of QPCs~\cite{ChenPRB09b,
MicolichJPCM11}. Finally, studies of electron ordering in 1D
systems~\cite{HewPRL09, SmithPRB09} are at an early stage, and
building in techniques such as dc conductance~\cite{ChenPRB09b} and
compressibility measurements~\cite{LuscherPRL07, SmithPRL11} may
yield some interesting new physics. Finally, the device fabrication
used in these electron ordering studies may be utilised studying how
the behaviour of the ZBA changes as the QPC potential is tuned using
an insulated top-gate.

There is also significant potential for further theoretical work
regarding the physics of QPCs below the last conductance plateau.
There is still clear disagreement amongst many of the density
functional theory models, and interesting possibilities such as
spontaneous electron ordering~\cite{KlironomosEPL06, GucluPRB09,
WelanderPRB10} and the coexistence of Kondo-like correlations and
exchange-driven spin-polarisation effects~\cite{SongPRL11} have only
just begun to be explored. Given the almost `Jekyll and Hyde' nature
of the conclusions drawn from the experimental data regarding
support for {\it either} dynamical Kondo-like processes {\it or}
static spin-polarization processes, and data that, when taken as a
whole, seems to suggest the possible coexistence of both, an
interesting direction would be to further consider how the
complexities of a real QPC (e.g., finite length, potential
mismatch/fluctuations, disorder, etc) may result in such a scenario.

Thinking further into the future, a major direction for research on
low dimensional systems are the novel states and physics predicted
to occur in the regime of very strong electron-electron
interactions. The first examples of these have already been seen in
2D systems with the fractional quantum Hall effect~\cite{TsuiPRL82}
and 0D systems with the Kondo effect~\cite{GoldhaberGordonNat98,
CronenwettSci98}, but in 1D systems, strong interactions are
expected to be particularly fruitful in terms of new physics
including spin-charge separation and Luttinger liquid behaviour.
Studies using cleaved edge overgrowth methods on GaAs
heterostructures~\cite{AuslaenderSci05} and carbon
nanotubes~\cite{DeshpandeNP08, DeshpandeSci09} hold particular
promise. This direction was the subject of an interesting recent
review by Deshpande {\it et al}~\cite{DeshpandeNat10}. Another
important direction where studies of spin-related phenomena such as
the $0.7$ plateau are important is spintronics, where QPCs may find
a role as injectors and detectors for spin. Several studies have
already been made in this direction~\cite{FolkSci03, DebrayNN09,
FrolovNat09, FrolovPRL09}, and an important goal is the
electrostatic manipulation of spin via the spin-orbit interaction in
order to realise spintronic devices that do not rely on magnetism
for operation~\cite{AwschalomPhys09}.

\ack The author acknowledges financial support from an Australian
Research Council (ARC) Future Fellowship (FT0990285), and the ARC
Discovery Projects Scheme (DP0877208 and DP110103802). I thank A.M.
Burke for proof-reading of this review, R. Newbury and O. Klochan
for helpful comments on the final draft. I thank K.-F. Berggren,
J.P. Bird, A.M. Burke, J.C.H. Chen, T.-M. Chen, W.R. Clarke, R.
Danneau, R. Fazio, D.K. Ferry, S. Fricke, D. Goldhaber-Gordon, A.R.
Hamilton, W.K. Hew, Y. Hirayama, M.J. Kay, O. Klochan, P.E.
Lindelof, H. Linke, S. L\"{u}scher, T.P. Martin, Y. Meir, R.
Newbury, D.J. Reilly, S. Reimann, L. Samuelson, F. Sfigakis, L.W.
Smith, O.P. Sushkov, R.P. Taylor, A. Wacker, R. Wirtz, P. Wu and U.
Z\"{u}licke for interesting discussions and helpful clarifications
on this topic, both large and small, over many years. Part of this
review was written whilst visiting the University of Oregon, and
their hospitality is gratefully acknowledged.

\end{document}